# Mineralogy and Geology of asteroid (4) Vesta from Dawn Framing Camera

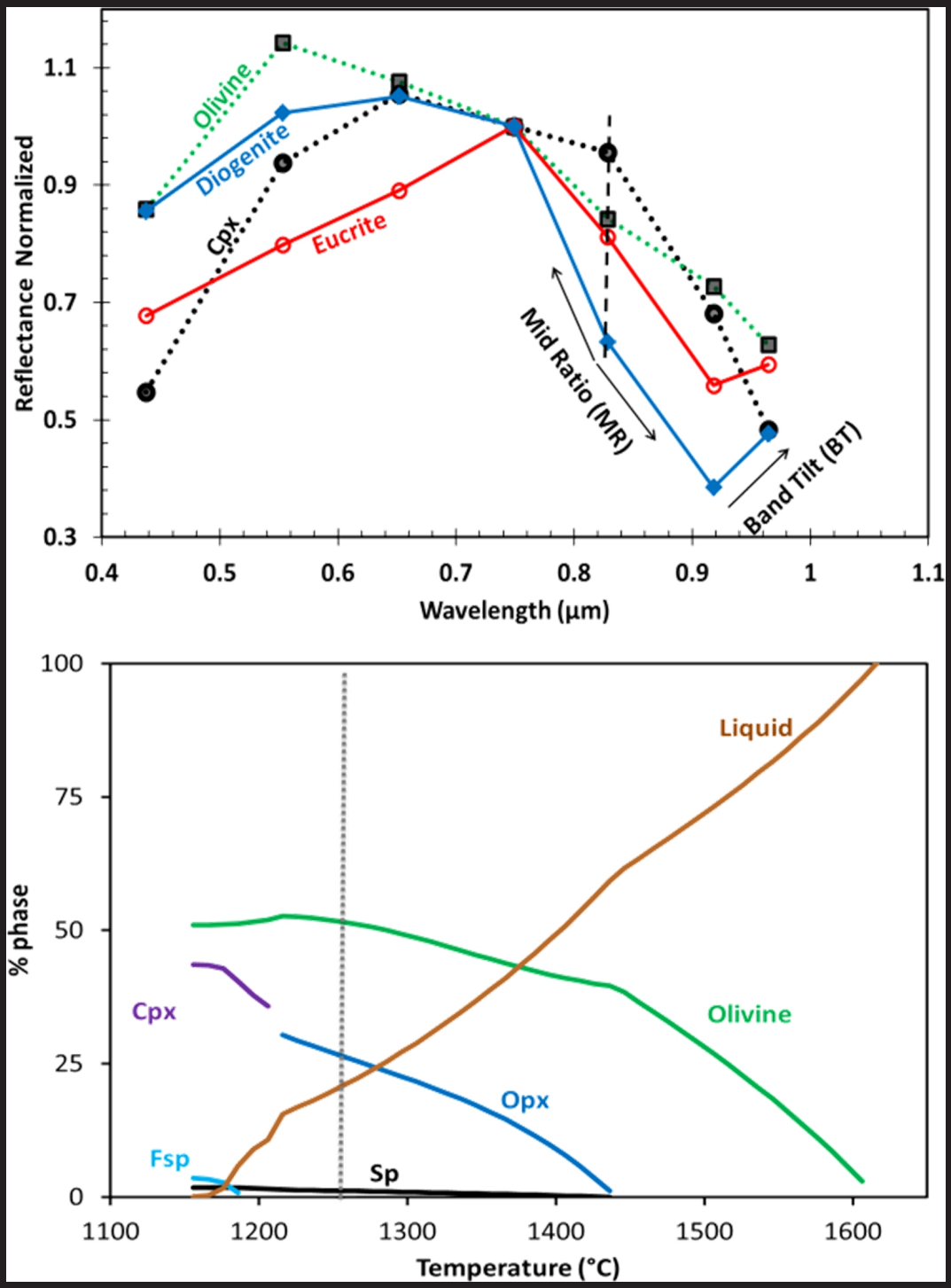

## Guneshwar Thangjam

International Max Planck Research School
for Solar System Science
at the University of Göttingen

# Mineralogie und Geologie des Asteroiden (4) Vesta von Dawn Framing Camera

## Dissertation

zur Erlangung des Doktorgrades
der Naturwissenschaften
(Dr. rer. nat.)

vorgelegt von

**Guneshwar Thangjam**
aus Manipur, Indien

genehmigt von der
Fakultät für Energie- und Wirtschaftswissenschaften
der Technischen Universität Clausthal
Tag der mündlichen Prüfung
07 September 2015

**Dekan**
Prof. Dr. Wolfgang Pfau (TUC)

**Vorsitzender der Promotionskommission**
Prof. Dr. Hans-Jürgen Gursky (TUC)

**Erster Gutachter**
Prof. Dr. Kurt Mengel (TUC)

**Weitere Gutachter**
Dr. Andreas Nathues (MPS)
Prof. Dr. W. van Berk (TUC)

***-To my mother and eldest brother-***

# Contents

# Zusammenfassung

Der Asteroid (4) Vesta (~ 520 km Durchmesser) wird als analog zu einem kleinen terrestrischen Planeten angesehen. Seine magmatischen Prozesse kamen höchstwahrscheinlich in den ersten 10 bis 100 Ma nach seiner Akkretion zum Erliegen. Die Untersuchung eines solch einzigartigen Objektes ist essentiell um die frühe geologische Evolution eines Planeten zu verstehen. In dieser Arbeit werden Analysen der Oberflächenzusammensetzung von Vesta mittels hochauflösender Bilder der Dawn Framing Camera präsentiert. Die Framing Camera (FC) besitzt sieben Farbfilter im Wellenlängenbereich von 0,4 bis 1,0 µm und einen ‚clear filter'. Die Kartierung der lithologischen Variationen auf Vesta's Oberfläche wurde mittels FC Farbfilter Daten (~ 480 m/Pixel) vorgenommen, welche in der frühen Phase der Mission aufgenommen wurden. Basierend auf diesen Beobachtungen kommt Thangjam et al. (2013) zu dem Schluss, dass der Großteil der Oberfläche eine howarditische Zusammensetzung aufweist.

Es wurden erstmals olivinreiche Regionen im Arruntia und Bellicia Krater mittels FC Farbfilter Daten mit einer Auflösung von ~ 60 m/Pixel identifiziert (Thangjam et al., 2014). Die Entdeckung von Olivin von Ammannito et al. (2013b) mittels Datensätzel des VIR Spektrometers im Bereich von 0,4-2,5 µm wird hier bestätigt. Die verwendeten Bandparameter der FC Farbfilter Daten sind, ‚Band-Tilt' ($BT=R_{0.92\ \mu m}/R_{0.96\mu m}$) und ‚Mid Ratio' ($MR=(R_{0.75\mu m}/R_{0.83\mu m})/(R_{0.83\mu m}/R_{0.92\mu m})$), wobei $R_{\lambda}$ die Reflektanz in dem entsprechenden Wellenlängenbereich darstellt. Hier wird zum ersten Mal eine dreidimensionale spektrale Herangehensweise vorgestellt, um unterschiedliche Zusammensetzungen zu analysieren und zu kartieren (Thangjam et al., im Press). Die Arruntia-Region wurde als Fallstudie ausgewählt und anhand der Bandparameter BT, MR und Reflektanz bei 0,55 µm ($R_{0,55\mu m}$) analysiert. Die Untersuchungen zeigen, dass diese Region überwiegend von Eukrit-dominierten Howarditen bedeckt ist. Olivinreiche Aufschlüsse wurden im Vergleich zu früheren Arbeiten in relativ großem Umfang entdeckt.

Um die petrologische Evolution von Vesta nachvollziehen zu können, wurde in dieser Studie eine chondritische Zusammensetzung von 80 % H Chondriten plus 20 % CV Chondriten gewählt. Das Sauerstoffisotopenverhältnis dieser Mischung entspricht jenem der HED Meteorite. Um Aussagen über Vestas interne Geometrie zu machen, wurden thermodynamische Berechnungen mittels MELTS verwendet. Dies ergab einen harzburgitischen Mantel mit einer durchschnittlichen Mächtigkeit von ~137 km und eine Kruste von ~15 km. Wenn ein Absenken der früh gebildeten Olivinphase angenommen wird, sind die Mächtigkeiten eines dunitschen Mantels, gefolgt von einem orhtopyroxenischen Mantel und einer Kruste ~46, ~84 und ~22 km. Diese Resultate können erklären, warum Dawn nicht in der Lage war, signifikante Olivin-Mengen im Rheasilvia Becken zu finden und lassen vermuten, dass Olivin auf Vestas Oberfläche einen exogenetischen Ursprung hat.

# Summary

Asteroid (4) Vesta (~ 520 Km diameter) is thought to serve as an analog for a small terrestrial planet. Its magmatic processes probably ceased in the first few 10's to 100's of Ma after accretion. The study of such an object is a key in understanding the early geologic evolution of planetary bodies. This thesis presents surface compositional analysis using high spatial resolution images obtained by the Dawn Framing Camera. The Framing Camera (FC) houses seven color filters in the wavelength range between 0.4 and 1.0 µm along with a clear filter. Lithologic variations on Vesta's surface are mapped using FC color data (~480 m/pixel) acquired during the early phase of the mission. Based on these observations, Thangjam et al. (2013) suggested that the majority of the surface is howarditic in composition.

Olivine-rich sites are identified for the first time using FC color data at a resolution of ~60m/pixel at Arruntia and Bellicia craters (Thangjam et al., 2014). The discovery of olivine by Ammannito et al. (2013b) by the VIR spectrometer data cubes in the range 0.4-2.5 µm is confirmed. The band parameters employed in FC color data are Band Tilt ($BT=R_{0.92 \mu m}/R_{0.96\mu m}$) and Mid Ratio ($MR=(R_{0.75\mu m}/R_{0.83\mu m})/(R_{0.83\mu m}/R_{0.92\mu m})$), where $R_{\lambda}$ is the reflectance at the given wavelength. A three-dimensional spectral approach is introduced for the first time for analyzing and mapping compositional heterogeneities (Thangjam et al., in press). The Arruntia region is selected for a case study and is analyzed by combining the band parameters BT, MR and reflectance at 0.55 µm ($R_{0.55\mu m}$). This study reveals that this region is mostly covered by eucrite-dominated howardite. Olivine-rich exposures are found in rather wide extents when compared to earlier works.

In an attempt to understand the petrological evolution of Vesta, a chondritic composition of 80% H chondrite plus 20% CV chondrite is presented as a case study. This mixture meets observed HED oxygen isotope relations. The results of thermodynamic calculation by MELTS are used to predict Vesta's internal geometry. The thicknesses of a harzburgitic mantle and crust are ~137 and ~15 km, respectively. If segregation of early-crystallized olivine phase is considered, the thicknesses of dunitic mantle followed upward by orthopyroxenitic mantle and crust are ~46, ~84, and ~22 km, respectively. These results may explain why Dawn failed to detect significant amounts of olivine in the Rheasilvia basin and emphasize that olivine on Vesta's surface is likely exogenic in origin.

# Outline of the thesis

This is a cumulative thesis consisting of an introductory part and three manuscripts of which two are published and one has been submitted (now accepted and available online).

Chapter 1 presents an introduction to Vesta, its geological and petrological evolution as well as a brief outline of the HED clan of meteorites. The latter is commonly believed to represent material from Vesta and the Vestan family. An evolutionary model of Vesta is presented as a case study in which the mineralogical composition of a silicate mantle is calculated by thermodynamic modelling (MELTS). Rock masses are then recalculated to thicknesses of mantle and crust at a given core diameter of 110 km and an assumed spherical shape (r=262 km). An outline is given for the use of spectral data acquired by the Framing Camera (FC) onboard the Dawn spacecraft with application to two crater regions.

Chapter 2 (Thangjam et al., 2013) presents a lithological mapping of Vesta during the earlier phase of the spacecraft orbit. The spatial resolution of the data is approximately 480 m/pixel that covered regions from approximately 75° S to 26° N.

Chapter 3 (Thangjam et al., 2014) presents the identification of olivine-rich sites in the Arruntia and Bellicia regions, for the first time using Dawn FC color data. The application of band parameters based on archived laboratory-derived spectra of HEDs and the rock-forming minerals olivine and orthopyroxene including their mixtures are discussed.

Chapter 4 (Thangjam et al., in press) presents an innovative three-dimensional spectral approach to analyze surface compositional heterogeneities. It employs an additional parameter along with commonly used two-dimensional approach. The effectiveness and the reliability of this approach are evaluated.

# 1. Asteroid (4) Vesta in the light of Dawn

## 1.1 (4) Vesta, HEDs, and the Dawn mission

### 1.1.1 Some general aspects of the minor planet Vesta

Vesta is the third largest and the second most massive asteroid in the Main Asteroid Belt (MAB). It is described as a small differentiated planetary body because of its overall resemblance to the terrestrial planets (Keil, 2002; Jaumann et al., 2012; Russell et al., 2013). The geologic evolution of this differentiated object in the first few 10's or 100's of Ma of the solar system formation is well documented (e.g., McSween et al., 2011; Mittlefehldt, 2015). A direct link between Vesta and the Howardite-Eucrite-Diogenite clan of achondrites (HED) has been established over the past 4 decades (e.g., McCord et al., 1970; Gaffey, 1997; Russell et al., 2012, 2013; Pieters et al., 2012). Despite the overwhelming evidence of Vesta as the parent body of the HED meteorites, Consolmagno et al. (2015) and Wasson (2013) challenged this view: Consolmagno et al. (2015) put forward mainly astronomic and mass-balance arguments against the HED's origin from Vesta; Wasson (2013) argued basically on the basis of the distribution of some siderophile elements (Au, Co, Ir) as well as O- and Cr-isotopic ties which are not in accordance with the widely accepted idea of HED being ejected from Vesta. However, there are undoubtedly strong lines of evidences to support a genetically direct link of many, if not all, of the HEDs with Vesta or Vestan family (McCord et al., 1970; Mittlefehldt et al., 1998; McSween et al., 2011, 2013; Russell et al., 2013; Greenwood et al., 2014; Mittlefehldt, 2015; McCoy et al., 2015).

The study of Vesta is essential in understanding the evolution of planetary bodies in the early solar system.  It is one of the primary scientific goals of the historic Dawn mission launched in September 2007. Dawn spacecraft was in orbit around Vesta for more than a year (July 2011- September 2012) and imaged the entire surface. Higher resolution images obtained by the onboard instruments allow us to analyze and study the mineralogy and geology of this body.

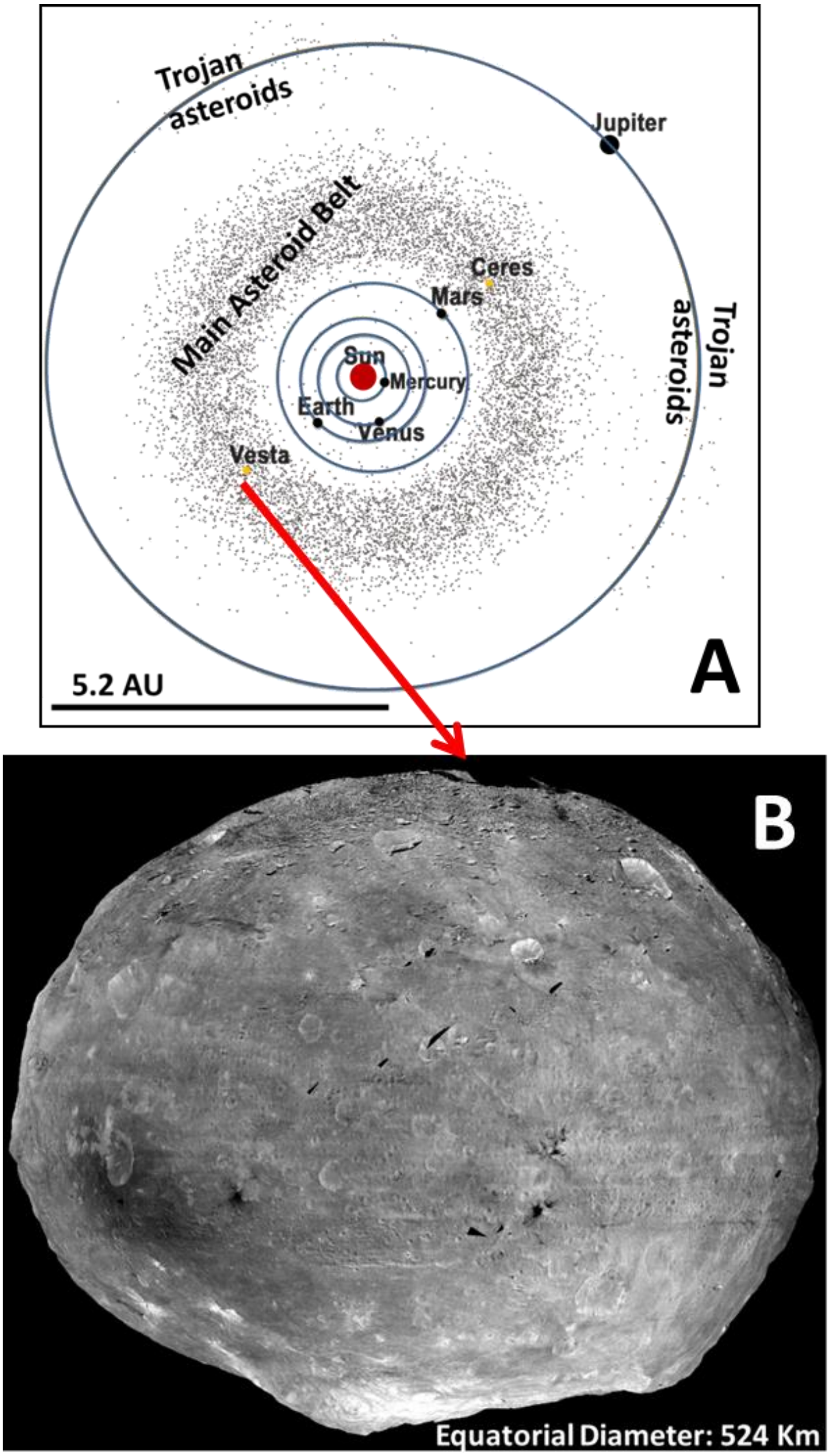

Fig. 1: (A) Location of (4) Vesta in the Main Asteroid Belt (adapted from McBride, 2011). (B) A view of Vesta from Dawn Framing Camera (image credit: NASA/JPL-Caltech/UCLA /MPS/DLR/IDA).

Asteroid (4) Vesta was discovered in 1807, by Wilhem H. Olbers in Bremen, Germany (Pilcher, 1979). The name was given by Carl Friedrich Gauss (Schmadel, 2003). The geometric albedo as revealed by the Dawn mission is in the range of 0.10 to 0.67 (Reddy et al., 2012a). Vesta is among the brightest objects in the MAB. The physical parameters of Vesta are given in Table 1.

Table 1: Some physical parameters of (4) Vesta

| **Physical parameters** | | **Source** |
|---|---|---|
| Major axes (radii) | (286/278/223) km | Russell et al., 2012 |
| Mean radius | 262 km | Russell et al., 2012 |
| Volume | $74.97 \times 10^6$ km$^3$ | Russell et al., 2012 |
| Mass | $2.59 \times 10^{20}$ kg | Russell et al., 2012 |
| Bulk density | 3456 kg /m$^3$ | Russell et al., 2012 |
| Geometric albedo | 0.1 to 0.67 | Reddy et al., 2012a |
| Semi-major axis | 2.36 AU | Russell and Raymond, 2011 |

### 1.1.2 HED meteorites

Before discussing the surface mineralogy and evolutionary aspects, it is worth understanding about Howardite, Eucrite and Diogenite (HED) suite of meteorites since they serve as analogues for Vesta's composition. The worldwide meteorite record contains ~ 8% of achondrites amongst the group of stony meteorites (McSween, 1999). Within the group of achondrites, the group of HEDs comprises 60% abundance (Voigt, 2012). A total of 1450 HED meteorites has been reported to date (Mittlefehldt, 2015). Falls and finds are known to be distributed unevenly in hot and cold deserts from systematic meteorite surveys over the decades. Most probably, these fragments have been transferred from Vesta to Earth in a multi-step process. There is convincing evidence that in a first step lumps of rock have left Vesta's gravitational field as a result of larger impacts. After ejection from the planetary body, these rock fragments might have exposed to cosmic irradiation before smaller pieces were mechanically removed and transferred to 3:1 orbital resonance space eventually allowing colliding with Earth (Binzel and Xu, 1993; Thomas et al., 1997). Cosmic exposure 'ages' of HED range from 3 to 110 Ma where 80 % of the HEDs investigated so far clustered in 6±1, 12±2, 21±4, 38±8, and 73±3 Ma intervals (McSween et al., 2011).

#### 1.1.2.1 Eucrites

Eucrites are known to have crystallized as lavas on Vesta's early surface or within relatively shallow level in the form of dikes or plutons (McSween et al., 2011). Eucrites are dominated by monoclinic pyroxenes, pigeonite and plagioclase plus subordinate orthopyroxene with traces of metal, troilite, chromite, ilmenite and silica (McSween et al., 2011). They are classified as either basaltic or cumulate eucrites (Stolper, 1977; McSween et al., 2011). Most eucrites are brecciated and polymict due to impacts (McSween et al., 2011). Pyroxenes in basaltic eucrites are Fe-rich and commonly contain low Ca-pigeonite hosts with finely exsolved augite lamellae (Mittlefehldt et al., 1998). The pyroxenes are crystallized as zoned-pigeonites (Takeda and Graham, 1991). The original Mg-Fe zonation is quite often obliterated by thermal metamorphism (Yamaguchi et al., 1996). Both low Ca-pyroxene and augite are present because of exsolution from pigeonite (Yamaguchi et al., 1996; Mittlefehldt et al., 1998; McSween et al., 2011). Plagioclase is more calcic ($An_{96-75}$) in basaltic eucrites than the cumulate eucrites ($An_{90-96}$) (Mittlefehldt et al., 1998; Mayne et al., 2009). Silica minerals occur in proportions ranging from 0 to >10% in basaltic eucrites, but cumulate eucrites contain in rather less amounts (Mayne et al., 2009).

### 1.1.2.2 Diogenites

Diogenites are categorized as coarse-grained cumulates that originated from a plutonic layer deep inside the crust, although sophisticated thermal histories and brecciation have obscured the primary features (McSween et al., 2011). Most diogenites are nearly monomineralic, composed of orthopyroxene (87-99%) with accessory chromite and olivine, and subordinate troilite, plagioclase, diopside, silica (all <2%) and rare metal (Sack et al., 1991; Bowman et al., 1997; McSween et al., 2011). Pyroxene compositions are generally in the range $Wo_{1-3}En_{71-77}Fs_{22-24}$, and the homogeneity in major element (Mg and Fe) composition is believed to be the result of post crystallization equilibration rather than igneous fractionation (Fowler et al., 1994; Mittlefehldt, 1994). Olivine compositions exhibit a narrow range, $Fo_{70-73}$ (Sack et al., 1991; Fowler et al., 1994; Mittlefehldt, 1994) but more ferroan and magnesian olivines are observed occasionally. Plagioclases are rather calcic in composition, i.e., $An_{86-82}$ (Mittlefehldt et al., 1998). A few specimens of olivine-rich diogenites have been reported. One specimen is dunitic in composition, and others show harzburgitic mineralogy (e.g., Beck and McSween, 2010; Beck et al., 2011). As reported by Tkalcec et al. (2013) and Tkalcec and Brenker (2014), few harzburgitic diogenite specimens reveal subgrain deformation features which are similar to what is known from peridotite xenoliths entrained in basaltic rocks of the Earth. Such subgrain boundary deformation is commonly interpreted as a result of plastic flow within the Earth's upper mantle. This is a challenging idea in terms of interpreting diogenites a potentially deep-seated mantle rock of Vesta. However, Yamaguchi et al. (2015) suggest that one of the samples of Tkalcec and coworkers' (diogenite NWA 5480) is most probably not a mantle rock because this specimen shows peculiar features of melts that are typical of cumulate rocks.

### 1.1.2.3 Howardites

Howardites are best described as a type of regolith breccia (Bunch, 1975; Dymek et al., 1976; Warren 1985) formed by impact-induced mixing on the surface. Warren et al. (2009) divide howardites into two types; regolithic (well-mixed regolith suggested to have originated from older-impacts) and fragmental (originated from more recent impacts). Delaney et al. (1983) define howardites as polymict basaltic achondrite breccia containing less than 90% of a single identifiable lithic component. Mason (1983) classifies howardites as a polymict basaltic achondrite breccia containing identifiable diogenite material. However, Delaney et al. (1983) encourage using howardite and polymict eucrites as end-members of a continuum of basaltic achondrites. Howardites and polymict eucrites are considered as polymict breccias (McSween et al., 2011). Impact mixing of eucritic and diogenitic lithologies has produced polymict breccias of eucrite-dominated and diogenite-dominated composition (Mittlefehldt et al., 1998). According to McSween et al. (2011), a distinction between howardites and polymict eucrites is arbitrary and the current sampling suggests that there is a wide range from unbrecciated eucrites to diogenites; unbrecciated eucrites to monomict eucrite to polymict eucrite to howardites to polymict diogenite to unbrecciated diogenites. Various exogenic chondritic materials are known to occur in howardites, for example, howardite PRA 04401 contains ~60% CM2 clasts (e.g., Mittlefehldt et al., 1998; Herrin et al., 2011). Individual olivine grains claimed to be the Vestan mantle in origin are found in howardites (Lunning et al., 2015; Hahn et al., 2015). As pointed out by Lunning et al. (2015), the oxygen three isotope and trace element signatures of the olivine grains are clearly in accordance with a depleted peridotite mantle left behind after melt extraction.

## 1.1.3 The Dawn mission and the Framing Camera

Dawn is the ninth NASA discovery mission, and the first mission to orbit the most massive asteroids in MAB (Russell and Raymond, 2011; Russell et al., 2012, 2013). The mission was launched in 2007, and the targets of this spacecraft are (4) Vesta and (1) Ceres. The name 'Dawn' is because these objects are believed to be the witnesses to the events at the 'dawn' of the solar system (Russell et al., 2013). The primary goal of this mission is to study the two minor planets to understand the processes occurred in the early solar system (Russell and Raymond, 2011). The mission is managed and operated by JPL/CALTECH (Jet Propulsion Laboratory/ California Institute of Technology) with the overall mission science responsibilities from UCLA (University of California, Los Angeles).

The spacecraft has three instruments onboard i.e., Framing Camera (FC), Visible and Infrared Spectrometer (VIR), Gamma Ray and Neutron Detector (GRaND). The VIR is a hyperspectral instrument operating in the wavelength range from near ultra-violet to infra-red (0.2 to 5 µm). It has two data channels, the visible channel (0.25-1.07 µm) and the infrared channel (0.95-5.1 µm) (De Sanctis et al., 2011, 2012; Ammannito et al., 2013a). The main objective of VIR is to map and analyze surface mineralogy/composition, thermal behavior of the surface and their interaction with the extreme space weather in an atmosphere-free environment (De Sanctis et al., 2011). The GRaND is able to derive information on surface elemental abundances in oxide and silicate minerals like O, Mg, Al, Si, K, Ca, Ti, Fe, Ni, U, Th, ices (H, C, N) and volcanic exhalation or aqueous alteration products (Prettyman et al., 2011). Of the three instruments, the Framing Camera is of very interesting and important to the entire mission, because apart from the scientific objectives, this camera serves for orbit navigation and control (Sierks et al., 2011).

The FC was developed and built at the Max Planck Institute for Solar System Research, Katlenburg-Lindau (now in Goettingen), with contributions from DLR/IPR (German Aerospace Centre / Institute for Planetary Science) and TUB/IDA (Technical University Braunschweig / Institute of Computer and Communication Network Engineering). The advantage of FC data over VIR cubes is its higher spatial resolution that exceeds ~ 2.3 times (Sierks et al., 2011). It is equipped with seven color filters in the wavelength range from 0.4 to 1.0 µm, and a broadband clear filter (Sierks et al., 2011). The center wavelengths of the color filters are chosen to characterize major absorption bands of HED meteorites (Sierks et al., 2011). The FC color data is used for surface compositional analysis and mapping while the clear data is for morphological analysis and geologic mapping. A view of the Framing Camera used onboard the Dawn spacecraft is shown in Fig. 2, with a detail of its color filters in Table 2.

Table 2: Characteristics of FC filters (Sierks et a., 2011)

| Channel No. | Center wavelength (nm) | Bandwith (nm) | Transmission (%) |
|---|---|---|---|
| 1 | 430±2 | 40±5 | >75 |
| 2 | 550±2 | 40±5 | >75 |
| 3 | 650±2 | 40±5 | >75 |
| 4 | 750±2 | 40±5 | >75 |
| 5 | 830±2 | 40±5 | >75 |
| 6 | 920±2 | 40±5 | >75 |
| 7 | 980±2 | 80±5 | >75 |

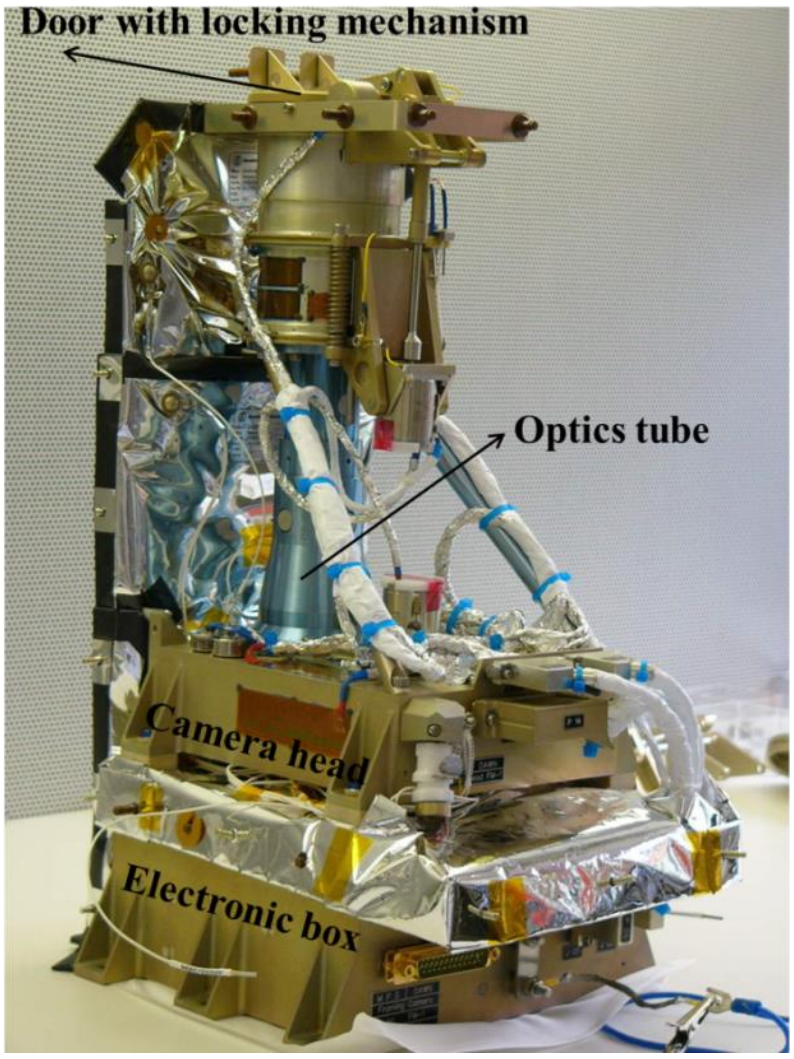

Fig. 2 Framing Camera (after Sierks et al., 2011)

## 1.2 Geology and petrology of Vesta

The density of Vesta (3.45 $kg/m^3$, Russell et al., 2012) is compatible with an origin from ordinary chondrite material that requires a composite silicate/FeNi structure. In contrast to the large inner planets like Venus, Earth, and Mars, the interior of Vesta did not preserve enough primordial heat that is the driving force for long-lived internal differentiation and whole-planet evolution. The rather limited size of early Vesta (~260 to 270 km radius) is responsible for enhanced cooling of the body, and as a result, the internal differentiation in terms of magmatic activity probably ceased some 10's or 100's of million years after accretion of the body (Mittlefehldt et al., 1998; McSween et al., 2011; Mittlefehldt, 2015). Consequently, the early formed rocks have been preserved largely in their respective petrologic compartments. Therefore, the bulk silicate mantle remained as a restitic component around the inner FeNi/FeNiS core while the magmatic products of early differentiation, basaltic and gabbroic rocks are located at or near the surface (Mittlefehldt et al., 1998; McSween et al., 2011; Mittlefehldt, 2015).

Eucrites are generally used to date the magmatic activities of Vesta. By means of the Sm/Nd method of dating ($^{147}Sm/^{143}Nd$), a crystallization age of 4.56 Ga has been reported by Wadhwa and Lugmaier (1995); a Pb/Pb isochrone ($^{207}Pb/^{206}Pb$, Wadhwa et al., 2009) yield an age of 4.566 Ga. Short lived isotope chronometers suggest that the time span of crystallizing basaltic eucrites on Vesta's outer shell ranges from 5 Ma after CAI formation ($^{26}Al/^{26}Mg$, Srinivasan et al., 1999; Nyquist et al., 2003). According to $^{53}Mn/^{53}Cr$ chronometry (Trinquier et al., 2008, and references therein) basaltic eucrite crystallization took place between 3 and 10 Ma after formation of CAI. The formation of cumulate eucrites took place within the first 150 Ma of Vesta's history and these eucrites reveal ages of formation of 4.43 Ga (McSween et al., 2011). The radioactive isotope record in basaltic HED rocks indicates that the petrologic evolution of the minor planet is confined to the first 10's or 100's of Ma after condensation of the solar nebula, i.e. the formation of CAIs (Mittlefehldt et al., 1998; McSween et al., 2011; Mittlefehldt, 2015).

This petrologic scenario is in contrast with the long/lived or ongoing geochemical evolution of the larger inner planets whose early constitution and structure has almost completely been overprinted by active processes of internal differentiation, e.g. plate tectonics on Earth. Thus, Vesta is a study object to learn about the early composition and evolution of the larger planets. However, although the near surface rocks have resided on Vesta's surface since they were formed; continuous bombardment by minor and larger impacts has led to a mechanical reworking commonly known as 'gardening'. Similar to the other almost atmosphere-free objects, at least the outer few meters or ten to hundreds of meters of Vesta's surface consists of very fine-grained dusty material or regolith (e.g., Jaumann et al., 2012). The dominant regolith size is estimated in the order of less than 45 µm (e.g., Hiroi et al, 1994; Palomba et al. 2014; Zambon et al., 2014). Figure 3 depicts processes including impacts and gardening of the regolith that have formed and reshaped the outer parts of Vesta.

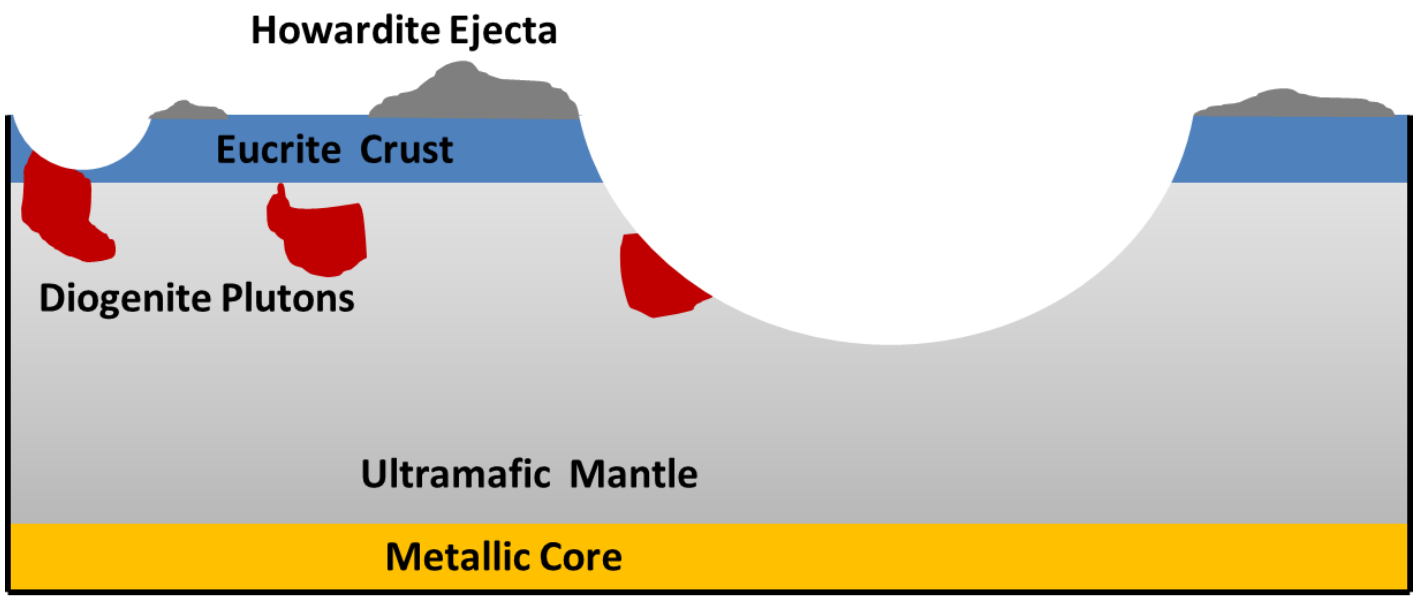

Fig. 3: A sketch of crust forming and reshaping processes including tentative positions of HED lithologies (adapted from McSween and Huss, 2010).

The current morphology and crater distribution suggests at least two larger impact events that gave the present shape: the Rheasilvia and the Veneneia basins. The former is roughly dated to 1 Ga, and it produced a deep basin around the modern South Pole including a prominent central peak (Schenk et al. 2012). The Veneneia basin is also located in the South Pole and partially overlaps with the Rheasilvia basin; it is believed to have formed around 2 Ga ago (Schenk et al., 2012). Beside these large craters, Vesta exhibits numerous smaller craters and structures, some of which have been discussed in more detail elsewhere (Jaumann et al., 2012; Schenk et al., 2012; Marchi et al., 2012).

The current Vestan surface hosts material which was ‘imported’ by the continuous influx of exogenic material from the Main Belt (e.g., Reddy et al., 2012; McCord et al., 2012; Nathues et al., 2014, 2015). Within the group of howardite meteorites there is not only evidence of impact induced glass but also of chondritic material that suggests addition of primitive material to Vesta’s surface after completion of endogenic geologic processes (e.g., Wee et al., 2010; Herrin et al., 2011). The early Vestan chondritic materials are unlikely to have survived the high-temperature geologic history of Vesta, and therefore the present chondritic contaminants or remnants should be exogenic in origin.

### 1.2.1 Petrologic models:

#### 1.2.1.1 Fractional crystallization

As early as in the sixties of the last century, Mason (1962, 1967) suggested a model based on sequential fractional crystallization to explain the evolution of the parent body of the achondrite clan of meteorites today known as HEDs. He considered pallasites as an associate to the HEDs and proposed a Howardite-Eucrite-Diogenite-Pallasite parent body (HEDP-PB). In his model, pallasites (metallic iron and olivine) form the core, and the silicates form an orthopyroxene-rich mantle (diogenite material); pigeonite and plagioclase seen in eucrites and howardites make up the crust. He suggested that the Mg/Fe ratio of the parent body (HEDP-PB) is similar to H chondrite, although there were dissimilarities in other aspects (e.g., alkali content, S content). The consideration of pallasites as a member of the HED clan was supported by Clayton and Mayeda (1996, 1999) who observed similar fractionation lines of HEDs and pallasites in their oxygen isotopic ratios. Following Mason's HEDP-PB model, Dreibus and Wänke (1980) estimated bulk compositions based on HED and pallasite meteorites' geochemical analysis.

#### 1.2.1.2 Partial Melting

Stolper (1975, 1977) introduced the partial melting model on the basis of melting experiments to explain the formation of eucrites. His experimental results showed that olivine, pyroxene, and plagioclase components are at the peritectic point of the silica-olivine-anorthite system. He suggested that eucrites are formed by partial melting of the source region that consists of olivine, pyroxene, plagioclase, chromite and metallic iron; while cumulate eucrites and diogenites are formed at varying degrees of partial melting. He further argued that fractional crystallization fails to generate the melt composition at the peritectic point. During the last three decades, this model was supported by many researchers (e.g., Consolmagno and Drake, 1977; Hertogen et al., 1977; Morgan et al., 1978; Jones, 1984).

#### 1.2.1.3 Magma Ocean

Ikeda and Takeda (1985) proposed a model of Magma Ocean based on their study of howardite Y-7308's lithic clasts and mineral fragments. In their model, primary magmas were produced from carbonaceous or LL chondritic material by partial or batch melting under reducing conditions. Fractional crystallization in an open system was introduced to explain the formation of lithic clasts and mineral fragments in their howardite sample. However, in this model a dunite layer exist as cumulate and, therefore, the model is termed as HED-Dunite parent body (HEDD-PB). Hewins and Newsome (1988) suggested that a maximum fractionation could also generate eucritic magma on the olivine-pyroxene-plagioclase peritectic point rather than the equilibrium process in Stolper's model. Ikeda (1989) improved the model of Ikeda and Takeda (1985) considering the bulk composition of Dreibus and Wänke (1980). Warren (1985, 1997) proposed magma ocean crystallization based on mass-balance constraints from major and trace elements. Magma ocean model was favored over partial melting model by many researchers based on major and trace element fractionations (Righter and Drake, 1997; Ruzicka et al., 1997). Ruzicka et al. (1997) chose fractional crystallization followed by equilibrium crystallization to explain the petrogenesis of HEDs while Righter and Drake (1997) supported equilibrium crystallization followed by fractional crystallization. Righter and Drake (1997) considered a convective lock-up in the magma chamber with a conductive lid. Mandler and Elkins-Tanton (2013) improved the approach of Righter and Drake' model and they claimed that their revised magma ocean model can explain eucrites and diogenites petrogenesis. The model is a two-step fractionation sequence (equilibrium crystallization up to 60-70% followed by fractional crystallization) with subsequent magmatic recharge from the underlying sub-crustal magmatic-mush. Figure 6 shows four stages of evolution of Vesta from their model. Mandler and Elkins Tanton (2013) challenged the idea of fractional crystallization because such models fail to explain HED petrogenesis because the residual liquid is too rich in iron to produce basaltic eucrites. In addition they stated that the partial melting model seems to be unrealistic to explain the petrogenesis of diogenites together with that of cumulate eucrites. Toplis et al. (2013) proposed a model based on thermodynamic and mass-balance constraints combining 1/3 H chondrite and ¾ CM2. Based on their model as well as observations from Dawn, they suggested that the mantle of Vesta (or HED parent body) is enriched in orthopyroxene rather than olivine.

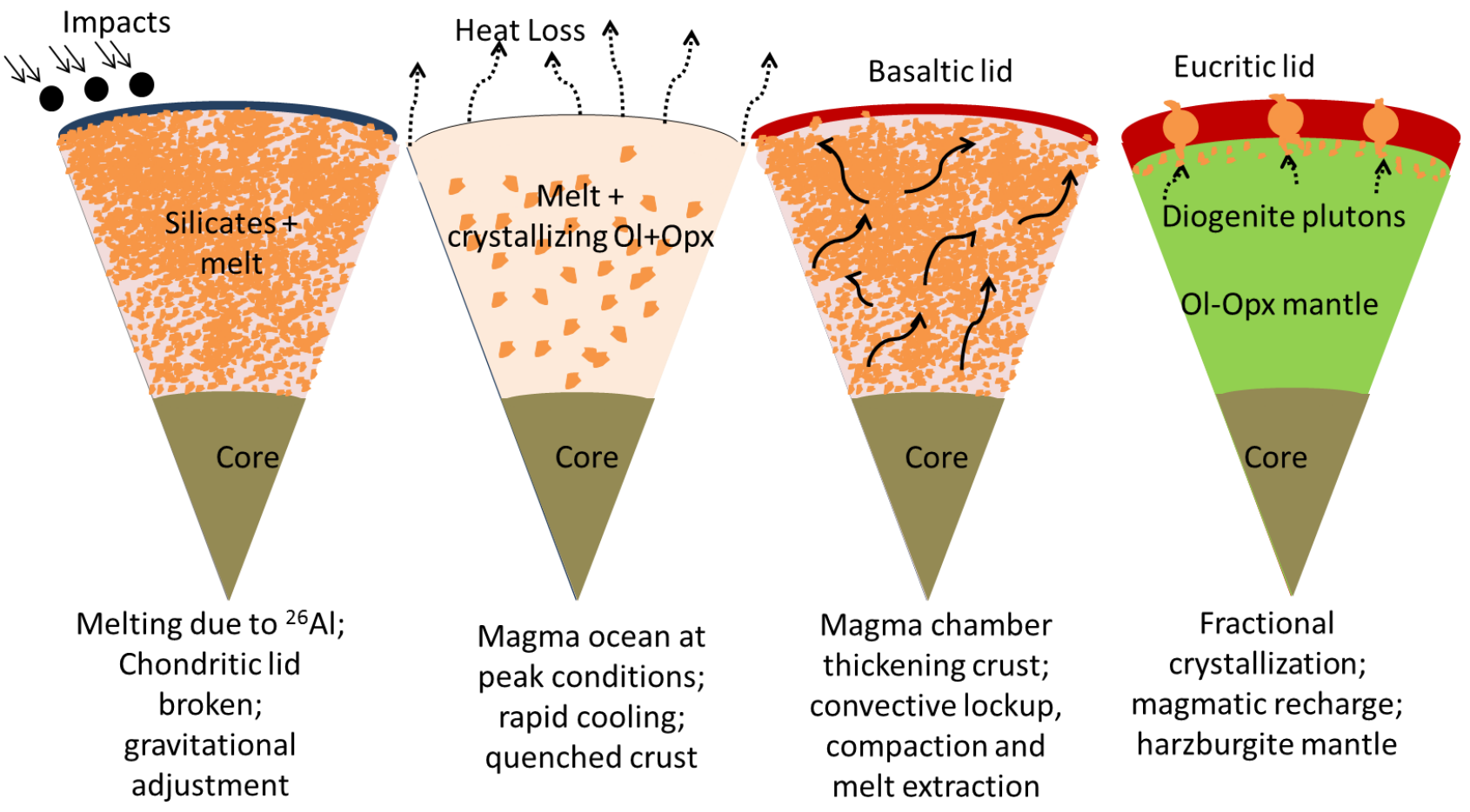

Fig. 4: Commonly accepted view of Vesta's internal differentiation from a composite chondritic source according to Mandler and Elkins-Tanton (2013). Formation of magma ocean by early heat production (left); near-complete melting and heat loss (center left); crystallization of olivine and orthopyroxene plus accessory oxides, and formation of an early basaltic outer shell (center right); evolution of plutonic and basaltic rocks (right). (adapted from Mandler and Elkins-Tanton, 2013).

#### 1.2.1.4 Serial-magmatism

Yamaguchi et al. (1996, 1997) proposed a model to explain the evolution of eucritic crust by subsequent eruption and rapid cooling followed by burial and reheating of lava flows. The model was based on their petrologic and thermal observations of eucrites. They suggested that metamorphosed or reequilibrated-eucrites in the hot lower crust could allow even small dikes to cool slowly thereby forming cumulate eucrites. Their model did not consider the formation of diogenites at that time, and they stated that eucritic magmas might have undergone re-melting to produce diogenites. Later on, Barrat et al. (2010) found deep negative Eu anomaly in some of the diogenites that could be explained by the mixing of these diogenites' parental melts with melts derived by eucritic crusts' partial melting. They concluded that occurrence of local eruptions of such lavas, i.e., diogenitic magmatism seems to be realistic. The serial-magmatism model is supported by the observation of a large range of incompatible trace elements in diogenites (Mittlefehldt, 1994; Fowler et al., 1995; Shearer et al., 1997) and in Stannern-trend eucrites (Barrat et al., 2007) as well as by the existence of variable Mg#s in diogenites (Beck and McSween, 2010; Shearer et al., 2010). Recently,

Cheek and Sunshine (2014) suggested that the co-existence of olivine and chromium in Arruntia and Bellicia craters imply an intra-crustal pluton in origin. It is worth mentioning that Barrat and Yamaguchi (2014) refuted the model of Mandler and Elkins-Tanton's magma ocean model claiming that multiple magmatic melt sources are needed to explain the HED petrogenesis.

## 1.2.2 Towards a model of Vesta's internal structure

Despite the unprecedented progress in understanding the geology of Vesta from the observations of Dawn mission and the study of HED meteorites, existing models fail to reveal an equivocal petrologic evolution and internal structure of Vesta. Based on impact modelling and the obvious lack of significant abundances of olivine in the Rheasilvia basin, Clenet et al. (2014) claimed that the crustal and mantle transition zone of Vesta is extended up to a depth of 80-100 km. This scenario is in contrast to widely accepted magma-ocean models proposed by Righter and Drake (1997), Ruzicka et al. (1997), and Mandler and Elkins-Tanton (2013). An orthopyroxene-rich mantle of Vesta has been favored rather recently (e.g., Toplis et al., 2013). This new idea is also incompatible to the afformentioned models. Meanwhile, the model composition of Toplis et al. (2013) is unlikely to meet the $\delta^{17}O$-$\delta^{18}O$ relations of HEDs as shown in Fig. 4.

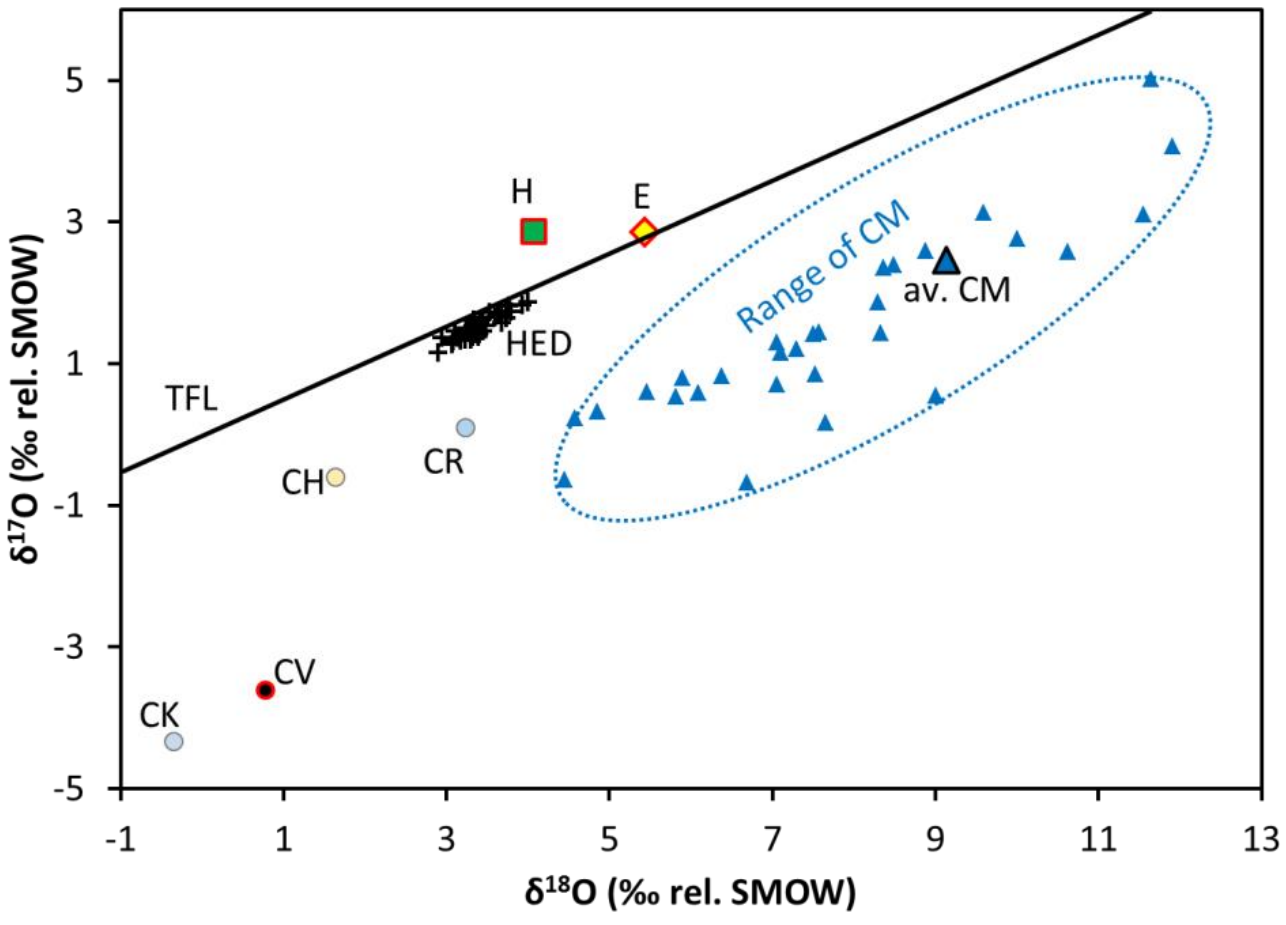

Fig. 5: $\delta^{17}O$-$\delta^{18}O$ relations of common chondrites and HEDs. TFL: Terrestrial Fractionation Line; H: Ordinary chondrites; E: Enstatite chondrites; CM, CV, CK, CH, CR: carbonaceous chondrites. Mixing of H plus CM chondrite seems to be unrealistic to match

HED signatures. Oxygen isotope data compiled from Clayton and coworkers' (Clayton and Mayeda, 1996, 1999; Clayton, 1993; Clayton et al., 1976).

Motivated by this current problem, a revised magma ocean model is presented. A starting chondritic composition is used that: (i) matches the three oxygen isotope signatures of HEDs, (ii) is rich enough in metal plus sulfide to produce a proper core size and (iii) reveals a bulk silicate composition that is able to produce an ultramafic mantle followed by a pyroxenitic (diogenite) and an gabbroic/basaltic (eucrite) outer shell.

The first basic requirement, i.e., appropriate $\delta^{17}O$-$\delta^{18}O$ relations is quite important because oxygen is the most abundant chemical element in the silicate fraction in all chondrites and in the bulk silicate Vesta composition. Therefore, a mixture of average ordinary chondrite (H) composition with average carbonaceous chondrite type, for example CV chondrite, which is low enough in metallic iron, is chosen. The latter is important in order to reduce the enormously high metal contents of H chondrites, and the CV carbonaceous chondrites can fulfill this requirement. The mixing line of average H chondrite with average CV meet the $\delta^{17}O$-$\delta^{18}O$ relations of HEDs is shown in Fig. 6 (red dashed line), and a mixture of 80% H plus 20% CV chondrite matches the $\delta^{17}O$-$\delta^{18}O$ features on HEDs (red/green box). This major oxide composition is used in this case study. Other mixtures of ordinary chondrite with compatible carbonaceous chondrites are also possible.

In the next step, the chemical major element data of the non-metallic i.e., the oxidized portion for the selected H and CV chondrites' composition are compiled. There are two reliable sources of data in the literature, one is from Wasson and coworkers (e.g., Wasson and Kalleymen, 1988; Hutchinson, 2007), the other stems from Jarosewich (1990, 2006). Jarosewich provides his data in major element oxides while the group of Wasson presents wt.% of elements. A recalculation of the Wasson (W) dataset to oxide values reveals a certain difference compared to that of Jarosewich (J), particularly in FeO and MgO (Table 3). Since there is no measure at hand to prefer one data set or the other, we use both data sets in parallel. It is worth mentioning that alkali depleted chondritic precursor is adopted in the existing models, and sodium, potassium and phosphorus are regarded as trace elements.

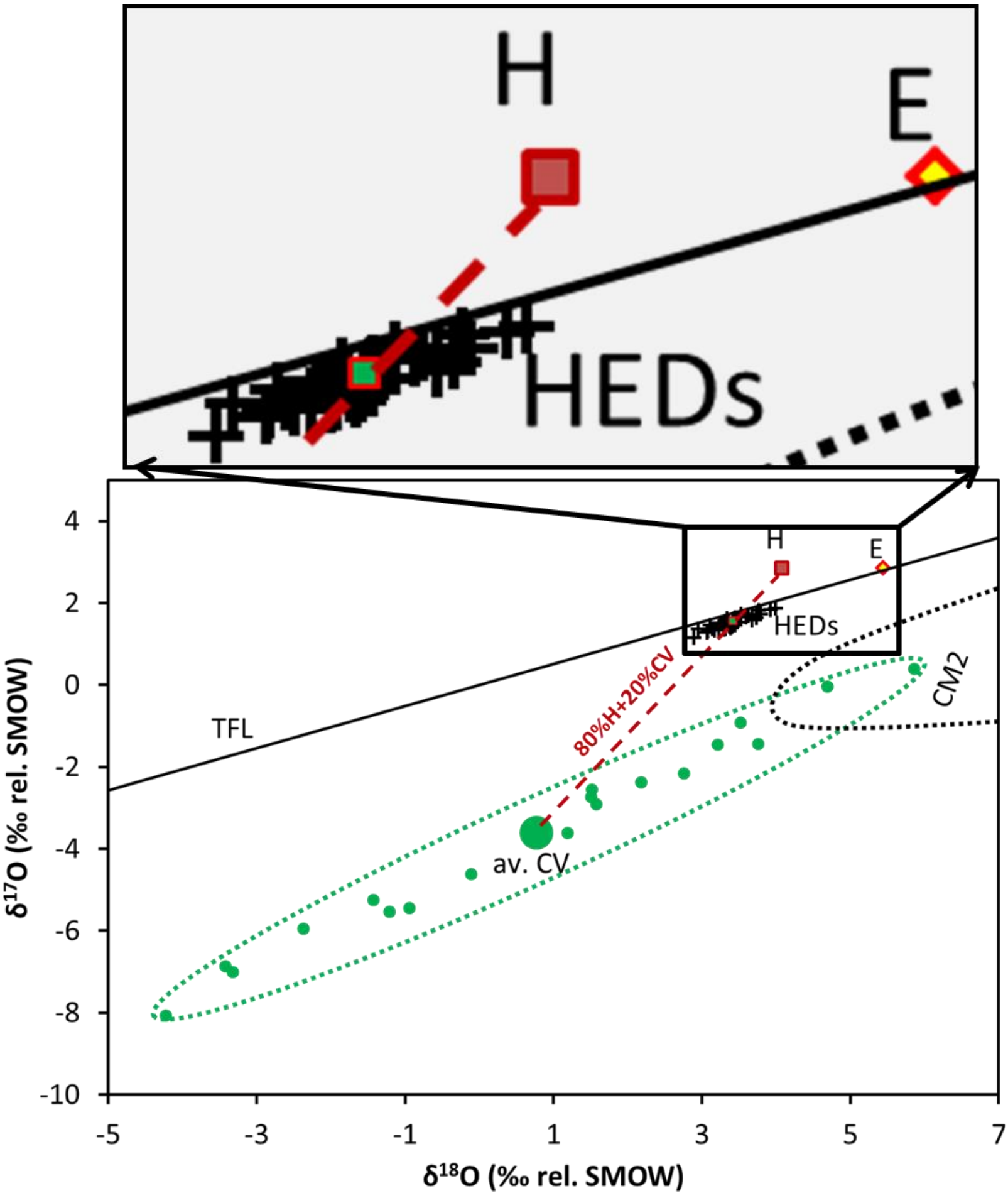

Fig. 6: Oxygen isotope relations of average H chondrite and average CV carbonaceous chondrite. Red dashed line: Binary mixing of CV and H; small red/green box in the HED field refers to 80% H + 20% CV mixture. Back dashed outline shows part of CM2 region. Data points of HEDs, and the mixing line is shown along with the location of 80% H+20% CV mixture (inset).

Table 3: Starting compositions: Upper part, 80% H - 20% CV based on chondrite compositions from Jarosewich (1990, 2006); lower part, Wasson and coworkers' data compiled by Hutchinson (2007).

| Data from Jarosewich, 1990, 2006 | | | | | | | | | | | |
|---|---|---|---|---|---|---|---|---|---|---|---|
| wt.% | $SiO_2$ | $TiO_2$ | $Al_2O_3$ | $Cr_2O_3$ | $Fe_2O_3$ | FeO | MnO | MgO | CaO | $Na_2O$ | Sum |
| CV | 37.7 | 0.2 | 3.6 | 0.6 | 0.0 | 27.7 | 0.2 | 26.9 | 2.8 | 0.4 | 100 |
| H | 48.3 | 0.2 | 2.9 | 0.7 | 0.0 | 13.6 | 0.4 | 30.5 | 2.3 | 1.1 | 100 |
| **80%H plus 20% CV** | **46.2** | **0.2** | **3.0** | **0.6** | **0.0** | **16.4** | **0.4** | **29.8** | **2.4** | **0.9** | **100** |

| Data from Hutchinson, 2007 based on Wasson and colleagues' analyses | | | | | | | | | | | |
|---|---|---|---|---|---|---|---|---|---|---|---|
| wt.% | $SiO_2$ | $TiO_2$ | $Al_2O_3$ | $Cr_2O_3$ | $Fe_2O_3$ | FeO | MnO | MgO | CaO | $Na_2O$ | Sum |
| CV | 35.8 | 0.2 | 3.5 | 0.6 | 0.0 | 30.5 | 0.2 | 25.8 | 2.9 | 0.5 | 100 |
| H | 44.9 | 0.1 | 2.7 | 0.7 | 0.0 | 19.0 | 0.4 | 28.9 | 2.2 | 1.1 | 100 |
| **80%H plus 20% CV** | **43.1** | **0.1** | **2.8** | **0.6** | **0.0** | **21.3** | **0.3** | **28.2** | **2.3** | **1.0** | **100** |

The MELTS software of Ghiorso and Sack (1995) is widely used to model the crystallization or partial melting of silicate systems. The program is used here to evaluate the crystallization sequence of liquids with compositions given in Table 3. The assumptions made are: $fO_2$= iron-wustite (IW); pressure: 500 bars; no water; range of T: 1600 to 1000°C. The results of our calculations with MELTS on the two starting compositions (W) and (J) are graphically shown in Fig. 7.

The solid line starts at 100% liquid (initially no solids). At this stage, the liquid has the compositions of the respective starting composition. Upon cooling, olivine is the first mineral to appear in both liquids. Between 1450 and 1480°C orthopyroxene along with minor spinel appears. At this stage, the liquids have crystallized approximately ***35 to 40% magnesian olivine.***

Further cooling gives rise to the simultaneous formation of olivine plus orthopyroxene (with traces of spinel). Feldspar and clinopyroxene are not in equilibrium with the liquid at this stage. The question arises to which degree the equilibrium crystallization stops so that the remaining liquid is allowed to form crustal rocks (basaltic/gabbroic mineral assemblages) upon fractional crystallization. Righter and Drake (1997) consider 80% of equilibrium crystallization while it is 60-70% for Mandler and Elkins-Tanton (2013) model. Experimental evidence on terrestrial basaltic rocks suggests that at a $K_d$ value of 0.31, a basaltic liquid is in equilibrium with olivine and orthopyroxene at low pressures (Takahashi and Kushiro, 1983). Here, the degree of crystallization at this point is referred for this transition in this crystallization sequence. This experimentally derived $K_d$ value is defined as $(MgO/FeO_{liquid})/(MgO/FeO_{olivine})$.

At the stage where the FeO/MgO ratio of the liquid – olivine + orthopyroxene arrives at $K_d$=0.31, around 20 to 25 % of evolved liquid is present. This value is compatible with independently derived estimates of 20 % liquid by Righter and Drake (1997) while there seems to be an apparent contradiction to the results of Mandler and Elkins-Tanton (2013). Assuming further that some 20 to 25 % of the remaining liquid will form the crustal rocks, it is easy to give the proportions of the olivine plus orthopyroxene solids. These are ***25-30 % orthopyroxene plus 5-10 % olivine.***

These orthopyroxene-olivine relations do ***not*** consider the early-crystallized olivine i.e., 35 to 40%.

Summing up the results of the MELTS calculation for a 80% H plus 20% CV chondrite source, it reveals two distinctly different scenarios:

***First,*** a nearly pure assemblage of olivine is formed. This volume of solid may constitute the inner solid shell of bulk silicate Vesta, and is dunitic in composition. The liquid remained after the extraction of this olivine proportion produces an orthopyroxene-dominated olivine-bearing solid. The evolution of the remaining liquid is not considered any further since it will form gabbroic and basaltic rocks. This model assumes separation of olivine and orthopyroxene from the respective liquids. As a consequence, the lower part of Vesta's silicate shell has a nearly pure dunitic composition while the layer above this peridotite sequence is of orthopyroxenitic composition.

***Second***, any segregation of early crystallized minerals is not considered, i.e., the crystallized phase is assumed to form a continuous sequence of pure olivine (a dunite rock) followed by harzburgite compositions of olivine-rich orthopyroxene-bearing assemblage or relatively similar proportions of olivine and orthopyroxene. The volume proportions discussed above can be recalculated into spherical geometries using the mass-balance equations considering the required density information of all minerals which are readily available (either from the MELTS results or from literatures e.g., Deer at al., 1997). Thus, the results yield from this revised model is used to predict an early intact spherical shape of Vesta. For simplicity, a spherical shape (radius ~ 262 km) of the body having a core radius of 110 Km (Russell et al., 2012) suggested by Dawn observation is used.

For the ***first*** scenario:

The thicknesses of the dunite layer and the orthopyroxenite layer are about 48 km and 81 km thickness (W), respectively (45 km and 86 km for J) (Fig. 8, showing average thicknesses).

For the ***second*** scenario:

The layer of gradually increasing orthopyroxene underlying an olivine-dominated assemblage gives rise to a very similar estimation of about 135 km (W) and (139 km for J) thickness of the mantle (Fig. 8, showing average thickness).

It is beyond the scope of this thesis to get deeper into the fractionation of the liquid in equilibrium with Ol and Opx at $K_d$=0.31. For now, it is assumed that the remaining liquid fraction will form the outer shell of Vesta comprising eucrite and diogenite rock assemblages. The thickness of the 'crustal' layer is estimated to ~23 km (W) or ~21 km (J) for the ***first*** scenario, and ~17 km (W) or ~13 km (J) for the ***second*** scenario. These volume relations are graphically shown in Fig. 8. Given the excavation depth of ~ 80 km by the Rheasilvia impact in the South Pole basin (Ivanov and Melosh, 2013; Jutzi et al., 2013; Clenet et al., 2014), the olivine-rich mantle material of Vesta is unlikely to excavate and expose on the surface.

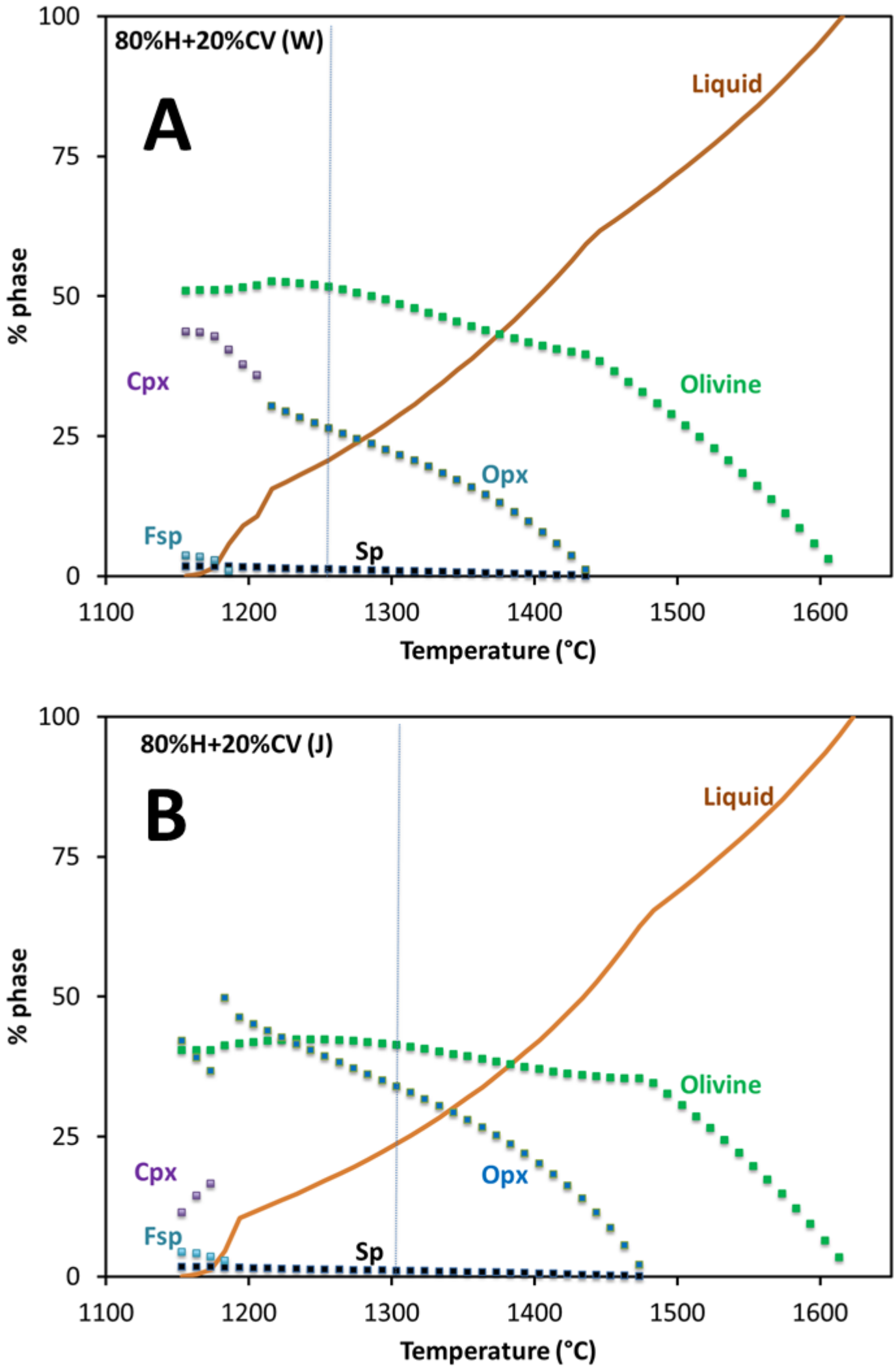

Fig. 7: Equilibrium crystallization of a chondritic precursor (80% H + 20% CV); compositions from Wasson and coworkers (W) and from Jarosewich (J). The solid straight line shows the temperature where the $K_d$ value (defined as $(MgO/FeO_{liquid})/(MgO/FeO_{olivine})$ is 0.31.

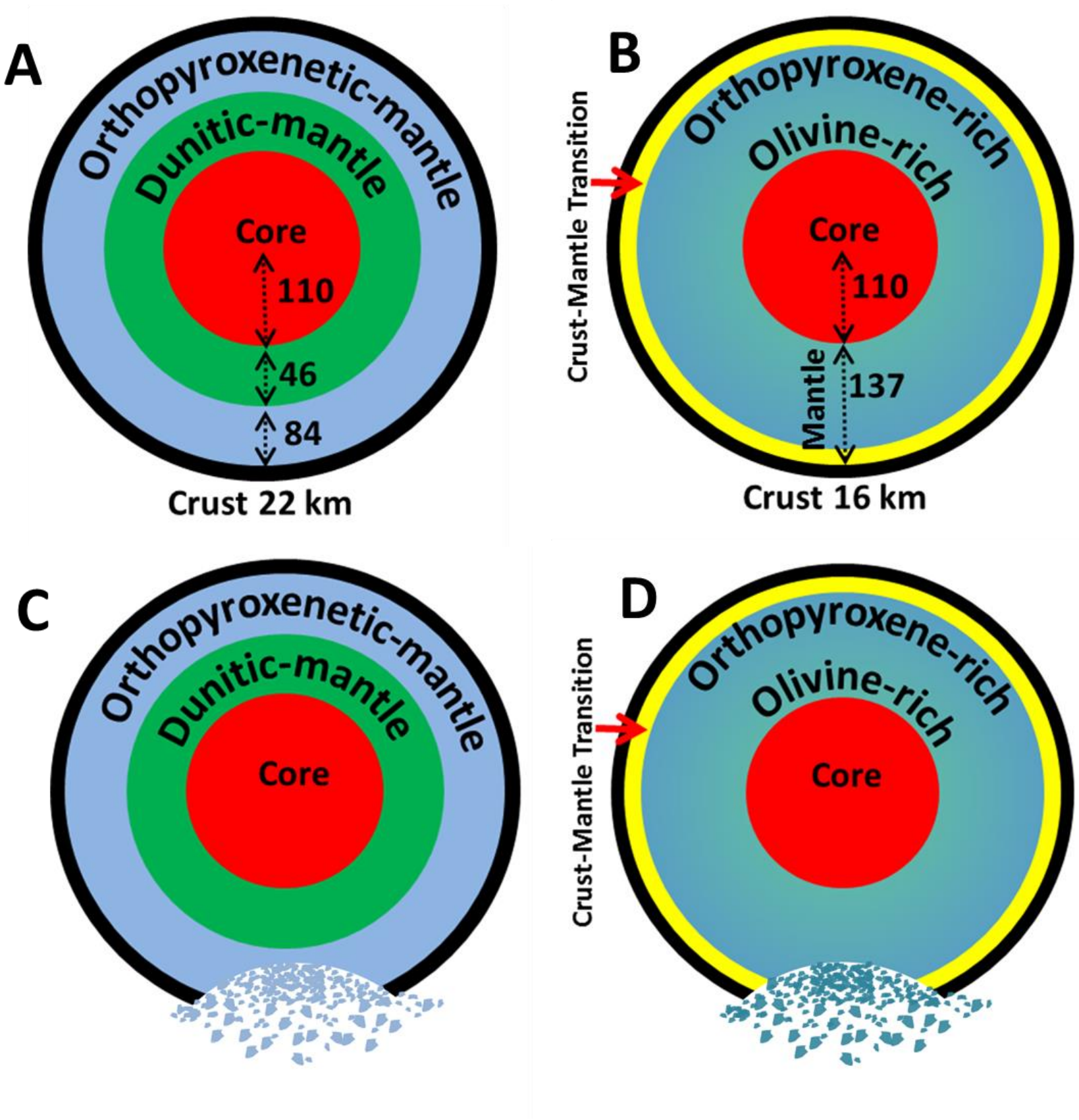

Fig. 8: Thickness of mantle and crust derived using the results of MELTS thermodynamic calculator, for 80% H + 20% CV composition, (W) for Wasson's composition, and (J) for Jarosewich. The crust-mantle transition in *scenario (ii)* is hypothetical. A spherical shape of the body (radius ~ 262 km) with a core radius of ~110 km suggested by Dawn observations is adopted (Russell et al., 2012). (A) *scenario (i)*, (B) *scenario (ii)*. Given an excavation depth of ~ 80 km of the Rheasilvia impact, it is unlikely to excavate the olivine-rich mantle materials (C, D).

## 1.3 Laboratory spectra of HEDs, and implications for Vesta

The link between the surface mineralogy of Vesta and the constituent minerals of HEDs is provided by an integrated study of laboratory derived-spectra of HEDs and spectral data obtained by Dawn spacecraft. As mentioned earlier, the FC color data has seven channels in the visible and near-infrared wavelength range (0.4-1.0 µm). The robustness of FC color data in analyzing mineralogy and geology of Vesta is well documented elsewhere (Reddy et al., 2012a, b; Thangjam et al., 2013, 2014; Nathues et al., 2014, 2015; Schaefer et al., 2014). Various FC band parameters are used for surface compositional analysis and mapping (Reddy et al., 2012a, b; Thangjam et al., 2013, 2014; Nathues et al., 2014a, b; Le Corre et al., 2011, 2013; Schaefer et al., 2014). The band parameters are derived by investigating the laboratory spectra. A large number of spectra of HEDs and individual rock forming minerals, archived at (i) the RELAB spectral library (Brown University, USA), (ii) the HOSERLab (University of Winnipeg, Canada), and (iii) the USGS spectral libraries (Flagstaff, Arizona), have been collected. The spectra usually cover the wavelength range between 0.3 and 2.5 µm (Fig. 9A). The whole rock and mineral spectra are resampled to the FC bandpasses (Sierks et al., 2011; Le Corre et al., 2011) and they are used in the analysis (Fig. 9B). Figure 9A shows an example of spectra of average HEDs, olivine and high-Ca pyroxene in the wavelength range from 0.4 to 2.5 µm at full resolution, whereas the resampled spectra to FC bandpasses in the wavelength range between 0.4 and 1.0 µm are displayed in Fig. 9B. Detailed discussions of many laboratory spectra and band parameters are presented in Chapter 2, 3 and 4.

The identification and mapping of olivine-rich sites at Arruntia and Bellicia craters on Vesta using FC color data is one of the significant contribution in this thesis. Despite the FC color data's narrow wavelength range (0.4-1.0 µm) to cover the full 1-µm absorption feature, diagnostic FC band parameters derived from various laboratory spectra allows identification of the olivine-rich exposures (Thangjam et al., 2014). Besides, another significant contribution is the three-dimensional spectral approach that is introduced for the first time for analyzing surface compositional heterogeneity (Thangjam et al., in press). A three-dimensional approach can constrain an additional parameter over the generally used two-dimensional spectral analysis, and it can yield rather effective results.

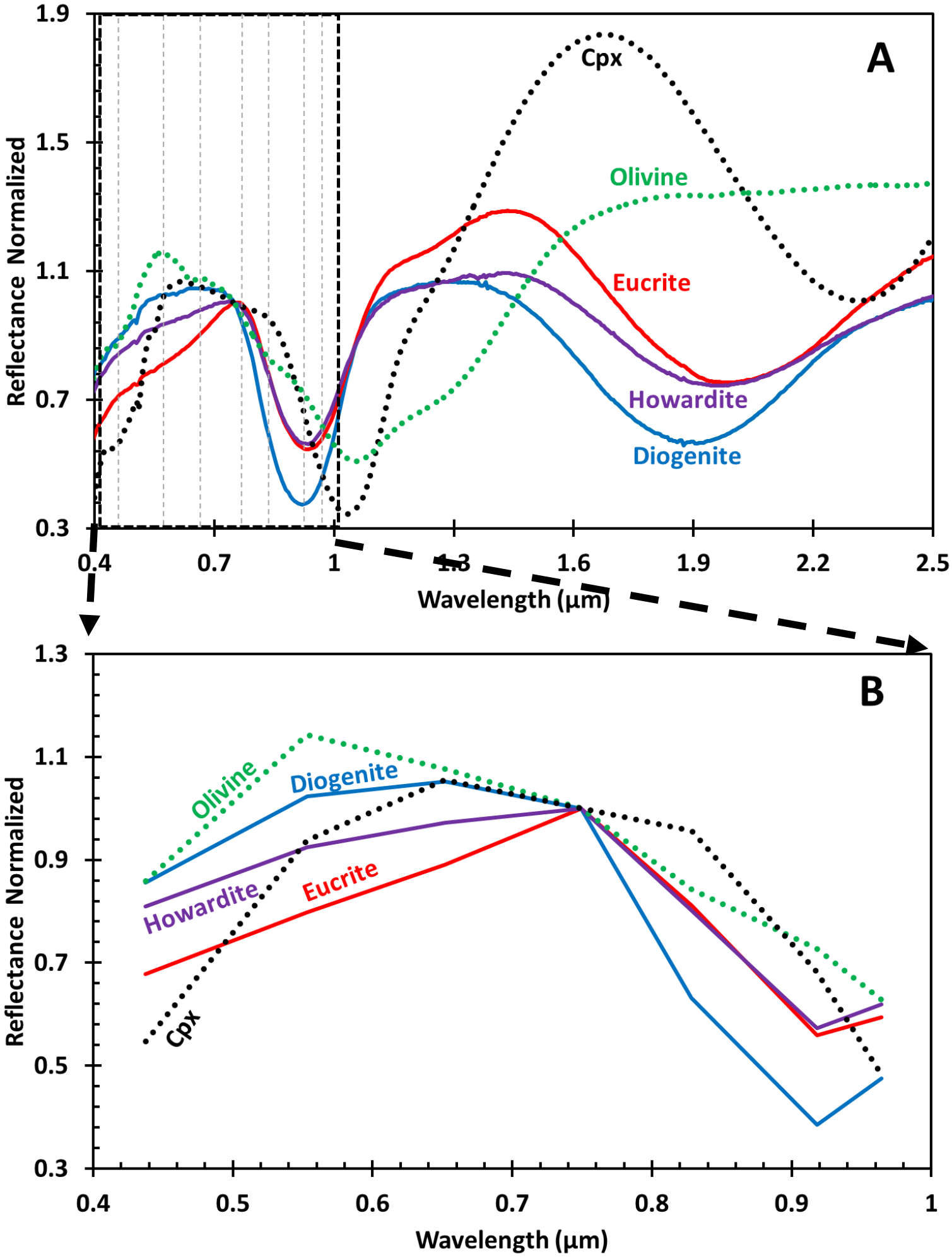

Fig. 9: Reflectance spectra in the visible to near infrared wavelength range of average eucrite, diogenite, and howardite along with a monoclinic high-Ca pyroxene (Cpx) and a magnesian olivine; spectra are normalized to 0.75 µm. (A) Wavelength range up to 2.5 µm, dashed box refers to FC spectral range, and dashed lines show the center wavelengths of seven FC color filters. (B) Spectra resampled to FC band passes.

For a brief summary of this work, the applications of some of the FC band parameters are outlined in the following section as a case study of the Arruntia region on Vesta (see Chapter 2, 3, 4 for more details). This impact crater is about 12 Km in diameter located in the northern hemisphere (approximately 36-42°N, 46-74°E). It is one of the freshest features on Vesta that is approximately 2 – 18 Ma old based on crater counting methodology (Ruesch et al., 2014a). This region shows important lithologic compositions and geologic features, i.e., olivine-rich sites (e.g., Ammannito et al., 2013b; Thangjam et al., 2014; Ruesch et al., 2014a, b; Nathues et al., 2015), dark and bright materials (e.g., Zambon et al., 2014; Ruesch et al., 2014a, b; Thangjam et al., in press), potential impact melts (Le Corre et al., 2014; Ruesch et al., 2014a; Thangjam et al., in press). Figure 10 shows a global mosaic of Vesta from FC HAMO data (~60 m/pixel, reflectance image at 0.55 µm) highlighting the location of Arruntia crater. A perspective view of the region draped over HAMO DTM (~ 62m/pixel) is also displayed in Fig. 10. Two-dimensional and three-dimensional analyzes are briefly presented in the following sub-sections.

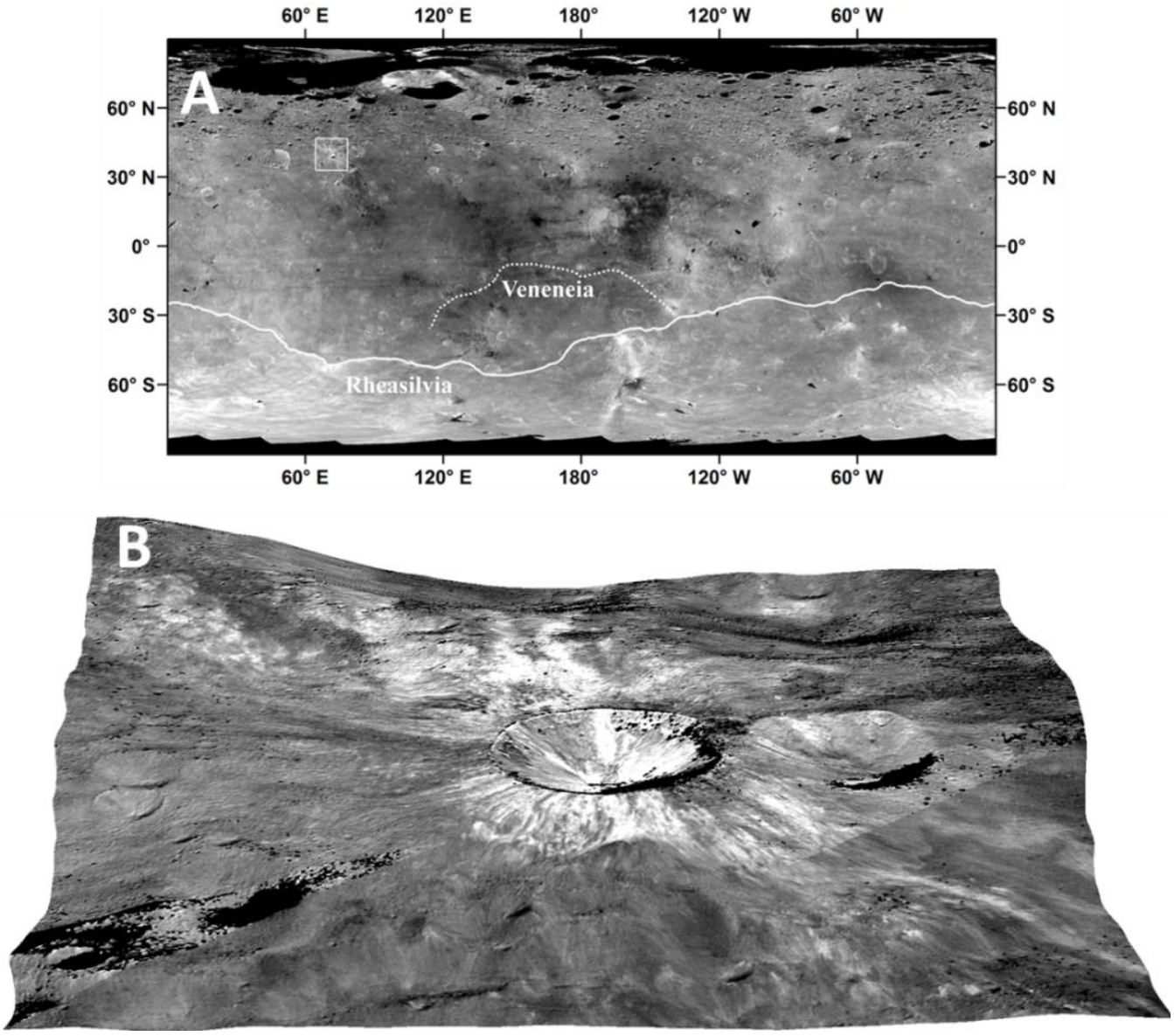

Fig. 10: (A) Global mosaic of Vesta (reflectance at 0.55 µm, ~60m/pixel) from FC HAMO data showing the location of Arruntia region in the northern hemisphere. Outlines of two basins are indicated by solid line (Rheasilvia basin) and dashed line (Veneneia basin).

(B) Perspective view of Arruntia region draped over the HAMO DTM (~62 m/pixel). The radius of the crater is ~ 12 km.

### 1.3.1 Two-dimensional spectral analysis: Mid Ratio and Band Tilt

The Band Tilt (BT) and Mid Ratio (MR) parameters are diagnostic for identifying olivine (Thangjam et al., 2014; Nathues et al., 2015). The BT and MR parameters are defined as:

$$BT= R_{0.92\mu m}/R_{0.96\mu m}$$

$$MR= (R_{0.75\mu m}/R_{0.83\mu m})/(R_{0.83\mu m}/R_{0.92\mu m})$$

where, $R_{\lambda}$ represents reflectance values at the given wavelengths.

Figure 11 shows normalized spectra of average eucrite, diogenite, olivine and high-Ca pyroxene, with a sketch of the band parameters used here. Based on the laboratory spectral data, two-dimensional polygons are defined for each lithologic group in BT-MR band parameters' space (Fig. 11B). The Ol-Opx and harzburgites represent olivine and orthopyroxene mixtures less than and more than 40 wt.% of olivine, respectively (Thangjam et al., 2014). Dunites are nearly pure olivine specimens. Two polygons of monoclinic clinopyroxene or high-Ca pyroxene is shown; Cpx/HED represents those minerals compatible with clinopyroxene or high-Ca clinopyroxene found in eucrite meteorites; High-Ca Cpx is nearly the whole range of clinopyroxenes (~ up to 50% wollastonite). As is seen from Fig. 11b, harzburgitic and dunitic polygons are separable from eucrites and howardites, almost distinguished from Cpx/HED. This two-dimensional band parameter space is used to identify olivine-rich sites on Vesta. Figure 12A shows a perspective view of Arruntia crater showing olivine-rich sites in red. Locations of data points of some of these exposures (Fig. 12A) in the BT and MR band parameters space are displayed in Fig. 12B. Many of the olivine-rich exposures are found in the ejecta nearby the inner rim of Arruntia crater and few sites in the floor as well as on the wall of the crater. The exposures extend from a few 100's of meters up to few kilometers. The sites with olivine-rich material show higher reflectance compared to the average Vesta surface. It is worth mentioning that within a region of 2.5 crater radii, olivine-rich sites cover ~ 1.5 % of the surface area. The results presented in this work confirmed the discovery of olivine-rich exposures by Ammannito et al. (2013b) using the VIR cubes in the wavelength range from 0.3 to 2.5 µm. This spectral approach is further extended on the entire Vesta's surface to identify and map global olivine-rich sites using FC color data (Nathues et al., 2015). Nathues et al., (2015) suggest that olivine on Vesta is likely exogenic in origin. The Arruntia region is found to be dominantly eucritic/howarditic in composition. Diogenite units are not observed in this region. Dark and bright areas display a notable difference in reflectance though brighter regions are relatively larger in extent.

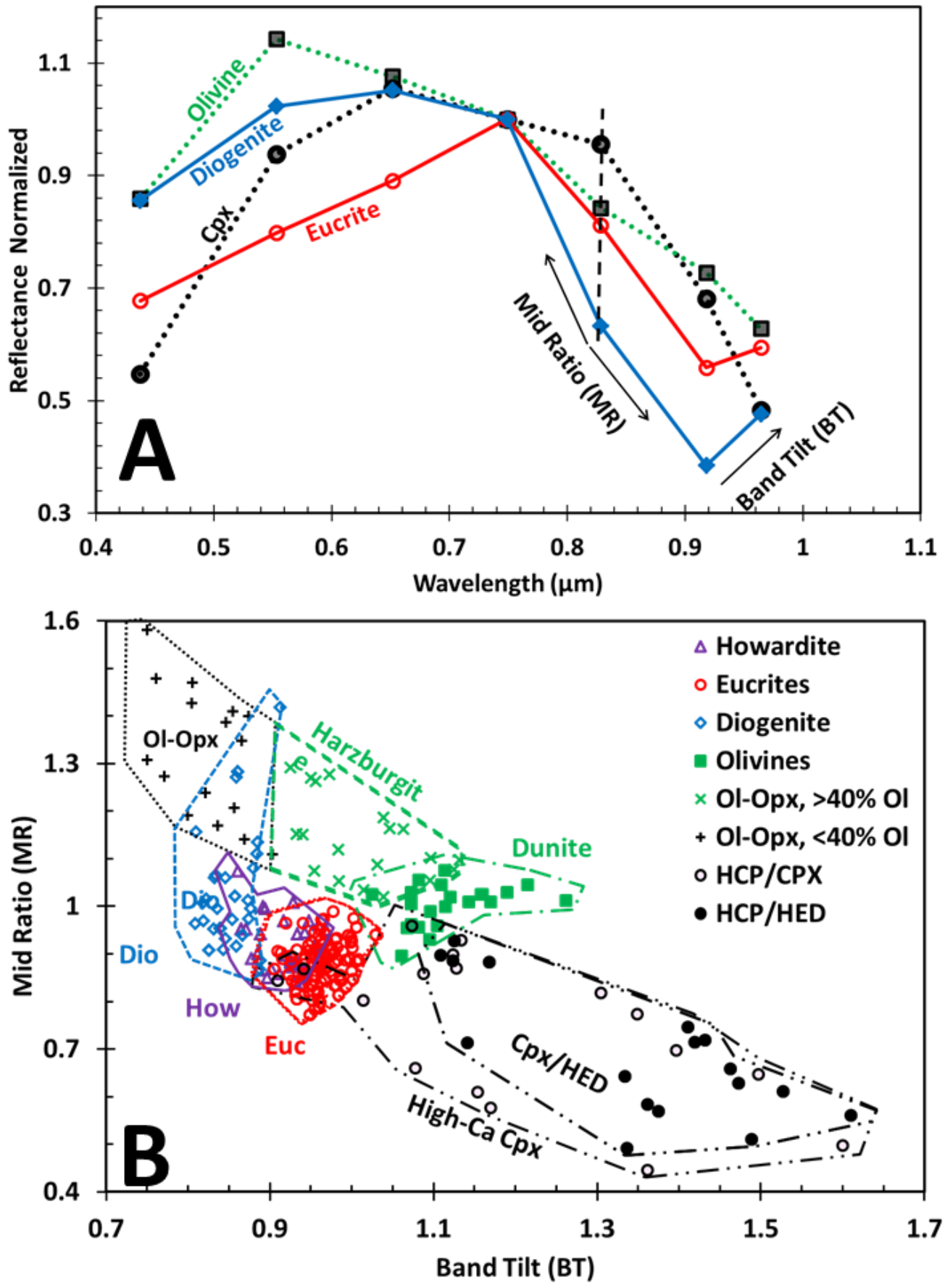

Fig. 11: (A) A sketch showing band parameters, BT and MR, with average spectra of eucrite, diogenite, olivine, and high-Ca pyroxene. (B) Band parameter space using derived band parameter values of BT and MR from various laboratory spectra. (See details in Chapter 3).

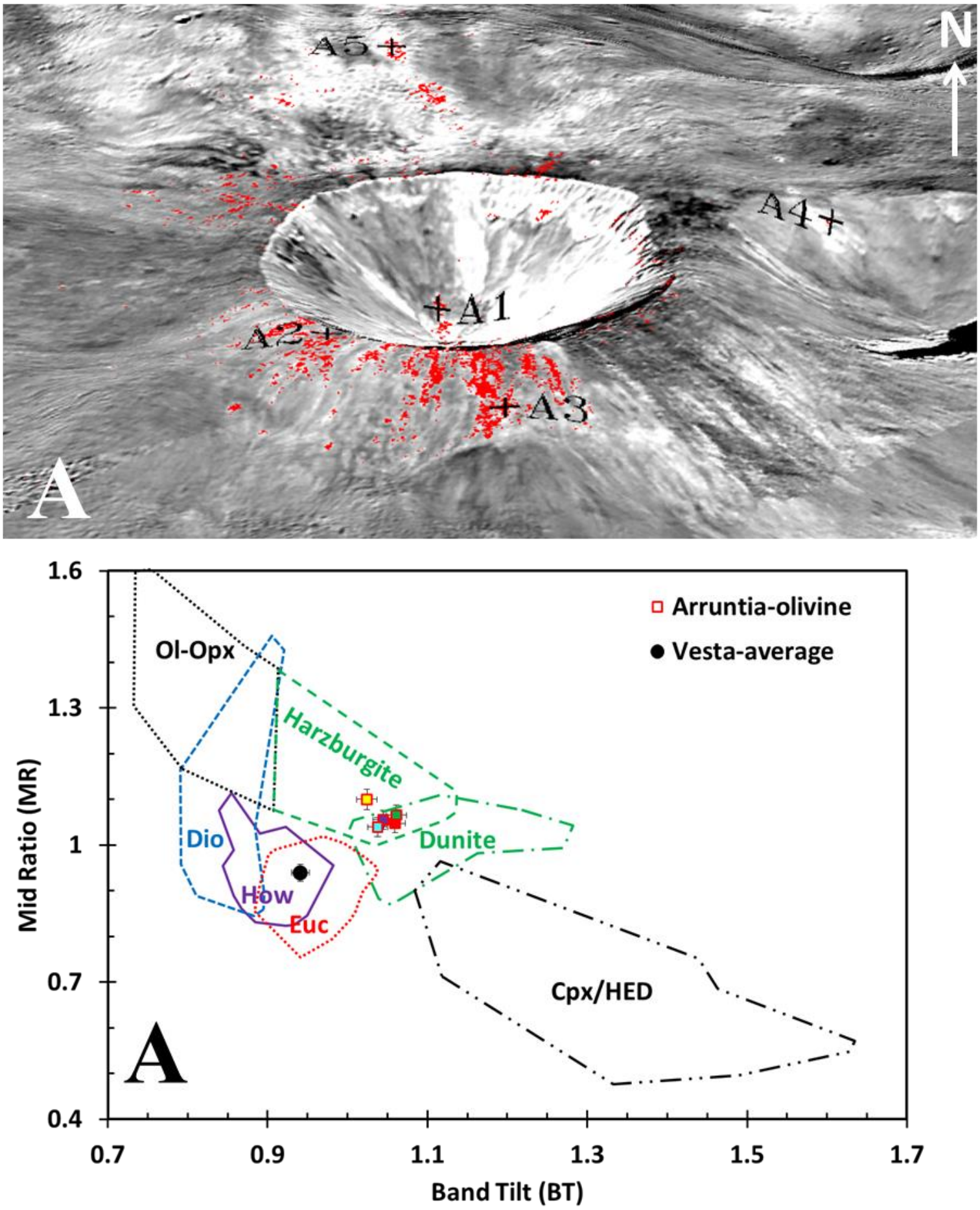

Fig. 12: (A) A perspective view of reflectance image ($R_{0.55\mu m}$) of Arruntia crater, projected on HAMO-DTM showing potential olivine-rich exposures marked in red. (B) Locations of the data points in BT-MR band parameters space, for the sites shown in (A). (See details in Chapter 3).

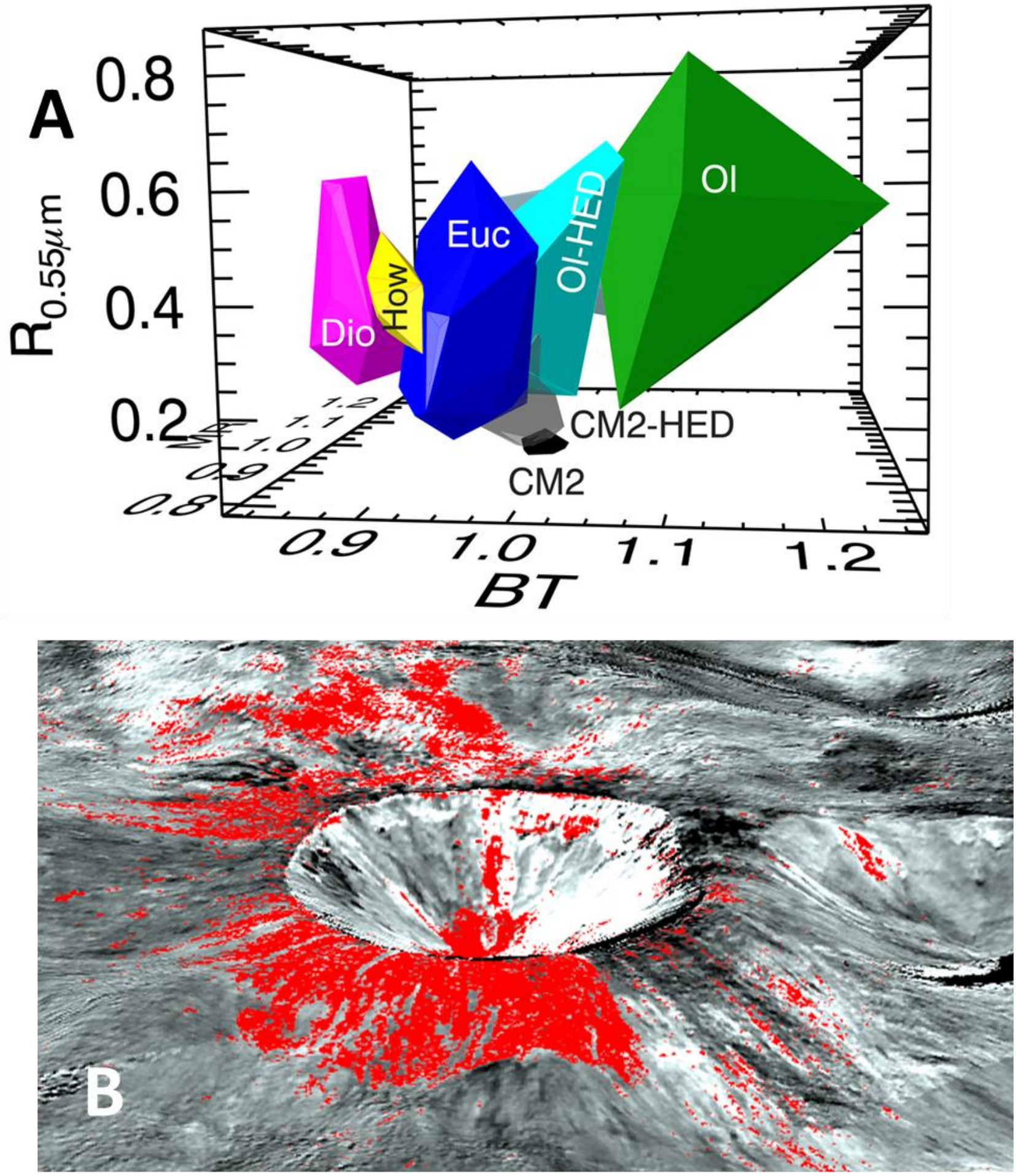

Fig. 13: (A) Three-dimensional polyhedrons defined in BT, MR, and $R_{0.55\mu m}$ band parameters space. (B) A perspective view showing olivine-rich exposures in red. The distribution of olivine-rich exposures is rather large compared to Fig. 12A.

### 1.3.2 Three-dimensional spectral analysis: Mid Ratio, Band Tilt and $R_{0.55\mu m}$

In continuation of the earlier work, a three-dimensional spectral approach is applied for the Arruntia region. In addition to the Mid Ratio (MR) and Band Tilt (BT) parameters of the two dimensional spectral analysis discussed in the previous section, a third parameter the reflectance at 0.55 µm ($R_{0.55\mu m}$) is employed here. The definitions of the band parameters are already illustrated in Fig. 11A. New spectra measured in the laboratory (two howardites, an olivine, a CM2-chondrite, and various mixtures of howardites with olivine or CM2 in 10 wt.% intervals) are also included in this work. Using band parameters derived from the laboratory spectra (of eucrites, diogenites, howardites, olivine, CM2 chondrite, and mixtures of HEDs with either olivine or CM2 chondrite), three-dimensional polyhedrons are defined that represent particular lithologies. Figure 13A shows the three-dimensional polyhedrons. The FC data points are assigned into these polyhedrons using an algorithm in IDL programming language. A detail on this algorithm is provided in Chapter 4. Based on these three-dimensional polyhedrons, surface lithologic units of Arruntia region are mapped. The region is found to be eucrite-dominated howardite in composition. Lithologies with nearly pure olivine, diogenite, or CM2 are not observed. Olivine-rich areas (more than 40-60 wt.% olivine) found to be more abundant than the areas identified using two-dimensional spectral analysis. Within an area of 2.5 crater radii, the olivine-rich exposures occupy about 14% which is about 12% more than the previous finding. Figure 13B shows a perspective view of Arruntia crater displaying olivine-rich lithologic units in red. The olivine-rich material shows higher reflectance than the surrounding areas, and they are located mainly on the ejecta. Olivine-poorer regions (less than 40-60% olivine) are difficult to separate from eucrites/howardites. Lithologic units with CM2 (less than 20-30% CM2) are indistinguishable from eucrites/howardites while distinct CM2-rich areas (more than 20-30% CM2) are found in few localities.

## 1.4 Concluding remarks

The lack of significant amount of olivine in the Rheasilvia and Veneneia basin and the identification of olivine-rich exposures far away from these basins raises some critical questions on the origin of this mineral as well as on the evolution of this planetary body. Petrologic modelling presented here suggests that the orthopyroxene-rich layer extends to considerable depths beneath a eucritic/diogenitic crust. This makes it particularly difficult to believe that an olivine-dominated mantle material has been excavated by the large impact, followed by an ambiguous process of redistribution of material to Vesta's northern hemisphere. Therefore, olivine in this region is likely exogenic in origin as suggested by Nathues et al. (2015) and Le Corre et al. (2015). However, an intra-crustal plutonic origin suggested by Cheek and Sunshine (2014) and Ruesch et al. (2014b) cannot be ruled out, though it seems rather unlikely in the contexts of petrological considerations presented above.

The presence of exogenic carbonaceous chondritic material (probably CM2), has been recognized (Reddy et al., 2012; McCord et al., 2012; Nathues et al. 2014; Palomba et al., 2014; Turrini et al., 2014) to explain the dark material on Vesta's surface which certainly is exogenic in origin.

The application of three-dimensional spectral approach for analyzing surface lithologic heterogeneity demonstrates that this method bears a large potential for tackling further questions on the mineralogy and geology of planetary bodies. It remains, however, to define spectral parameters to identify impact melts and glass-bearing material. Howardites quite often contain remnants of glassy material along with recrystallized melt/impact-melt. Such material needs to be investigated in terms of its spectral reflectance in the visible and the near-infrared wavelength range. Furthermore, a reliable method of identifying and quantifying calcic plagioclase or feldspar among mafic minerals needs to be established. These aspects seem to be of particular importance too, for characterizing the surface heterogeneities on Vesta. Thus, a better understanding of the geology and petrology of Vesta (the HED parent body) is a key to reveal the early evolution of planetary bodies as well as to unlock the processes during the early solar system formation.

## 1.5 References

Ammannito, E., et al., 2013a. Vestan lithologies mapped by the visual and infrared spectrometer on Dawn. Meteorit. Planet. Sci. 48, 2185-2198.

Ammannito, E., et al., 2013b. Olivine in an unexpected location on Vesta's surface. Nature 504, 122-125.

Barrat, J.A., Yamaguchi, A., Greenwood, R.C., Bohn, M., Cotten, J., Benoit, M., Franchi, I.A., 2007. The Stannern trend eucrites: contamination of main group eucriticmagmas by crustal partial melts. Geochim. Cosmochim. Acta 71, 4108–4124.

Barrat, J.-A., Yamaguchi, A., Zanda, B., Bollinger, C., Bohn, M., 2010. Relative chronol-ogy of crust formation on asteroid Vesta: insights from the geochemistry ofdiogenites. Geochim. Cosmochim. Acta 74, 6218–6231.

Barrat, J.-A., and Yamaguchi, A., 2014. Comment on "The origin of eucrites, diogenites, and olivine diogenites: Magma ocean crystallization and shallow magma processes on Vesta" by B. E. Mandler and L. T. Elkins-Tanton. Meteoritics & Planetary Science 49, 468-472.

Beck, A.W., McSween, H. Y., 2010. Diogenites as polymict breccias composed of orthopyroxenite and harzburgite. Meteorit. Planet. Sci. 47, 850-872.

Beck, A.W., Mittlefehldt, D.W., McSween, H.Y., Rumble, D., Lee, C.-T.A., Bodnar, R.J., 2011. MIL 03443, a Dunite from asteroid 4 Vesta: Evidence for its classification and cumulate origin. Meteoritics & Planetary Science 48:1133-1151.

Binzel, R.P., Xu, S., 1993. Chips off of asteroid 4 Vesta: Evidence for the parent body of basaltic achondrite meteorites. Science 260, 186-191.

Bowman, L.E., Spilde, M.N., Papike, J.J., 1997. Automated EDS modal analysis applied to the diogenites. Meteoritics & Planetary Science 32, 869-875.

Bunch, T.E., 1975. Petrography and petrology of basaltic achondrite polymict breccias (howardites). Proceedings, 6th Lunar Science Conference 469-492.

Cheek, L.C., and Sunshine, J.M., 2014. Evidence of differentiated near-surface plutons on Vesta in integrated Dawn color images and spectral datasets (abstract #2735). Asteroids, Comets, Meteors Conference.

Clayton, R.N., et al., 1976. A classification of meteorites based on oxygen isotopes. Earth and Planetary Science Letters 30, 10-18.

Clayton, R.N., 1993. Oxygen isotopes in meteorites. Annual Review of Earth and Planetary Sciences 21, 115–149.

Clayton, R.N., and Mayeda, T.K., 1996. Oxygen isotope studies of achondrites. Geochimica et Cosmochimica Acta 60, 11, 1999-2017.

Clayton, R.N. and Mayeda, T.K., 1999. Oxygen isotope studies of carbonaceous chondrites. Geochimica et Cosmochimica Acta 63, 2089–2104.

Clenet, H., et al., 2014. A deep crust–mantle boundary in the Asteroid 4 Vesta. Nature 511, 303–306. http://dx.doi.org/10.1038/nature13499.

Consolmagno, G.J., and Drake, M.J., 1977. Composition and evolution of eucrite parent body-Evidence from rare-earth elements. Geochimica et Cosmochimica Acta 41, 1271–1282.

Consolmagno, G.J., et al., 2015. Is Vesta an intact and pristine protoplanet? Icarus 254, 190–201. doi:10.1016/j.icarus.2015.03.029.

De Sanctis, M.C., et al. 2011. The VIR spectrometer. Space Sci. Rev. 163, 329-369.

De Sanctis, M.C., et al., 2012. Spectroscopic characterization of mineralogy and its diversity across Vesta. Science 336, 697-700.

Deer, W.A., Howie, R.A., Zussman, J., 1997. Rock Forming Minerals. Vol. 2A, Single Chain Silicates. Geological Society London.

Delaney, J.S., H. Takeda, Prinz, M., Nehru, C.E., Harlow, G.E., 1983. Meteoritics 18, 103.

Dreibus, G., and Wänke, H., 1980. The bulk composition of the eucrite parent asteroid and its bearing on planetary evolution. Zeitschrift für Naturforschung 35a, 204-216.

Dymek, R.F., Albee, A.L., Chodos, A.A., Wasserburg, G.J., 1976. Petrography of isotopically-dated clasts in the Kapoeta howardite and petrologic constraints on the evolution of its parent body. Geochimica et Cosmochimica Acta 40, 1115-1116.

Fowler, G.W., Papike, J.J., Spilde, M.N., Shearer, C.K., 1994. Diogenites as asteroidal cumulates: Insights from orthopyroxene major and minor element chemistry. Geochimica et Cosmochimica Acta 58, 3921-3929.

Fowler, G.W., Shearer, C.K., Papike, J.J., Layne, G.D., 1995. Diogenites as asteroidal cumulates: Insights from orthopyroxene trace element chemistry. Geochimica et Cosmochimica Acta 59, 3071-3084.

Gaffey, M.J., 1997. Surface lithologic heterogeneity of asteroid 4 Vesta. Icarus 127, 130-157.

Ghiorso, M.S. and Sack, R.O., 1995. Chemical mass transfer in magmatic processes. IV. A revised and internally consistent thermodynamic model for the interpolation and extrapolation of liquid–solid equilibria in magmatic systems at elevated temperatures and pressures. Contributions to Mineralogy and Petrology 119, 197–212.

Greenwood, R.C., Barrat, J.-A., Yamaguchi, A., Franchi, I.A., Scott, E.R.D., Bottke, W.F.,Gibson, J.M., 2014. The oxygen isotope composition of diogenites: evidence forearly global melting on a single, compositionally diverse, HED parent body. EarthPlanet. Sci. Lett. 390, 165–174.

Hahn, T.M., McSween, H.Y., and Taylor, L.A., 2015. Vesta's Missing Mantle: Evidence from new harzburgite components in Howardites. Lunar Planet. Sci. 46. Abstract #1964.

Hertogen, J.J., Vizgirda, J., Anders, E., 1977. Composition of the parent body of the eucrite meteorites. Bull. Amer. Astron. Soc. 9, 458-459.

Herrin, J.S., Zolensky, M.E., Cartwright, J.A., Mittlefehldt, D.W., Ross, D.K., 2011. Carbonaceous chondrite-rich howardites: The potential for hydrous lithologies on the HED parent. Lunar Planet. Sci. 42. Abstract #2806.

Hewins, R., Newsom, H., 1988. Igneous activity in the early solar system. In: Kerridge, J.F., Matthews, M.S. (Eds.), Meteorites and the early solar system. University ofArizona Press, Tucson, AZ, USA, pp. 73–101.

Hiroi, T., Pieters, C.M., Hiroshi, T., 1994. Grain size of the surface regolith of asteroid 4 Vesta estimated from its reflectance spectrum in comparison with HED meteorites. Meteoritics 29:394-396.

Hutchinson, R., 2007. Meteorites a petrologic, chemical and isotopic synthesis. Cambridge University Press. pp. 524.

Ikeda, Y., 1989. Igneous activity in early solar system based on HED and mesosiderite meteorites. Abstract for 28th International Geological Congress, Washington, 2, 92-93.

Ikeda, Y., and Takeda, H., 1985. A model for the origin of basaltic achondrites based on the Yamoto 7308 howardite. Proceedings, 15th Lunar and Planetary Science Conference. pp. C649–C663.

Ivanov, B.A., Melosh, H.J., 2013. Two-dimensional numerical modeling of the Rheasilvia impact formation. J. Geophys. Res. Planets 118, 1545-1557.

Jarosewich, E., 1990. Chemical analyses of meteorites: A compilation of stony and iron meteorite analyses. Meteoritics, 25, 323–337.

Jarosewich, E., 2006. Chemical analyses of meteorites at the Smithsonian Institution: An update Meteorit. Planet. Sci., 41, 1381–1382.

Jaumann, R., et al., 2012. Vesta's Shape and Morphology. Science 336, 687-690.

Jones, J. H., 1984. The composition of the mantle of the eucrite parent body and the origin of eucrites. Geochim. Cosmochim. Acra 48, 641-648.

Jutzi, M., et al., 2013. The structure of the asteroid 4 Vesta as revealed by models of planet-scale collisions. Nature 494, 207-210.

Keil, K., 2002. Geological history of asteroid 4 Vesta: The "Smallest Terrestrial Planet". In Asteroids III, ed. by W. Bottke, A. Cellino, P. Paolicchi, R. P. Binzel (University of Arizona Press, Tucson, 2002), 573-584.

Le Corre, L., Reddy, V., Nathues, A., Cloutis, E.A., 2011. How to characterize terrains on 4 Vesta using Dawn Framing Camera color bands? Icarus 216, 376-386.

Le Corre, L., et al., 2015. Exploring exogenic sources for the olivine on Asteroid (4) Vesta. Icarus, 258, 483-499.

Lunning, N.G., McSween, H.Y., Tenner, T.J., and Kita, N.T., 2015. Olivine and pyroxene from the mantle of asteroid 4 Vesta. Earth Planet. Sci. Let. 418, 126–135.

Mandler B. E., and Elkins-Tanton L. T. 2013. The origin of eucrites, diogenites, and olivine diogenites: Magma ocean crystallization. Meteoritics & Planetary Science 48, 2333-2349.

Marchi, S., McSween, H. Y., O'Brien, D. P., Schenk, P., De Sanctis, M. C., Gaskell, R., Jaumann, R., Mottola, S., Preusker, F., Raymond, C. A., and Russell, C. T., 2012. The violent collisional history of asteroid 4 Vesta. Science 336, 690-694.

Mason, B., 1962. Meteorites. Wiley, New York.

Mason, B., 1967. Meteorites. American Scientist, 55, 4, 429-455.

Mayne, R.G., McSween, H.Y., McCoy, T.J., Gale, A., 2009. Petrology of the unbrecciated eucrites. Geochimica et Cosmochimica Acta 73, 794-819.

McCord, T.B., Adams, J.B., Johnson, T.V., 1970. Asteroid Vesta: Spectral reflectivity and compositional implications. Science 178, 745-747.

McCord, T. B. et al. 2012. Dark material on Vesta: Adding carbonaceous volatile-rich materials to planetary surfaces. Nature 491, 83-86.

McCoy, T.J., Beck, A.W., Mittlefehdt, D.W., 2015. Asteroid (4) Vesta II: Exploring a geologically and geochemically complex world with the Dawn Mission. Chemie der Erde-Geochemistry 75, 3, 273-285.

McSween, H.Y., 1999. Meteorites and their Parent Planets. Cambridge University Press. pp. 324.

McSween, H.Y., and Huss, G.R., 2010. Cosmochemistry. pp. 565, Cambridge Univ. Press.

McSween, H.Y., Mittlefehldt D.W., Beck A.W., Mayne, R.G., McCoy T.J., 2011. HED meteorites and their relationship to the geology of Vesta and the Dawn mission. Space Sci. Rev. 163, 141-174.

McSween, H. Y., et al. 2013. Composition of the Rheasilvia basin, a window into Vesta's interior. Journal of Geophysical Research 118, 335-346.

Mittlefehldt, D.W., 1994. The genesis of diogenites and HED parent body petrogenesis. Geochim. Cosmochim. Acta 58, 1537-1552.

Mittlefehldt, D.W., McCoy, T.J., Goodrich, C.A., Kracher A., 1998. Non-chondritic meteorites from asteroidal bodies. In Planetary Materials, vol. 36 (ed. J. J. Papike). Mineralogical Society of America, Chantilly, Virginia, pp. 4-1 to 4-195.

Mittlefehldt, D.W., 2015. Asteroid (4) Vesta: I. The howardite-eucrite-diogenite (HED) clan of meteorites. Chemie Erde-Geochem. 75, 2, 155–183.

Morgan, J.W., Higuchi, H., Takahashi H. and Hertogen, J., 1978. A "chondritic" eucrite parent body: Inference from trace elements. Geochim. Cosmochim Acta 42, 27-38.

Nathues, A., et al., 2014. Detection of serpentine in exogenic carbonaceous chondrite material on Vesta from Dawn FC data. Icarus 239, 222-237.

Nathues, A., et al., 2015. Exogenic olivine on Vesta from Dawn Framing Camera color data. Icarus 258, 467-482.

Nyquist, L. E., Reese, Y., Weismann, H., Shih, C.-Y., and Takeda, H., 2003. Fossil $^{26}$Al and $^{53}$Mn in the Asuka 881394 eucrite: Evidence of the earliest crust on asteroid 4 Vesta. Earth and Planetary Science Letters 214, 11–25.

Palomba, E., et al. 2014. Composition and mineralogy of dark material deposits on Vesta. Icarus, 240, 58–72.

Pieters, C.M., et al., 2012. Distinctive space weathering on Vesta from regolith mixing processes. Nature 491, 79-82.

Pilcher, F., 1979. Circumstances of minor planet discovery. In Asteroids (T. Gehrels, ed.), pp. 1130–1154. Univ. of Arizona, Tucson.

Reddy, V. et al., 2012b. Delivery of dark material to Vesta via carbonaceous chondritic impacts. Icarus 221, 544-559.

Reddy, V., et al., 2012a. Color and albedo heterogeneity of Vesta from Dawn. Science 336, 700-704.

Righter, K., Drake, M.J., 1997. A magma ocean on Vesta: Core formation and petrogenesis of eucrites and diogenites. Meteorit. Planet. Sci. 32, 929-944.

Ruesch, O., et al., 2014b. Detections and geologic context of local enrichments in olivine on Vesta with VIR/Dawn data. J. Geophys. Res.: Planets 119, 2078–2108. http://dx.doi.org/10.1002/2014JE004625.

Ruesch, O., et al. 2014a. Geologic map of the northern hemisphere of Vesta based on Dawn Framing Camera (FC) images, Icarus 244, 41-59. doi:10.1016/j.icarus.2014.01.035.

Russell, C. T. and Raymond, C.A. 2011. The Dawn mission to Vesta and Ceres. Space Science Reviews 163, 3–23.

Russell, C.T., et al., 2012. Dawn at Vesta: testing the protoplanetary paradigm. Science 336, 684-686.

Russell, C.T., et al., 2013. Dawn completes its mission at 4 Vesta. Meteorit. Planet Sci. 48, 2076-2089.

Ruzicka, A., Snyder, G. A., Taylor, L. A., 1997. Vesta as the howardite, eucrite and diogenite parent body: Implications for the size of a core and for large-scale differentiation. Meteoritics & Planetary Science 32, 825-840.

Sack, R.O., Azeredo, W.J., Lipschutz, M.E., 1991. Olivine diogenites: The mantle of the eucrite parent body. Geochimica et Cosmochimica Acta 55, 1111-1120.

Schenk, P., et al., 2012. The geologically recent giant impact basins at Vesta's south pole. Science 336, 694-697.

Schaefer, M., et al., 2014. Imprint of the Rheasilvia impact on Vesta – Geologic mapping of quadrangles Gegania and Lucaria. Icarus 244, 60–73.

Sierks, H., et al., 2011. The Dawn Framing Camera. Space Sci. Rev. 163, 263-327.

Shearer, C.K., Fowler, G.W., Papike, J.J., 1997. Petrogenetic models for magmatism on the eucrite parent body: Evidence from orthopyroxene in diogenites. Meteoritics & Planetary Science 32, 877-889.

Shearer, C.K., Burger, P., and Papike, J.J., 2010. Petrogenetic relationships between diogenites and olivine diogenites: Implications for magmatism on the HED parent body. Geochimica et Cosmochimica Acta 74, 4865-4880.

Schmadel, Lutz D., 2003. Dictionary of minor planet names: Prepared on behalf of commission 20 under the auspices of the International Astronomical Union. Springer. p. 15. ISBN 3-540-00238-3.

Stolper, E, 1975. Petrogenesis of eucrite, howardite and diogenite meteorites. Nature 258, 220-222.

Stolper, E., 1977. Experimental petrology of eucritic meteorites. Geochimica et Cosmochimica Acta 41, 587-611.

Srinivasan, G., Goswami, J. N., and Bhandari, N., 1999. 26Al in eucrite Piplia Kalan: Plausible heat source and formation chronology. Science 284, 1348-1350.

Takahasi, E., Kushiro, I., 1983. Melting of a dry peridotite at high pressures and basalt magma genesis. American Mineralogist. 68, 859-879.

Takeda, H., Graham, A.L., 1991. Degree of equilibration of eucritic pyroxenes andthermal metamorphism of the earliest planetary crust. Meteoritics 26, 129–134.

Thangjam, G., et al., 2013. Lithologic mapping of HED terrains on Vesta using Dawn Framing Camera color data. Meteorit. Planet. Sci. 48, 2199-2210.

Thangjam, G., et al., 2014. Olivine-rich exposures at Bellicia and Arruntia craters on (4) Vesta from Dawn FC. Meteorit. Planet. Sci. 49, 1831–1850.

Thangjam, G., Nathues, A., Mengel, K., Schäfer, M., Hoffmann, M., Cloutis, E.A., Mann, P., Müller, C., Platz, T., Schäfer, T., in press. Three-dimensional spectral analysis of compositional heterogeneity at Arruntia crater on (4) Vesta using Dawn FC, Icarus, doi: http://dx.doi.org/10.1016/j.icarus.2015.11.031.

Thomas, P.H., Binzel, R.P., Gaffey, M.J., Storrs, A.D., Wells, E.N., and Zellner, B.H., 1997. Impact excavation on asteroid 4 Vesta: Hubble Space Telescope results. Science 277, 1492–1495.

Tkalcec, B.J., and Brenker, F., 2014. Plastic deformation of olivine-rich diogenites and implications for mantle processes on the diogenite parent body. Meteorit. Planet. Sci. 49, 1202–1213.

Tkalcec, B.J., Golabek, G.J., Brenker, F.E., 2013. Solid-state plastic deformation in the dynamic interior of a differentiated asteroid. Nature Geosci. 6, 93-97.

Toplis, M.J., Mizzon, H., Monnereau, M., Forni O., McSween, H.Y., Mittlefehldt, D.W., McCoy, T.J., Prettyman, T.H., De Sanctis, M.C., Raymond, C.A., and Russell, C.T., 2013. Chondritic models of 4-Vesta: Implications for geochemical and geophysical properties. Meteoritics & Planetary Science 48, 2300-2315.

Trinquier, A., Birck, J.L., Allègre, C.J., Göpel, C., Ulfbeck, D., 2008. 53Mn-53Cr systematics of the early Solar System revisited. Geochim. Cosmochim. Acta 72, 5146–5163.

Turrini, D., et al., 2014. The contamination of the surface of Vesta by impacts and thedelivery of the dark material. Icarus 240, 86–102.

Voigt, H.-H., Abriss der Astronomie. 2012. Wiley-VCH, Berlin. pp519.

Wadhwa, M., Amelin, Y., Bogdanovski, O., Shukolyukov, A., Lugmair, G.W., Janney, P., 2009. Ancient relative and absolute ages for a basaltic meteorite: implications fortimescales of planetesimal accretion and differentiation. Geochim. Cosmochim.Acta 73, 5189–5201.

Wadhwa, M., Lugmair, G.W., 1995. Lunar Planet. Sci. XXVI, 1453.

Warren, P.H., 1985. Origin of howardites, diogenites and eucrites: a mass balanceconstraint. Geochim. Cosmochim. Acta 49, 577–586.

Warren, P.H., 1997. Magnesium oxide-iron oxide mass balance constraints and a more detailed model for the relationship between eucrites and diogenites. Mete-orit. Planet. Sci. 32, 945–963.

Warren, P.H., Kallemeyn, G. W., Huber, H., Ulff-Møller, F., Choe, W., 2009. Siderophile and other geochemical constraints on mixing relationships among HED meteorite breccias. Geochimica et Cosmochimica Acta 73, 5918-5943.

Wasson, J.T., and Kallemeyn, G.W., 1988. Compositions of chondrites. Philosophical Transactions of the Royal Society of London 325, 535–544.

Wasson, J.T., 2013. Vesta and extensively melted asteroids: why HED meteorites areprobably not from Vesta. Earth Planet. Sci. Lett. 381, 138–146.

Wee, B.S., Yamaguchi, A., Ebihara, M., 2010. Platinum group elements in howardites and polymict eucrites: Implications for impactors on the HED parent body. Lunar Planet. Sci. 41. Abstract #1533.

Yamaguchi, A., Taylor, G.J., Keil, K., 1996. Global crustal metamorphic of the eucrite parent body. Icarus 124, 97-112.

Yamaguchi, A., et al., 2015. Petrology and geochemistry of Northwest Africa 5480 diogenite and evidence for a basin-forming event on Vesta. Meteoritics & Planetary Science 1–11. doi: 10.1111/maps.12470.

Yamaguchi, A., Taylor, G.J., Keil, K., 1997. Metamorphic history of the eucritic crust of 4 Vesta. J. Geophys. Res. 102, 13381-13386.

Zambon, F., et al., 2014. Spectral analysis of the bright materials on the asteroid Vesta. Icarus 240, 73-85.

# 2. Lithologic Mapping of HED Terrains on Vesta using Dawn Framing Camera Color Data

Guneshwar Thangjam[1], Vishnu Reddy[1,2], Lucille Le Corre[1,3], Andreas Nathues[1], Holger Sierks[1], Harald Hiesinger[4], Jian-Yang Li[5], Robert Gaskell[3], Juan A. Sanchez[1], Christopher T. Russell[6],Carol Raymond[7]

[1]Max-Planck-Institute for Solar System Research, Justus-von-Liebig-Weg 3, 37077 Göttingen, Germany

[2]Department of Space Studies, University of North Dakota, Grand Forks, USA

[3]Planetary Science Institute, 1700 East Fort Lowell, Suite 106, Tucson, AZ 85719-2395, USA

[4]Institut für Planetologie, Westfälische Wilhelms-Universität Münster, Munster, Germany

[5]Department of Astronomy, University of Maryland, College Park, Maryland, USA

[6]Institute of Geophysics and Planetary Physics, University of California, 3845 Slitcher Hall, 603 Charles, USA

[7]Jet Propulsion Laboratory, California Institute of Technology, Pasadena, California, USA

## 2.0 Abstract

The surface composition of Vesta, the most massive intact basaltic object in the asteroid belt, is interesting because it provides us with an insight into magmatic differentiation of planetesimals that eventually coalesced to form the terrestrial planets. The distribution of lithologic and compositional units on the surface of Vesta provides important constraints on its petrologic evolution, impact history and its relationship with Vestoids and howardite-eucrite-diogenite (HED) meteorites. Using color parameters (band tilt and band curvature) originally developed for analyzing lunar data, we have identified and mapped HED terrains on Vesta in Dawn Framing Camera (FC) color data. The average color spectrum of Vesta is identical to that of howardite regions, suggesting an extensive mixing of surface regolith due to impact gardening over the course of solar system history. Our results confirm the hemispherical dichotomy (east-west and north-south) in albedo/color/composition that has been observed by earlier studies. The presence of diogenite-rich material in the southern hemisphere suggests that it was excavated during the formation of the Rheasilvia and Veneneia basins. Our lithologic mapping of HED regions provides direct evidence for magmatic evolution of Vesta with diogenite units in Rheasilvia forming the lower crust of a differentiated object.

## 2.1 Introduction

Vesta is the most massive intact basaltic object in the asteroid belt and the only surviving member of its class of protoplanets (Thomas et al. 1997; Keil 2002; McSween et al. 2011). Like the Moon, the minerals plagioclase and pyroxene, indicators of its volcanic past dominate Vesta's surface (McCord et al. 1970, 1981; Pieters 1993; Takeda 1997; Gaffey 1997). However, Vesta and the Moon have several important differences despite similar surface mineralogy. On the Moon, albedo, topography and composition are strongly related (Metzger et al. 1977; Lucey et al. 1994; Hiesinger and Head 2006). Feldspathic lunar highlands have higher albedo and are topographically higher than the lunar mare which are pyroxene-rich and have lower albedo and topography (Pieters 1993; Lucey et al. 1994, 2006). In contrast, Vesta does not show this clear correlation between albedo, composition and topography (Reddy et al. 2012b; Jaumann et al. 2012). While the topographically lower Rheasilvia basin in the southern hemisphere has higher albedo and has a predominantly diogenitic composition, several topographically higher regions also show diogenite signatures (Reddy et al. 2012b; Schenk et al. 2012). Unlike the Moon, where late stage volcanism played an important role in creating its albedo, topography and composition link (Metzger et al. 1977; Lucey et al. 1994, 2006; Hiesinger and Head 2006), Vesta's basaltic activity ceased within the first 10 million years after solar system formation (Lugmair and Shukolyukov 1998; Srinivasan et al. 1999; Coradini et al 2011; McSween et al. 2011). Eons of impacts have erased any morphological traces of its volcanic past, creating a global regolith blanket of howarditic composition (Chapman and Gaffey 1979; Bell et al. 1988; Hiroi et al. 1994; Gaffey 1997; Reddy et al. 2012b).

The NASA Dawn mission entered orbit around Vesta in July 2011, for a year-long mapping mission (Russell et al. 2011, 2012). During this period, the entire visible surface of the asteroid was imaged using a clear filter and seven color filters (Sierks et al. 2011). Recent observations (Reddy et al. 2012b; De Sanctis et al. 2012) have revealed the compositional/mineralogical diversity across Vesta, confirming several observations made from ground-based telescopes and the Hubble Space Telescope, including rotational color and albedo variations (e.g., Gaffey 1983, 1997; Dumas and Hainaut 1996; Binzel et al. 1997; Zellner et al. 2005; Carry et al. 2010; Li et al. 2010; Reddy et al. 2010). Using the Framing Camera (FC) multispectral data (0.44-0.98 μm), Reddy et al. (2012b) analyzed the color properties and albedo of Vesta. They also used 1-μm band depth and a spectral eucrite-diogenite ratio to identify regions rich in eucrites and diogenites. Using Visible and Infrared

Spectrometer (VIR) hyperspectral data (0.25-5.01 µm), De Sanctis et al. (2012) analyzed regional as well as local compositional variations across the surface of Vesta. They used both 1-µm and 2-µm pyroxene band parameters, i.e., band center and band depth to interpret units in terms of diogenitic and eucritic mineralogy. These recent works from the Dawn mission show the presence of diogenites in the southern hemisphere and the eucrites in the equatorial and northen hemisphere (Reddy et al. 2012b; De Sanctis et al. 2012).

Here we present our analysis and results for identification and mapping of HED (howardite, eucrite, diogenite) terrains on Vesta using a different approach from the initial studies. We expanded our analysis by using 1-µm pyroxene band parameters that have been successfully applied to classify terrains on the Moon (Tompkins and Pieters 1999; Pieters et al. 2001; Dhingra 2007; Isaacson and Pieters 2009). Our goals are to identify and map the lithological units of the Vestan surface as constrained by HEDs, and to confirm the earlier observations, i.e., hemispherical dichotomy, albedo and band strength variations, and lithological heterogeneity, as well as an overall howarditic lithological characteristic of Vesta.

## 2.2 HED analyses

### 2.2.1 1-µm Pyroxene Band Parameters and HEDs:

Pyroxene is spectrally the most ubiquitous mineral on the surface of Vesta (e.g., McCord et al. 1970; Gaffey 1997). Eucrites are basaltic in composition, crystallized closer to the surface and contain equal amounts of low-Ca pyroxene (pigeonite) often with exsolution lamellae, and plagioclase (Mittlefehldt et al. 1998; Mayne et al. 2009; McSween et al. 2011). Diogenites are coarse-grained cumulates formed at depth and are composed mostly of orthopyroxene (Sack et al. 1991; Bowman et al. 1997; McSween et al. 2011). Howardites are regolith breccias formed by impact mixing of eucrites and diogenites (Bunch 1975; Dymek et al. 1976; Warren 1985; Warren et al. 2009). Near-IR spectra of HEDs have prominent absorption features around 1-µm and 2-µm (Gaffey 1976; Feierberg and Drake 1980; McFadden and McCord 1978; Gaffey 1997) due to the mineral pyroxene (Adams 1974, 1975; Burns 1993). The 1-µm pyroxene band center shifts to longer wavelength as the calcium content increases (Adams 1974; Cloutis and Gaffey 1991). However, it is the presence of iron ($Fe^{2+}$) in the pyroxene structure that is the cause of the two absorptions (Burns 1970, 1993; Mayne et al. 2009). Eucrites have higher calcium and iron content in their pyroxenes (54-60 mol%) than diogenites (20-33 mol%) (Mittlefehldt et al. 1998), and hence have band centers at longer wavelength than diogenites (Adams 1974; Gaffey 1976; Cloutis and Gaffey 1991).

Several authors (Tompkins and Pieters 1999; Pieters et al. 2001; Dhingra 2007; Isaacson and Pieters 2009) have successfully used pyroxene 1-µm band parameters (band curvature, band tilt and band strength) for mineralogical/compositional analyses of the lunar surface with Clementine multispectral data. These parameters can define the 1-µm absorption feature's overall shape, strength and position (Tompkins and Pieters 1999; Pieters et al. 2001; Dhingra 2007; Isaacson and Pieters 2009). Band curvature is the spectral curvature due to the 1-µm absorption feature at around 0.9 µm. Band tilt is the reflectance ratio of 0.9- and 1-µm filters while band strength is a ratio of 0.75- and 1-µm filters. Band strength that is a proxy for 1-µm band depth corresponds to the abundance of ferrous iron in a mineral (Lucey et al. 1995; Blewett et al. 1995; Tompkins and Pieters 1999; Pieters et al. 2001). Band curvature and band tilt are sensitive to mineral chemistry (Fe-rich/Ca-rich). Higher band curvature and lower band tilt are indicative of orthopyroxene, and vice versa for clinopyroxene (Pieters et al. 2001; Dhingra 2007; Isaacson and Pieters 2009). In this work, we apply these three parameters to the color spectra from the FC to constrain the surface mineralogy of Vesta and to identify terrains rich in howardites, eucrites and diogenites.

For our study we used visible near-IR spectra of 239 HED meteorites, i.e., 17 howardite, 44 eucrite, and 13 diogenite samples from the RELAB (Reflectance Experiment Laboratory) spectral database. The majority of the spectra are acquired in the wavelength range between 0.3 and 2.5 µm at 30° source angle and 0° detect angle with a spectral resolution of 5 or 10 nm. The laboratory spectra were resampled to FC filter bandpasses using the responsivity functions of the instrument (Sierks et al. 2011; Le Corre et al. 2011). Figure 1 shows the average spectra of HEDs from RELAB (1A) along with their resampled FC color spectra (1B), and 1-µm band parameters (1C) that will be used for this study. Band parameters were extracted from RELAB spectra using the following equations from Isaacson and Pieters (2009) that have been adapted for Dawn FC data:

Band Curvature (BC) = $(R_{0.749}+R_{0.964})/R_{0.918}$ Eq. 1

Band Tilt (BT) = $R_{0.918}/R_{0.964}$ Eq. 2

Band Strength (BS) = $R_{0.964}/R_{0.749}$ Eq. 3

R is the reflectance value at the corresponding wavelengths.

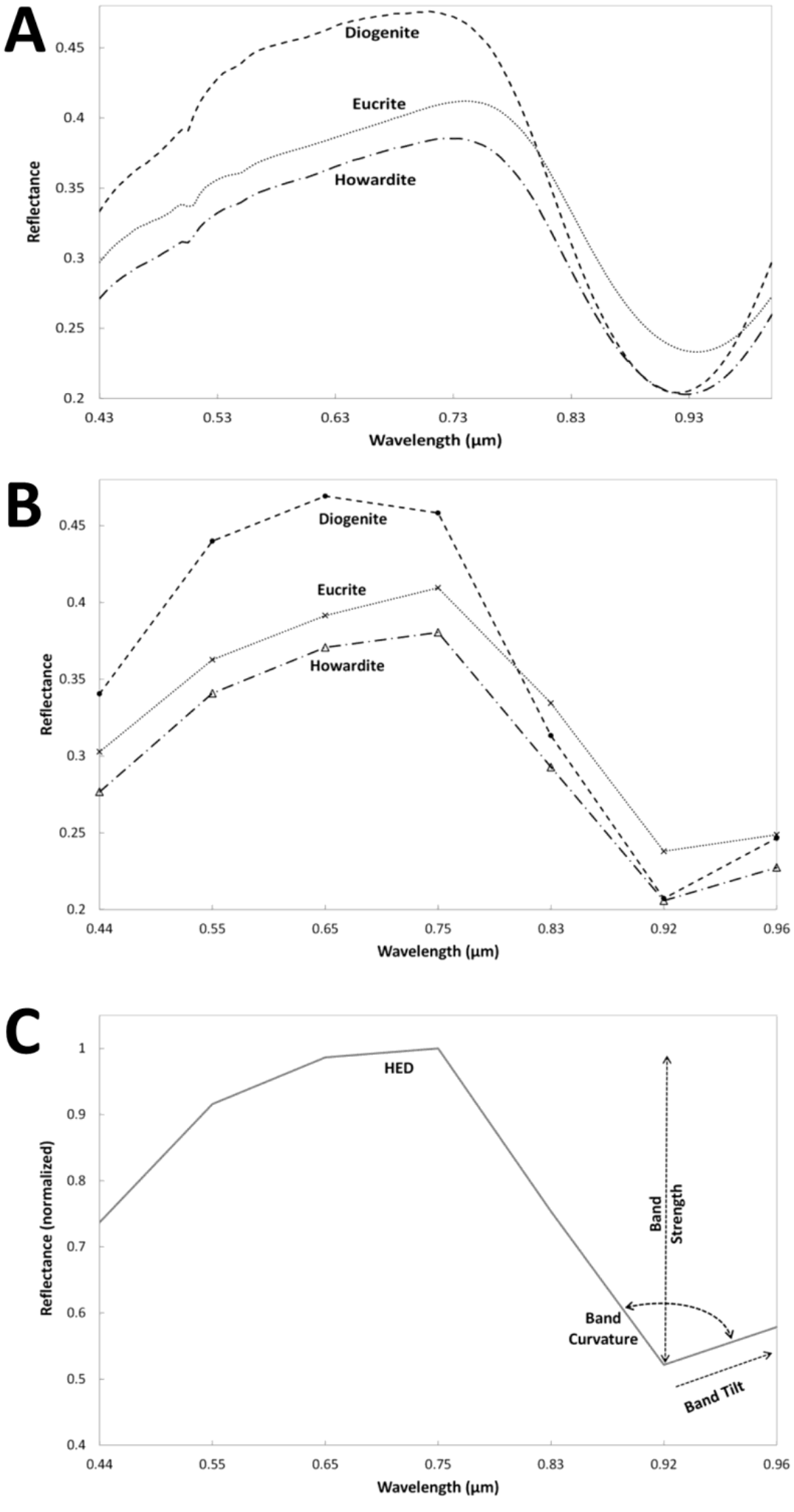

Figure 1. (A) Average spectra of HED meteorites obtained from RELAB spectral database in the visible/near-infrared wavelength range showing 1-µm pyroxene absorption feature. (B) Average HED spectra resampled to Dawn Framing Camera (FC) wavelength range using FC bandpasses and filter responsivity. (C) Average spectra of the HEDs normalized to unity at 0.75-µm showing the three 1-µm band parameters, i.e., band curvature, band tilt and band strength.

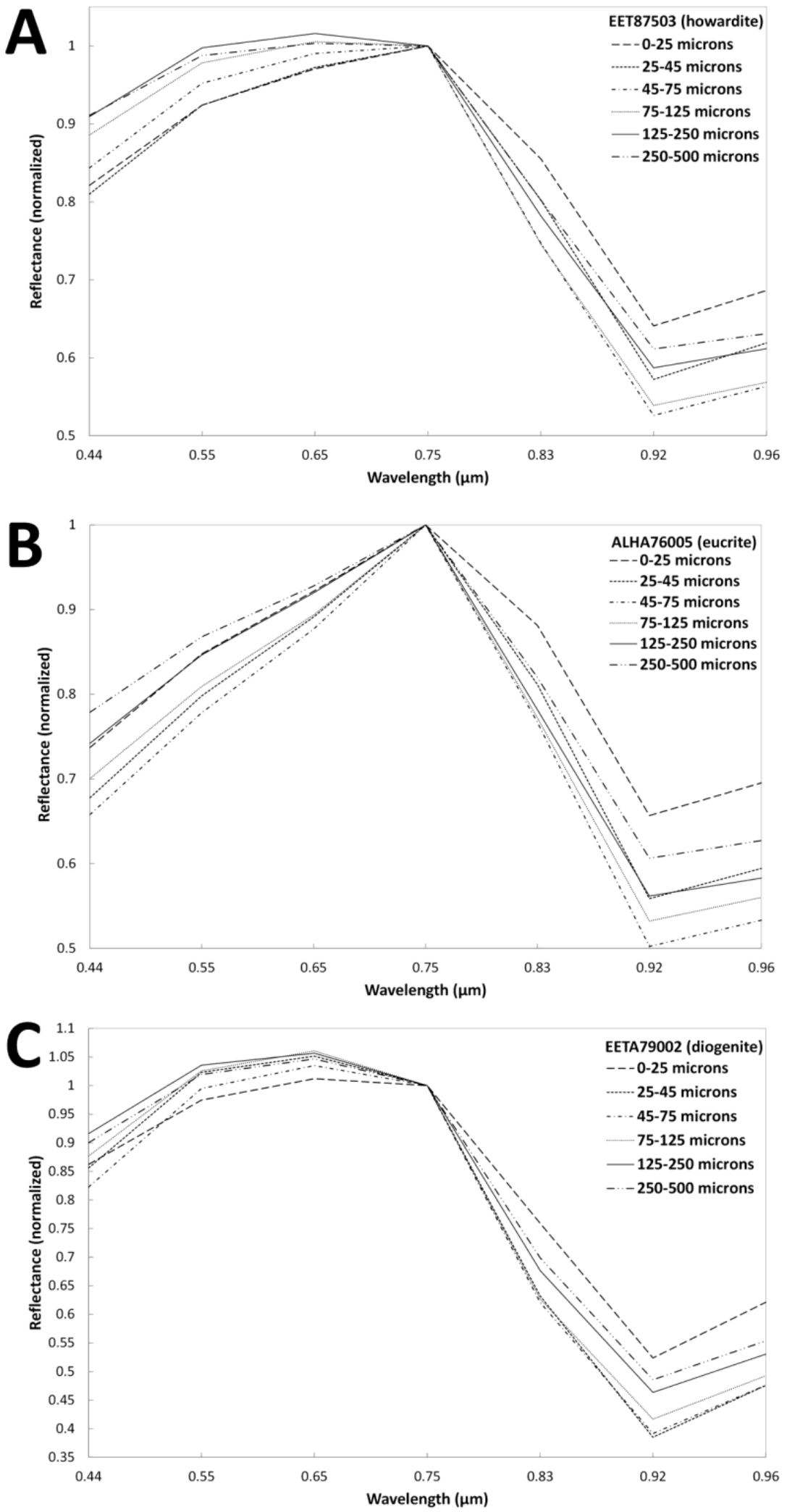

Figure 2. Normalized spectra of a (A) howardite (EET87503), (B) eucrite (ALHA76005, and (C) diogenite (EETA79002) showing the effect of particle size. Spectra normalized to unity at 0.75 µm and samples with particle size ranges of ≤25 µm, 25 to 45 µm, 45 to 75 µm, 75 to 125 µm, 125 to 250 µm, and 250 to 500 µm are shown.

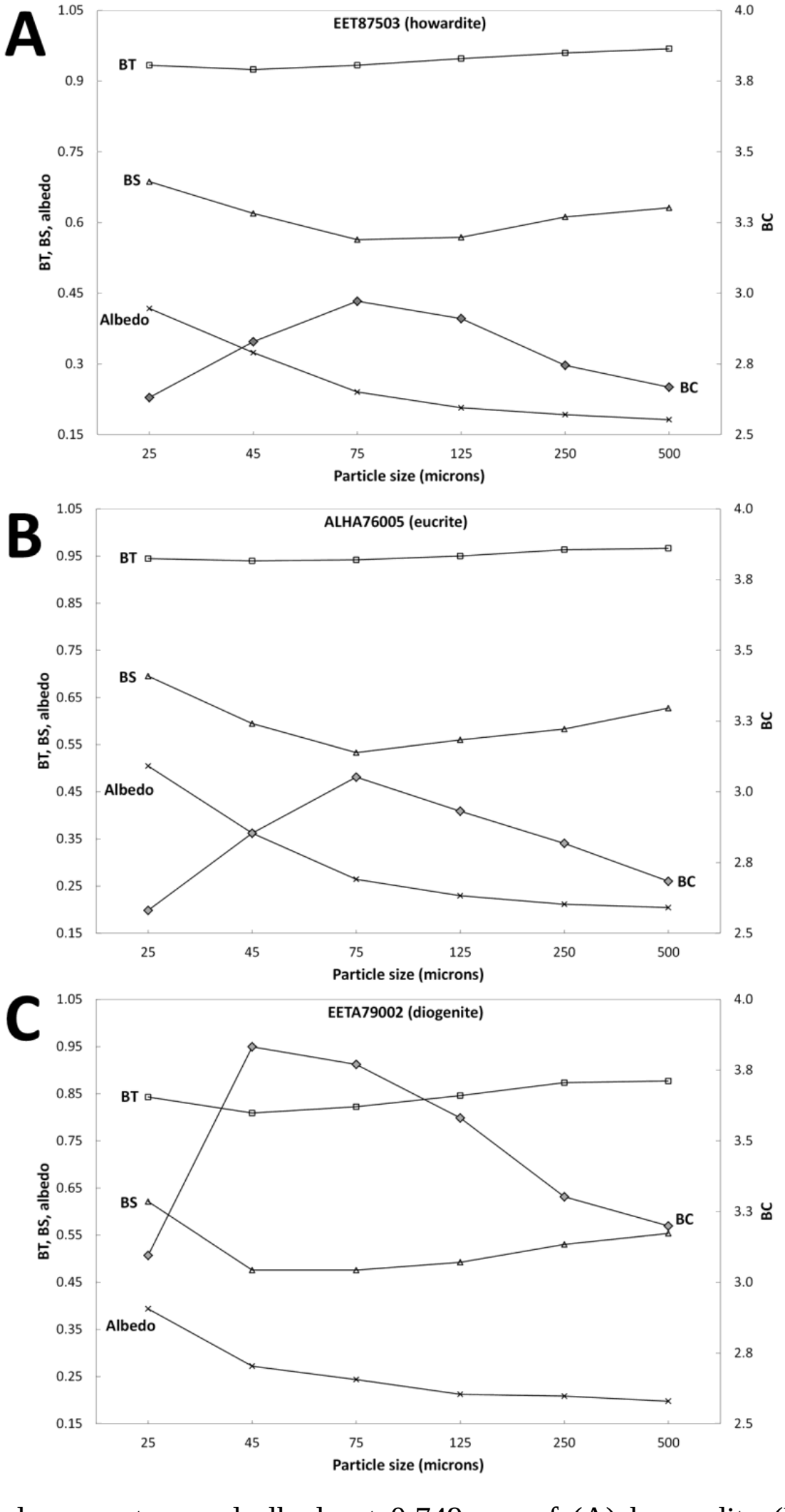

Figure 3. Spectral parameters and albedo at 0.749-µm of (A) howardite (EET87503), (B) ALHA76005 (eucrite), (C) EETA79002 (diogenite) showing the effect of particle size on the band parameters and albedo.

### 2.2.2 Band Curvature (BC):

We applied the band curvature equation (Eq. 1) to RELAB spectra of HEDs in an effort to differentiate the three meteorite types. According to our analyses, eucrites have a band curvature range of 2.18-3.4 with an average of 2.75±0.22, while diogenites are in the range of 2.78-4.44 (80% in the range of 3.4-4.5) with an average of 3.39±0.32. Howardites have intermediate values (2.48-3.33) with an average value of 2.90±0.22. BCs for the samples with ≤25 µm particle size (howardites-13, eucrites-27, diogenites-11 samples) show a similar trend (92% of eucrites in the range of 2.6-3.1 and 63% of diogenites in the range of 3.4-3.8). Based on these analyses, eucrites have lower BCs, while diogenites have higher values, with intermediate values for howardites. The higher band curvature in diogenites is due to 1-µm absorption features located at shorter wavelength due to lower total iron/calcium abundance while lower band curvature in eucrites is due to higher iron/calcium abundance that shifts the 1-µm absorption feature to longer wavelength.

To study the effect of particle size on BC, we analyzed color spectra of three samples ALHA76005 (eucrite), EETA79002 (diogenite) and EET87503 (howardite) in different particle size ranges (≤25, 25-45, 45-75, 75-125, 125-250, 250-500 µm). Figure 2 (A, B, C) shows the spectra normalized to unity at 0.75 µm for the three samples and figure 3 (A, B, C) shows the observations for the effects of grain-size for the three samples. We found that the band curvature of HEDs is affected by particle size. Depending on the particle size range, BCs varies between 11-19% with reference to the minimum value in their size ranges. The samples in the 25-45 and 45-75 µm particle size range are most affected while ≤25 µm size range is least affected (Fig. 3). Despite particle size effects, BC is still a reliable indicator for iron abundance and can be used for differentiating eucrites and diogenites as shown in Figure 4A.

### 2.2.3 Band Tilt (BT):

Applying the band tilt equation (Eq. 2) to our samples, we found that the band tilt of eucrites ranges between 0.90-1.1 with an average of 0.96±0.02. The band tilts of diogenites range between 0.73-0.89 (95% in the range of 0.81-0.89) with an average of 0.84±0.03. Howardites have intermediate values (0.83-0.97) with an average value of 0.92±0.03. For the samples with particle sizes ≤25 µm, 81% of eucrites are in the range of 0.93-0.99 while 81% of diogenites are in the range of 0.82-0.87. Based on our analysis, eucrites have stronger BTs while diogenites have lower values and howardites have intermediate values. Higher band tilts in eucrites are due to 1-µm absorption features located at longer wavelengths due to higher iron/calcium abundances while lower band tilts in diogenites are due to lower iron/calcium abundances shifting the 1-µm absorption feature to shorter wavelength. Higher band tilt is observed in olivines too (Pieters et al. 2001; Isaacson and Pieters 2009), but it is rare in HED meteorites (Sack et al. 1991; Fowler et al. 1994; Mittlefehldt 1994; McSween et al. 2011) and not yet reported on Vesta. Therefore higher band tilt is referred to eucrites for the present analyses.

Unlike band curvature, band tilt is less affected by particle size and varies only 2-7%. The samples with the largest particle size-range (250-500 µm) show the widest variations and the effect decreases as the particle size decreases (Fig. 3). Because of the negligible particle size effect, band tilt can be used as a very effective parameter for differentiating eucrites from diogenites.

## 2.2.4 Band Strength (BS):

Applying band strength equation (Eq. 3), howardites, eucrites and diogenites are observed to have wide ranges of BS values. Eucrites are in the range of 0.44-0.84 with an average of 0.62±0.08, while diogenites are between 0.35-0.68 with an average of 0.49±0.07. Howardites have intermediate values (0.50-0.61) with an average value of 0.61±0.06. For the samples with ≤25 µm particle size-range, similar distributions are observed. Despite overlapping BS ranges, diogenites have overall lower BS values, implying deeper band depths compared to eucrites and howardites.

Depending on the particle size, BS varies between 17% and 23%. In the observed range, samples with 25-45 and 45-75 µm particle size-ranges have the deepest band depth and the depths decrease as the particle sizes increase or decrease (Fig. 2, 3). Therefore, band depth is not uniquely affected by composition, but also by particle size. The band depth on Vesta is also affected by phase angle, temperature, iron abundance and the presence of opaques (Hiroi et al. 1994; Duffard et al. 2005; Reddy et al. 2012a, 2012b, 2012c).

*2.2.5 Combined approach of band curvature (BC) and band tilt (BT):*

Based on our analysis of spectra of HED meteorites, we determined that band curvature and band tilt are the most robust spectral parameters to distinguish eucrites from diogenites. A scatter plot of BC vs. BT with an approximate outline of HED-regions shows distinct spatial distributions for eucrites and diogenites with howardites falling in between (Fig. 4A, B). Irrespective of particle size, 19% and 52% of howardites fall in the diogenite-region and the eucrite-region, respectively (Fig. 4A). A similar trend is observed when we use spectra of samples with a particle size-range of ≤25 µm (Fig. 4B). Here 25% and 68% of howardites are in the diogenite region and eucrite-region, respectively. This suggests that howardites, in general, are more eucrite-rich (50-58%) than diogenite-rich (19-25%). This result is consistent with observed compositional variations in Vestoids (Binzel and Xu 1993; Burbine et al. 2009; Reddy et al. 2011).

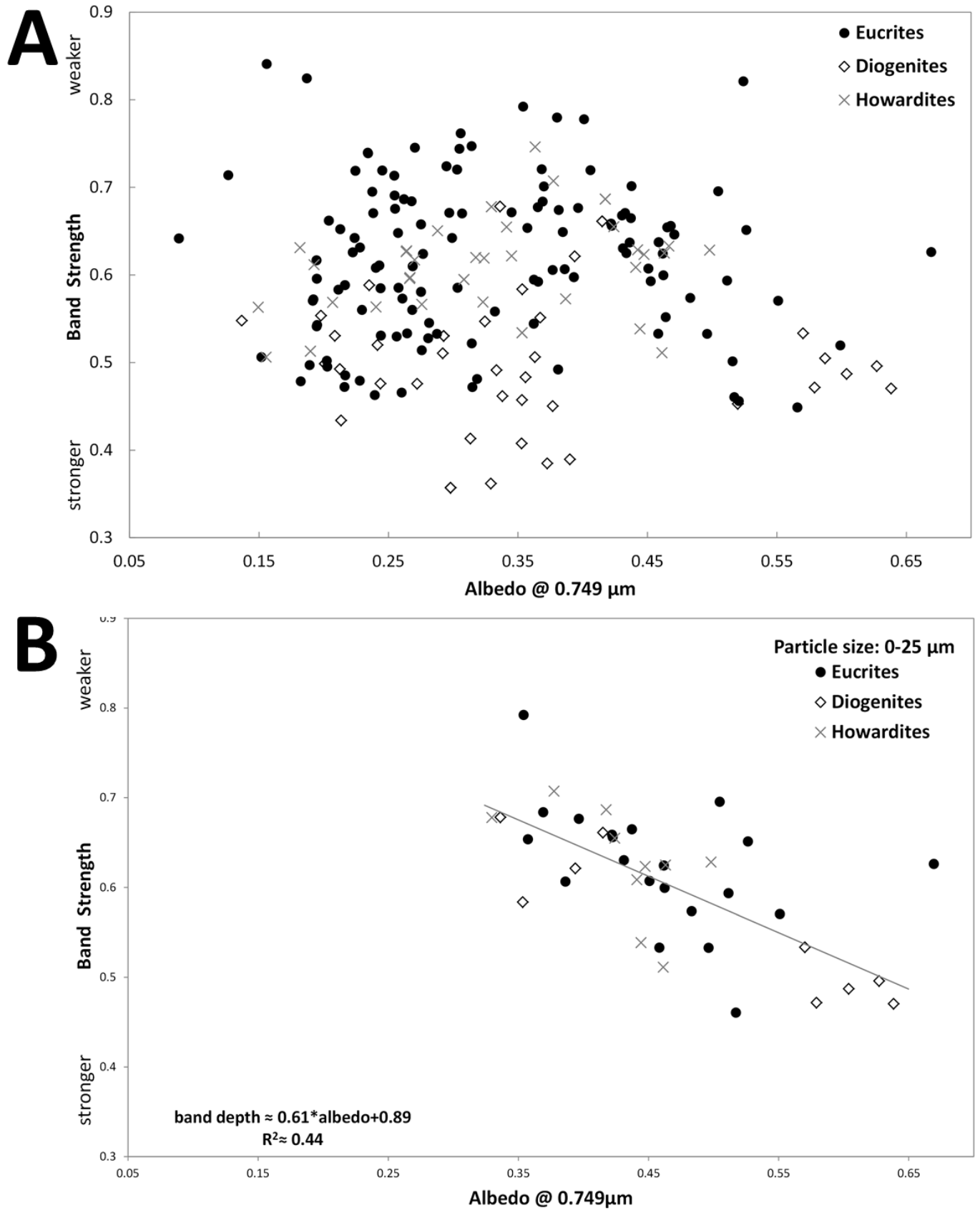

Figure 5. (A) Scatter plot of band strength (BS) vs. albedo at 0.749-µm derived from the resampled spectra of the HEDs (41 howardites, 157 eucrites, and 41 diogenites) irrespective of particle size. (B) Samples with particle size ≤25 µm (13 howardites, 27 eucrites, 11 diogenites).

### 2.2.6 Band Strength (BS) and Albedo at 0.749-µm:

Despite variations in albedo at 0.75-µm and band depth of HEDs, a correlation between band depth and albedo exists (Reddy et al. 2012a, 2012b, 2012c). The scatter plot of BS and albedo for the HEDs irrespective of the particle size (Fig. 5A) shows the variability of band depth as well as albedo with an overall deeper band depth for the diogenites. However, the scatter plot for the samples considered in the particle size-range ≤25 µm shows a linear correlation with an r-factor (correlation coefficient) of 0.66 (Fig. 5B). We also analyzed the effect of particle size on albedo for the three samples with different grain size-ranges and found that albedo decreases with increasing particle size, i.e., from ≤25 µm (54-70%) to a minimum for the 250-500 µm grain size-range. In contrast, band depth does not directly correlate with particle size, but the maximum band depth is observed between 25-45 and 45-75 µm, and decreases as the particle size increases or decreases. The smallest particle size range (≤25 µm) has the weakest band depth but highest albedo variation (54-70%) of the three samples. Given the lack of clear correlation between albedo and band strength with composition and the observed effect of opaques in the form of dark material on band depth on Vesta (Reddy et al. 2012c; McCord et al. 2012), we are not using these observations for the present study.

## 2.3 Data Reduction and Analysis

The Dawn FC acquired multispectral images using a clear filter and seven band-pass filters in the wavelength range of 0.43-µm to 0.96-µm with high spatial resolution up to 16 m/pixel (Sierks et al. 2011; Reddy et al. 2012b). Basic calibration (bias, dark, and flats) is accomplished prior to the generation of higher-level data products in the Integrated Software for Imagers and Spectrometers (ISIS) developed and maintained by the USGS (United States Geological Survey) (Anderson et al. 2004). The data are photometrically corrected using Hapke functions derived from disk integrated, ground-based telescopic and Dawn spacecraft observations of Vesta and Vestoids (Li et al. 2012). The data are map-projected and co-registered to align the seven color frames to create the color cubes. A detailed discussion of the FC data processing pipeline is presented in the supplementary section of Reddy et al. (2012b). For our study, data acquired during the approach phase (more specifically during the Rotational Characterization 3b, i.e., RC3b) of the mission were used with a pixel scale of ~480 m/pixel. All maps shown here are in the Claudia coordinate system (Russell et al. 2011, 2012) in simple cylindrical projection and span the latitudinal range from 75°S to 26°N.

We used the image analysis software ENVI 4.8 to analyze the global maps. The three band parameters equations (band curvature, band tilt and band strength) are calculated with the band math application and are color-coded. Apart from the three band parameter maps (Fig. 6B-D), an albedo map (the reflectance image at 0.75-µm) was created to provide the context for the other maps (Fig. 6A). To analyze and map the HED regions on Vesta, 2D scatter plot of band curvature vs. band tilt and the band parameters' global images as well as the trend of HEDs spectral observations were used. Using the Regions of Interests (ROIs) tool in ENVI 4.8 software, HED regions identified in the previous step were exported separately for further analysis.

## 2.4 FC data analyses

The global albedo mosaic at 0.75-µm (Fig. 6A) shows wide variations across the surface of Vesta. The geometric albedo varies between 0.10 and 0.67 (Reddy et al. 2012b). Most of the regions in the eastern hemisphere and few regions in the western hemisphere have lower albedo while regions in the southern hemisphere and few in the northern hemisphere have higher albedos. Such albedo variations on hemispheric scale are consistent with the north-south and east-west dichotomies (Bobrovnikoff 1929; Haupt 1958; Gehrels 1967; Dumas and Hainaut 1996; Gaffey 1997; Binzel et al. 1997; Rivkin et al., 2006; Li et al. 2010; Reddy et al. 2010, 2012b).

### 2.4.1 Band Curvature (BC):

The band curvature global mosaic from the RC3b phase is shown in Figure 6B. Band curvature across the surface of Vesta has a range of 1.61-3.96, with an average value 2.59±0.09. Regions with the weak BCs (blue) have lower albedos and are largely confined to the eastern hemisphere and a few limited areas in the western hemisphere. In contrast, the southern hemisphere and a few regions in the northern hemisphere (probably recent impact craters and their ejecta) have higher values of BC (red/yellow/green). These regions have higher albedos, suggesting a link between BC and albedo. As observed from the analysis of HED meteorites, higher values of BC are indicative of diogenites, while lower values are typical of eucrites. Our BC map of Vesta suggests that the southern hemisphere and few limited regions in the northern hemisphere are dominated by diogenite-rich material, while equatorial regions and the northern hemisphere are dominated by eucrite-rich materials. The intermediate values of BC are indicative of howardites, which are the result of collisional mixing of eucrite-rich and diogenite-rich regions. However, since BC is also affected by particle size besides the composition, it is worthy of cautious in interpreting compositions.

### 2.4.2 Band Tilt (BT):

Figure 6C shows the BT map of Vesta from the RC3b phase of the mission. The color-coding and values of BT are inverted (Fig. 6C) to be consistent with the BC map (Fig. 6B). Band tilt measured on the surface of Vesta has a range between 0.71-1.69, with an average value of 1.04±0.01. The band tilt map shows the hemispherical dichotomy with the lower albedo and lower BT regions in the eastern hemisphere, while the higher albedo regions in the southern hemisphere have higher BT values. As observed from the analysis of HED meteorites, BT is a very robust parameter to distinguish eucrites from diogenite with negligible effects of particle size differences. Therefore we conclude that the equatorial and the northern hemisphere regions are dominated by eucrite-rich material, while the regions in the southern hemisphere are more diogenitic.

### 2.4.3 Band Strength (BS):

The band strength map of Vesta is shown in Figure 6D. Similar to the band tilt map (Fig. 6C), the BS color-code and values are inverted to show areas of deeper 1-µm band depth, in red. Band strength ranges between 0.56-2.76 with an average value of 1.49±0.07. The BS map shows variations in the strength of the 1-µm pyroxene absorption band across the surface of Vesta. The southern hemisphere has higher albedos and deeper band strengths than the northern hemisphere, which has lower albedos and weaker band strengths. Similarly, the low albedo regions in the eastern hemisphere and a few regions in the western hemisphere also show weaker band strengths. This suggests a correlation between band strength and albedo. The presence of carbonaceous chondrite materials in the northern hemisphere, largely in the western hemisphere causes weaker band depths and lower albedos (Reddy et al. 2012b, 2012c; McCord et al. 2012). As observed from the analysis of HED meteorites, diogenites typically show stronger BSs than eucrites. However, particle size and the presence of opaques and metal have significant impact on this parameter (Hiroi et al. 1994; Duffard et al. 2005; Reddy et al. 2012b, 2012c; McCord et al. 2012). Our BS analysis coupled with BT and BC data suggests that diogenite is the dominant material in the southern hemisphere of Vesta (Li et al. 2010; De Sanctis et al. 2012; Reddy et al. 2010, 2012b).

### 2.4.4 Lithological mapping using pyroxene band parameters:

We qualitatively analyzed the Vestan surface using the pyroxene band parameters (Fig. 6B-D) in an effort to identify howardite-, eucrite- and diogenite-rich regions. The lithological units are mapped over the global albedo mosaic (Fig. 6E) considering the observations of the ratio images (preferentially band tilt, and then band curvature) as well as the observations of the HED meteorites. The mapped HED units are also marked in the scatter plot of band curvature vs. band tilt (Fig. 4C). Although the band parameter values observed from HED meteorites and FC data are slightly mismatched, the trends observed in HED meteorites (Fig. 4 A, B) and FC global mosaic (Fig. 4C) are generally similar. An exact comparison of values from the laboratory and actual observations may not be possible because of the uncertainties and the difference primarily in the spectra of laboratory samples and actual surface materials (e.g., Mustard et al. 1993; Tompkins and Pieters 1999). Considering the uncertainty limits of FC filters spectral responses and their calibration in the RC3b data (filters with the center wavelengths 0.43-, 0.65-, 0.75-, 0.96-μm could be affected upto 2% and filters with the center wavelengths 0.83-, 0.92-μm upto 4%), deviations from the RC3b values are calculated to find the maximum probable error for the band parameters (Fig. 4C). The mismatch in the values remains to be answered in the near future with the availability of better higher resolution FC data with entire coverage of Vesta. As seen from the scatter plot and the lithological map, the howardite-rich regions fall between eucrite-rich and diogenite-rich regions. Like in the HED data, we also observed a larger overlap between the eucrite- and howardite-rich units compared to the diogenite- and howardite-rich regions.

We compared color spectra of the HED regions and the average color spectrum of Vesta from the RC3b data. Figure 7 shows the average spectra for each lithology normalized to unity at 0.75 μm. These spectra show variations in 1-μm band parameters due to composition. The average color spectrum of Vesta is consistent with howardite-rich regions, confirming the howarditic nature of the Vestan surface (Chapman and Gaffey 1979; Bell et al. 1988; Hiroi et al. 1994; Gaffey 1997; Zellner et al. 2005; Delaney 2009; Carry et al. 2010; De Sanctis et al. 2012). The average band tilt for eucrite-, diogenite- and howardite-rich regions is 0.97±0.02, 0.93±0.02, and 0.96±0.01, respectively while the band tilt value for the entire surface is 0.96±0.01. Similarly, the average value of band curvature is 2.5±0.15, 2.79±0.07, 2.6±0.18 for eucrite-, diogenite-, and howardite-rich regions and 2.6±0.2 for the entire surface. Thus BT and BC values for surface average are consistent with howardite-rich regions, confirming the earlier spectral match (Fig. 7). As derived from the band parameters, diogenite-rich regions are generally in the southern hemisphere while eucrite- and howardite-rich regions are in the equatorial and northern hemisphere (Li et al. 2010; De Sanctis et al. 2012; Reddy et al. 2010, 2012b). Again, the diogenite-rich regions are generally characterized by relatively deeper band depth and higher albedo as compared to eucrite-rich and howardite-rich regions (Li et al. 2010; De Sanctis et al. 2012; Reddy et al. 2010, 2012b). Thus, the hemispherical dichotomy in albedo and composition is also observed in the band parameters (Bobrovnikoff 1929; Haupt 1958; Gehrels 1967; Dumas and Hainaut 1996; Gaffey 1997; Binzel et al. 1997; Rivkin et al. 2006; Li et al. 2010; Reddy et al. 2010, 2012b). The presence of diogenite-rich materials dominantly in the southern hemisphere suggests that it was excavated during the formation of the giant south pole basins, particularly the Rheasilvia basin (Reddy et al. 2012b; McSween et al. 2013).

## 2.5. Conclusions

Our analysis of Dawn FC color data using 1-µm pyroxene band parameters has confirmed several findings from previous works as discussed above and has provided new insight into the distribution of eucrite- and diogenite-rich material on the surface of Vesta. Our study reveals the following:

- We have successfully applied lunar band parameter analysis technique to Dawn FC data to identify terrains rich in eucrites and diogenites.
- HED laboratory spectra show distinct variations in pyroxene absorption spectral parameters (band strength, band tilt and band curvature) that can be used to interpret surface mineralogy.
- Band tilt and band curvature are strongly influenced by pyroxene chemistry (Fe-rich/Ca-rich) and are robust indicators to identify eucrite- or diogenite-rich regions.
- Band strength is strongly influenced by particle size and abundance of opaques (carbonaceous chondrite materials) on the surface and is a less reliable indicator for the abundance of ferrous iron.
- Vesta shows hemispherical dichotomies (north-south and east-west) in albedo and color that is strongly related to the excavation of diogenite-rich lower mantle material during the formation of the Rheasilvia and Veneneia basins.

The average color spectrum of Vesta is similar to the average spectrum of howardite regions and the band tilt and band curvature parameters are also identical to those of howardites.

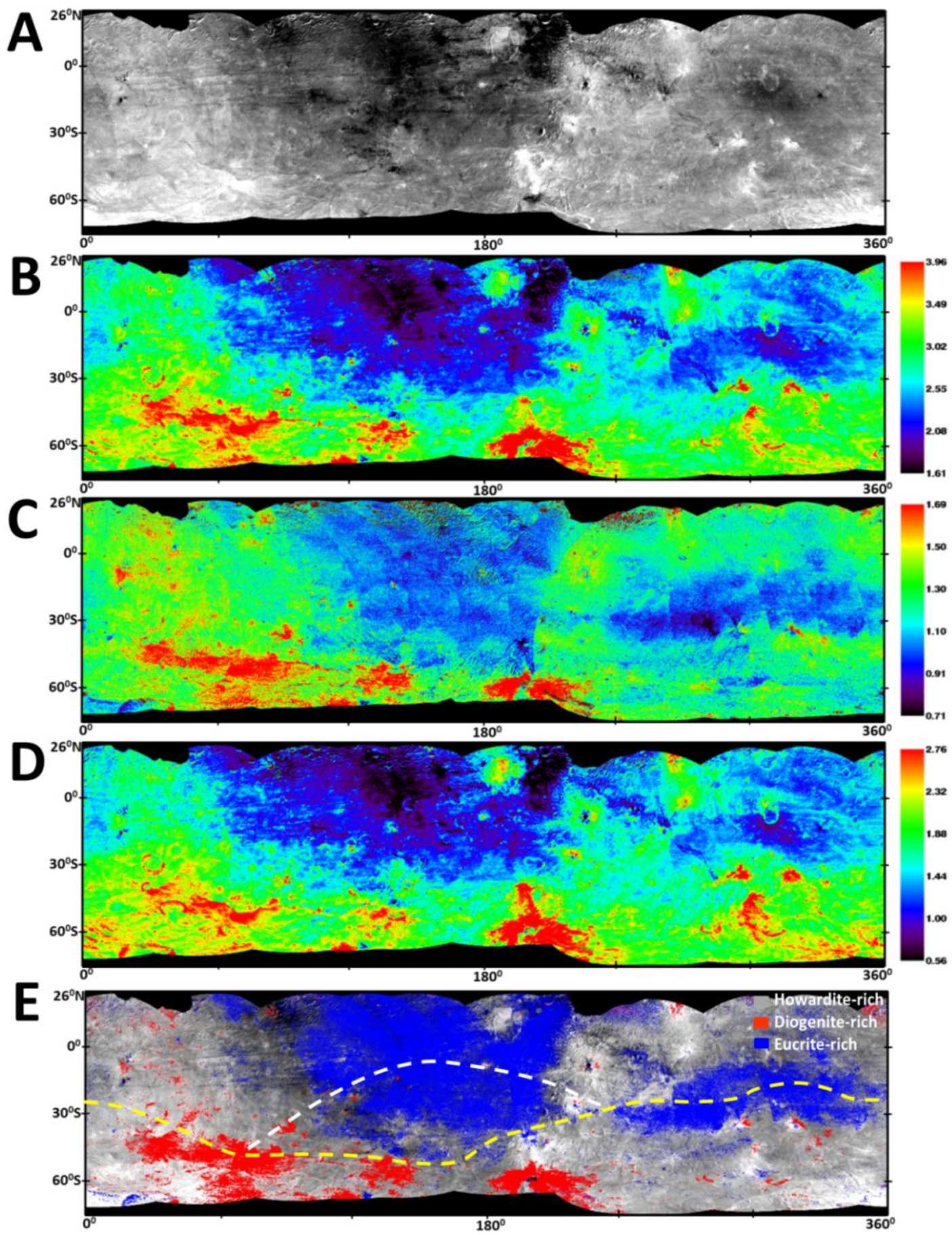

Figure 6. Global mosaic of Vesta acquired during the approach phase (RC3b) at a resolution of 480 m/pixel in the Claudia coordinate system with a simple cylindrical projection. (A) Albedo image at 0.75-µm shows variations in albedo across the surface and 1-µm band parameter maps i.e., (B) band curvature (BC), (C) band tilt (BT) and (D) band strength (BS) with rainbow-color code and their values. The original values of BT and BS and the color ramp are inverted to show diogenite-rich regions as reddish/yellowish and eucrite-rich/howardite-rich regions as bluish/greenish color in the band parameter's image. (E) Lithological units mapped as eucrite-rich, diogenite-rich and howardite-rich. Yellow lines (dashed and dotted) are approximate outlines for the south pole basins (the Rheasilvia and the Veneneia).

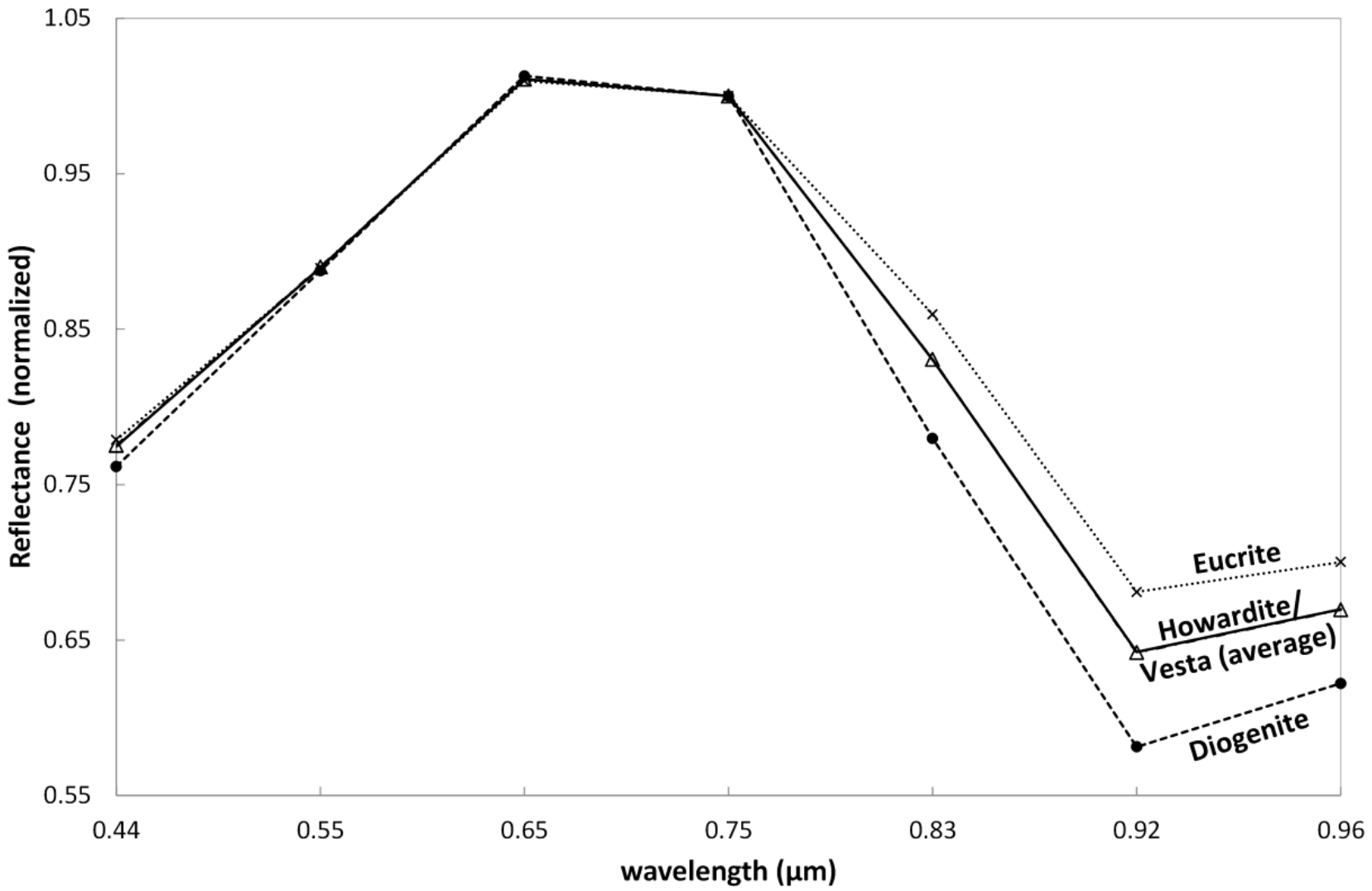

Figure 7. Average color spectra of howardite-rich, eucrite-rich and diogenite-rich regions from Fig. 6E normalized to unity at 0.749-µm. The average color spectrum of the howardite-rich region and the one for the whole Vestan surface are overlapped and similar in nature.

## 2.6 Acknowledgment

We thank the Dawn team for the development, cruise, orbital insertion, and operations of the Dawn spacecraft at Vesta. The Framing Camera project is financially supported by the Max Planck Society and the German Space Agency, DLR. We also thank the Dawn at Vesta Participating Scientist Program for funding the research. A portion of this work was performed at the Jet Propulsion Laboratory, California Institute of Technology, under contract with NASA. Dawn data is archived with the NASA Planetary Data System. This study uses 239 HED meteorites spectra from the RELAB spectral database at Brown University and we acknowledged the concerned PIs and the RELAB team for their effort.

## 2.7 References

Adams J. B. 1974. Visible and near-infrared diffuse reflectance spectra of pyroxenes as applied to remote sensing of solid objects in solar system. *Journal of Geophysical Research* 79:4829-4836.

Adams J. B. 1975. Interpretation of visible and near-infrared diffuse reflectance spectra of pyroxenes and other rock forming minerals. In *Infrared and Raman Spectroscopy of Lunar and Terrestrial Materials* (C. Karr, ed.), Academic, New York, pp. 91-116.

Anderson J. A., Sides S. C., Soltesz D. L., Sucharski T. L, Becker K. J. 2004. Modernization of the Integrated Software for Imagers and Spectrometers (abstract#2039). 35th *Lunar and Planetary Science Conference*.

Bell J. F. 1988. An earth-crossing source body for the basaltic achondrites: Vesta's son or Vesta's nephew? (abstract). 19$^{th}$ *Lunar and Planetary Science Conference,* 55-56.

Binzel R. P., Xu S. 1993. Chips off of asteroid 4 Vesta: Evidence for the parent body of basaltic achondrite meteorites. *Science* 260:186-191.

Binzel R. P., Gaffey M. J., Thomas P. C., Zellner B., Storrs A. D., Wells E. N. 1997. Geologic mapping of Vesta from 1994 Hubble Space Telescope images. *Icarus* 128:95-103.

Blewett D. T., Lucey P. G., Hawke B. R., Jolliff B. L. 1997. Clementine images of the lunar sample-return stations: Refinement of FeO and $TiO_2$ mapping techniques. *Journal of Geophysical Research* 102:16319-16325.

Bobrovnikoff N. T. 1929. The spectra of minor planets. *Lick Observatory Bulletin* 14 (407):18-27.

Bowman L. E., Spilde M. N., Papike J. J. 1997. Automated EDS modal analysis applied to the diogenites. *Meteoritics & Planetary Science* 32:869-875.

Bunch T. E. 1975. Petrography and petrology of basaltic achondrite polymict breccias (howardites). Proceedings, 6$^{th}$ *Lunar Science Conference* 469-492.

Burbine T. H., Buchanan P. C., Dolkar T., Binzel R. P. 2009. Pyroxene mineralogies of near-Earth Vestoids. *Meteoritics & Planetary Science* 44:1331-1341.

Burns R. G. 1970. Mineralogical applications of crystal field theory. Cambridge Univ., London. 224 pp.

Burns R. G. 1993. Origin of electronic spectra of minerals in the visible to near-infrared region. In *Remote Geochemical Analysis: Elemental and Mineralogical Composition* edited by C. M. Pieters and P. A. J. Englert (Eds) on Cambridge Univ. Press, New York. pp. 3-29.

Carry B., Vernazza P., Dumas C., Fulchignoni M. 2010. First disk-resolved spectroscopy of 4 Vesta. *Icarus* 205:473-482.

Chapman C. R. and Gaffey M. J. 1979. Reflectance spectra for 277 asteroids. In *Asteroids* (T. Gehrels and M. S. Matthews, Eds.), pp. 655–687. Univ. of Arizona Press, Tucson.

Cloutis E. A. and Gaffey M. J. 1991. Pyroxene spectroscopy revisited: Spectral-compositional correlations and relationship to geothermometry. *Journal of Geophysical Research* 96:22809-22826.

Coradini A., Turrini D., Federico C. 2011. Vesta and Ceres: Crossing the history of the solar system. *Space Science Reviews* 164:25-40.

De Sanctis M. C. et al. 2012. Spectroscopic characterization of mineralogy and its diversity across Vesta. *Science* 336:697-700.

Delaney S. J. 2009. The surface of 4 Vesta: A petrologist's view (abstract #1600). 40$^{th}$ *Lunar and Planetary Science Conference*.

Dhingra D. 2007. Exploring links between crater floor mineralogy and layered lunar crust. *Advances in Space Research* 42 (2):275-280.

Duffard R., Lazzaro D., De Leon J. 2005. Revisiting spectral parameters of silicate-bearing meteorites. *Meteoritics & Planetary Science* 40 (3):445-459.

Dumas C. and Hainaut O. R. 1996. Ground-based mapping of the asteroid 4 Vesta. *The Messenger* 84:13-16.

Dymek R. F., Albee A. L., Chodos A. A., Wasserburg G. J. 1976. Petrography of isotopically-dated clasts in the Kapoeta howardite and petrologic constraints on the evolution of its parent body. *Geochimica et Cosmochimica Acta* 40:1115-1116.

Feierberg M. A. and Drake M. J. 1980. The meteorite-asteroid connection: The infrared spectra of eucrites, shergottites, and Vesta. *Science* 209:805-807.

Floran R. J., Prinz M., Hlava P. F., Keil K., Spettel B., Wänke H. 1981. Mineralogy, petrology, and trace element geochemistry of the Johnstown meteorite: A brecciated orthopyroxenite with siderophile and REE-rich components. *Geochimica et Cosmochimica Acta* 45:2385-2391.

Fowler G. W., Papike J. J., Spilde M. N., Shearer C. K. 1994. Diogenites as asteroidal cumulates: Insights from orthopyroxene major and minor element chemistry. *Geochimica et Cosmochimica Acta* 58:3921-3929.

Gaffey M. J. 1976. Spectral reflectance characteristics of the meteorite classes. *Journal of Geophysical Research* 81:905-920.

Gaffey M. J. 1983. First "map" of Vesta. *Sky & Telescope* 66:502.

Gaffey M. J. 1997. Surface lithologic heterogeneity of asteroid 4 Vesta. *Icarus* 127:130-157.

Gehrels T. 1967. Minor planets. I. The rotation of Vesta. *The Astronomical Journal* 72:929-938.

Haupt H. 1958. Sitz. Ber. österr. Akad. Wiss., Math.-Nat. Kl. 167:303.

Hiesinger H. and Head J. W. III. 2006. New views of lunar geoscience: An introduction and overview. In *New Views of the Moon* (B. Jolliff, M. Wiezcorek, editors.). *Reviews in Minearalogy and Geochemistry,* Mineralogical Society of America Geochemical Society 60:1-81.

Hiroi T., Pieters C. M., Hiroshi T. 1994. Grain size of the surface regolith of asteroid 4 Vesta estimated from its reflectance spectrum in comparison with HED meteorites. *Meteoritics* 29:394-396.

Isaacson P. J. and Pieters C. M. 2009. Northern Imbrium noritic anomaly. *Journal of Geophysical Research* 114:E09007.

Jaumann R. et al. 2012. Vesta's shape and morphology. Science 336:687-690.

Keil K. 2002. Geological history of asteroid 4 Vesta: The "Smallest Terrestrial Planet". In *Asteroids III*, ed. by W. Bottke, A. Cellino, P. Paolicchi, R. P. Binzel (University of Arizona Press, Tucson, 2002), pp. 573-584.

Le Corre L., Reddy V., Nathues A., Cloutis E. A. 2011. How to characterize terrains on 4 Vesta using Dawn Framing Camera color bands? *Icarus* 216 (2):376-386.

Li J.-Y., McFadden L. A., Thomas, P. C., Mutchler M. J., Parker J. W., Young E. F., Russell C. T., Sykes M. V., Schmidt B. E. 2010. Photometric mapping of asteroid (4) Vesta's southern hemisphere with Hubble Space Telescope. *Icarus* 208:238-251.

Li J.-Y. et al. 2012. Photometric properties of Vesta. Proceedings, *Asteroids, Comets, Meteors* 6387.

Lucey P. G., Spudis P. D., Zuber M., Smith D., Malaret E. 1994. Topographic-compositional units on the Moon and the early evolution of the lunar crust. *Science* 266:1855-1858

Lucey P. G., Taylor G. J., Malaret E. 1995. Abundance and distribution of iron on the Moon. *Science* 268:1150-1153.

Lucey P. et al. 2006. Understanding the lunar surface and space-moon interactions. *Reviews in Mineralogy and Geochemistry* 60 (1):83-219.

Lugmair G. W. and Shukolyukov A. 1998. Early solar system timescales according to 53Mn-53Cr systematics. *Geochimica et Cosmochimica Acta* 62:2863-2886.

Mayne R. G., McSween H. Y., McCoy T. J., Gale A. 2009. Petrology of the unbrecciated eucrites. *Geochimica et Cosmochimica Acta* 73:794-819.

McCord T. B., Adams J. B., Johnson T. V. 1970. Asteroid Vesta: Spectral reflectivity and compositional implications. *Science* 178:745-747.

McCord T. B., Clark R. N., Hawke B. R., McFadden L. A, Owensby P. D., Pieters C. M., Adams J. B. 1981. Moon: Near-infrared spectral reflectance, a good first look. *Journal of Geophysical Research* 86:10883-10892.

McCord T. B. et al. 2012. Dark material on Vesta: Adding carbonaceous volatile-rich materials to planetary surfaces. *Nature* 491:83-86.

McFadden L. A. and McCord T. B. 1978. Prospecting for plagioclase on Vesta (abstract). *Bulletin of the American Astronomical Society* 10:601.

McSween H. Y., Mittlefehldt D. W., Beck A. W., Mayne R. G., McCoy T. J. 2011. HED meteorites and their relationship to the geology of Vesta and the Dawn mission. *Space Science Reviews* 163:141-174.

McSween H. Y. et al. 2013. Composition of the Rheasilvia basin, a window into Vesta's interior. *Journal of Geophysical Research* 118:335-346.

Metzger A. E., Haines E. L., Parker R. E., Radocinski R. G. 1977. Thorium concentrations in the lunar surface. I-Regional values and crustal content. Proceedings, 8$^{th}$ *Lunar Science Conference* 1:949-999.

Mittlefehldt D. W. 1994. The genesis of diogenites and HED parent body petrogenesis. *Geochimica et Cosmoschimica Acta* 58:1537-1552.

Mittlefehldt D. W. 2008. Meteorite dunite breccia MIL 03443: A probable crustal cumulate closely related to diogenites from the HED parent asteroid (abstract). 39$^{th}$ *Lunar and Planetary Science Conference* 1919.

Mittlefehldt D. W., McCoy T. J., Goodrich C. A., Kracher A. 1998. Non-chondritic meteorites from asteroidal bodies. In *Planetary Materials*, vol. 36 (ed. J. J. Papike). Mineralogical Society of America, Chantilly, Virginia, pp. 4-1 to 4-195.

Mustard J. F., Sunshine J. M., Pieters C. M., Hoppin A., Pratt S. F. 1993. From minerals to rocks: Toward modeling lithologies with remote sensing (abstract). 24$^{th}$ *Lunar and Planetary Science Conference* 1041-1042.

Pieters C. M. 1993. Compositional diversity and stratigraphy of the lunar crust derived from reflectance spectroscopy. In *Remote Geochemical Analyses: Elemental and Mineralogical Composition* (eds. C. M. Pieters and P. Englert), pp. 309-336. Cambridge Univ. Press, Houston, Texas, USA.

Pieters C. M., Gaddis L., Jolliff B., Duke M. 2001. Rock types of South Pole-Aitken basin and extent of basaltic volcanism. *Journal of Geophysical Research* 106:28001-28022.

Reddy V., Gaffey M. J., Kelley M. S., Nathues A., Li J-Y., Yarbrough R. 2010. Compositional heterogeneity of asteroid 4 Vesta's southern hemisphere: Implications for the Dawn mission. *Icarus* 210 (2):693-706.

Reddy V., Nathues A., Gaffey M. J. 2011. Fragment of asteroid Vesta's mantle detected. *Icarus* 212 (1):175-179.

Reddy V., Sanchez J. A., Nathues A., Moskovitz N. A., Li J-Y., Cloutis E. A., Archer K., Tucker R. A., Gaffey M. J., Mann J. P., Sierks H., Schade U. 2012a. Photometric, spectral phase and temperature effects on Vesta and HED meteorites: Implications for Dawn mission. *Icarus* 217:153-168.

Reddy V. et al. 2012b. Color and albedo heterogeneity of Vesta from Dawn. *Science* 336:700-704.

Reddy V. et al. 2012c. Delivery of dark material to Vesta via carbonaceous chondritic impacts. *Icarus* 221 (2):544-559.

Rivkin A., McFadden L., Binzel R., Sykes M. 2006. Rotationally-resolved spectroscopy of Vesta I: 2-4 µm region. *Icarus* 180:464-472.

Russel C. T., Raymond C. A. 2011. Dawn mission to Vesta and Ceres. *Space Science Reviews* 164:3-23.

Russell C. T. et al. 2012. Dawn at Vesta: testing the protoplanetary paradigm. *Science* 336:684-686.

Sack R. O., Azeredo W. J., Lipschutz M. E. 1991. Olivine diogenites: The mantle of the eucrite parent body. *Geochimica et Cosmochimica Acta* 55:1111-1120.

Schenk P. et al. 2012. The geologically recent giant impact basins at Vesta's south pole. *Science* 336:694-697.

Sierks H. et al. 2011. The Dawn Framing Camera. *Space Science Reviews* 163:263-327.

Srinivasan G., Goswami J. N., and Bhandari N. 1999. 26Al in eucrite Piplia Kalan: Plausible heat source and formation chronology. *Science* 284:1348-1350.

Takeda H. 1997. Mineralogical records of early planetary processes on the howardite, eucrite, diogenite parent body with reference to Vesta. *Meteoritics & Planetary Science* 32:841–853.

Thomas P. C., Binzel R. P., Gaffey M. J., Storrs A. D., Wells E. N., Zellner B. H. 1997. Impact excavation on asteroid 4 Vesta: Hubble Space Telescope results. *Science* 277:1492-1495.

Tompkins S. and Pieters C. M. 1999. Mineralogy of the lunar crust: Results from Clementine. *Meteoritics & Planetary Science* 34:25-41.

Warren P. H. 1985. Origin of howardites, diogenites and eucrites: A mass balance constraint. *Geochimica et Cosmochimica Acta* 49:577-586.

Warren P. H., Kallemeyn G. W., Huber H., Ulff-Møller F., Choe W. 2009. Siderophile and other geochemical constraints on mixing relationships among HED meteorite breccias. *Geochimica et Cosmochimica Acta* 73:5918-5943.

Zellner N. E. B., Gibbard S., de Pater I., Marchis F., Gaffey M. J. 2005. Near-IR imaging of asteroid 4 Vesta. *Icarus* 177:190-195.

# 3. Olivine-rich exposures at Bellicia and Arruntia craters on (4) Vesta from Dawn FC

Guneshwar Thangjam[1], Andreas Nathues[1], Kurt Mengel[2], Martin Hoffmann[1], Michael Schäfer[1], Vishnu Reddy[1,3], Edward A. Cloutis[4], Ulrich Christensen[1], Holger Sierks[1], Lucille Le Corre[3], Jean-Baptiste Vincent[1], Christopher T. Russell[5]

[1]Max-Planck-Institute for Solar System Research, Justus-von-Liebig-Weg 3, 37077 Göttingen, Germany

[2]Clausthal University of Technology, Adolph-Roemer-Straße 2a, 38678 Clausthal-Zellerfeld, Germany

[3]Planetary Science Institute, 1700 East Fort Lowell, Suite 106, Tucson, AZ 85719-2395, USA

[4]Department of Geography, University of Winnipeg, 515 Portage Avenue Winnipeg, Manitoba, Canada R3B 2E9

[5]Institute of Geophysics and Planetary Physics, University of California, 3845 Slitcher Hall, 603 Charles, USA

## 3.0 Abstract:

We present an analysis of the olivine-rich exposures at Bellicia and Arruntia craters using Dawn Framing Camera (FC) color data. Our results confirm the existence of olivine-rich materials at these localities as described by Ammannito et al. (2013a) using Visual Infrared Spectrometer (VIR) data. Analyzing laboratory spectra of various Howardite-Eucrite-Diogenite meteorites, high-Ca pyroxenes, olivines and olivine-orthopyroxene mixtures, we derive three FC spectral band parameters that are indicators of olivine-rich materials. Combining the three band parameters allows us, for the first time, to reliably identify sites showing modal olivine contents >40%. The olivine-rich exposures at Bellicia and Arruntia are mapped using higher spatial resolution FC data. The exposures are located on the slopes of outer/inner crater walls, on the floor of Arruntia, in the ejecta, as well as in nearby fresh small impact craters. The spatial extent of the exposures ranges from a few hundred meters to few kilometers. The olivine-rich exposures are in accordance with both the magma ocean and the serial magmatism model (e.g., Righter and Drake 1997; Yamaguchi et al. 1997). However, it remains unsolved why the olivine-rich materials are mainly concentrated in the northern hemisphere (~36-42° N, 46-74° E) and are almost absent in the Rheasilvia basin.

## 3.1 Introduction

Asteroid (4) Vesta is geologically the most diverse differentiated chondritic body that remained intact surviving the catastrophic collisional events in the Solar System (e.g., Keil 2002; Russell et al. 2012, 2013). The exploration of such a proto-planetary body enriches the understanding of the geological conditions prevalent in the early Solar System. The observational, meteoritic and dynamical evidences so far suggest that Vesta is the parent body of many of the Howardite-Eucrite-Diogenite (HEDs) meteorites (McCord et al. 1970; Thomas et al. 1997; Migliorini et al. 1997; Schenck et al. 2012; Reddy et al. 2012b; Russell et al. 2012, 2013). Models based on the petrogenesis of HEDs (Ruzicka et al. 1997; Righter and Drake 1997; Warren 1997) favor the evolution of Vesta by an extensive melting (magma ocean). The evolution by sequential development of eruptions in shallow multiple magma chambers (serial magmatism) is also postulated (Yamaguchi et al. 1996, 1997). The serial magmatism on Vesta is consistent with the existing variations of incompatible trace element abundances in diogenites (Mittlefehldt 1994; Fowler et al. 1995; Shearer et al. 1997), and the wide range of Mg- compositions in pyroxene or olivine among olivine-bearing diogenites (Beck and McSween 2010; Shearer et al. 2010). Mandler and Elkins-Tanton (2013) proposed a two-step model of magmatic evolution from a bulk mantle composition based on major and minor elements estimated from earlier studies by Righter and Drake (1997), Dreibus and Wänke (1980), Ruzicka et al. (1997), Lodders (2000), Boesenberg and Delaney (1997). They claimed that their magma ocean model (60-70% equilibrium crystallization followed by fractional crystallization of the residual liquid in shallow magma chambers) can explain the evolution of Vesta in terms of the diverse lithologies/petrogenesis among HEDs. The assumption of an olivine rich mantle of Vesta seems to be justified regardless to the above mentioned models by which various olivine-bearing lithologies like dunites (>90% olivines), harzburgites (40-90% olivines) and olivine-orthopyroxenites (<40% olivine) could be formed in the Vestan mantle or deeper crust (Mandler and Elkins-Tanton, 2013). The exposures of olivine-rich mantle materials were expected in the huge Rheasilvia basin (Thomas et al. 1997; Gaffey 1997; Reddy et al. 2010, 2011a; Beck and McSween 2010; McSween et al. 2011, 2013; Tkalcec et al. 2013; Le Corre et al. 2013). The Rheasilvia basin (~500 km in diameter) superimposes the older Veneneia basin (~400 km in diameter) to a large extent in the southern hemisphere (Schenk et al. 2012).

Recently, olivine has been identified in Bellicia and Arruntia craters by Ammannito et al. (2013a) using Visible and Infrared Spectrometer (VIR) data. VIR is a hyperspectral instrument, which operates in the wavelength range between 0.2 and 5 µm (De Sanctis et al. 2011). The finding of olivine-rich sites in the northern hemisphere (~36-42° N, 46-74° E) despite the absence of such sites in the huge Rheasilvia basin complicates the understanding of the geological evolution of Vesta. Therefore, we mapped and investigated the potential olivine exposures using FC color data to understand the origin and nature of the olivine (Thangjam et al. 2014; Nathues et al. 2014a). The FC instrument (Sierks et al. 2011) acquired color images in 7 filters between 0.44 and 0.96 µm. The spatial resolution of FC exceeds the VIR resolution threefold. Despite FC's limited wavelength range and limited number of filters, the robustness of color parameters in constraining surface composition and mineral identification on Vesta data is well demonstrated (e.g., Le Corre et al. 2011; Reddy et al. 2012b; Thangjam et al. 2013; Nathues et al. 2014b)

### 3.1.1 Olivine in HEDs and pre-Dawn Vesta background

Among HEDs, olivine has been primarily found in diogenites, commonly associated with orthopyroxene and some accessory minerals like troilite, chromite, silica, iron-nickel (e.g., Mittlefehldt 1994; Bowman et al. 1997; Irving et al. 2009; Beck and McSween 2010; Beck et al. 2011, 2012, 2013; McSween et al. 2011). A survey of olivine-rich HED meteorites shows that there are about 30 diogenites with ≤25 wt.-% olivine (generally less than 10%), 8 diogenites with 40-68% olivine, and 4 dunites with >90% olivine (Floran et al. 1981; Sack et al. 1991; Mittlefehldt 1994; Bowman et al. 1997; Bunch et al. 2006, 2010; Irving et al. 2009; Beck and McSween 2010; Beck et al. 2011, 2012, 2013; McSween et al. 2011). The abundance of olivine in diogenites is typically very heterogeneous, and the estimates could be the result of a sampling bias (e.g., Bowman et al. 1997; Irving et al. 2009; Beck et al. 2011, 2012). Eucrites, which are one of the main components in howardites, normally do not contain olivine (Delaney et al. 1980). It is because of the fact that eucritic components have been removed from the parental melt upon fractional crystallization of basaltic magmas (Mason 1962; Stolper 1977; Grove and Bence, 1979; Delaney et al. 1980). Mikouchi and Miyamoto (1997) also suggested that eucrites don't contain olivine except the late crystallized Fe-rich olivines. Very rarely, a few eucrites have been reported with fayalitic olivine veinlets (Barrat et al. 2011; Zhang et al. 2011). Olivine has also been found in a few howardites, in most cases at the level of less than a few percent (Delaney et al. 1980; Beck et al. 2011, 2012, 2014; Lunning et al. 2014). Olivine-rich impact melts in the range ~0-30 vol.-% (containing ~50-75 vol.-% olivine in the melt) are observed in PCA 02 and GRO 95574 howardites (Beck et al. 2011, 2014).

Prior to the arrival of the Dawn spacecraft at Vesta, various attempts were made to detect olivine. McFadden et al. (1977) suggested either the presence of little olivine (<10%) or no olivine on the Vestan surface from ground-based spectra taken in the wavelength range 0.5-1.06 µm. Their conclusion was based on the apparent symmetry of 1 µm absorption feature. Larson and Fink (1975) and Feierberg et al. (1980) didn't find any spectral indication of olivine in their near-infrared spectra. Gaffey (1997) reported an olivine-bearing unit based on rotationally resolved ground based spectra. Binzel et al. (1997) also suggested olivine-bearing regions based on the observations of four-band spectra (0.43-1.04 µm) from Hubble Space Telescope (HST). Shestopalov et al. (2010) predicted up to 6.8% olivine by simulating the spectra from Binzel et al. (1997) and examining the available ground based spectra. However, Li et al. (2010) and Reddy et al. (2010) didn't confirm olivine on Vesta from their four-band HST spectra and ground based near-infrared spectra, respectively. More recently, Le Corre et al. (2013) and Reddy et al. (2013) revisited sites reported to be olivine rich from ground-based and HST observations using data from Dawn FC and VIR instruments. Le Corre et al. (2013) concluded that the olivine-rich unit reported by Gaffey (1997) called Leslie formation corresponds to the ejecta around the crater Oppia. While the observations of this feature from Dawn instruments and Gaffey (1997) agree well with each other, data from Dawn suggests that the feature is likely to be of impact melt origin rather than olivine (Le Corre et al. 2013; Reddy et al. 2013).

### 3.1.2 Olivine versus High-Ca pyroxene in 1 & 2 µm:

Pyroxene and olivine are common rock-forming minerals of mafic/ultramafic terrestrial bodies. The presence of olivine in any asteroid or meteorite can signify its origin and evolution tracing the igneous or nebular history (Sunshine et al. 2007), while high-calcium pyroxenes (HCPs) can be used to trace the degree of melting and differentiation of the body (Sunshine et al. 2004). The visible and near-infrared wavelength region (0.4-2.5 µm) is widely applied to detect and analyze these mineralogical compositions. In general, olivine spectra show a composite of three overlapping absorption bands in the 1 µm region. It is basically due to $Fe^{2+}$ in which the major broad absorption feature (around 1.05 µm) is attributed to the M2 cation site in the crystal structure, while the weaker absorption features (around 0.85 and 1.25 µm) are attributed to the M1 cation site (Adams 1974; Burns 1970, 1993; Singer 1981; Cloutis et al. 1986; King and Ridley 1987; Reddy et al. 2011b; Sanchez et al. 2014). The spectral features depend not only on the olivine chemistry, where 1 µm absorption moves to longer wavelength with increasing Fe content, but they also depend on physical parameters of the regolith, like grain size and temperature as well as observational parameters such as phase angle (Adams 1975; Burns 1970, 1993; Cloutis et al. 1986; King and Ridley 1987; Sanchez et al. 2014). Pyroxenes have prominent 1 µm and 2 µm absorption features with varying absorption band centers and band depths. They depend on $Fe^{2+}$-$Ca^{2+}$-$Mg^{2+}$ chemistry and the asymmetry of the cations (crystallographic sites) as well as grain size, temperature and phase angle (Adams 1974, 1975; Burns 1993; Singer 1981; Cloutis and Gaffey 1991; Klima et al. 2007; Schade et al. 2004). The low-Ca pyroxenes have absorption band centers near 0.9 µm and 1.9 µm while high-Ca pyroxenes have band centers near 0.98 µm and 2.15 µm (Pieters 1986). The missing absorption feature of olivine in 2 µm is the key to distinguish olivine from pyroxenes using the band area ratio (BAR) approach (Gaffey 1983; Cloutis et al. 1986; Cloutis and Gaffey 1991). It should be noted that very high-Ca pyroxenes, termed spectral type A by Adams (1974) can have reflectance spectra superficially similar to olivine (Schade et al. 2004), however, such pyroxenes have not been detected in HEDs (e.g., Mayne et al. 2009, 2010; McSween et al. 2011). It is also worth to mention that many authors (Duffard et al. 2005; Moroz et al. 2000; Sanchez et al. 2012) discussed how temperature and grain size could affect mafic silicate reflectance spectra and spectral parameters. They suggested caution in implementing the spectral parameters. Several spectral parameters like the HCP index, forsterite index, fayalite index, and olivine index have been developed and applied to the Martian surface (Poulet et al. 2007; Pelkey et al. 2007; Carrozzo et al. 2012). De Sanctis et al. (2013), Palomba et al. (2012a, b, 2013a, b), Ruesch et al. (2013, 2014) adapted the Martian spectral parameters to identify potential olivine-rich sites on Vesta using the VIR data.

Distinguishing olivine from high Ca-pyroxene using datasets without having full 1-µm absorption band coverage (e.g., Clementine Ultraviolet/Visible or UVVIS, HST Wide Field Planetary Camera/WFPC, Dawn FC) is rather challenging. It is because of their close spectral similarity in the 1 µm absorption band minima. Clementine UVVIS multispectral data has five bands in the wavelength range between 0.41 and 1 µm. Tompkins and Pieters (1999) and Pieters et al. (2001) suggested olivine-bearing lithologies on the lunar surface using Clementine UVVIS data, however it was difficult to distinguish them from high-Ca pyroxene bearing lithologies. The four band HST/WFPC data covering the wavelength range between 0.43 and 1.04 um were used by Binzel et al. (1997), Shestopalov et al. (2008) and Li et al. (2010) to analyze the likely presence or absence of olivine on Vesta. The Dawn FC is comparable to Clementine/UVVIS and HST/WFPC in terms of their wavelength coverage, but the FC has more spectral bands with better spatial resolution.

## 3.2 Laboratory-derived Spectra

In this work, spectra in the visible and near-infrared wavelength range have been compiled from available data sets of Reflectance Experiment Laboratory (RELAB) at Brown University/USA (http://www.planetary.brown.edu/relab/), Hyperspectral Optical Sensing for Extraterrestrial Reconnaissance Laboratory (HOSERlab) at University of Winnipeg/Canada (http://psf.uwinnipeg.ca/Home.html), and Unites States Geological Survey (USGS) Spectroscopy Lab (http://speclab.cr.usgs.gov/spectral-lib.html). The compilation includes:

(1) 241 spectra of HEDs (45 eucrite, 13 diogenite, and 17 howardite samples) of various grain sizes and of bulk rock samples from RELAB. A few spectra were excluded in this analysis because of inconsistencies/exceptions observed in the FC spectral range (Appendix-1).

(2) 43 spectra of terrestrial olivine (~$Mg_{90.4}Fe_{9.6}$) - orthopyroxene (~ $Mg_{86.8}Ca_{0.4}Fe_{12.8}$) mixtures (Ol-Opx, 10-90% olivine) at 10% intervals in various grain size ranges (<38, 38-53, 63-90, 90-125 µm) from HOSERLab;

(3) Nearly pure olivines (Ol) having various forsterite contents $Fo_{10-90}$ (<45 µm) from RELAB, $Fo_{11}$-$Fo_{91}$ (<65 µm) from USGS and $Fo_{86.8}$ (<38, 38-53, 63-90, 90-125 µm) from HOSERLab;

(4) 46 spectra of synthetic low/high Ca-pyroxenes (Klima et al. 2011) with various wollastonite contents ($Wo_{2-51}$, <45 µm) from RELAB. Only those HCPs have been considered in our analysis that shows Wo-En-Fs compositions, which are indeed observed in eucrites (Mayne et al. 2009, 2010; McSween et al. 2011). HCPs outside the eucritic compositional range has been discarded (Appendix-2). The selected HCPs (>$Wo_{20}$) are termed HCP/HED in this analysis. All Wo-En-Fs sample spectra of synthetic low/high Ca-pyroxenes (Klima et al. 2011) are termed HCP/CPX ($Wo_{2-51}$).

The laboratory spectra are resampled to FC filter band passes as presented in Sierks et al. (2011). Fig. 1A shows normalized spectra of eucrite (ALHA76005, 25-45 µm), diogenite (EETA79002, 25-45 µm), HCP ($Wo_{45}En_{14}Fs_{41}$, ≤45 µm), Ol ($Fo_{90}$, 38-53 µm) and Ol-Opx (60 wt.-% olivine, 38-53 µm). The filter band passes and center wavelengths of the FC are also presented in this figure. The same spectra resampled to the FC band passes are displayed in Fig. 1B. The figure also visualizes the three band parameters, which are defined as follows:

Band Tilt (BT) = $(R_{0.92\mu m} / R_{0.96\mu m})$

Mid Ratio (MR) = $(R_{0.75\mu m} / R_{0.83\mu m}) / (R_{0.83\mu m} / R_{0.92\mu m})$

Mid Curvature (MC) = $(R_{0.75\mu m} + R_{0.92\mu m}) / R_{0.83\mu m}$

; where R(λ) is the reflectance in the corresponding filter.

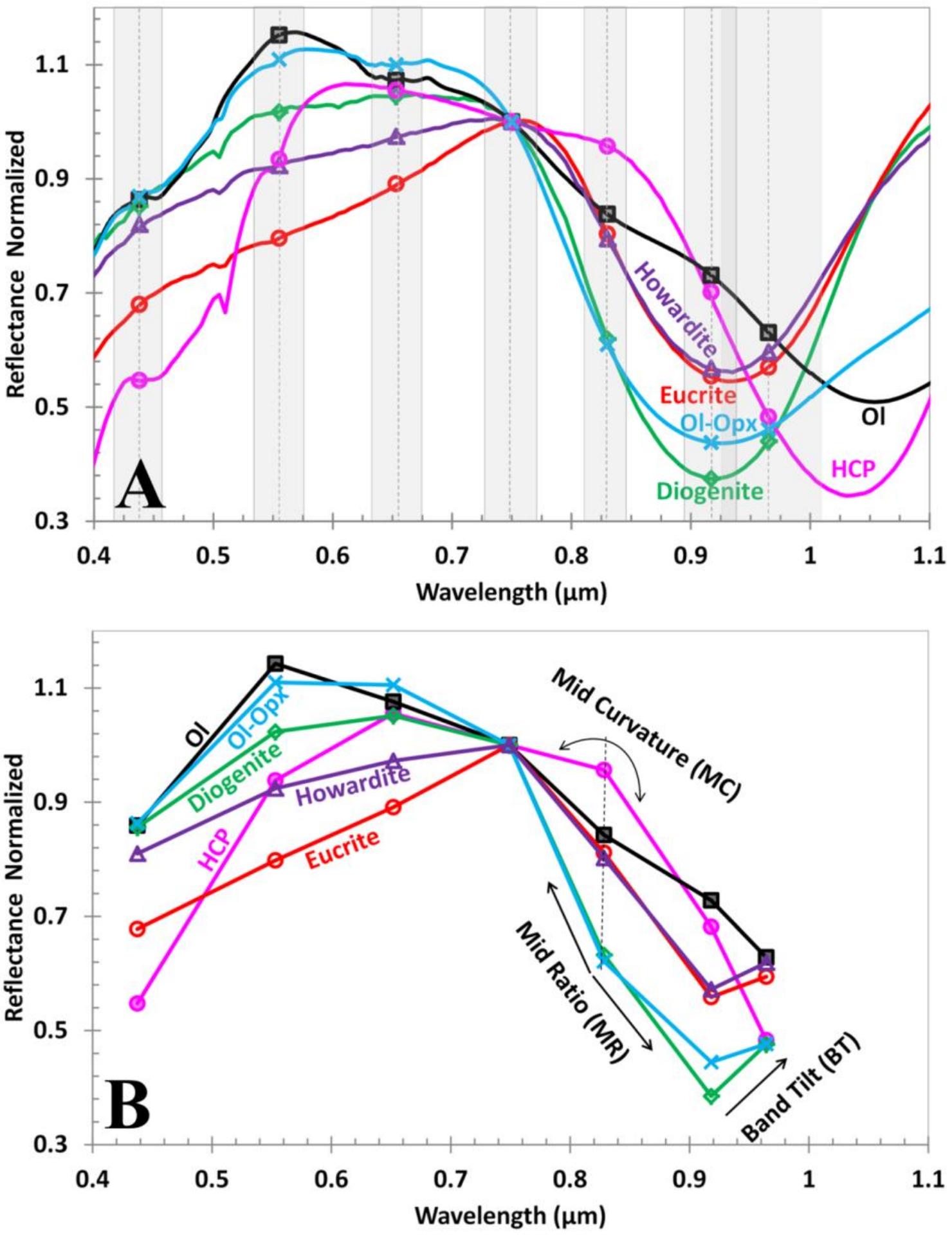

Fig. 1: Reflectance spectrum of howardite (EET875003, 25-45 µm), eucrite (ALHA76005, 25-45 µm), diogenite (EETA79002, 25-45 µm), high-Ca pyroxene (Wo45En14Fs41, ≤45 µm), olivine (Fo90, 38-53 µm) and olivine-orthopyroxene mixture (60 wt.% olivine, 38-53 µm) normalized to unity at 0.75 µm. Framing Camera filter band passes and center wavelengths are marked. (B) Spectra are resampled to FC filter band passes. A sketch defines the band parameters Mid Curvature (MC), Mid Ratio (MR) and Band Tilt (BT).

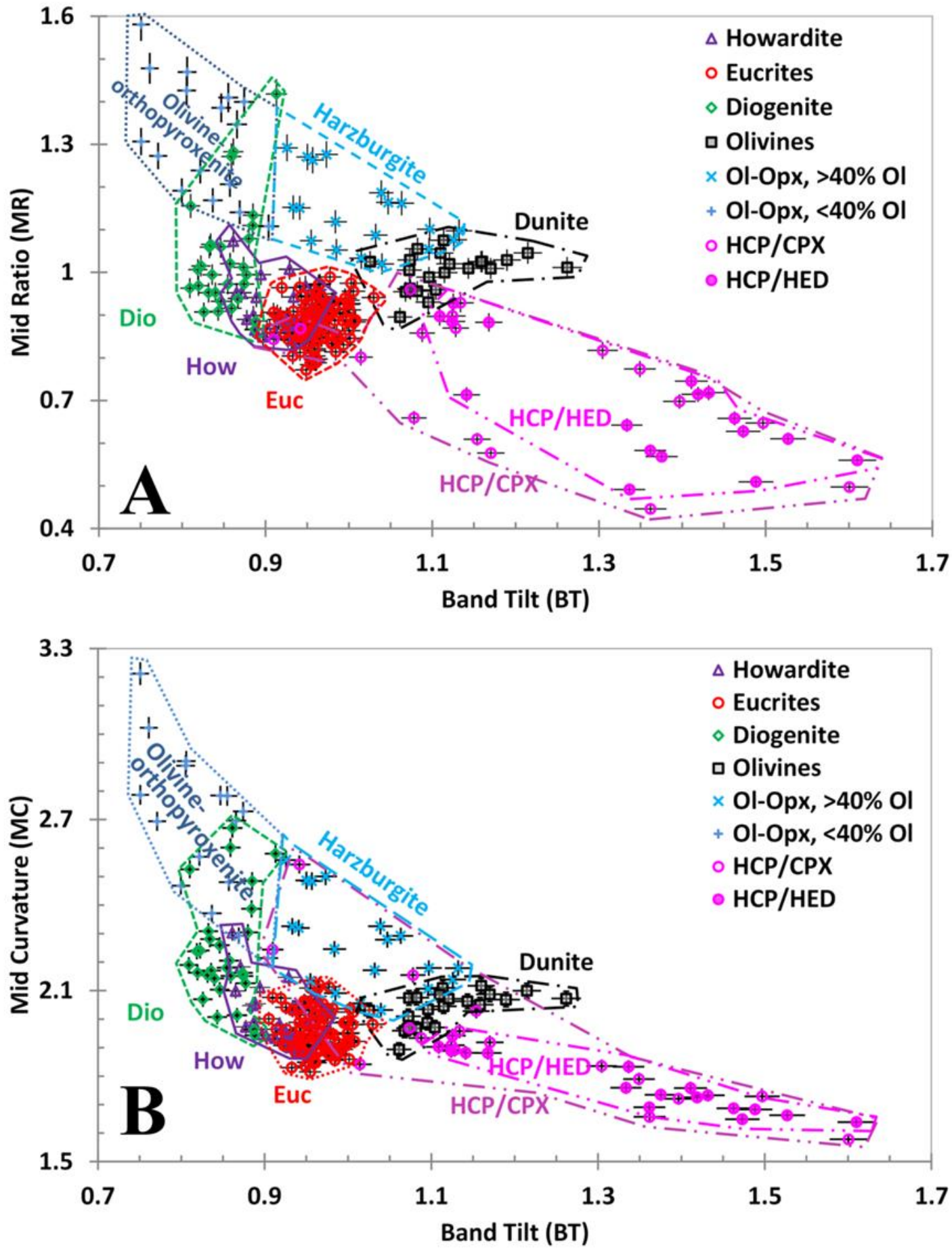

Fig. 2: (A) Mid Ratio versus Band Tilt, and (B) Mid Curvature versus Band Tilt for eucrites, diogenites, howardites, Ol, HCP and Ol-Opx mixtures. The HED samples are in various grain sizes/bulk from RELAB. HCP/CPX samples (<45 µm) are synthetic clinopyroxenes with compositional range Wo2-51, while HCP/HED are selected samples compatible with the existing HCPs among eucrites. Ol-Opx spectra (38-53, 63-90, 90-125 µm) are from HOSERLab. Ol spectra (terrestrial olivines) are from RELAB (<45 µm, Fo10-90), USGS (<65 µm, Fo11-91), and HOSERLAB (Fo90 in various grain sizes). Ol and HCP/HED are highlighted by filled symbols.

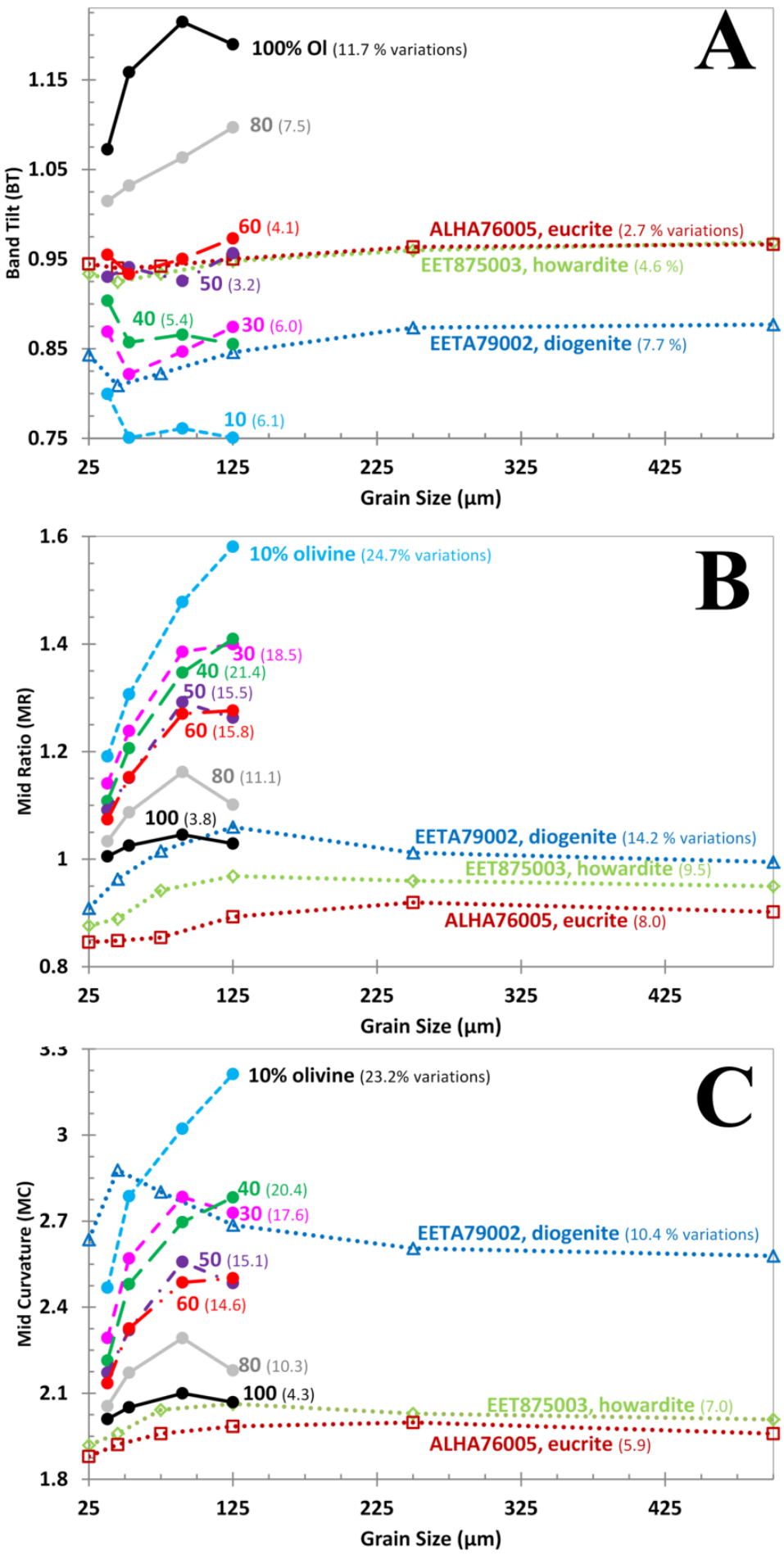

Fig. 3: Influence of grain size on (A) BT, (B) MR and (C) MC parameters, for HEDs in size ranges <25, 25-45, 45-75, 75-125, 125-250, 250-500 µm, and olivine-orthopyroxene mixtures (10-90% Olivine) and 100% olivine in size ranges <38, 38-53, 63-90, 90-125 µm. The maximum variations (%) of each sample over the whole grain size ranges are given in brackets. Data points for some of the mixtures are not shown to enhance readability.

### 3.2.1 Band Tilt (BT)

The BT parameter was found to be well suited to distinguish eucrites from diogenites (Thangjam et al. 2013). Here, we will discuss the relevance of this parameter for distinguishing Ol, Ol-Opx and HCPs. The BT parameter values of the individual samples mentioned in section 2 are plotted along the X-axis in Fig. 2. The values of olivine are in the range 1.02-1.26. The values of HCP/CPX samples lie in the range 0.90-1.61, with the majority (95%) falling in the range 1.01-1.61. The values of HCP/HED samples are in the range 1.11-1.61. Most of the olivine (95%) and HCP samples (89%) have larger BT values than eucrites (0.91-1.03). Eucrites, olivines and HCP samples have higher BT values than diogenites (0.80-0.91). The olivine-orthopyroxene mixtures have BT values in the range 0.75-1.13, while those samples above 40 wt.-% olivine have higher values (0.92-1.13) than diogenites. In this study, we followed the nomenclature of olivine-orthopyroxene mixtures/rock assemblages in accordance with the IUGS system (Streckeisen 1974; Wittke et al. 2011; Mandler and Elkins-Tanton 2013), i.e. olivine-orthopyroxenites (<40%), harzburgites (40-90% olivine), and dunites (>90% olivine). We use the term diogenites to denote the olivine-free orthopyroxenites (Appendix-3). Based on our analyses, the BT parameter is effective in separating peridotites, HCPs and eucrites from diogenites as well as from olivine-orthopyroxenites. HED sample spectra seldom reach a BT value larger than 1.03.

The influence of grain size on the BT parameter is shown in Fig. 3A. The BT values of eucrite ALHA76005, howardite EET875003, diogenite EETA79002 (over the size intervals ≤25, 25-45, 45-75, 75-125, 125-250, 250-500 µm), and Ol-Opx mixtures (over the size intervals ≤38, 38-53, 63-90, 90-125 µm) are presented. The trend of the BT values with increasing grain size of the three HED samples is similar. The BT values of the grain size range 25-45 µm and 125-250 µm are the extremes. There is no significant variation of the values from 125-250 µm to 250-500 µm, while the variations below 125 µm are larger. The BT values of Ol-Opx mixtures do not show a parallel systematic trend. The maximum variations over the whole range of grain sizes are given in brackets (see Fig. 3A). The maximum variation is 7.7% for the three HEDs, 8.2% for Ol-Opx mixtures and 11.7% for pure olivine. The influence of grain size on the BT parameter of HEDs and Ol-Opx mixtures is found to be similar

### 3.2.2 Mid Ratio (MR)

The MR parameter values for the individual samples given in section 2 are plotted along the Y-axis in Fig. 2A. Eucrites have MR values between 0.77 and 0.99, while diogenites range between 0.86 and 1.42. Howardites (0.84-1.07) lie between eucrites and diogenites. The values of olivines range between 0.89 and 1.08, while the range is 1.01-1.58 for Ol-Opx mixtures. The values of HCP/CPX samples are in the range 0.44-0.96, while HCP/HED values lie in the range 0.5-0.93. The olivine-orthopyroxene mixtures have larger MR values than HCPs, and therefore these two are distinguishable.

The influence of grain size on the MR parameter is shown in Fig. 3B. The MR parameter of the three HED samples increases with increasing grain size up to 90-125 µm, but decreases thereafter. The values for grain size ranges ≤25 µm and 90-125 µm are the extremes. There is no significant change of the MR values from 125-250 µm to 250-500 µm. A systematic trend is not noticeable for Ol-Opx mixtures (see Fig. 3B). The MR parameter of Ol-Opx mixtures increases with increasing grain size (up to 63-90 µm), while the values are increasing or decreasing for the size range 90-125 µm. The overall influence of grain size on MR parameter decreases with increasing olivine content. The effect is the least (3.8%) for pure olivine, whereas the maximum effect (24.7%) is observed for 10 wt.-% olivine Ol-Opx mixture. Among HEDs, the effect increases from the eucrite sample (8.0%) to the diogenite sample (14.2%).

### 3.2.3 Mid Curvature (MC)

The MC parameter values for the samples given in section 2 are plotted along the Y-axis in Fig. 2B. Eucrites have MC values in the range 1.81-2.12 while the values of diogenites range between 1.94 and 2.67. Howardites have an intermediate range (1.88-2.30) between eucrites and diogenites. The MC values of olivine are in the range 1.89-2.11 while the range of olivine-orthopyroxene mixtures is 2.03-3.21. The values of HCP/CPX are between 1.57 and 2.54, while the values of HCP/HED lie in the range 1.64-1.94. The majority of eucrites (83%) and HCP/CPX (87%) have lower MC values than Ol-Opx mixtures. However, HCP/HED is distinguishable from Ol-Opx mixtures.

The influence of grain size on the MC parameter is shown in Fig. 3C. Although the values of the three HED samples vary over the whole grain size ranges, there is no significant change from 125-250 µm to 250-500 µm. The MC values of Ol-Opx mixtures increase with increasing grain size (from ≤38 μm to 63-90 μm), but the trend for the grain size range 90-125 μm are different. The overall influence of grain size on the MC parameter decreases with increasing olivine content. The effect is the least (4.3%) for pure olivine, while it is maximal (23.2%) for 40 wt.-% olivine Ol-Opx mixture. Among HEDs, the effect increases from the eucrite sample (5.9%) to the diogenite sample (10.4%).

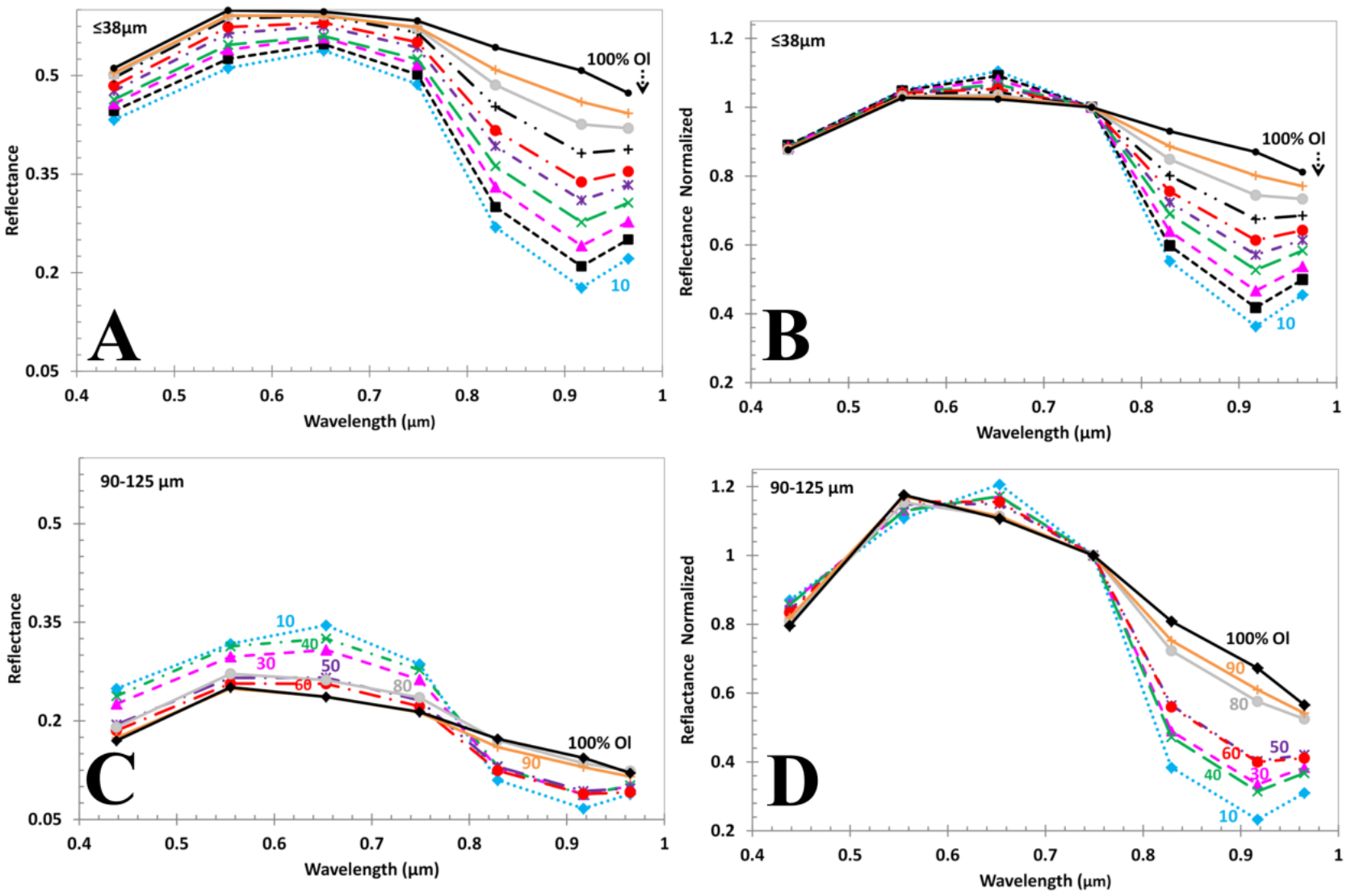

Fig. 4: (A) Absolute and (B) normalized spectra of Ol-Opx mixtures (≤38 μm). (C) Absolute and (D) normalized spectra of Ol-Opx mixtures (90-125 μm). Spectra of a few mixtures are not shown to enhance readability (B, D).

### 3.2.4. Band depth and albedo at 0.75 μm

The optical parameters, band depth and albedo, can characterize the abundance of mafic minerals (Pieters et al. 2001). However, these parameters are highly affected by grain size, temperature, viewing geometry and the presence of opaques (Nathues 2000; Reddy et al. 2012a; Hiroi et al. 1994; Duffard et al. 2005; Thangjam et al. 2013). The ratio of the reflectance values at 0.75 and 0.96 (or 1.0) μm is used as a proxy to 1-μm absorption band depth for datasets like lunar Clementine/UVVIS (Tompkins and Pieters 1999; Pieters et al. 2001; Isaacsson and Pieters 2009). In the present work, the ratio of reflectance values of the filters 0.75 and 0.92 μm is used to define the apparent band depth. Figure 4 shows spectra of olivine-orthopyroxene mixtures at 10 wt.-% intervals for the grain size ranges ≤38 and 90-125 µm. For grain sizes ≤38 µm, the band depth and reflectance at 0.75 µm behaves linearly, i.e. with increasing olivine, the albedo gradually increases while the band depth decreases (Fig. 4A, B). For larger grain sizes (90-125 µm), the trend is less systematic (Fig. 4C, D). The influence of grain size on band depth and albedo for the three HED samples (≤25, 25-45, 45-75, 75-125, 125-250, 250-500 µm) and Ol-Opx mixtures (≤38, 38-53, 63-90, 90-125 µm) are analyzed similarly to the band parameters discussed above (Figure not shown). The band depth values vary up to 23.4% for the HED samples and 37% for Ol-Opx mixtures, while the albedo values vary up to 23.3% for the HED samples and 63.4% for Ol-Opx mixtures. Figure 5 shows a scatter plot of band depth and reflectance values for all the samples described in section 2. It is obvious that band depth versus reflectance at 0.75 µm is not suited to distinguish the samples in this analysis.

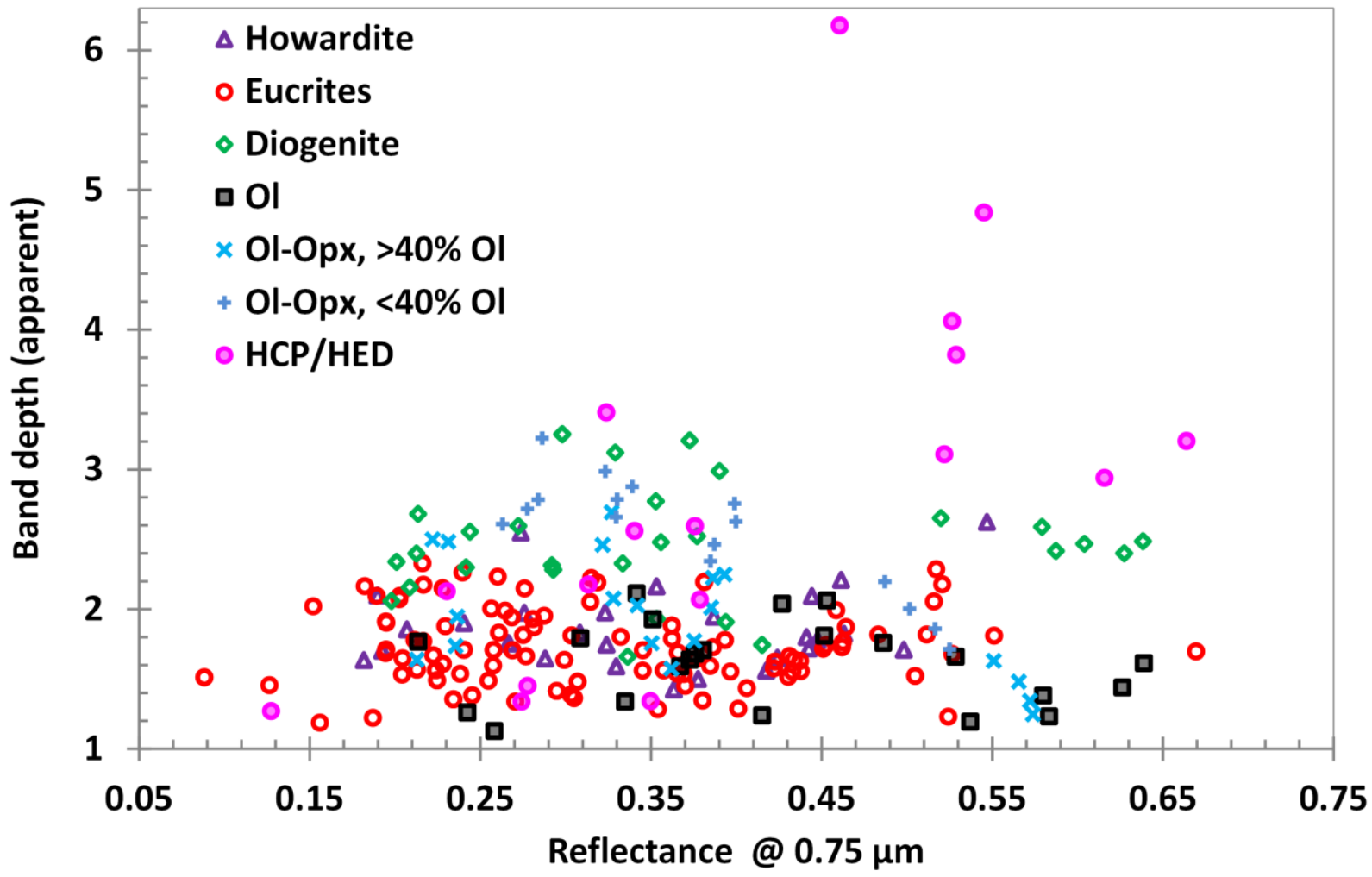

Fig. 5: Band depth (apparent) and reflectance values at 0.75 µm for howardites, eucrites, diogenites, olivines, olivine-orthopyroxene mixtures, and HCP/HED.

### 3.2.5 Band parameter approach

Based on the laboratory data analysis, we conclude:

(1) The BT parameter is effective in distinguishing eucrites from olivine-rich Ol-Opx mixtures (harzburgites), HCPs and eucrites from diogenites and olivine-poor Ol-Opx mixtures (olivine-orthopyroxenites). The dunites (>90% olivine) and HCPs have often larger BT values than eucrites. For BT values above 1.03, dunites and HCPs are obviously separated from HEDs.

(2) The MR values of Ol-Opx mixtures are larger than HCP/HED and HCP/CPX, which mean that the MR parameter is suited to distinguish them from HCPs.

(3) The majority of eucrites (83%) and HCP/CPX (87%) have lower MC values than Ol-Opx mixtures, while HCP/HED samples have lower values than Ol-Opx mixtures.

The combination of the above band parameters is found to be useful to identify olivine-rich Ol-Opx mixtures (peridotites) in the geological context of Vesta (Fig. 2). The band parameter space BT versus MR (BT-MR polygons, Fig. 2A) is more appropriate than BT versus MC (BT-MC polygons, Fig. 2B), because the separation of peridotites from HCP/CPX and HCP/HED are clearer in BT-MR parameter space. Since uncertainties of the laboratory datasets are unavailable, we assumed a 1% standard deviation error. The error propagation for each band parameter is computed statistically (Appendix-4), and the polygons were drawn accordingly (Fig. 2). A quantitative analysis of the influence of grain size on the band parameters for HEDs, Ol, and Ol-Opx mixtures is also presented (Fig. 3). It is worth to mention that the influence of grain size on the BT and MR parameters are lower when compared to the MC parameter. This is one of the reasons why we prefer the use of BT-MR polygons rather than BT-MC polygons. The polygons defined in band parameter space consider various grain sizes and bulk samples (see section 2). Given the fact that the spectra of the laboratory samples used in our analyses are not entirely representing the whole compositional range on Vesta, there may be changes in the polygons defined here.

## 3. 3 FC data analyses

The Dawn Framing Camera (Sierks et al. 2011) acquired images of the entire visible surface of Vesta in three different orbits at spatial resolutions of ~250 m/pixel, ~60 m/pixel, and ~20 m/pixel. There are three standard levels of FC images from which level 1c is processed correcting the “in-field” stray light component (Kovacs et al. 2013). Level 1c I/F data is used for processing in the Integrated Software for Imagers and Spectrometers/ISIS (Anderson et al. 2004) pipeline, where the photometric corrections of the FC color data are performed to standard viewing geometry using Hapke functions. The resulting reflectance data are then map-projected in various steps, and co-registered aligning the color frames to create the color cubes. For the photometric correction, the Vesta shape model derived from FC clear filter images (Gaskell 2012) is used. Further descriptions of the data processing method and the photometric corrections are presented in Nathues et al. (2014b). The FC mosaics generated by the ISIS pipeline were analyzed using ENVI software. For the present analysis, FC color data having ~60 m/pixel spatial resolution from HAMO and HAMO-2 phase are used.

The global mosaic of Vesta in the Claudia-Coordinate system is shown in Fig. 6 using the cylindrical projection. The approximate outlines of the Rheasilvia and the Veneneia basins are marked. Uncertainties of each individual FC color filters were estimated from homogeneous, small areas of different size (2 x 2 to 7 x 7 pixels) at Bellicia and Arruntia. We observed that the relative statistic error for the 4 x 4 pixel sized area is reliable, and these values are used to compute the error propagation of the band parameters (Appendix-4).

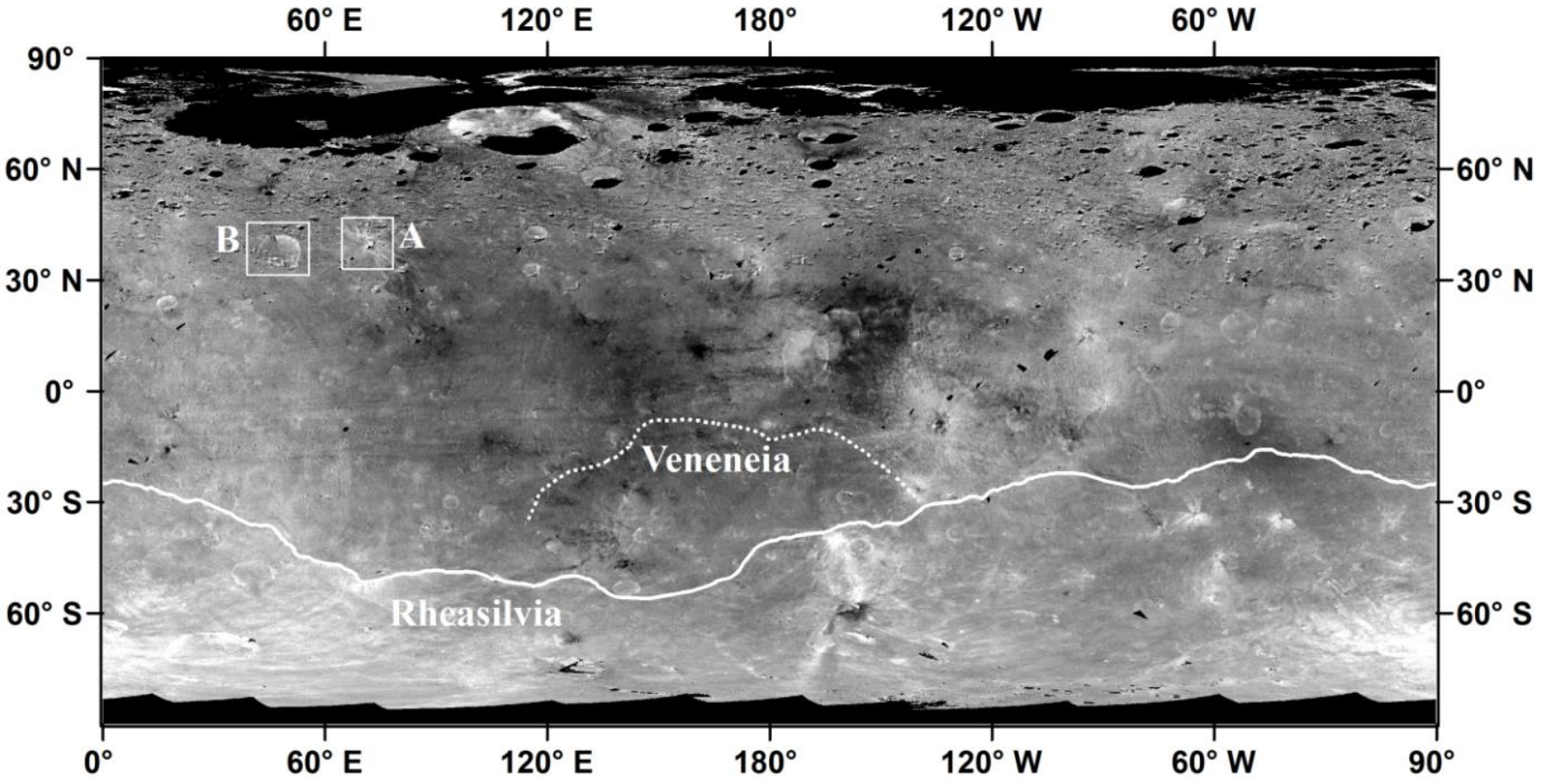

Fig. 6: HAMO global mosaic of Vesta in the Claudia-Coordinate system at ~60 m/pixel resolution in simple cylindrical projection. Arruntia (A) and Bellicia (B) craters are in the northern hemisphere. The approximate outlines of the Rheasilvia and Veneneia basins are marked in bold and dashed lines, respectively.

### 3.3.1 Arruntia crater

Arruntia is an impact crater of ~12 km diameter and 2.5 km depth in the northern hemisphere (Fig. 6). A perspective view of the reflectance image at 0.55 µm, projected on HAMO DTM (~62 m/pixel resolution) is shown in Fig. 7A. Potential olivine-rich exposures are highlighted in red by selecting those pixels that have band parameters in the peridotitic field. A few sites are selected for illustration (A1-A5; Fig. 7A) and their average absolute and normalized reflectance spectra over a region of 2 x 2 pixels are presented in Figs. 8A and 8B, respectively. The Vesta average spectrum from FC HAMO-1 & -2 is also displayed. The band parameter space of the identified sites is illustrated in Fig. 9. Olivine-rich exposures are located on the ejecta blanket nearby the outer rim, and a few of them are located on the inner wall and the crater floor. Many of the exposures extend a few hundreds of meters in length, and the exposure marked A3 extends up to few kilometers. The olivine-rich exposures cover ~1.6 % of the area within 2.5 crater radii from the center of the crater. The exposures have higher reflectance value than the average surface of Vesta. In general, the band depth of the olivine exposures is similar to that of the average Vesta, and sometimes slightly deeper. However, a few sites exhibit shallower band depth than the average Vesta. The exposures exhibit a redder visible slope compared to the average Vesta spectrum, which could be due to the associated lithological background materials in the regolith. Dark material is observed in the ejecta blanket nearby the crater rim, and on the slopes of inner crater wall. The lithological background of the olivine-rich exposures is investigated in the band parameter space. The band parameter space of the olivine-rich exposure A3 (located in the ejecta nearby the outer rim, Fig. 10A, B) is presented using the BT-MR polygons (Fig. 10C). The Arruntia region is howarditic/eucritic in composition.

### 3.3.2 Bellicia crater

Bellicia is an impact crater located westward of Arruntia, having a diameter of ~35 km and 5.9 km depth (Fig. 6). A perspective view of the reflectance image at 0.55 µm is displayed in Fig. 7B. The potential olivine-rich exposures are marked in red. Some sites have been selected (B1-B5, Fig. 7B), and their average absolute and normalized reflectance spectra over a region of 2 x 2 pixels are presented in Figs. 8C and 8D, respectively. The average Vesta spectrum is also shown. Many of the olivine-rich exposures are located on the slopes of the inner crater wall, and a few of them are possibly on the crater floor and in nearby small fresh craters. The exposures extend few hundreds of meters, while some of the exposures (e.g., B1 and B4) are up to few kilometers. The exposures cover ~0.7 % of the crater area within 1 crater radius from the center of the crater. The exposures in Bellicia exhibit higher reflectance values than the average Vesta surface. The exposures in general have similar to slightly higher band depth than the average Vesta. Dark material nearby the olivine exposure B1 are apparently moving along the slope of the inner crater wall (Fig. 7B). The exposures exhibit a redder visible slope compared to the average Vesta spectrum, while two of the sites (B1 and B2) have both higher reflectance values and redder visible slopes than the rest of the exposures (Fig. 8C, D). The locations of the data points in the band parameter space (BT-MR polygons, and BT-MC polygons) are displayed in Fig. 9. The background materials of olivine-rich exposures are analyzed in the band parameter space. The band parameter space of the olivine-rich site B1 (located along the slope of the inner crater wall, Fig. 11A, B) is presented using the BT-MR polygons (Fig. 11C). The majority of the data points are in the howarditic/eucritic field.

## 3.4 Discussion

Our analysis using FC data suggests olivine-rich exposures at Bellicia and Arruntia. The exposures have higher reflectance values, and similar or slightly higher/lower 1-µm band depths than the average Vesta spectrum. The exposures at Arruntia have redder visible slope than that of Bellicia, which is likely due to the background lithology. The red slope of the ejecta materials at Arruntia could also be caused by an association of impact melt component (Le Corre et al. 2013). The spatial extent of the olivine-rich exposures is found in the range of a few hundred meters up to few kilometers. The exposures are located on the inner crater walls, on the floor of Arruntia, in the ejecta, and in nearby fresh small impact craters. It is to be noted that smaller planetary bodies reveal deep seated minerals, like olivine and spinel, on inner crater walls, central peaks, crater floors, ejecta, and in the vicinity of basins (e.g., Koeppen and Hamilton 2008; Pieters et al. 2011; Yamamoto et al. 2010, 2012). A quantitative analysis of mineralogy and olivine abundance of the exposures seems difficult using FC color data. Moreover, the influence of other factors like grain size has to be considered. However, based on the locations of the data points over the band parameter space (Fig. 9), the identified sites suggest peridotitic lithologies with modal olivine contents above 60%. Such an olivine content is in accordance with the predicted abundance of olivine (60-80%) in the mantle material of Vesta by Mandler and Elkins-Tanton (2013). The exposures at Bellicia and Arruntia could be potential mantle material, but the abundance of olivine is not a sufficient criterion for a mantle origin. The exposures are in general associated with a howarditic/eucritic environment.

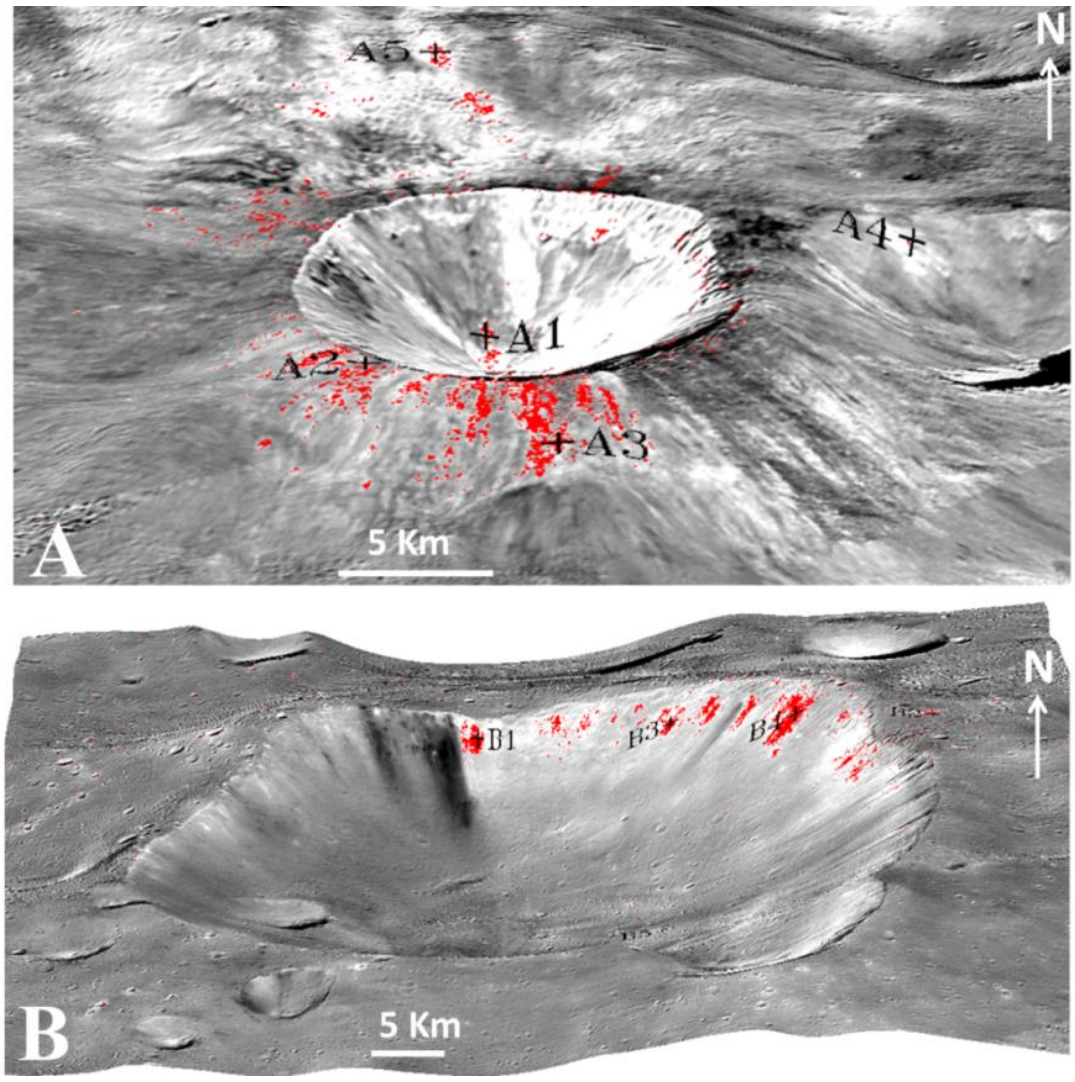

Fig. 7: Perspective view of reflectance image of (A) Arruntia and (B) Bellicia crater, projected on HAMO DTM. Potential olivine-rich exposures are marked in red.

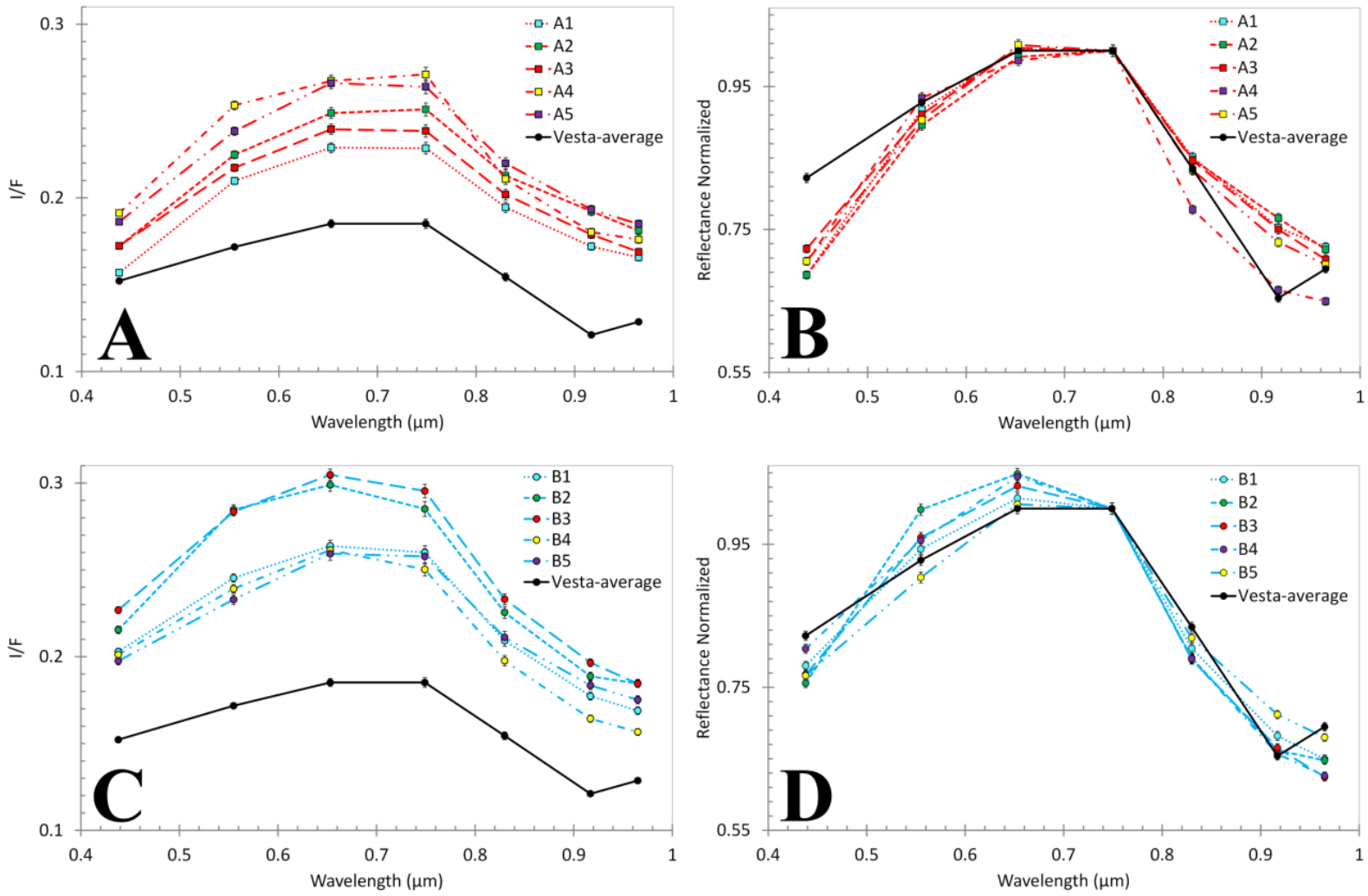

Fig. 8: Spectra of olivine-rich sites as indicated in Fig. 7. (A) Absolute and (B) normalized spectra from sites at Arruntia. (C) Absolute and (D) normalized spectra from Bellicia. Each spectrum is an average of 2 by 2 pixels. The spectrum of the average Vesta surface (black solid line) is also shown.

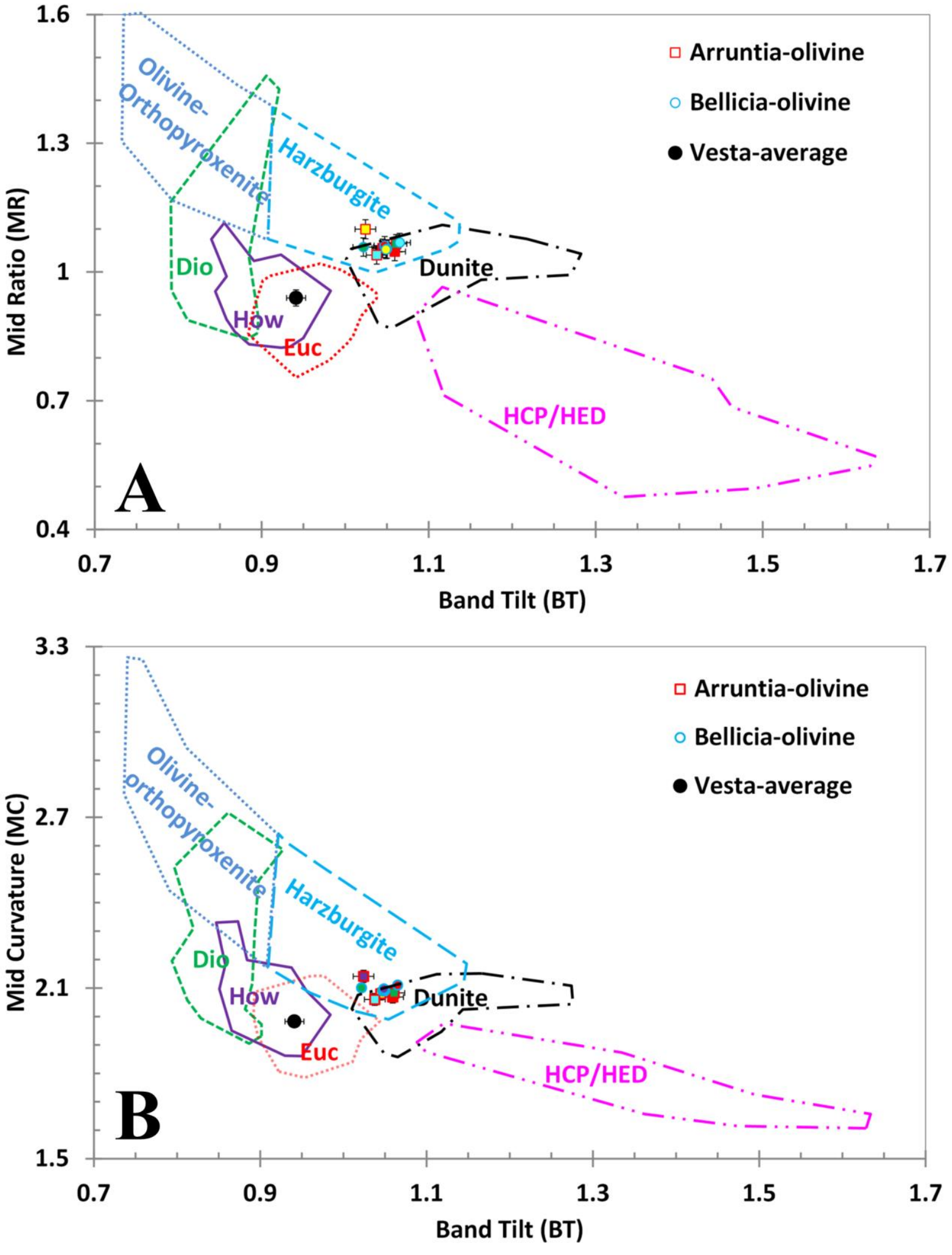

Fig. 9: Location of data points of olivine-rich exposures at Bellicia and Arruntia crater, projected on (A) BT-MR polygons, and (B) BT-MC polygons. The polygons are based on our laboratory spectral analyses (see Fig. 2).

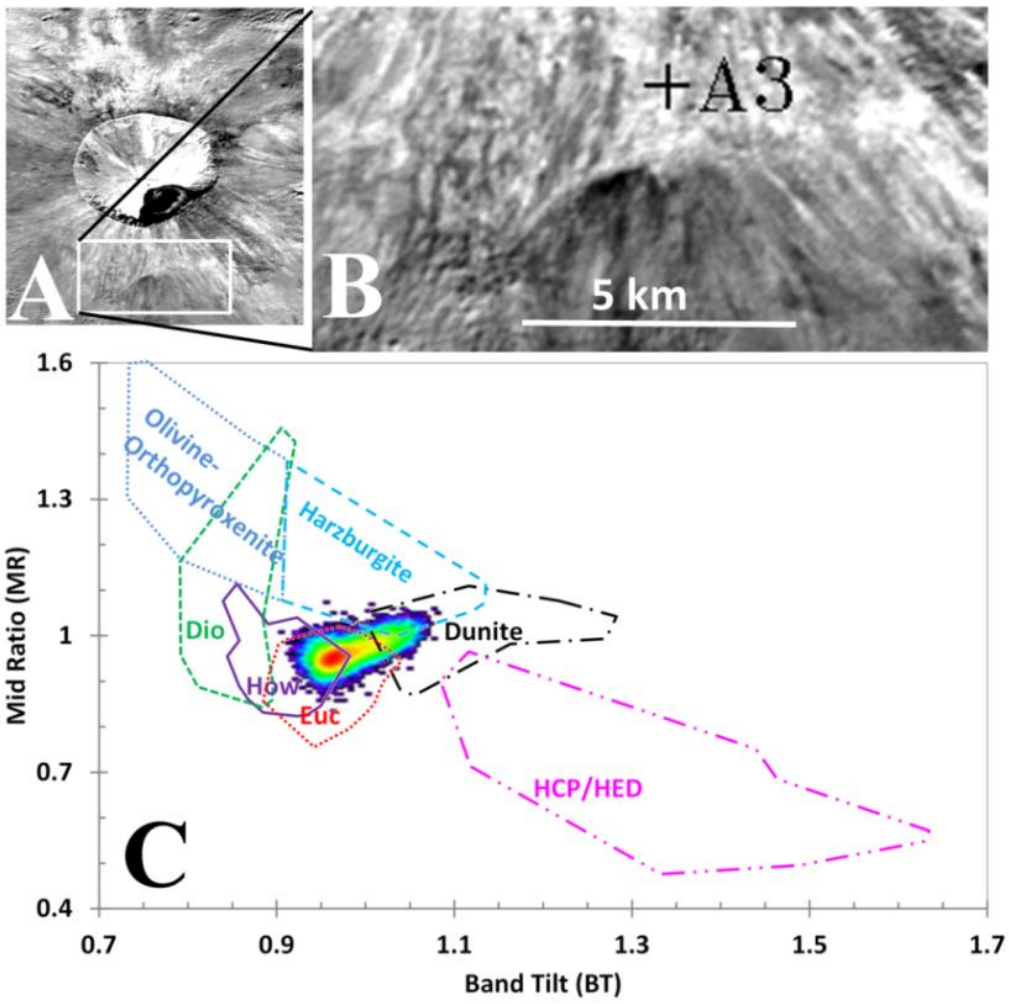

Fig. 10: (A) Arruntia crater, and (B) selected region near the olivine-rich exposure A3. (C) Band parameter values of the selected region plotted over the BT-MR polygons.

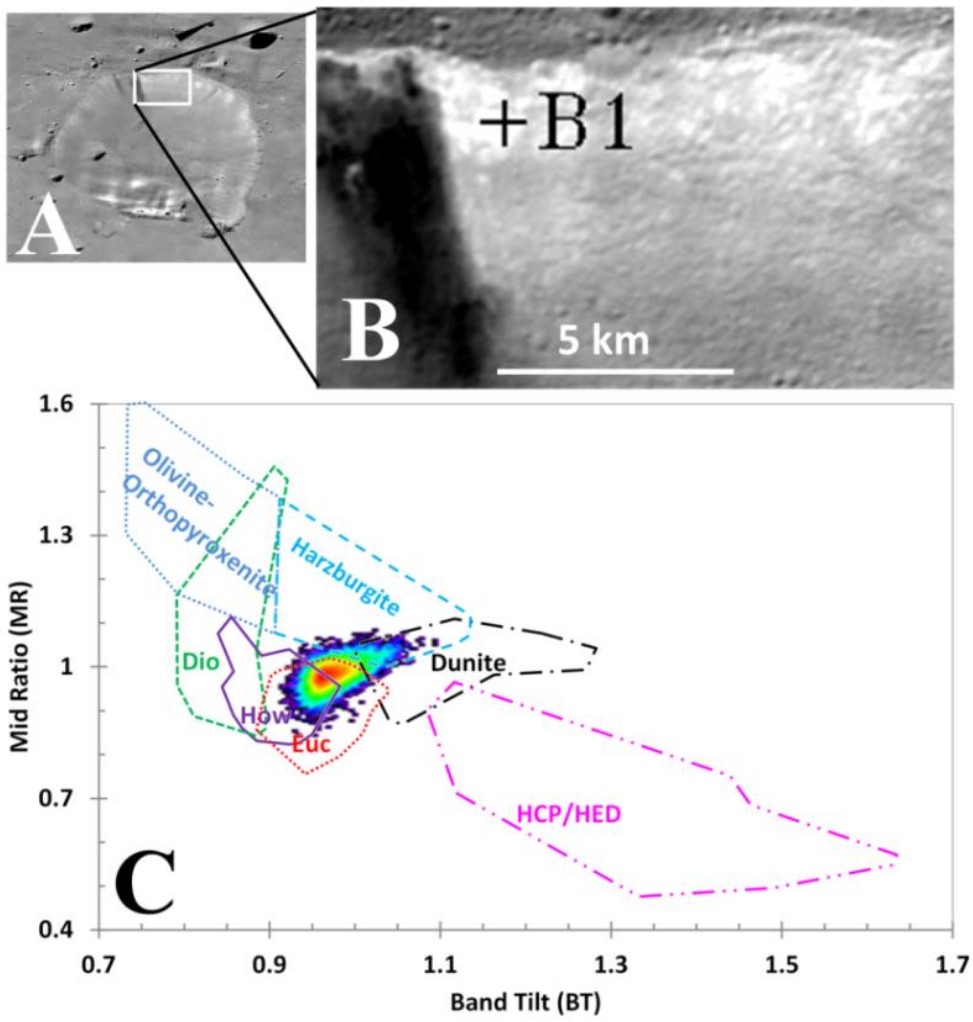

Fig. 11: (A) Bellicia crater, and (B) selected region near the olivine-rich exposure B1. (C) Band parameter values of the selected region plotted over the BT-MR polygons.

### 3.4.1 Source of the olivine-rich material:

#### 3.4.1.1 Excavations from a nearby old impact basin

The peridotitic exposures at Bellicia and Arruntia are close to the rim of an old basin identified by Marchi et al. (2012). They interpreted this old basin as one of the largest impact structures in the northern hemisphere. It extends ~180 km across with a relative depth of ~10-15 km. They also suggested another large basin nearby the old basin. Assuming an excavation depth in the order of 10-15 km (Marchi et al. 2012) and a crustal thickness of 15-20 km (McSween et al. 2013a), the olivine-rich materials could be excavates of such a basin followed by recent impacts. Meanwhile, olivine is almost absent in the huge Rheasilvia basin (Nathues et al. 2014a; Ruesch et al. 2014), and it seems to be unlikely that even smaller impacts excavated mantle materials. However, Cheek and Sunshine (2014) suggested that the olivine-rich exposures at Bellicia and Arruntia support shallow crustal origin, probably signifying a late stage serial magmatism. The idea of crustal thickness or density variations as well as the petrogenetic model of serial magmatism on Vesta is strengthened by the recent observations of Dawn geophysical data (Raymond et al. 2014b). On the other hand, De Sanctis et al. (2014) suggested that the olivine-rich exposures at Bellicia's crater wall are hard to explain in terms of crustal pluton origin.

#### 3.4.1.2 Exogenic origin

The olivine-rich exposures in the northern hemisphere could be of exogenic origin, delivered by an olivine-rich impactor. The survival of impactor remnants particularly in oblique impacts coupled with low velocity is possible (Bland et al. 2008; Pierazzo and Melosh 2000; Yue et al. 2013). However, the olivine-rich asteroids (A-type) are rare in the main belt (Burbine et al. 1996; Bus and Binzel 2002; Reddy et al. 2011b; Sanchez et al. 2014). The 'missing mantle problem' in the main belt, i.e. the scarcity of asteroidal bodies having composition similar to mantle materials of differentiated and disrupted bodies (Burbine et al. 1996; Sanchez et al. 2014) is a well-known dilemma. Meanwhile, olivine could be an endogenic lithological component of Vesta, which is believed to be abundant in the mantle, as hypothesized from the geochemical/petrological evolution model (e.g., Mandler and Elkins-Tanton 2013; Ruzicka et al. 1997; Righter and Drake 1997; Sack et al. 1991), and the geophysical and thermal models (e.g., Fu et al. 2012; Gupta and Sahijpal 2010; Formisano et al. 2012). Moreover, olivine is comparatively more susceptible to weathering and alteration than pyroxenes when exposed to solar winds and micrometeorite bombardments (Duffard et al. 2005; Yamada et al. 1999). In accordance with these observations, Ammannito et al. (2013a) argued that the exogenic origin of the olivine-rich exposures is unlikely. However, the exogenous origin of olivine cannot be ruled out, and will be discussed in an upcoming paper.

### 3.4.2 Consequences on the geology of Vesta

Ammannito et al. (2013a) summarized that the finding of olivine-rich exposures at Bellicia and Arruntia in the northern hemisphere and the absence of such materials in the huge basins in the southern hemisphere suggest a more complex evolutionary history compared to pre-Dawn models. Recently, Nathues et al. (2014a) and Ruesch et al. (2014) presented their preliminary observation and mapping of the global distribution of olivine-rich exposures on Vesta using higher spatial resolution FC and VIR HAMO data, respectively. Many of their identified olivine sites are in the northern hemisphere, and only a few are in the Rheasilvia basin. The lack of olivine in the Rheasilvia basin is in contrary to what was expected. However, Beck et al. (2013) suggested that an upper limit of olivine abundance of 30% can be expected on the surface of Vesta within ~60 m exposures (6-16% olivine abundance as more realistic). Furthermore, olivine abundance in diogenites is quite heterogeneous ranging from ~0 to >90 vol.-% (e.g., Bowman et al. 1997; Beck and McSween 2010; Tkalcec et al. 2013). Assuming the olivine abundance suggested by Beck et al. (2013) and the heterogeneity of the distribution of olivine in the regolith of Vesta, it might be possible that the FC and the VIR instruments are not able to spectrally detect them (e.g., Beck et al. 2013; Jutzi et al. 2013).

Meanwhile, Mandler and Elkins-Tanton (2013) argued against the excavation of olivine-rich mantle materials by the Rheasilvia impact. They suggested that the excavation depth in the order of 40 km will excavate all the HED lithologies without the olivine-rich mantle materials assuming their model's crustal thickness of 30-41 km. The non-excavation of olivine-rich mantle materials during the Rheasilvia (and Veneneia impact) was also opined by Jutzi et al. (2013) as an alternative reason to explain the lack of olivine in Rheasilvia. They suggested that Vesta might have a thicker eucritic crust (~100 km) with ultramafic (diogenitic) inclusions.

Contrarily, the chondritic model for Vesta's origin and evolution proposed by Toplis et al. (2013) predicts a relatively orthopyroxene-rich mantle, which is further supported by the observations from Dawn (e.g., Prettyman et al. 2013; Yamashito et al. 2013; Park et al. 2014). Fe abundances (Yamashito et al. 2013) and thermal neutron absorptions (Prettyman et al. 2013) in the Rheasilvia basin and its ejecta observed from the Gamma Ray and Neutron Detector (GRaND) indicate that orthopyroxene-rich lithologies are the excavated materials by the Rheasilvia impact. McSween et al. (2013b) also suggested that the Rheasilvia impact is supposed to excavate the mantle materials, and therefore the occurrence of diogenites in this basin floor (observed from FC and VIR) implies that the mantle materials appear to be excavated and mixed in the ejecta blanket extending across almost half the Vestan surface. McSween et al. (2014) further predicted an olivine-free upper mantle of Vesta because of the lack of spectrally detectable olivine in the Rheasilvia basin.

The evolution of Vesta by serial magmatism (shallow magmatic plutons) is fostered to explain the olivine-rich exposures in the northern hemisphere (Ammannito et al. 2013a; Cheek and Sunshine 2014; Ruesch et al. 2014). The thermo-chemical evolution model of Neumann et al. (2014) predicted the possibility of a shallow magma ocean on Vesta. Again, Barrat and Yamaguchi (2014) argue that the recent magma ocean model proposed by Mandler and Elklins-Tanton (2013) fails to explain the diversity of trace elements observed in diogenites. They suggested that the most likely explanation for the diversity in trace elements in diogenites is by multiple parental melts on Vesta, but not by a magma ocean model. Ruesch et al. (2014) proposed a local enrichment of olivine on Vesta to explain some of the unusual distributions of olivine-rich exposures in the northern hemisphere. De Sanctis et al. (2014) noted that the apparent absence of olivine in Rheasilvia is probably due to the heterogeneity of the Vestan crustal-mantle depths. The observations from Dawn geophysical data reveal significant gravity anomalies that may reflect crustal thickness and/or density variations on Vesta (Raymond et al., 2014b). The observed gravity anomaly and the density variations strengthen the idea of the heterogeneity in the primordial crust and mantle of Vesta favoring multiple plutons within the deep crust or upper mantle (Raymond et al. 2014a, b; Park et al. 2014). Raymond et al. (2014a, b) also suggested that the olivine-rich exposures of Arruntia and Bellicia are part of a northward extension of a strong positive anomaly observed on the eastern equatorial troughs.

Despite all these complexities, the finding of solid-state plastic deformation in olivine grains of diogenite NWA 5480 (57% olivine) and diogenite NWA 5784 (92% olivine) show that they are likely formed in the mantle of Vesta (Tkalcec et al. 2013; Tkalcec and Brenker 2014). Again, Lunning et al. (2014) claimed that the Mg-rich olivine grains found in paired

GRO 95 howardites are also of potential Vesta mantle origin. Such findings strengthen the idea of an olivine-rich mantle, which are in accordance with various evolution models (e.g., Ruzicka et al. 1997; Righter and Drake 1997; Sack et al. 1991; Mandler and Elkins-Tanton 2013). The finding of olivine grains in HEDs, which are typical of mantle origin, signifies that the mantle materials were excavated and/or exposed on the surface somehow during Vesta's geological history.

## 3. 3 Conclusion

Dawn FC data can be used to distinguish olivine-rich materials from the howardite-eucrite-diogenite (HED) lithologies on Vesta, despite the FC's limited wavelength range, covering about half of the pyroxene/olivine 1-µm absorption band. Our results are consistent with the recent findings from the VIR data by Ammannito et al. (2013a). The olivine-rich exposures at these localities are mapped using higher spatial resolution Dawn FC color data. The olivine exposures are located on the slopes of the inner/outer crater walls, floor of Arruntia, ejecta, and nearby young small impact craters. The identified sites are potential olivine-rich exposures with modal olivine contents of more than 60%. Both craters are located in howarditic/eucritic background. The recent observation and mapping of globally-distributed potential olivine-rich exposures, including some regions in the Rheasilvia basin (Nathues et al. 2014a; Ruesch et al. 2014) could be a key to better understand the source of the olivine-rich exposures, and consequently the geological evolution of Vesta. However, it remains unsolved why olivine-rich mantle materials were not excavated and/or detected in the Rheasilvia basin as predicted by various models discussed above. Therefore, it calls for in-depth observation and understanding of the composition of the Vestan lithology on a global and local scale from Dawn mission datasets, integrating information from HEDs and the evolution models of Vesta.

## 3.4 Acknowldegment

We acknowledge the entire Dawn mission team, the RELAB and USGS spectral library. E.A.C. thanks the University of Winnipeg, the Canadian Space Agency (CSA), the Canada Foundation for Innovation and the Manitoba Research Innovations Fund for establishment of HOSERLab, and CSA and the Natural Science and Engineering Research Council of Canada (NSERC) for supporting the laboratory spectral data acquisitions and analysis. T.G. is grateful to International Max-Planck Research School, Max-Planck Institute for Solar System Research Institute (IMPRS/MPS). We would like to thank the reviewers (P. Harderson and an anonymous reviewer) and A.E. (Carle Pieters) for their constructive comments and suggestions.

## 3.5 Appendix

1. Among HEDs, few spectra with the RELAB file names CBMS48, CCMS48, CBMS49, CCMS49, C3MS52, L3MS52, L6MS52, C9MS52, L9MS52, CAMS52, LAMS52, C1MS52, L1MS52, C4MS52, L4MS52, C7MS52, L7MS52, CATB20, LATB20, C1TB27, LATB27, MGP023, MGP025, MGP027, MGP029, MGP031, MGP033, MGP035, MGP037, MGP039, MGP041, MGP043, MGP045, MGP047, MGP049, MGP051, MGP053, MGP055, MGP057, MGP059, MGP061, MGP063, MGP065, MGP067, MGP069, CAMP76, CAMP70, CAMP84, CAMP71 are not considered in our analyses. i) Laser irradiated spectra for Millbillillie eucrite (MS-JTW-048-B/CBMS48, MS-JTW-048-C/CCMS48, MS-JTW-052-3/C3MS52, MS-JTW-052-3/L3MS52, MS-JTW-052-6/L6MS52, MS-JTW-052-9/C9MS52, MS-JTW-052-9/L9MS52, MS-JTW-052-0/CAMS52, MS-JTW-052-0/LAMS52, MS-JTW-052-1/C1MS52, MS-JTW-052-1/L1MS52, MS-JTW-052-4/C4MS52, MS-JTW-052-4/L4MS52, MS-JTW-052-7/C7MS52, MS-JTW-052-7/L7MS52) and Johnstown diogenite (MS-JTW-049-B/CBMS49, MS-JTW-049-C/CCMS49) are excluded because the impulse laser treatment yields quite different alteration products, and after all, the correlation of such experimental results with the observed spectra of HEDs are poorly understood (Wasson et al. 1997, 1998). On the other hand, Vesta exhibits a distinctive style of space weathering that differs from other airless bodies since there is no evidence for (lunar-like) nano-phase iron on its regolith (Pieters et al. 2012). It is worth mentioning that if we include all these irradiated spectra, the BT-MR and BT-MC polygons don't change significantly (Fig. 2). ii) The Macibini clast 3 melted/quenched spectra TB-RPB-027/C1TB27 and TB-RPB-027/LATB30 are excluded since the spectra look quite unusual and are significantly different from Macibini clast 3 (TB-RPB-027/C1TB27 and TB-RPB-027/LATB27). iv) C1TB07 and LATB07 (TB-RPB-007; Y792510, ≤1000 μm), C1TB15 and LATB15 (TB-RPB-015; ALHA85001, ≤1000 μm) are excluded because of inconsistencies observed in the individual measurements of the same sample. iii) 24 datasets with the index MGP- are excluded because spectra are not wavelength corrected; instead the corresponding wavelength corrected spectra having file names with the index CGP- are used. iv) The HED samples (EET90020 eucrite (MP-TXH-076-A/CAMP76, TB-RPB-020/C1TB20/LATB20, Cachari eucrite (MP-TXH-084-A/CAMP84), Petersburg howardite (MP-TXH-070-A/CAMP70), and ALHA77256 diogenite (MP-TXH-071-A/CAMP71) which are likely weathered/rusted are excluded.

The spectra used in this analysis are EET87503 (CAMB68, LAMB68A, NCMB68A, CBMB68, CCMB68, CDMB68, CEMB68, CFMB68, CGMB68, CHMB68), Kapoeta (C1MP53, LAMP53, CAMP53), GRO95535 (CAMP67), QUE94200 (CAMP69), EET83376 (CAMP73), EET87513 (CAMP74), Binda (CAMP82), Bununu (CAMP83), Frankfort (CAMP85, CGP049), Le Teilleul (CAMP93, CGP051),Y-7308 (CAMP97), Y-790727 (CAMP98), Y-791573 (CAMP99), GRO95574 (BKR1MP125, C1MP125), QUE97001 (BKR1MP126, C1MP126), Pavlovka (CGP047), Petersburg (CGP053, CGP055), EETA79002, (CAMB67, CBMB67, CCMB67, CDMB67, CEMB67, CFMB67), Y-74013 (CAMB73, CBMB73), Y-75032 (CAMB74, CBMB74), Johnstown (CAMB95, LAMB95A, CBMB95, CGP057, CGP059, CGP061, CAMS49, C1MS51), Ellemeet (BKR1MP112, C1MP112, BKR1MP113, C1MP113), LAP91900 (CAMP77, C1TB18, LATB18), Aioun el Atrouss (CAMP81), Tatahouine (CAMP88, CGP065, CGP067), A-881526 (CAMP95), Roda (CGP063), Shalka (CGP069), GRO95555 (CAMP68), ALHA76005 (CAMB66, CBMB66, CCMB66, CDMB66, CEMB66, CFMB66, C1TB23, LATB23, C1TB24, LATB24), Millbillillie (C1HH03, C1MB69, C2MB69, C3MB69, CAMB69, LAMB69A, CBMB69, CCMB69, CDMB69, CAMS48, C1MS50, C1RK116A, C1RK116F2, C1RK116G, C1RK116I, C1RK116L), Juvinas (C1MB70, C2MB70, CAMB70, CBMB70, CCMB70, CDMB70, CEMB70, CGP035, CGP037), Y-74450 (CAMB71, CBMB71, CCMB71, CDMB71), ALH-78132 (CAMB72, CBMB72, CCMB72), Padvarninkai (CAMB96, CBMB96, CCMB96, CDMB96, CGP025), Stannern (CAMB97, CBMB97, CGP039, CGP041, CGP043), ALH85001 (CDMB99, CWMB99), Moore County (CAMP86, CAMP86), Pasamonte (CAMP87, CGP033), Bereba (CAMP89, CGP023), Bouvante (CAMP90, C1TB28, LATB28, C1TB29, LATB29, BKR1TB118, C1TB118), Jonzac (CAMP91, CGP029), Serra de Mage (CAMP92), A-881819 (CAMP96), Sioux County (CGP027), Haraiya (CGP031), Nobleborough (CGP045), EET87520 (C1MT29), PCA91078 (C1MT31), Y-792510 (CAMT41), Y-792769 (CAMT42), Y-793591 (CAMT43), Y-82082 (CAMT44), Macibini (C1TB27, LATB27), GRO95533 (CAMP66), PCA82501 (C1TB12, LATB12, BKR1MP124, C1MP124), PCA82502 (C1TB21, LATB21, CAMP80), ALHA85001 (C1TB15, LATB15), ALHA81011 (C1TB14, LATB14, BKR1MP122,

C1MP122), ALHA81001 (BKR1MP121, C1MP121, BKR1MT030, C1MT30), LEW87004 (C1TB19, LATB19, CAMP79), Y-75011 (C1TB08, LATB08), EETA79005 (CAMP72, C1TB26, LATB26), EETA79006 (BKR1MP123, C1MP123), LEW85303 (CAMP78), EET83251 (C1TB22, LATB22), EET92003 (BKR1MP118, C1MP118), PCA91006 (BKR1MP119, C1MP119), PCA91007 (BKR1MP120, C1MP120, C1TB16, LATB16), EET87542 (CAMP75, C1TB14, LATB14), Y-791186 (C1TB09, LATB09), A-87272 (CAMP94).

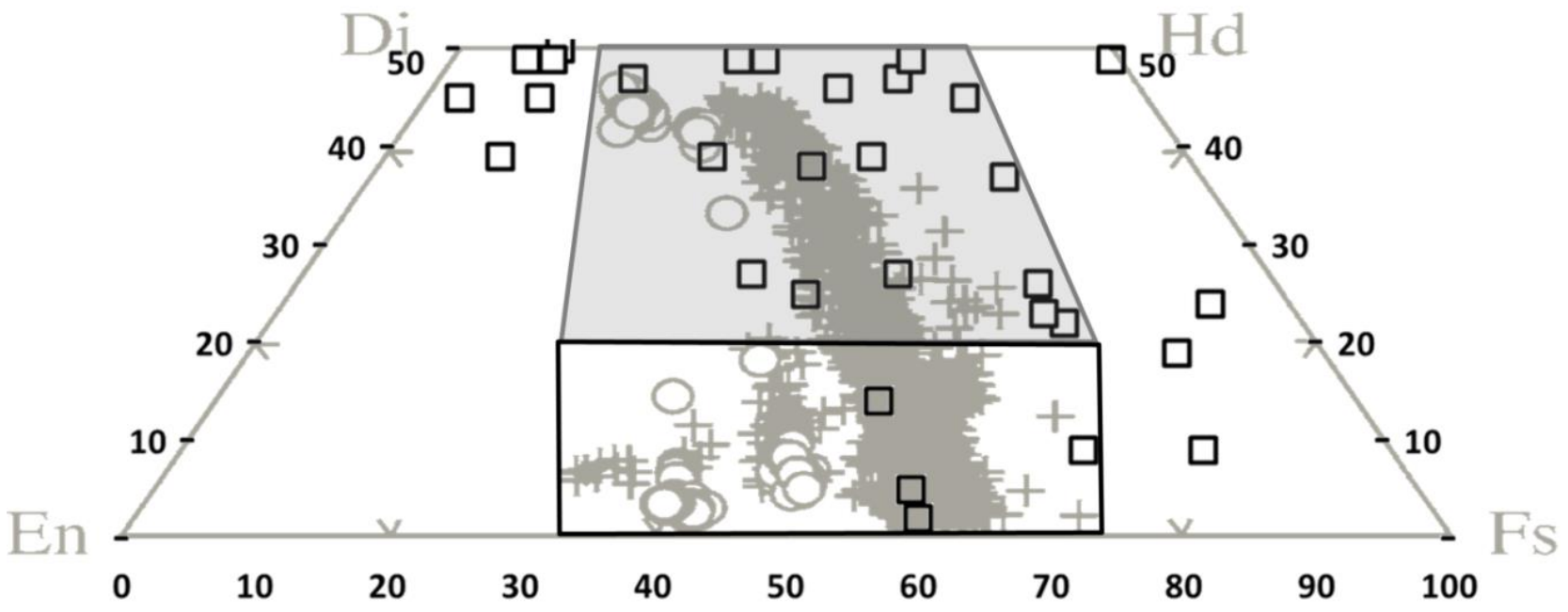

Fig. A: Pyroxene composition in eucrites (open circles/plus), adapted from McSween et al. (2011) and Mayne et al. (2009). Markers in open squares are synthetic pyroxenes (Klima et al. 2011). We restrict high-Calcium pyroxene compositions (HCP/HED) to Wo- contents above 20 mol-%. The gray box indicates those clinopyroxenes considered in our analysis, while open box represents clinopyroxene compositions below 20 mol-% Wo.

Out of 46 spectra of synthetic low/high Ca-pyroxenes ($Wo_{2-51}$, <45 µm) from RELAB (Klima et al. 2011), we selected samples that have reasonable compositions in the contexts of geology of Vesta. i) We consider the calcium bearing pyroxenes free of Fe and Mg as unreasonable. Such pyroxenes neither exist in common basaltic and gabbroic rocks on Earth (Deer et al. 1997), nor in HEDs (Mayne et al. 2009; Mittlefehldt et al. 1998, 2012). Besides, the basaltic Vesta surface in general exhibits prominent 1 µm absorption feature that implies ubiquitous presence of mafic mineralogy. Thus, synthetic pyroxenes without Fs- or without En- components ($Wo_2Fs_{98}$, $Wo_9En_{91}$, $Wo_7Fs_{93}$, $Wo_{10}En_{90}$, $Wo_{17}En_{83}$, $Wo_{25}Fs_{75}$, $Wo_{29}Fs_{71}$, $Wo_{30}En_{70}$, $Wo_{46}En_{54}$, $Wo_{35}Fs_{65}$, $Wo_{38}En_{63}$, $Wo_{39}Fs_{61}$, $Wo_{46}En_{54}$, $Wo_{51}Fs_{49}$) are omitted. ii) We follow Sunshine et al. (2004) and Klima et al. (2011) in restricting the high-Calcium clinopyroxene (HCP) compositions to Wo- contents above 20 mol-%. Clinopyroxene compositions below that value are very rare in eucrites and diogenites and probably result from exsolved HCP lamellae from pigeonite hosts (e.g., McSween et al. 2011). This compositional field is marked by the open box in the Figure given below (Fig. A). This Figure (open circles/plus markers) is based on the compilation of pyroxene compositions in eucrites given by Mayne et al. (2009). Comparing the compositional variations of pyroxenes in eucrites with that of the synthetic pyroxenes investigated by Klima et al. (2011) and Sunshine et al. (2004) reveals that a small number of syntheitic pyroxenes plots outside the compositional variation of diogenites and eucrites (indicated by the gray box). iii) All synthetic pyroxenes plotting outside that compositional field ($Wo_{23}En_6Fs_{70}$, $Wo_{39}En_{52}Fs_9$, $Wo_{45}En_{52}Fs_3$, $Wo_{45}En_{46}Fs_9$, $Wo_{49}En_{45}Fs_6$, $Wo_{49}En_{42}Fs_8$, $Wo_{49}En_{43}Fs_8$, and $Wo_{49}En_1Fs_{50}$) are not considered here.

2. The existing nomenclature of diogenites and/or olivine bearing diogenites by different researchers is ambiguous (Sack et al. 1991; Bunch et al. 2010; Beck and Mcsween, 2010; Wittke et al. 2011; Ammannito et al. 2013a; Mandler and Elkins-Tanton 2013). Here, we follow the nomenclature that is in accordance with IUGS system (Streckeisen 1974; Wittke et al. 2011; Mandler and Elkins-Tanton 2013).
3. The error propagations are computed using the statistical formulations given below-

$$\partial|BT| = |BT|\sqrt{\left(\frac{\partial|0.92\mu m|}{|0.92\mu m|}\right)^2 + \left(\frac{\partial|0.96\mu m|}{|0.96\mu m|}\right)^2}$$

$$\partial|MR| = |MR|\sqrt{\left(\frac{\partial|0.75\mu m|}{|0.75\mu m|}\right)^2 + \left(\frac{\partial|0.92\mu m|}{|0.92\mu m|}\right)^2 + 2\left(\frac{\partial|0.83\mu m|}{|0.83\mu m|}\right)^2}$$

$$\partial|MC| = |MC|\sqrt{\left(\frac{\sqrt{(\partial|0.75\mu m|)^2+(\partial|0.92\mu m|)^2}}{|0.75\mu m|+|0.92\mu m|}\right)^2 + \left(\frac{\partial|0.83\mu m|}{|0.83\mu m|}\right)^2}$$

Where, $\partial|\lambda|$ is the standard deviation, and $|\lambda|$ is the value of the band parameters or reflectance at their respective wavelength. The uncertainties of the band parameters for laboratory spectra (BT- 1.4%, MR- 2%, MC- 1.2%) are shown in Figure 2. In the same way, the uncertainty limits of the band parameters for FC spectra are also computed by selecting nearly homogenous areas, based on similar ranges/values of the BT parameter, topography and reflectance. Four such sites are selected, and spectra are collected for various pixel sizes (2 x 2, 3 x 3, 4 x 4, 5 x 5, 6 x 6 and 7 x 7). The average spectra and their standard deviation are used to statistically compute the error propagation for the band parameters. The uncertainty limits for 4 x 4 pixels are selected after examining the trend of the values for consistency. Therefore, we observed 0.79, 0.78, 0.74, 0.81, 0.81, 0.87, 0.84% uncertainties for the seven filters (in ascending center wavelengths of the filters), and then, the uncertainties for the band parameters (BT- 1.19%, MR- 2.02%, MC- 1.02%) are also estimated (Fig. 9).

## 3.6 References

Adams J. B. 1974. Visible and near-infrared diffuse reflectance spectra of pyroxenes as applied to remote sensing of solid objects in solar system. *Journal of Geophysical Research* 79:4829-4836.

Adams J. B. 1975. Interpretation of visible and near-infrared diffuse reflectance spectra of pyroxenes and other rock forming minerals. In: *Infrared and Raman Spectroscopy of Lunar and Terrestrial Materials* (C. Karr, ed.), Academic, New York, pp. 91-116.

Ammannito E., De Sanctis M. C., Palomba E., Longobardo A., Mittlefehldt D. W., McSween H. Y., Marchi S., Capria M. T., Capaccioni F., Frigeri A., Pieters C. M., Ruesch O., Tosi1 F., Zambon F., Carraro F., Fonte S., Hiesinger H., Magni G., McFadden L. A., Raymond C. A., Russell C. T., Sunshine J. M. 2013a. Olivine in an unexpected location on Vesta's surface. *Nature* 504:122-125.

Ammannito E., De Sanctis M. C., Capaccioni F., Capria M. T., Carraro F., ombe J.-P., Fonte S., Frigeri A., Joy S., Longobardo A., Magni G., Marchi S., McCord T. B., McFadden L. A, McSween H. Y., Palomba E., Pieters C. M., Polanskye C. A., Raymond C. A., Sunshine J., Tosi F., Zambon F., and Russell C. T. 2013b. Vestan lithologies mapped by the visual and infrared spectrometer on Dawn. *Meteoritics & Planetary Science* 48:2185-2198.

Anderson J. A., Sides S. C., Soltesz D. L., Sucharski T. L., Becker K. J. 2004. Modernization of the Integrated Software for Imagers and Spectrometers. *35th Lunar and Planetary Science Conference* (abstract#2039).

Barrat J. A.,Yamaguchi T. E., Bunch M., Bohn C., Bollinger C., Ceuleneer G. 2011. Fluid-rock interactions recorded in unequilibrated eucrites. *42nd Lunar and Planetary Science Conference* (abstract#1306).

Barrat J.-A., and Yamaguchi A. 2014. Comment on "The origin of eucrites, diogenites, and olivine diogenites: Magma ocean crystallization and shallow magma processes on Vesta" by B. E. Mandler and L. T. Elkins-Tanton. *Meteoritics & Planetary Science* 49:468-472.

Beck A. W., McSween H. Y. 2010. Diogenites as polymict breccias composed of orthopyroxenite and harzburgite. *Meteoritics & Planetary Science* 47:850-872.

Beck A. W., Mittlefehldt D. W., McSween H. Y., Rumble D., Lee C. -T. A., Bodnar R. J. 2011. MIL 03443, a Dunite from asteroid 4 Vesta: Evidence for its classification and cumulate origin. *Meteoritics & Planetary Science* 48:1133-1151.

Beck A. W., Welten K. C., McSween H. Y., Viviano C. E., Caffee M. W. 2012. Petrologic and textural diversity among the PCA 02 howardite group, one of the largest pieces of the Vestan surface. *Meteoritics & Planetary Science* 47:947-969.

Beck A. W., McCoy T. J., Sunshine J. M., Viviano C. E., Corrigan C. M., Hiroi T., Mayne R. G. 2013. Challenges in detecting olivine on the surface of 4 Vesta. *Meteoritics & Planetary Science* 48:2166-2184.

Beck A. W., Lunning N. G., De Sanctis M. C., Hiroi T., Plescia J., Viviano-Beck C. E., Udry A., Corrigan C. M., McCoy T. J. 2014. A meteorite analog for olivine-rich terrain in unexpected locations on Vesta. *45th Lunar and Planetary Science Conference* (abstract # 2499).

Binzel R. P., Gaffey M. J., Thomas P. C., Zellner B., Storrs A. D., Wells E. N. 1997. Geologic mapping of Vesta from 1994 Hubble Space Telescope images. *Icarus* 128:95-103.

Bland P. A., Artemieva N. A., Collins G. S., Bottke W. F., Bussey D. B. J., Joy K. H. 2008. Asteroids on the moon: projectile survival during low velocity impact. *39$^{th}$ Lunar and Planetary Science Conference* (abstract # 2045).

Boesenberg J. S., and Delaney J. S. 1997. A model composition of the basaltic achondrite planetoid. *Geochimica et Cosmochimica Acta* 61:3205-3225.

Bowman L. E., Spilde M. N., Papike J. J. 1997. Automated EDS modal analysis applied to the diogenites. *Meteoritics & Planetary Science* 32:869-875.

Bunch T. E., Irving A. J., Wittke J. H., Kuehner S. M., Rumble D. III., and Sipiera P. P. 2010. Northwest Africa 5784, Northwest Africa 5968 and Northwest Africa 6157: More Vestan dunites and olivine diogenites. *Meteoritics & Planetary Science* 45:A27 (abstract).

Bunch T. E., Wittke J. H., Rumble D. III., Irving A. J., and Reed B. 2006. Northwest Africa 2968: A dunite from 4 Vesta. *Meteoritics & Planetary Science* 41:A31 (abstract).

Burbine T. H., Meibom A., Binzel R. P. 1996. Mantle material in the main belt: battered to bits? *Meteoritics & Planetary Science* 31:607-620.

Burns R. G. 1970. Mineralogical applications of crystal field theory. Cambridge Univ., London. 224 pp.

Burns R. G. 1993. Origin of electronic spectra of minerals in the visible to near-infrared region. In: C.M., Pieters, and P.A.J., Englert, (Eds.), *Remote Geochemical Analysis: Elemental and Mineralogical Composition*. Cambridge Univ. Press, New York. pp. 3-29.

Bus S. J. and Binzel R. P. 2002. Phase II of the small main-belt asteroid spectroscopic survey: A feature-based taxonomy. *Icarus* 158:146-177.

Carrozzo F. G., Altieri F., Bellucci G., Poulet F., D'Aversa E., Bibring J. -P. 2012. Iron mineralogy of the surface of Mars from the 1 µm band spectral properties. *Journal of Geophysical Research* 117:E00J17, doi:10.1029/2012JE004091.

Cloutis E. A., Gaffey M. J., Jackowski T. L., Reed K. L. 1986. Calibrations of phase abundance, composition and particle size distribution for olivine-orthopyroxene mixtures from reflectance spectra. *Journal of Geophysical Research* 91:11641-11653.

Cloutis E. A., and Gaffey M. J. 1991. Pyroxene spectroscopy revisited: Spectral-compositional correlations and relationship to geothermometry. *Journal of Geophysical Research* 96:22809-22826.

De Sanctis M. C., Coradini A., Ammannito E., Filacchione G., Capria M. T., Fonte S., Magni G., Barbis A., Bini A., Dami M., Ficai-Veltroni I., and Preti G. 2011. The VIR spectrometer. *Space Science Reviews* 163:329-369.

De Sanctis M. C., Ammannito E., Capria M. T., Tosi F., Capaccioni F., Zambon F., Carraro F., Fonte S., Frigeri A., Jaumann R., Magni G., Marchi S., McCord T. B., McFadden L. A., McSween H. Y., Mittlefehldt D. W., Nathues A., Palomba E., Pieters C. M., Raymond C. A., Russell C. T., Toplis M. J., and Turrini D. 2012. Spectroscopic characterization of mineralogy and its diversity across Vesta. *Science* 336:697-700.

De Sanctis M. C., Ammannito E., Palomba E., Longobardo A., Mittlefehldt D. W., McSween H. Y., Marchi S., Capria M. T., Capaccioni F., Frigeri A., Pieters C. M., Ruesch O., Tosi F., Zambon F., Hiesinger H., Magni G., McFadden L. A., Raymond C. A., Russell C. T., Sunshine J. M. 2014. Vesta evolution from surface mineralogy: mafic and ultramafic mineral distribution. *45th Lunar and Planetary Science Conference* (abstract # 1748).

Deer W. A., Howie R. A., Zussman J. 1997. Rock Forming Minerals. Vol. 2A, Single Chain Silicates. Geological Society London.

Delaney J. S., Nehru C. E., Prinz M. 1980. Olivine clasts from mesosiderites and howardites: Clues to the nature of achondrite parent bodies. Proc. *11th Lunar and Planetary Science Conference* 1073-1087.

Dreibus G., and Wänke H. 1980. The bulk composition of the eucrite parent asteroid and its bearing on planetary evolution. *Zeitschrift für Naturforschung* 35a:204-216.

Duffard R., Lazzaro D., De Leon J. 2005. Revisiting spectral parameters of silicate-bearing meteorites. *Meteoritics & Planetary Science* 40:445-459.

Feierberg M. A., Larson H. P., Fink U., Smith H. A. 1980. Spectroscopic evidence for two achondrite parent bodies: Asteroids 349 Dembowska and 4 Vesta. *Geochimica et Cosmochimica Acta* 44:513-524.

Floran R. J., Prinz M., Hlava P. F., Keil K., Spettel B., Wänke H. 1981. Mineralogy, petrology, and trace element geochemistry of the Johnstown meteorite: A brecciated orthopyroxenite with siderophile and REE-rich components. *Geochimica et Cosmochimica Acta* 45:2385-2391.

Formisano M., Federico C., Coradini A. 2012. Vesta thermal models. *Memorie della Società Astronomica Italiana Supplementary* 20: 90.

Fowler G. W., Shearer C. K., Papike J. J., Layne G. D. 1995. Diogenites as asteroidal cumulates: Insights from orthopyroxene trace element chemistry. *Geochimica et Cosmochimica Acta* 59:3071-3084.

Fu R. R, Weiss B. P., Shuster D. L., Gattacceca J., Grove T. L., Suavet C., Lima E. A., Li L., Kuan A. T. 2012. An ancient core dynamo in asteroid Vesta. *Science* 338:238-241.

Gaffey M. J. 1983. The asteroid (4) Vesta: Rotational spectral variations, surface material heterogeneity, and implications for the origin of the basaltic achondrites. *24$^{th}$ Lunar and Planetary Science Conference* 231–232 (Abstract).

Gaffey M. J. 1997. Surface lithologic heterogeneity of asteroid 4 Vesta. *Icarus* 127:130-157.

Gaskell R. W. 2012. SPC shape and topography of Vesta from Dawn imaging data. *American Astronomical Society, DPS meeting 44, 209.03.*

Gupta G., Sahijpal S. 2010. Differentiation of Vesta and the parent bodies of other achondrites. *Journal of Geophysical Research* 115:E08001.

Grove T. L., and Bence A. E. 1979. Crystallization kinetics in a multiply saturated basalt magma: An experimental study of Luna 24 ferrobasalt. Proc. *10th Lunar and Planetary Science Conference* 439-478.

Hiroi T., Pieters C. M., Hiroshi T. 1994. Grain size of the surface regolith of asteroid 4 Vesta estimated from its reflectance spectrum in comparison with HED meteorites. *Meteoritics* 29:394-396.

Irving A. J., Bunch T. E., Kuehner S. M., Wittke J. H., Rumbleet D. 2009. Peridotites related to 4 Vesta: Deep crustal igneous cumulates and mantle samples. *15th Lunar and Planetary Science Conference* (abstract # 2466).

Isaacson P. J. and Pieters C. M. 2009. Northern Imbrium Noritic anomaly. *Journal of Geophysical Research* 114:E09007.

Jutzi M., Asphaug E., Gillet P., Barrat J.-A., Benz W. 2013. The structure of the asteroid 4 Vesta as revealed by models of planet-scale collisions. *Nature* 494:207-210.

Keil K. 2002. Geological history of asteroid 4 Vesta: The "Smallest Terrestrial Planet". In *Asteroids III*, ed. by W. Bottke, A. Cellino, P. Paolicchi, R. P. Binzel (University of Arizona Press, Tucson, 2002), pp. 573-584.

King T. V. V., and Ridley I. 1987. Relation of the spectroscopic reflectance of olivine to mineral chemistry and some remote sensing implications. *Journal of Geophysical Research* 92:11457-11469.

Klima R. L., Pieters C. M., Dyar M. D. 2007. Spectroscopy of synthetic Mg-Fe pyroxenes I: Spin allowed and spin-forbidden crystal field bands in the visible and near-infrared. *Meteoritics & Planetary Science* 42:235-253.

Klima R. L., Dyar M. D., Pieters C. M. 2011. Near-infrared spectra of clinopyroxenes: Effects of calcium content and crystal structure. *Meteoritics & Planetary Science* 46:379-395.

Koeppen W. C., and Hamilton V. E. 2008. Global distribution, composition, and abundance of olivine on the surface of Mars from thermal infrared data. *Journal of Geophysical Research* 113:E05001, doi:10.1029/2007JE002984.

Kovacs G., Sierks H., Richards M. L., Gutierrez-Marqués P., Nathues A. 2013. Stray light calibration of the DAWN framing camera. *SPIE Remote Sensing Conference, Proc. SPIE 8889, Sensors, Systems, and Next-Generation Satellites* XVII, 888912.

Larson H. P., and Fink U. 1975. Infrared spectral observations of asteroid 4 Vesta. *Icarus* 26:420-427.

Le Corre L., Reddy V., Nathues A., Cloutis E.A. 2011. How to characterize terrains on 4 Vesta using Dawn Framing Camera color bands? *Icarus* 216:376-386.

Le Corre L., Reddy V., Schmedemann N., Becker K. J., O'Brien D. P., Yamashita N., Peplowski P. N., Prettyman T. H., Li J-Y., Cloutis E.A., Denevi B., Kneissl T., Palmer E., Gaskell R., Nathues A., Gaffey M. J., Garry B., Sierks H., Russell C. T., Raymond C.

2013. Olivine vs. Impact Melt: Nature of the Orange Material on (4) Vesta from Dawn Observations. *Icarus* 226:1568–1594.

Li J.-Y., McFadden L. A., Thomas P. C., Mutchler M. J., Parker J. W., Young E. F., Russell C. T., Sykes M. V., Schmidt B. E. 2010. Photometric mapping of asteroid (4) Vesta's southern hemisphere with Hubble Space Telescope. *Icarus* 208:238-251.

Lodders K. 2000. An oxygen isotope mixing model for the accretion and composition of rocky planets. *Space Science Reviews* 92:341-354.

Lunning N. G., McSween H. Y., Tenner T. J., and Kita N. T. 2014. Olivine from the mantle of 4 Vesta identified in howardites. *48th Lunar and Planetary Science Conference* (abstract#1921).

Mandler B. E., and Elkins-Tanton L. T. 2013. The origin of eucrites, diogenites, and olivine diogenites: Magma ocean crystallization. *Meteoritics & Planetary Science* 48:2333-2349.

Marchi S., McSween H. Y., O'Brien D. P., Schenk P., De Sanctis M. C., Gaskell R., Jaumann R., Mottola S., Preusker F., Raymond C. A., and Russell C. T. 2012. The violent collisional history of asteroid 4 Vesta. *Science* 336:690-694.

Mason B. 1962. *Meteorites* (Wiley, New York) pp. 274.

Mason, B., 1967. Meteorites. American Scientist, 55, 4, 429-455.

Mason, B., 1983. The definition of a howardite. Meteoritics, 18: 245. doi: 10.1111/j.1945-5100.1983.tb00825.x.

Mayne, R.G., McSween, H. Y., McCoy, T. J., Gale, A., 2009. Petrology of the unbrecciated eucrites. *Geochimica et Cosmochimica Acta* 73:794-819.

Mayne R. G., Sunshine J. M., McSween H. Y., McCoy T. J., Corrigan C. M., Gale A. 2010. Petrologic insights from the spectra of the unbrecciated eucrites: Implications for Vesta and basaltic asteroids. *Meteoritics & Planetary Science* 45:1074-1092.

McCord T. B., Adams J. B., Johnson T. V. 1970. Asteroid Vesta: Spectral reflectivity and compositional implications. *Science* 178:745-747.

McFadden L. A., Pieters C. M., McCord T. B. 1977. Vesta - The first pyroxene band from new spectroscopic measurements. *Icarus* 31:439-446.

McSween H. Y., Mittlefehldt D. W., Beck A. W., Mayne R. G., McCoy T. J. 2011. HED meteorites and their relationship to the geology of Vesta and the Dawn mission. *Space Science Reviews* 163:141-174.

McSween H. Y., Ammannito E., Reddy V., Prettyman T. H., Beck A. W., De Sanctis M. C., Nathues A., Le Corre L., O'Brien D. P., Yamashita N., McCoy T. J., Mittlefehldt D. W.,

Toplis M. J., Schenk P., Palomba E., Tosi F., Zambon F., Longobardo A., Capaccioni F., Raymond C. A., and Russell C. T. 2013a. Composition of the Rheasilvia basin, a window into Vesta's interior. *Journal of Geophysical Research* 118:335-346.

McSween H. Y. Binzel R. P., DeSanctis M. C., Ammannito E., Prettyman T. H., Beck A. W., Reddy V., Le Corre L., Gaffey M. J., McCord T. B., Raymond C. A., and Russell C. T. 2013b. Dawn; the Vesta-HED connection, and the geologic context for eucrites, diogenites, and howardites. *Meteoritics & Planetary Science* 48:2090-2104.

McSween H. Y., Mittlefehldt D. W., and the Dawn Science Team. 2014. Vesta in the light of Dawn, but without HEDs?. *Vesta in the light of Dawn: First exploration of a protoplanet in the asteroid belt* (abstract # 2016).

Migliorini F., Morbidelli A., Zappala V., Gladman B. J., Bailey M. E., Cellino A. 1997. Vesta fragments from υ6 and 3:1 resonances: Implications for V-type near-Earth asteroids and howardite, eucrite and diogenite meteorites. *Meteoritics & Planetary Science* 32:903-916.

Mikouchi T., and Miyamoto M. 1997. Forsteritic olivines from angrites and howardites. *28th Lunar and Planetary Science Conference* (abstract # 1733).

Mittlefehldt D. W. 1994. The genesis of diogenites and HED parent body petrogenesis. *Geochimica et Cosmochimica Acta* 58:1537-1552.

Mittlefehldt D. W., McCoy T. J., Goodrich C. A., Kracher A. 1998. Non-chondritic meteorites from asteroidal bodies. In: *Planetary Materials*, vol. 36 (ed. J. J. Papike). Mineralogical Society of America, Chantilly, Virginia, pp. 4-1 to 4-195.

Mittlefehldt D. W., Beck A. W., Lee C.-T. A., McSween H. Y., and Buchanan P. C. 2012. Compositional constraints on the genesis of diogenites. *Meteoritics & Planetary Science* 47:72–98.

Moroz L., Schade U., Wäsch R. 2000. Reflectance spectra of olivine-orthopyroxene bearing assemblages at decreased temperatures: Implications for remote sensing of asteroids. *Icarus*147:79-93.

Nathues A. 2000. Spectroscopic study of Eunomia asteroid family. Ph.d. dissertation, University of Berlin, Germany.

Nathues A., Hoffmann M., Schäfer M., Thangjam G., Reddy V., Cloutis E. A., Mengel K., Christensen U., Sierks H., Le Corre L., Vincent J. B., Russell C. T., Raymond C. 2014a. Distribution of potential olivine sites on the surface of Vesta by Dawn FC. *45$^{th}$ Lunar and Planetary Science Conference* (abstract # 1740).

Nathues A., Hoffmann M., Cloutis E. A., Schäfer M., Reddy V., Christensen U., Sierks H., Thangjam G. S., Le Corre L., Mengel K., Vincent J.-B., Russell C. T., Prettyman T., Schmedemann N., Kneissl T., Raymond C., Gutierrez-Marques P., Hall I. Büttner I. 2014b. Detection of serpentine in exogenic carbonaceous chondrite material on Vesta from Dawn FC data. *Icarus* (in press, doi: 10.1016/j.icarus.2014.06.003).

Neumann W., Breuer D., and Spohn T. 2014. Differentiation of Vesta: Implications for a shallow magma ocean. *Earth & Planetary Science Letters* 395:267-280.

Palomba E., De Sanctis M. C., Ammannito E., Capaccioni F., Capria M. T., Farina M., Frigeri A., Longobardo A., Tosi F., Zambon F., McSween H. Y., Mittlefehldt D. W., Russell C. T., Raymond C. A., Sunshine J., McCord T. B. 2012a. Search for olivine spectral signatures on the surface of Vesta. *EGU* vol. 14, EGU2012-10814-1 (abstract).

Palomba E., Longobardo A., De Sanctis M. C., Ammannito E., Capaccioni F., Capria M. T., Farina M., Frigeri A., Tosi F., Zambon F., McSween H. Y., Mittlefehldt D. W., Russell C. T., Raymond C. A., Sunshine J., McCord T. B., McCoy T., Beck A., Toplis M. J., McFadden L., Cloutis E. 2012b. Search for olivine in the Vesta surface from the Dawn-VIR hyperspectral data. *European Planetary Science Congress* Vol. 7 (abstract # EPSC2012-652-1).

Palomba E., Longobardo A., De Sanctis M. C., Ammannito E., Capaccioni F., Capria M. T., Frigeri A., Tosi F., Zambon F., Russell C. T., Raymond C. A., Cloutis E. 2013a. Distribution of olivine rich materials on Vesta: application of spectral indices to the VIR-Dawn data. *European Planetary Science Congress* vol. 8, (abstract # EPSC2013-604-2).

Palomba E., Longobardo A., De Sanctis M. C., Ammannito E., Capaccioni F., Capria M. T., Frigeri A., Tosi F., Zambon F., Russell C. T., Raymond C. A., Toplis M. J., Cloutis E. 2013b. Calibration of spectral indexes suitable for olivine detection on Vesta. *44$^{th}$ Lunar and Planetary Science Conference* (abstract#1922).

Park R. S., Konopliv A. S., Asmar S. W., Bills B. G., Gaskell R. W., Raymond C. A., Smith D.E., Toplis M. J., Zuber M. T. 2014. Gravity field expansion in ellipsoidal harmonic and polyhedral internal representations applied to Vesta. *Icarus* (in press, doi: 10.1016/j.icarus.2013.12.005).

Pelkey S. M., Mustard J. F., Murchie S., Clancy R. T., Wolff M., Smith M., Milliken R.,. Bibring J.-P, Gendrin A., Poulet F., Langevin Y., Gondet B. 2007. CRISM multispectral summary products: Parameterizing mineral diversity on Mars from reflectance. *Journal of Geophysical Research* 112:E08S14, doi:10.1029/2006JE002831.

Pierazzo E., and Melosh H. J. 2000. Hydrocode modeling of oblique impacts: The fate of the projectile. *Meteoritics & Planetary Science* 35:117-130.

Pieters C. M. 1986. Composition of the lunar highland crust from near infrared spectroscopy. *Rev. Geophys.* 24:557-578.

Pieters C. M., Gaddis L., Jolliff B., Duke M. 2001. Rock types of South Pole-Aitken basin and extent of basaltic volcanism. *Journal of Geophysical Research* 106:28001-28022.

Pieters C. M., Besse S., Boardman J., Buratti B., Cheek L., Clark R. N., Combe J. P., Dhingra D., Goswami J. N., Green R. O., Head J. W., Isaacson P., Klima R., Kramer G., Lundeen S., Malaret E., McCord T., Mustard J., Nettles J., Petro N., Runyon C., Staid M., Sunshine J., Taylor L. A., Thaisen K., Tompkins S., Whitten J. 2011. Mg-spinel lithology: A new rock type on the lunar farside. *Journal of Geophysical Research* 116:E00G08, doi:10.1029/2010JE003727.

Pieters C. M., Ammannito E., Blewett D. T., Denevi B. W., De Sanctis M. C., Gaffey M. J., Le Corre L., Li J.-Y., Marchi S., McCord T. B., McFadden L. A., Mittlefehldt D. W., Nathues A., Palmer E., Reddy V., Raymond C. A., Russell C. T. 2012. Distinctive space weathering on Vesta from regolith mixing processes. *Nature* 491:79-82.

Poulet F., Gomez C., Bibring J.-P., Langevin Y., Gondet B., Belluci G., Mustard J. and the Omega Team 2007. Martian surface mineralogy from OMEGA/Mars Express: Global mineral maps. *Journal of Geophysical Research* E08S02, doi:10.1029/2006JE002840.

Prettyman T. H., Mittlefehldt D. W., Yamashita N., Beck A. W., Feldman W. C., Hendricks J. S., Lawrence D. J., McCoy T. J., McSween H. Y., Peplowski P. N., Reedy R. C., Toplis M. J., Le Corre L., Mizzon H., Reddy V., Titus T. N., Raymond C. A., and Russell C. T. 2013. Neutron absorption constraints on the composition of 4 Vesta. *Meteoritics & Planetary Science* 48:2211-2236.

Raymond C. A., Park R. S., Konopliv A. S., Asmar S. W., Jaumann R., McSween H. Y., De Sanctis M. C., Ammannito E., Buczkowski D. L., Russell C. T., Smith D. E., Toplis M. J., Zuber M. T. 2014a. Constraints on Vesta's interior evolution from Dawn geophysical data. *Vesta in the light of Dawn: First exploration of a protoplanet in the asteroid belt* (abstract#2033).

Raymond C. A., Park R. S., Konopliv A. S., Asmar S. W., Jaumann R., McSween H. Y., De Sanctis M. C., Ammannito E., Buczkowski D. L., Russell C. T., Smith D. E., Toplis M. J., Zuber M. T. 2014b. Geophysical constraints on the structure and evolution of Vesta’s crust and mantle. *45th Lunar and Planetary Science Conference* (abstract#2214).

Reddy V., Gaffey M. J., Kelley M. S., Nathues A., Li J-Y., Yarbrough R. 2010. Compositional heterogeneity of asteroid 4 Vesta's southern hemisphere: Implications for the Dawn mission. *Icarus* 210:693-706.

Reddy V., Nathues A., Gaffey M. J. 2011a. Fragment of asteroid Vesta's mantle detected. *Icarus* 212:175-179.

Reddy V., Nathues A., Gaffey M. J., Schaeff S. 2011b. Mineralogical characterization of potential targets for the ASTEX mission scenario. *Planetary and Space Science* 59: 772-778.

Reddy V., Sanchez J. A., Nathues A., Moskovitz N. A., Li J-Y., Cloutis E. A., Archer K., Tucker R. A., Gaffey, M. J., Mann J. P., Sierks H., Schade U. 2012a. Photometric, spectral phase and temperature effects on Vesta and HED meteorites: Implications for Dawn mission. *Icarus* 217:153-168.

Reddy V., Nathues A., LeCorre L., Sierks H., Li J.-Y., Gaskell R., McCoy T., Beck A. W., Schroeder S. E., Pieters C. M., Becker K. J., Buratti B. J., Denevi B., Blewett D. T., Christensen U., Gaffey M. J., Gutierrez-Marques P., Hicks M., Keller H. U., Maue T., Mottola S., McFadden L. A., McSween H. Y., Mittlefehldt D., O'Brien D. P., Raymond C., and Russell C. T. 2012b. Color and albedo heterogeneity of Vesta from Dawn. *Science* 336:700-704.

Reddy V., Li J.-Y., Le Corre L., Scully J. E., Gaskell R., Russell C. T., Park R. S., Nathues A., Raymond C. R., Gaffey M. J., Becker K. J., McFadden L. A. 2013. Comparing Dawn, Hubble Space Telescope and ground-based interpretations of (4) Vesta. *Icarus* 226:1103-1114.

Righter K., Drake M. J. 1997. A magma ocean on Vesta: Core formation and petrogenesis of eucrites and diogenites. *Meteoritics & Planetary Science* 32:929-944.

Ruesch O., Hiesinger H., De Sanctis M. C., Ammannito E., Palomba E., Longobardo A., Capria M. T., Capaccioni F., Frigeri A., Tosi F., Zambon F., Fonte S., Magni S., Raymond C. A., Russell C. T. 2014. Distribution of the near-IR spectral signature of olivine on Vesta with VIR/Dawn data: the ultramafic side of Vesta's surface. *45th Lunar and Planetary Science Conference* (abstract # 1715)

Ruesch O., Hiesinger H., De Sanctis M. C., Ammannito E., Palomba E., Longobardo A., Capria M. T., Capaccioni F., Frigeri A., Tosi F., Zambon F., Fonte S., Magni S., Raymond C. A., Russell C. T. 2013. Seeing through pyroxene: distribution of olivine-bearing material on Vesta using VIR/Dawn data. *EPSC* vol. 8, (abstract # 891).

Russell C. T., Raymond C. A., Coradini A., McSween H. Y., Zuber M. T., Nathues A., De Sanctis M. C., Jaumann R., Konopliv A. S., Preusker F., Asmar S. W., Park R. S., Gaskell R., Keller H. U., Mottola S., Roatsch T., Scully J. E. C., Smith D. E., Tricarico P., Toplis M. J., Christensen U. R., Feldman W. C., Lawrence D. J., McCoy T. J., Prettyman T. H., Reedy R. C., Sykes M. E., and Titus T. N. 2012. Dawn at Vesta: testing the protoplanetary paradigm. *Science* 336:684-686.

Russell C. T., Raymond C. A., Jaumann R., McSween H. Y., De Sanctis M. C., Nathues A., Prettyman T. H., Ammannito E., Reddy V., Preusker F., O'Brien D. P., Marchi S., Denevi B. W., Buczkowski D. L., Pieters C. M., McCord T. B., Li J.-Y., Mittlefehldt D. W., Combe J.-P., Williams D., Hiesinger H., Yingst R. A., Polanskey C. A., and Joy S. P. 2013. Dawn completes its mission at 4 Vesta. *Meteoritics & Planetary Science* 48:2076-2089.

Ruzicka A., Snyder G. A., Taylor L. A. 1997. Vesta as the howardite, eucrite and diogenite parent body: Implications for the size of a core and for large-scale differentiation. *Meteoritics & Planetary Science* 32:825-840.

Sack R. O., Azeredo W. J., Lipschutz M. E. 1991. Olivine diogenites: The mantle of the eucrite parent body. *Geochimica et Cosmochimica Acta* 55:1111-1120.

Sanchez J. A., Reddy V., Nathues A., Cloutis E. A., Mann P., Hiesinger H. 2012. Phase reddening on near-Earth asteroids: Implications for mineralogical analysis, space weathering and taxonomic classification. *Icarus* 220:36–50

Sanchez J. A., Reddy V., Kelley M. S., Cloutis E. A., Bottke W. F., Nesvorný D., Lucas M. P., Hardersen P. S., Gaffey M. J., Abell P. A., Le Corre L. 2014, Olivine-dominated asteroids: Mineralogy and origin. *Icarus* 228:288-300.

Schade U., Wäsch R., Moroz L. 2004. Near-infrared reflectance spectroscopy of Ca-rich clinopyroxenes and prospects for remote spectral characterization of planetary surfaces. *Icarus* 168:80-92.

Schenk P., O'Brien D. P., Marchi S., Gaskell R., Preusker F., Roatsch T., Jaumann R., Buczkowski D., McCord T., McSween H. Y., Williams D., Yingst A., Raymond C., Russell C. T. 2012. The geologically recent giant impact basins at Vesta's south pole. *Science* 336:694-697.

Shearer C. K., Fowler G. W., Papike J. J. 1997. Petrogenetic models for magmatism on the eucrite parent body: Evidence from orthopyroxene in diogenites. *Meteoritics & Planetary Science* 32:877-889.

Shearer C. K., Burger P., and Papike J. J. 2010. Petrogenetic relationships between diogenites and olivine diogenites: Implications for magmatism on the HED parent body. *Geochimica et Cosmochimica Acta* 74:4865-4880.

Shestopalov D. I., McFadden L. A., Golubeva L. F., Khomenko V. M., Gasanova L. O. 2008. Vestoid surface composition from analysis of faint absorption bands in visible reflectance spectra. *Icarus* 195:649–662.

Shestopalov D. I., McFadden L. A., Golubeva L. F., Orujova L. O. 2010. About mineral composition of geologic units in the northern hemisphere of Vesta. *Icarus* 209:575-585.

Sierks H., Keller H. U., Jaumann R., Michalik H., Behnke T., Bubenhagen F., Büttner I., Carsenty U., Christensen U., Enge R., Fiethe B., Gutierrez-Marques P., Hartwig H., Krüger H., Kühne W., Maue T., Mottola S., Nathues A., Reiche K.-U., Richards M. L., Roatsch T., Schröder S. E., Szemerey I., and Tschentscher M. 2011. The Dawn Framing Camera. *Space Science Reviews* 163:263-327.

Singer R. B. 1981. Near-infrared spectral reflectance of mineral mixtures: Systematic combinations of pyroxenes, olivine and iron oxides. *Journal of Geophysical Research* 86:7967-7982.

Stolper E. 1977. Experimental petrology of eucritic meteorites. *Geochimica et Cosmochimica Acta* 41:587-611.

Streckeisen A. 1974. Classification and nomenclature of plutonic rocks. *Geologische Rundschau* 63:773-785.

Sunshine J. M., Bus S. J., Corrigan C. M., McCoy T. J., Burbine T. H. 2007. Olivine-dominated asteroids and meteorites: Distinguishing nebular and igneous histories. *Meteoritics & Planetary Science* 42:155-170.

Sunshine J. M., Bus S. J., McCoy, T. J., Burbine T. H., Corrigan C. M., Binzel R. P. 2004. High-calcium pyroxene as an indicator of igneous differentiation in asteroids and meteorites. *Meteoritics & Planetary Science* 39:1343-1357.

Thangjam G., Reddy V., Le Corre L., Nathues A., Sierks H., Hiesinger H., Li J.-Y., Sanchez J. A., Russell C. T., Gaskell R., Raymond C. 2013. Lithologic mapping of HED terrains on Vesta using Dawn Framing Camera color data. *Meteoritics & Planetary Science* 48:2199-2210.

Thangjam G., Nathues A., Mengel K., Hoffmann M., Schäfer M., Reddy V., Cloutis E. A., Christensen U., Sierks H., Le Corre L., Vincent J. B., Russell C. T., Raymond C. 2014.

Olivine rich exposures in Bellicia and Arruntia craters on Vesta using Dawn FC. *45th Lunar & Planetary Science* (abstract # 1755).

Thomas P. C., Binzel R. P., Gaffey M. J., Storrs A. D., Wells E. N., Zellner B. H. 1997. Impact excavation on asteroid 4 Vesta: Hubble Space Telescope results. *Science* 277:1492-1495.

Tkalcec B. J., Golabek G. J., Brenker F. E. 2013. Solid-state plastic deformation in the dynamic interior of a differentiated asteroid. *Nature Geosciences* 6:93-97.

Tkalcec B. J., and Brenker F. E. 2014. Early dynamic mantle movements in young, semi-crystallized Vesta. *Vesta in the light of Dawn: First exploration of a protoplanet in the asteroid belt* (abstract # 2022).

Tompkins S., and Pieters C. M. 1999. Mineralogy of the lunar crust: Results from Clementine. *Meteoritics & Planetary Science* 34:25-41.

Toplis M. J., Mizzon H., Monnereau M., Forni O., McSween H. Y., Mittlefehldt D. W., McCoy T. J., Prettyman T. H., De Sanctis M. C., Raymond C. A., and Russell C. T. 2013. Chondritic models of 4-Vesta: Implications for geochemical and geophysical properties. *Meteoritics & Planetary Science* 48:2300-2315.

Warren P. H. 1997. Magnesium oxide iron oxide mass balance constraints and a more detailed model for the relationship between eucrites and diogenites. *Meteoritics & Planetary Science* 32:945-963.

Wasson J. T., Pieters C. M., Fisenko A. V., Semjonova L. F., and Warren P. H. 1998. Simulation of space weathering of eucrites by laser impulse irradiation. *29th Lunar and Planetary Science Conference* (abstract # 1940).

Wasson J. T., Pieters C. M., Fisenko A. V., Semjonova L. F., Moroz L.V., and Warren P. H. 1997. Simulation of space weathering of HED meteorites by laser impulse irradiation. *28th Lunar and Planetary Science Conference* (abstract # 1730).

Wittke J. H., Irving A. J., Bunch T. E., Kuehner S. M. 2011. A nomenclature system for diogenites consistent with the IUGS system for naming terrestrial ultramafic rocks. *74th Annual Metting of the Meteoritical Society* (abstract # 5233).

Yamada M., Sasaki S., Nagahara H., Fujiwara A., Hasegawa S., Yano H., Hiroi T., Ohashi H., Otake H. 1999. Simulation of space weathering of planet-forming materials: Nanosecond pulse laser irradiation and proton implantation on olivine and pyroxene samples. *Earth Planets Space* 51:1255-1265.

Yamaguchi A., Taylor G. J., Keil K. 1996. Global crustal metamorphic of the eucrite parent body. *Icarus* 124:97-112.

Yamaguchi A., Taylor G. J., Keil K. 1997. Metamorphic history of the eucritic crust of 4 Vesta. *Journal of Geophysical Research* 102:13381-13386.

Yamamoto S., Nakamura R., Matsunaga T., Ogawa Y., Ishihara Y., Morota T., Hirata N., Ohtake M., Hiroi T., Yokota Y., Haruyama J. 2010. Possible mantle origin of olivine around lunar impact basins detected by SELENE. *Nature Geosciences* 3:533-536.

Yamamoto S., Nakamura R., Matsunaga T., Ogawa Y., Ishihara Y., Morota T., Hirata N., Ohtake M., Hiroi T., Yokota Y., Haruyama J. 2012. Olivine-rich exposures in the South Pole-Aitken basin. *Icarus* 218:331-344.

Yamashita N., Prettyman T. H., Mittlefehldt D. W., Toplis M. J., McCoy T. J., Beck A. W., Reedy R. C., Feldman W. C., Lawrence D. J., Peplowski P. N., Forni O., Mizzon H., Raymond C. A., Russell C. T. 2013. Distribution of iron on Vesta. *Meteoritics & Planetary Science* 48:2237-2251.

Yue Z., Johnson B. C., Minton D. A., Melosh H. J., Di K., Hu W., Liu Y. 2013. Projectile remnants in central peaks of lunar impact craters. *Nature Geosciences* 6:435-437.

Zhang A., Sun Q., Hsu W., Wang R., Itoh S., Yurimoto H. 2011. Fe-rich olivine in brecciated eucrite Northwest Africa 2339: petrography and mineralogy. *34th Symposium on Antarctic Meteorite Program, National Institute of Polar Research* (abstract).

# 4. Three-dimensional spectral analysis of compositional heterogeneity at Arruntia crater on (4) Vesta using Dawn FC

Guneshwar Thangjam[1,2], Andreas Nathues[1], Kurt Mengel[2], Michael Schäfer[1], Martin Hoffmann[1], Edward A. Cloutis[3], Paul Mann[3], Christian Müller[2], Thomas Platz[1], Tanja Schäfer[1]

[1]Max-Planck-Institute for Solar System Research, Justus-von-Liebig-Weg 3, 37077, Göttingen, Germany

[2]Clausthal University of Technology, Adolph-Roemer-Straße 2a, 38678, Clausthal-Zellerfeld, Germany

[3]University of Winnipeg, 515 Portage Avenue Winnipeg, Manitoba, Canada R3B 2E9

## 4.0 Abstract

We introduce an innovative three-dimensional spectral approach (three band parameter space with polyhedrons) that can be used for both qualitative and quantitative analyses improving the characterization of surface heterogeneity of (4) Vesta. It is an advanced and more robust methodology compared to the standard two-dimensional spectral approach (two band parameter space). The Dawn Framing Camera (FC) color data obtained during High Altitude Mapping Orbit (resolution ~ 60 m/pixel) is used. The main focus is on the howardite-eucrite-diogenite (HED) lithologies containing carbonaceous chondritic material, olivine, and impact-melt. The archived spectra of HEDs and their mixtures, from RELAB, HOSERLab and USGS databases as well as our laboratory-measured spectra are used for this study. Three-dimensional convex polyhedrons are defined using computed band parameter values of laboratory spectra. Polyhedrons based on the parameters of Band Tilt ($R_{0.92µm}/R_{0.96µm}$), Mid Ratio (($R_{0.75µm}/R_{0.83µm}$)/($R_{0.83µm}/R_{0.92µm}$)) and reflectance at 0.55 µm ($R_{0.55µm}$) are chosen for the present analysis. An algorithm in IDL programming language is employed to assign FC data points to the respective polyhedrons. The Arruntia region in the northern hemisphere of Vesta is selected for a case study because of its geological and mineralogical importance. We observe that this region is eucrite-dominated howarditic in composition. The extent of olivine-rich exposures within an area of 2.5 crater radii is ~ 12% larger than the previous finding (Thangjam et al., 2014). Lithologies of nearly pure CM2-chondrite, olivine, glass, and diogenite are not found in this region. Although there are no unambiguous spectral features of impact melt, the investigation of morphological features using FC clear filter data from Low Altitude Mapping Orbit (resolution ~ 18 m/pixel) suggests potential impact-melt features inside and outside of the crater. Our spectral approach can be extended to the entire Vestan surface to study the heterogeneous surface composition and its geology.

## 4.1 Introduction

Vesta is one of the most geologically interesting small bodies in the Main Asteroid Belt that bears characteristics similar to the terrestrial planets (e.g., Keil, 2002; Jaumann et al., 2012; Russell et al., 2013). Vesta is an intact differentiated object that most probably survived catastrophic impact events in the early solar system (e.g., Russell et al., 2012, 2013). The geologic study of such a body allows us to understand the evolution of planetary bodies with a silicate-dominated shell. The knowledge about Vesta has been improved by an integrated study of HED meteorites (e.g., Mittlefehldt et al., 1998; Mittlefehldt, 2015; McSween et al., 2011) and ground-based telescopic observations (e.g., Thomas et al., 1997; Hicks et al., 2014; Hiroi et al., 1995). The investigations using Dawn spacecraft images mark a significant progress because of its higher spatial and spectral resolution data compared to all former observations. The spacecraft spent more than a year in orbit around Vesta (July 2011 - September 2012) and acquired images from various distances (e.g., Russell et. al., 2012, 2013). The Framing Camera (FC) and the Visible and Near-infrared Spectrometer (VIR) are imaging instruments onboard Dawn (e.g., Sierks et al., 2011; De Sanctis et al., 2011). The Framing Camera houses seven color filters in the wavelength range from 0.4 to 1.0 µm, and a clear filter. The spatial resolution of FC color data is about 2.3 times higher than the resolution of VIR cubes (Sierks et al., 2011). The FC color data is used in this study.

### 4.1.1 Composition and geology: pre- and post-Dawn

Before the arrival of Dawn at Vesta, many researchers have studied this object using Earth-based telescopes. Non-uniform reflectivity and albedo variations across the surface have long been recognized (e.g., Bobrovnikoff, 1929; Gehrels, 1967). McCord et al. (1970) observed the spectral similarity of Vesta with the eucrite Nuevo Laredo in the visible and near-infrared region (0.3-1.1 µm). Later on, the spectral similarity between eucrite meteorites and Vesta were confirmed (e.g., Feierberg and Drake, 1980). Since then, the evidence for a Vestan origin of the HED suite of meteorites has increased substantially. This hypothesis was supported by the finding of Vestoids (V-type asteroids) that are spectrally similar to Vesta and located in the 3:1 orbital resonance region (e.g., Binzel and Xu, 1993; Thomas et al., 1997). A pyroxene-distribution map (Dumas et al., 1996) indicated albedo and lithological variations. A schematic map of major lithologic units of diogenites and eucrites including an olivine-bearing unit was produced by Gaffey (1997) based on rotationally-resolved ground-based telescopic data. He noted hemispheric and sub-hemispheric compositional variations. A geologic map of Vesta was constructed by Binzel et al. (1997) using Hubble Space Telescope data/Wide Field Planetary Camera (HST/WFPC2). These observations suggested that the eastern hemisphere is spectrally more diverse than the western hemisphere, dominated by units similar to diogenites and eucrites, respectively. Li et al. (2010) presented albedo and color ratio maps of the southern hemisphere using HST/WFPC2 images. They suggested that the brighter eastern hemisphere has a deeper 1-µm absorption band than the darker western hemisphere. Reddy et al. (2010) investigated the southern hemisphere of Vesta using rotationally resolved ground-based telescopic data. These authors noted that the composition in the southern hemisphere ranges from diogenites to cumulate eucrites whereas the northern hemisphere is dominated by cumulate to basaltic eucrites. Olivine was initially not detected (Reddy et al., 2010; Li et al., 2010), despite the predictions of its presence by previous investigators (Gaffey, 1997; Binzel et al., 1997; McFadden et al., 1977).

Analyzes of higher resolution Dawn images have led to a remarkable progress. A general compositional and lithological heterogeneity are mapped in terms of HED lithologies (Reddy et al., 2012b; De Sanctis et al., 2012a; Ammannito et al., 2013b; Thangjam et al., 2013; Nathues et al., 2014, 2015; Schaefer et al., 2014). The surface of Vesta is found to be dominantly eucrite-rich howardite (e.g., Thangjam et al., 2013; Ammannito et al., 2013b). Diogenite-rich lithologies are spotted inside the Rheasilvia basin and in a few localized regions outside the basin, while the equatorial regions are rather eucritic in composition (e.g., Reddy et al., 2012b; De Sanctis et al., 2012a; Ammannito et al., 2013b; Thangjam et al., 2013). Pieters et al. (2012) suggest that space weathering differs from the lunar style and is minimal compared to other airless planetary bodies. The heterogeneous composition of Vesta includes several additional components mixed in with the HED lithology, i.e., dark material (McCord et al., 2012; Reddy et al., 2012c; Palomba et al., 2014; Nathues et al., 2014), bright material (Longobardo et al., 2014; Zambon et al., 2014), glass/shock/impact-melts or orange material (Le Corre et al., 2013; Schaefer et al., 2014; Ruesch et al., 2014b) and olivine-rich material (Ammannito et al., 2013a; Thangjam et al., 2014; Ruesch et al., 2014b; Poulet et al., 2014; Nathues et al., 2015; Palomba et al., 2015).

Dark material (DM) affects Vesta’s surface reflectivity significantly (e.g., McCord et al., 2012; Reddy et al., 2012c; Jaumann et al., 2012). Dark material is likely exogenic in origin, and its spectral parameters resemble carbonaceous chondrites, in particular the CM2-chondrites (McCord et al., 2012; Reddy et al., 2012; Nathues et al., 2014; Palomba et al., 2014). The CM2-chondrites along with other chondritic materials are found in variable proportions as xenoliths in HED meteorites, for instance the PRA 04401 howardite contains about 40-50% CM2-material (e.g., Herrin et al., 2011; Wee et al., 2010). Detection of OH-bearing minerals (De Sanctis et al., 2012b) in particular serpentine (Nathues et al., 2014) and elevated H-abundance (Prettyman et al., 2012) in DM regions further supports an exogenic source.

Bright material (BM) showing unusual high reflectivity is frequently encountered on Vesta (Zambon et al., 2014). Zambon et al. (2014) have analyzed sites of BM using VIR data and found that this material is spectrally consistent with pyroxenes in HEDs showing deeper band depths. Regions with BM are interpreted to be generally fresh which are neither contaminated by howarditic material through continuous impact gardening nor by exogenic carbonaceous chondritic material (Zambon et al., 2014; Reddy et al., 2012b).

Olivine-rich sites are identified for the first time at Arruntia and Bellicia craters in Vesta's northern latitudes using VIR data (Ammannito et al., 2013a). Further, such exposures are found in far northern regions (e.g., Pomponia crater) and in a few spots in the equatorial region, including potential sites at the Rheasilvia basin rim/wall (Matronalia Rupes), on crater floors (Severina crater, Tarpeia crater) and on crater central peaks (Nathues et al., 2015; Ruesch et al., 2014b; Palomba et al., 2015). Olivine was not detected initially during the early stages of Dawn observations despite the predictions of significant olivine exposures in the south polar basin (e.g., Thomas et al., 1997; Gaffey, 1997; Beck and McSween 2010; Jutzi and Asphaug, 2011; Jutzi et al., 2013; Ivanov and Melosh, 2013; Tkalcec et al., 2013), and therefore, concerns raised about the missing mantle material of Vesta (e.g., McSween et al., 2014). Based on linear spectral unmixing analysis, Combe et al. (2015) argue that a mixture of low- and high-Ca pyroxenes (hypersthene, pigeonite, and diopside) is an alternative to the presence of olivine in these regions.

An integrated analysis of Dawn FC, VIR and GRaND data, Le Corre et al. (2013) find that 'Orange Material' (OM) that appeared orange in the 'Clementine' RGB ratio image is analogue to material of impact-melt/shock/glass on Vesta's surface. They suggest that a steeper visible slope (red visible slope) is a diagnostic spectral parameter for this material. This parameter is referred elsewhere to detect and map regions with OM (e.g., Williams et al., 2014; Schaefer et al., 2014). However, Ruesch et al. (2014b) argue that this parameter is not uniquely diagnostic for impact-melt characterization.

To better understand the nature of compositional heterogeneity across Vesta, we undertake a comprehensive study of the spectral reflectance properties of HEDs and other types of materials known or inferred to be present on Vesta.

## 4.2 Samples and laboratory spectra

Reflectance spectra of two lines of mixtures between howardite and olivine or CM material are measured to understand how mixtures of Vesta-relevant materials vary spectrally. One mixing line is based on the diogenite-dominated howardite DaG 779, the other is based on the eucrite-dominated howardite NWA 1929. The carbonaceous chondrite material is from a specimen of the Jbilet-Winselwan CM2 carbonaceous chondrite. The olivine material is prepared from a forsterite-rich gemstone quality sample. Descriptions of the meteorites/samples are given in Appendix 1 with their compositional data in Appendix 2.

The meteorite samples are received in chips/bulks. The outer rims of each specimen are removed thoroughly before crushing to minimize the inclusion of fusion crust and terrestrial weathering products. They are ground to powders using an agate mortar either in a grinding mill or by hand at the Clausthal University of Technology, Germany (TUC). The fine-grained material are dry-sieved and further processed (milled) until all material passed through a 63 µm sieve. The sieved powders of the two howardites are used to prepare intimate mixtures in 10 wt.% intervals, each with olivine or Jbilet-Winselwan.

Individual powders are homogenized by stirring and shaking the powders in glass vials for about 1 minute before preparing the mixtures. The samples are mixed to weigh 250 mg in total using an analytical weighing balance in a clean laboratory at Max-Planck Institute for Solar System Research, Germany (MPS). Each mixture is again gently stirred for about 5 minutes using a pestle in a small agate mortar to produce homogeneous mixtures. The reflectance spectra (0.35-2.5 µm) of the particulate samples and their mixtures are measured at HOSERLab (University of Winnipeg, Canada) using an Analytical Spectral Devices (ASD) FieldSpec Pro HR spectrometer. The measurements are done relative to a Spectralon® standard using a 150W quartz-tungsten-halogen collimated light source. In each case, 200 spectra are acquired and averaged to provide sufficient signal-to-noise. The details of these measurements are discussed in Cloutis et al. (2013).

Apart from the measured spectra, available reflectance spectra of powdered HED meteorites, olivine, CM2, and olivine-orthopyroxene mixtures from the RELAB, USGS, and HOSERLab spectral libraries are used. Details on the spectra/samples are listed in Appendix 3. Figure 1 shows absolute (Fig. 1A) and normalized (Fig. 1B) reflectance spectra of the measured spectra (solid lines) of olivine, NWA 1929, DAG 779 and Jibilet Winselwan CM2. The Macibini-glass spectrum is from RELAB. The resampled data points at the effective wavelength of each filter are represented by the symbols in Fig. 1. The spectra are resampled to FC bandpasses using the instrument response function per filter (Sierks et al., 2011).

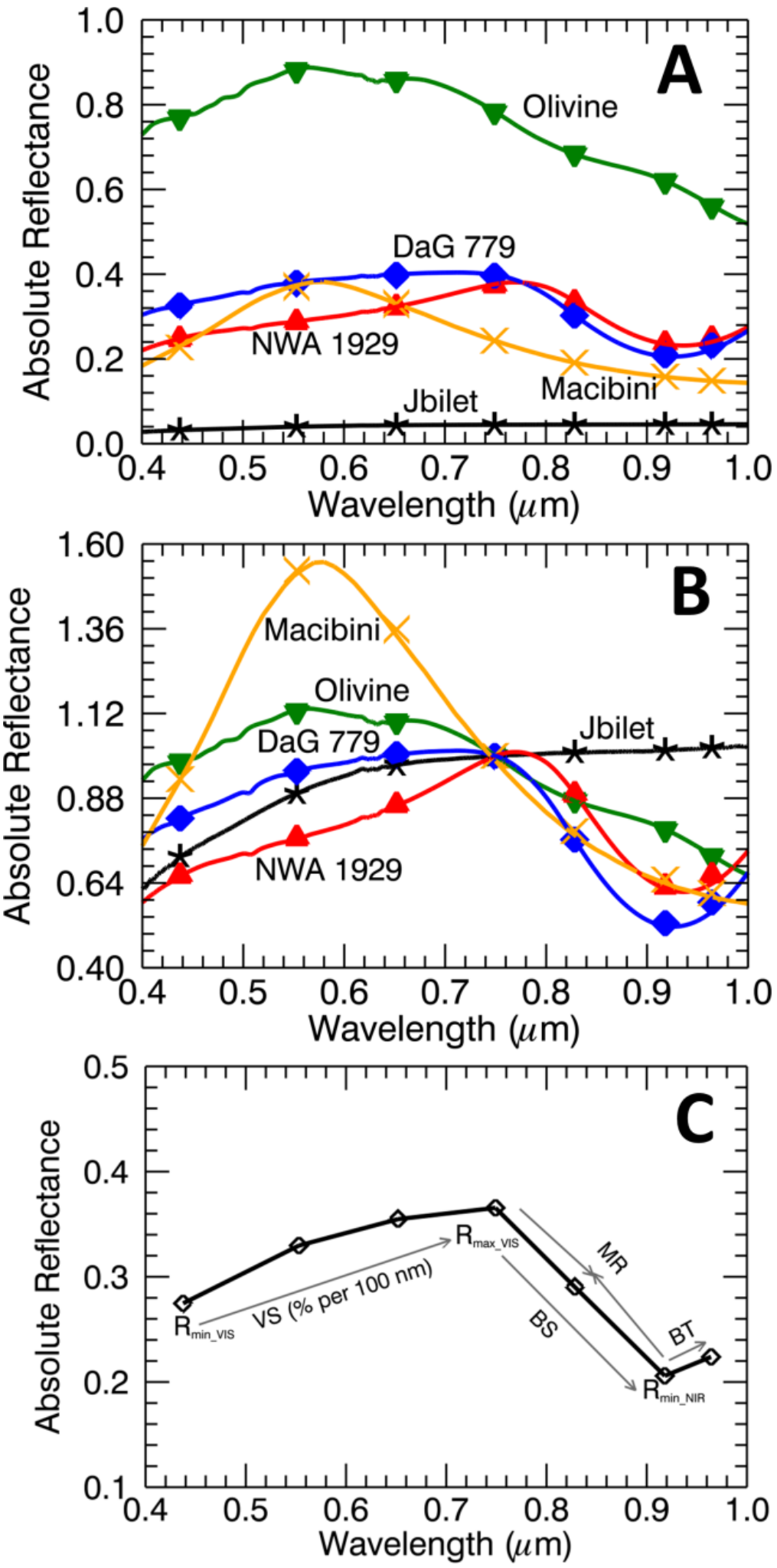

Fig. 1. (A) Absolute and (B) normalized laboratory spectra (line) resampled to FC bandpasses (symbols). The spectra of NWA 1929 howardite (eucrite-rich), DaG 779 howardite (diogenite-rich), Jbilet-Winselwan (CM2-chondrite), and olivine are measured in this work. The Macibini-eucrite glass spectrum is from RELAB. (C) Sketch illustrating the band parameters calculated from the absolute reflectance spectrum, Band Tilt (BT), Mid Ratio (MR), Visible Slope in % per 0.1 μm (VS) and Band Strength (BS). The acronyms $R_{min_VIS}$ and $R_{max_VIS}$ represent those FC color filters showing minimum and maximum reflectance in the visible wavelength range, while $R_{min_NIR}$ is the minimum reflectance in the near-infrared wavelength range.

Table 1: Band parameters used in this study.

| Band Parameters | Definition |
|---|---|
| Band Tilt (BT) | $R_{0.92}/R_{0.96}$ |
| Mid Ratio (MR) | $(R_{0.75}/R_{0.83})/(R_{0.83}/ R_{0.92})$ |
| Band Strength (BS) | $R_{max_VIS}/R_{min_NIR}$ |
| Visible Slope (VS) in % per 0.1µm | $(R_{\lambda 2}-R_{\lambda 1})*10/(\lambda_2- \lambda_1)$ |

where, R is the reflectance value at the given wavelengths in µm; $R_{max_VIS}$ and $R_{min_NIR}$ are the maximum reflectance in the visible wavelength range and the minimum reflectance in the near-infrared wavelength range, respectively.

## 4.3 Band Parameters

The robustness of FC color data in constraining Vestan composition and mineralogy has been investigated in a number of studies. Despite the fact that the FC data does not cover the entire 1-µm absorption feature, mafic mineral compositions can still be derived and linked to particular HED lithologies (e.g., Reddy et al., 2012b, c; Thangjam et al., 2013, 2014; Nathues et al., 2014, 2015). Various FC band parameters have been developed to analyze and map lithologic units (Reddy et al., 2012; Thangjam et al., 2013), dark material (Reddy et al., 2012c; Nathues et al., 2014), olivine-rich regions (Thangjam et al., 2014; Nathues et al., 2015), and impact-melt/glass/orange material (Le Corre et al., 2013). Band parameters used in this study are listed in Table 1.

The spectral slope in the visible wavelength range of FC color data, generally defined by a ratio of reflectance values at 0.43 and 0.75 µm (0.43/0.75 µm), has been used elsewhere (e.g., Reddy et al., 2012b; Buratti et al., 2013; Le Corre et al., 2013; Schaefer et al., 2014) to analyze composition and surface properties of Vesta. The spectral slope has also widely implemented in comparative studies of Vesta, Vestoids, and HED meteorites (e.g., Buratti et al., 2013; Hicks et al., 2014; Hiroi and Pieters, 1998; Burbine et al., 2001). Computation of the slope from spectra normalized to a particular wavelength is typical for ground-based telescopic data (e.g., Luu and Jewitt, 1990; Bus and Binzel, 2002; Hicks et al., 2014). The general FC visible slope definition is similar to what has been applied to lunar Clementine Ultraviolet/Visible Camera (UVVIS) data (e.g., Pieters et al., 2001; Isaacson and Pieters, 2009). However, Vesta's regolith differs in many ways from the lunar regolith: for example, different space weathering processes (Pieters et al., 2012), absence/insignificance of lunar-like agglutinates and Fe°-np (nano-phase iron) and TiO2 in HED meteorites (e.g., Buchanan et al., 2000; Singerling et al., 2013), dominantly mafic composition on Vesta unlike the prominent lunar anorthositic (felsic) highlands and mafic mare, etc. Again, the FC is equipped with 4 filters (0.43, 0.55, 0.65, 0.75 µm) in the visible wavelength range compared to 2 filters (0.41, 0.75 µm) for the Clementine UVVIS camera. We define visible slope (VS) with slight modifications from Luu and Jewitt (1990), Doressoundiram et al. (2008), and Hiroi and Pieters (1998). The slope is calculated in % per 0.1 µm from the absolute reflectance spectra, over a wavelength range framed by minimum reflectance and a maximum reflectance in the visible wavelength range.

Band strength or band depth of the 1-µm absorption feature is generally represented by a ratio of the reflectance values at 0.75 and 0.92 µm (0.75/0.92 µm) for FC color data (Reddy et al., 2012a; Le Corre et al., 2013; Thangjam et al., 2014). This is similar to the lunar Clementine parameter (e.g., Tompkins and Pieters 1999; Pieters et al. 2001; Isaacson and Pieters 2009). It is worth mentioning that Tompkins and Pieters (1999) defined this parameter by a ratio of 0.75 µm to any of the bands near 1 µm (0.75/0.90 µm, 0.75/0.95 µm, 0.75/1.0 µm). They used one of the ratios, depending on the closest wavelength to the 1-µm absorption band minimum, to estimate a first-order relative abundance of mafic minerals (and plagioclase). Though this parameter is diagnostic to analyze lunar surface's FeO abundance and soil maturity and their composition, it is important to be cautious in applying this parameter to Vesta's surface because of the aforementioned differences between the two bodies. Here, we define the parameter Band Strength (BS) as the ratio of maximum reflectance value in the visible wavelength range (λmax_VIS) to the minimum reflectance in the near-infrared range (λmin_NIR) (see Fig. 1C).

Band Tilt (BT) is one of the most diagnostic parameters in the near-infrared region to characterize the 1-µm absorption feature of mafic minerals. This is discussed in detail in Thangjam et al. (2013, 2014), and Nathues et al. (2015). This parameter is basically a lunar Clementine UVVIS parameter, which was introduced by Pieters et al. (2001), and then later modified by Dhingra (2008), and Isaacson and Pieters (2009).

Mid Ratio (MR) was introduced in Thangjam et al. (2014) and discussed in detail elsewhere (Thangjam et al., 2014; Nathues et al., 2015). This parameter is useful for distinguishing olivine-rich lithologies from HED lithologies, especially from High-Ca pyroxenes observed in HEDs (Thangjam et al., 2014).

A sketch illustrates our band parameters together with a typical HED spectrum (Fig. 1C). The wavelengths with minimum reflectance (λmin_VIS) and maximum reflectance (λmax_VIS) in the visible wavelength range, and minimum reflectance in the near-infrared region (λmin_NIR) are determined before calculating the band parameters. The band parameter values of the spectra used in this study are provided in Appendix 3.

## 4.4 Analysis - Laboratory spectra

More than 80% of eucrites and diogenites show λmax_VIS at 0.75 and 0.65 µm, respectively. The spectrum of eucrite-dominated howardite NWA 1929 shows λmax_VIS at 0.75 µm, while λmax_VIS of the diogenite-dominated howardite DaG 779 spectrum is at 0.65 µm. Both the olivine and the Macibini-eucrite glass spectra exhibit λmax_VIS at 0.55 µm. The spectrum with the steepest visible slope is the Macibini-eucrite glass, followed by NWA 1929, olivine, and DaG 779. Similarly, the glass spectrum shows the deepest band strength, followed by DaG 779, NWA 1929, and olivine. The Jbilet-Winselwan spectrum appears to be almost flat in the FC wavelength range though it displays a notably visible slope.

Figure 2 presents FC resampled spectra of Jbilet-Winselwan, howardites DaG 779, and NWA 1929, and their mixtures. Absolute reflectance spectra of these mixtures are shown in 10 wt.% intervals (Fig. 2A and 2B) whereas normalized spectra are displayed in steps of 10, 30, 50, 70 and 90 wt.% for clarity (Fig. 2C and 2D). The two howardites show significantly higher reflectance and band strength values compared to that of Jbilet-Winselwan. Both the parameter values are significantly reduced by adding 10 wt.% of Jbilet-Winselwan, while alternatively, rather a large amount of howardites (more than 30-40 wt.%) can increase the low reflectance and band strength values of Jbilet-Winselwan. Though a rapid change of reflectance and band strength is apparent in these mixtures (Fig. 2A, B), only slight variations are observed in the visible slope (Fig. 2C, D). The visible slope values of mixtures with DaG 779 increase with increasing CM2 abundance, whereas, in case of NWA 1929, this behavior is less systematic. This is probably due to differences in the reflectance in the visible wavelength range of the different samples that affects their spectral slope. The majority of the mixtures show λmax_VIS at 0.75 µm, and the spectral slope from 0.43 µm to either 0.65 or 0.55 µm drops very rapidly compared to the spectral slope from 0.65 to 0.75 µm.

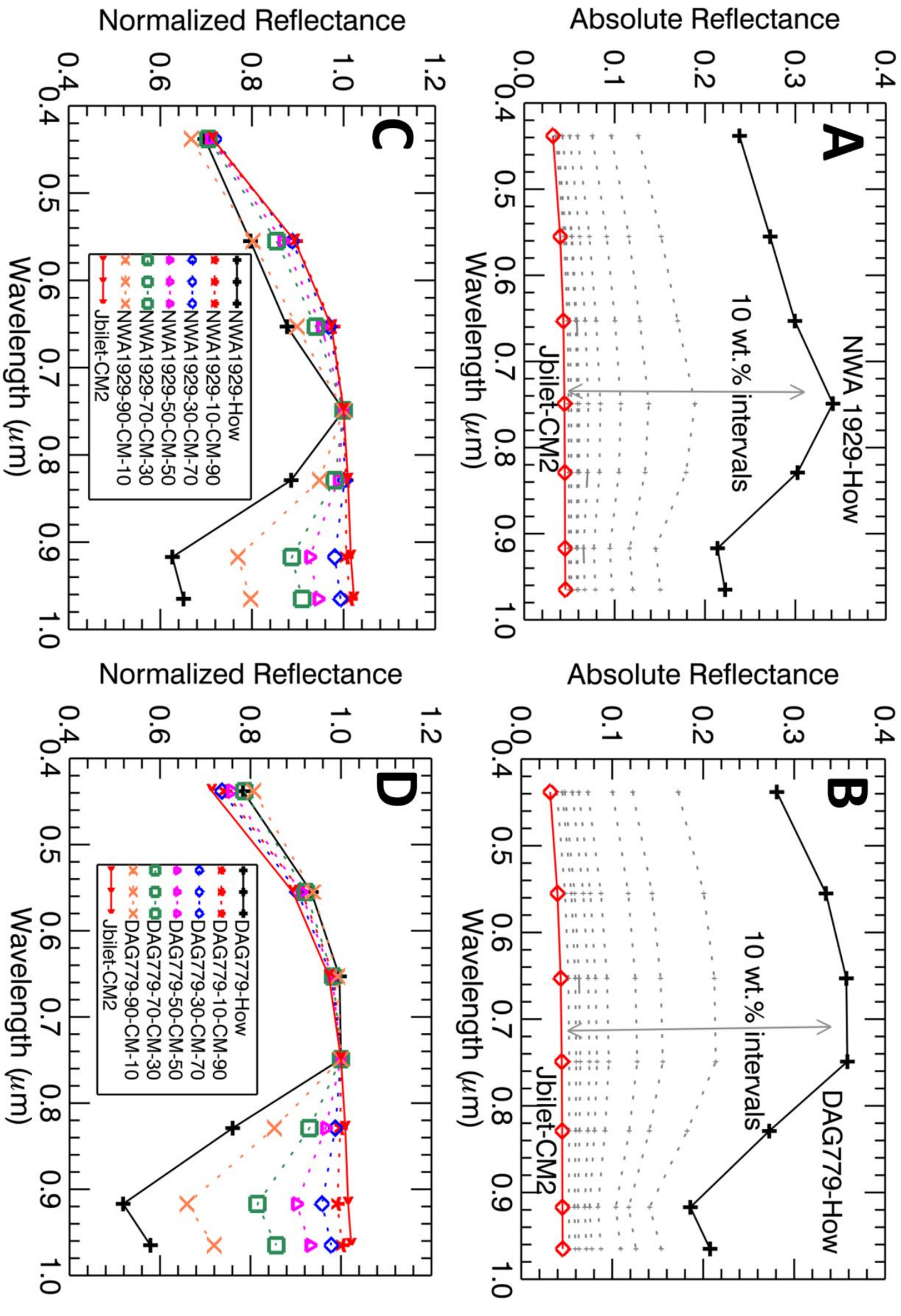

Fig. 2. Resampled spectra of Jbilet-Winselwan (CM2-chondrite) mixed with howardites DaG 779 and NWA 1929. (A, B) Absolute reflectance spectra in an interval of 10 wt.%, represented by the dotted lines. (C, D) Normalized spectra of samples and mixtures in steps of 10, 30, 50, 70, 90 wt.%.

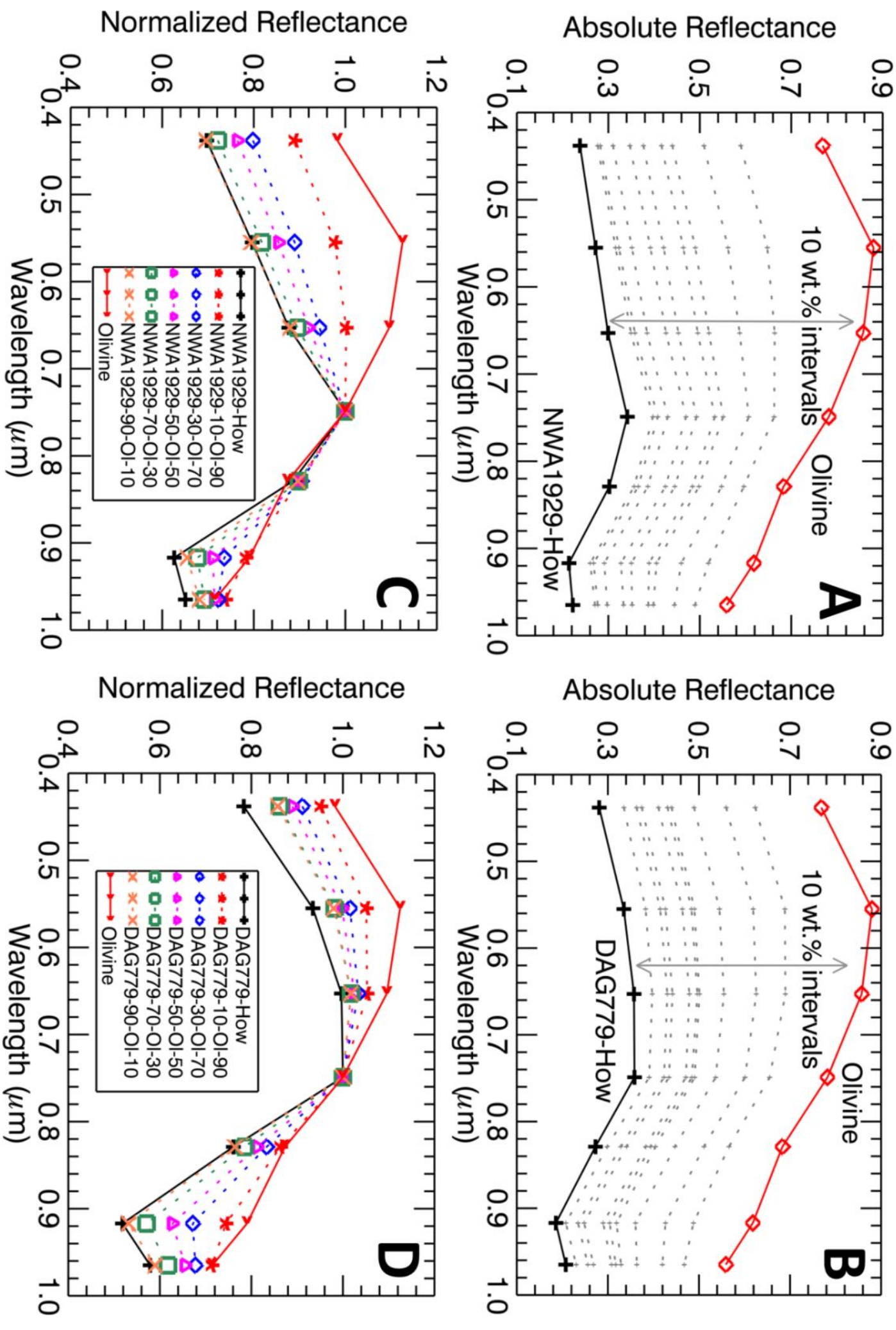

Fig. 3. Resampled spectra of olivine mixed with howardites DaG 779 and NWA 1929. (A, B) Absolute reflectance spectra in an interval of 10 wt.%, represented by the dotted lines. (C, D) Normalized spectra of samples and mixtures in steps 10, 30, 50, 70, 90 wt.%.

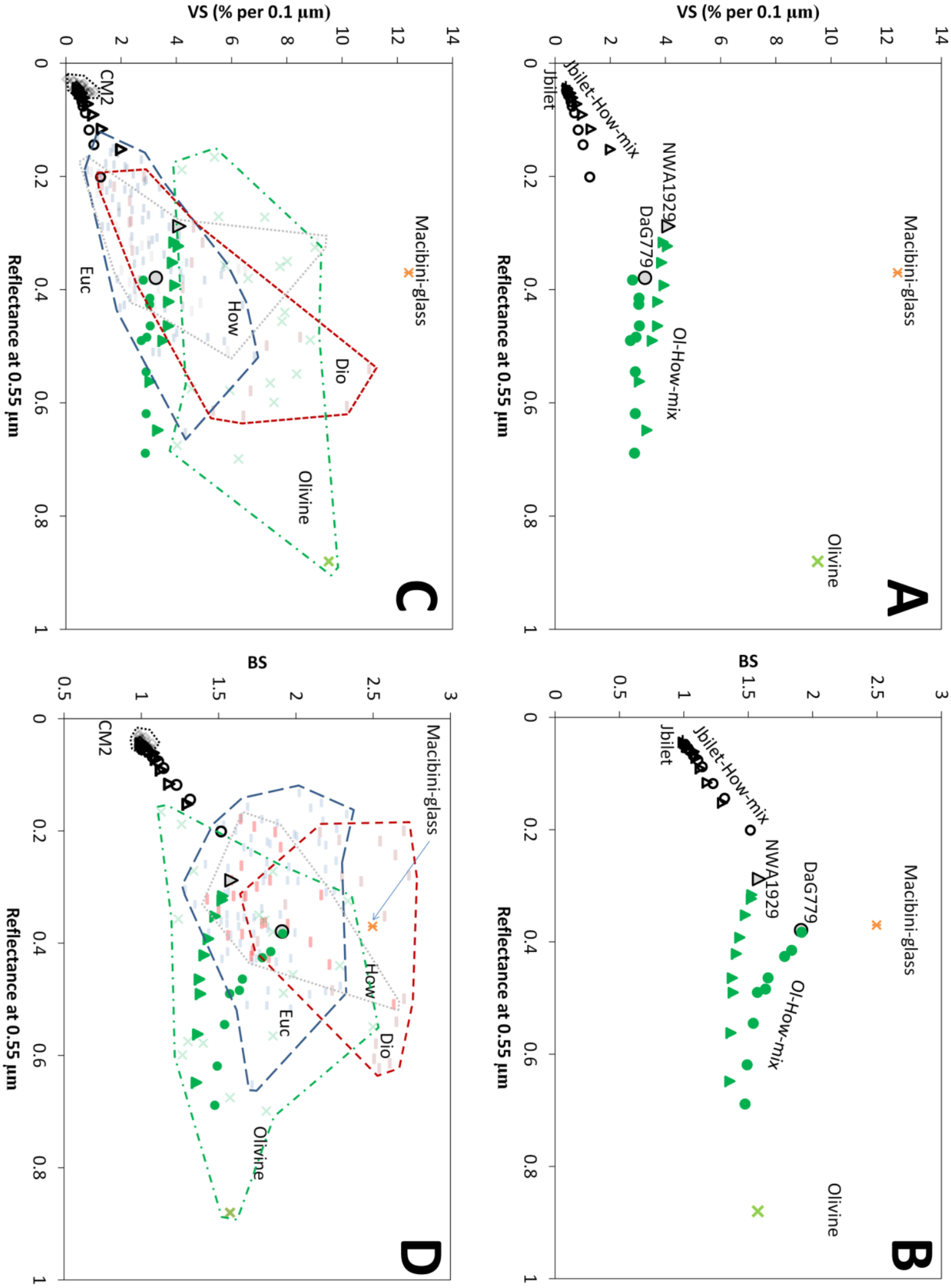

Fig. 4: Two-dimensional band parameter spaces for DaG 779, NWA 1929, olivine, Jbilet-Winselwan, Macibini-eucrite glass, and mixtures of howardites with olivine or Jbilet-Winsewlan. (A) VS versus R0.55μm, (B) BS versus R0.55μm, (C, D) polygons defined by all the spectra of eucrites, diogenites, howardites, olivine, and CM2 in the corresponding band parameter spaces.

Figure 3 shows resampled spectra of mixtures of olivine and howardite (either with DaG 779 or NWA 1929). Absolute reflectance spectra in 10 wt.% intervals (Fig. 3A, B) and normalized spectra at steps of 10, 30, 50, 70, 90 wt.% howardite content (Fig. 3C, D) are displayed. Changes in the spectral shape and their band parameters, such as visible slope, band tilt, and band strength are noticed with changes in end member abundances. Unlike the CM2 spectrum, the olivine spectrum exhibits a notable 1-µm feature. The olivine spectrum shows higher reflectance compared to the howardites. Visible slope and reflectance of mixtures gradually changes as a function of the components. The same is true for parameters BT and BS. Olivine shows higher values of BT but lower BS parameters than the two howardites. A significant change in reflectance, VS, BT and MR values of olivine is seen by adding even only 10 wt.% howardite. A howarditic amount of more than 30-40 wt.% in these mixtures produces spectra similar to HEDs. This is almost opposite to what we observed in CM2 mixtures where the howardites spectra are significantly affected by even 10 wt.% of CM2.

Eucrite and olivine show larger BT values than diogenites, and therefore, the BT parameter is the most effective for distinguishing diogenite from eucrite and olivine (Thangjam et al., 2013, 2014). The MR values of diogenite, olivine, and olivine-rich olivine-orthopyroxene mixtures are in general larger than eucrite and clinopyroxene found in HEDs (Thangjam et al., 2014). The CM2s spectra show much lower reflectances compared to HEDs and olivine, and thus, the reflectance serves a useful criterion to recognize the presence of CM2 material (Fig. 4). Macibini-eucrite glass spectrum displays a rather large value of VS compared to the HEDs, olivine, and CM2s, and hence, the VS parameter can be useful to identify and discriminate the glass spectrum (Fig. 4). However, the VS values of HEDs, olivine, and CM2s overlap. Similarly, the BS parameter shows a broad range of values for HEDs and olivine, though the BS parameter is generally used to assess abundance of mafic minerals (e.g., Reddy et al., 2012b; Thangjam et al., 2013). This indicates that multiple spectral parameters are required for more robust discrimination of the various materials included in this study.

### 4.4.1 Two-dimensional approach

Two-dimensional approaches (i.e., two band parameter spaces) are generally employed in earlier works to characterize surface lithologies, for example, Band Tilt versus Band Curvature parameter space to distinguish among eucrites and diogenites (Thangjam et al., 2013), and Band Tilt versus Mid Ratio to separate olivine from HEDs (Thangjam et al., 2014; Nathues et al., 2015), etc. Apart from the aforementioned band parameter spaces, other parameter spaces such as band strength or visible slope versus reflectance at 0.75µm need to be discussed, since they are suggested to be diagnostic for identifying impact melts, exogenic dark materials, etc., (e.g., Le Corre et al., 2013; Reddy et al., 2012b; McCord et al., 2012). A detailed analysis of these parameter spaces is presented here.

The parameters VS and BS are plotted versus reflectance at 0.55 µm (R0.55µm) for Macibini-eucrite glass, howardites DaG 779 and NWA 1929, olivine, Jbilet-Winselwan, and mixtures of howardites with olivine or CM2 (Fig 4). In the VS versus R0.55µm band parameter space (Fig. 4A), the Jbilet-Winselwan, olivine, and glass sample are well separated from each other and the two howardites. The CM2-chondrite mixtures and the olivine mixtures plot on either side of the howardite samples along the x-axis (R0.55µm). We do not have data points of mixtures of glass and howardite, but a trend more or less in between the howardites and the glass can be expected. The parameter VS of the olivine and howardite mixtures show similar values with a negligible change compared with the howardites though the values are significantly different from the nearly pure olivine. The samples and mixtures in Fig. 4A are not distinguishable if all the eucrite, diogenite, olivine, and CM2 are plotted together (Fig. 4C). Similar plots are displayed for BS versus R0.55µm (Fig. 4B, D). The BS parameter of Macibini-glass is not as distinct as for the VS parameter, and this data point is not separable from the eucrite, diogenite, and olivine. These analyzes suggest that the VS and BS parameters are not ideally suited to distinguish lithologies among HEDs, olivine, and their mixtures. Meanwhile, visible slope and band strength parameters are known to be significantly affected not only by the mineralogy but multiple parameters such as grain size, viewing geometry or phase angle, temperature, etc. (e.g., Nathues et al., 2000; Reddy et al., 2012c; Duffard et al., 2005; Izawa et al., in review; Cloutis et al., 2013; Moskovitz et al., 2010).

Though the two-dimensional parameter spaces are useful to identify particular lithologies of interest, they seem to be insufficient in dealing with multiple components. Therefore, an advanced spectral approach of three-dimensional analysis (i.e., three-band parameter space with polyhedrons) is introduced and applied for the first time in this study. It is worth mentioning that the three-dimensional parameter space used elsewhere (e.g.,

Filacchione et al., 2012) are scattered data points, however the polyhedrons defined in this study allow to assess both qualitative and quantitative information. Again, this approach is different from other methodologies employing multi-dimensional parameters defined through statistical analyses for example, principal component analysis (e.g., Roig and Gill-Hutton, 2006; Nathues et al., 2012), or cluster analysis (e.g., Pinilia-Alonso et al., 2011), etc.

### 4.4.2 Three-dimensional approach

A three-dimensional band parameter space uses a further parameter (information), and thus, the results can be more robust and precise. The band parameter values computed from laboratory spectra representing different lithologies are used to define a three-dimensional space spitted in several convex polyhedrons. The polyhedrons are constructed by many convex triangular facets, using the band parameter values as vertices and a connectivity array defined by the convex hull method. Plotting are done in IDL programming language. Various sets of combinations of the band parameters are discussed, and a particular parameter space of interest is chosen for further analysis and application.

The polyhedrons of the BT versus MR versus R0.55μm band parameter space are shown in Fig. 5A in two different views. The parameters BT, MR, and R0.55μm are plotted along the X, Y, and Z axis, respectively. Each polyhedron or data point is given a name for clarity (Euc: eucrite; Dio: diogenite; How: howardite; Ol: olivine; Ol-HED: olivine plus howardite mixtures, and olivine-bearing diogenites; Ol-rich-Opx: Olivine rich olivine-orthopyroxene mixtures (>40 wt% olivine); CM2: CM2 chondrite; CM2-HED: Jbilet-Winselwan plus howardite mixture, Murchison plus eucrite mixture, and CM2-rich howardite; Gl: Macibini eucrite glass). The polyhedrons of diogenite and eucrite are clearly distinct while the howardite polyhedron overlaps with both of them. The polyhedrons of olivine and CM2-chondrites are separable from HEDs while the CM2-HEDs and Ol-HEDs polyhedrons gradually merge with the howardite/eucrite polyhedron. The CM2-HEDs polyhedron overlaps with that of eucrite/howardite at a range of approximately 20-30 wt.% CM2 content, whereas the overlapping is at about 40-60 wt.% of olivine content for the Ol-HEDs polyhedron. The data point of the glass neither overlaps with any polyhedron of olivine nor Ol-HEDs, but it lies within the Ol-rich-Opx and Ol-HEDs polyhedrons. Thus, the glass sample is not uniquely distinguishable in this band parameter space.

Two different views of the polyhedrons in the BT versus MR versus VS band parameter space are shown in Fig. 5B. Diogenite, howardite, olivine, CM2, CM2-HEDs, Ol-HEDs display a broad range of VS values, and they overlap to varying extents with each other, making it difficult to distinguish them. However, CM2, olivine, Ol-rich-Opx mixtures are distinguishable from the HEDs. Eucrite and diogenite are separated from each other but overlap with howardite. Since the data point of Macibini-glass shows a rather large VS value, this data point is distinct. Thus, this parameter space can be used to identify eucrite-glass samples assuming that glasses in HEDs are similar to the Macibini eucrite glass. However, data points of two Padvarninkai eucrite impact melts, and a heavily shocked eucrite, JaH 626, do not follow the glass sample, and they overlap with other polyhedrons. A small amount (< 10-20 wt.%) of howardite mixed with glass can yield spectra similar to HEDs (e.g., Buchanan et al, 2014), and it may further complicate separation of such data points from HEDs.

The polyhedrons of the BT versus MR versus BS band parameter space are shown from two different views in Fig. 5C. Some of the polyhedrons are distinct, and the BS parameter is insensitive to distinguish among HEDs, olivine, and Ol-HEDs. Eucrite and diogenite are separable from each other. Olivine, CM2, and Ol-rich-Opx polyhedrons are distinguished from HEDs. The Ol-HEDs and CM2-HEDs polyhedrons overlap with the HEDs to a large extent. The data point of the eucrite glass sample does not overlap with either of the polyhedrons though it lies close to the polyhedrons of olivine and Ol-rich-Opx. However, impact-melt and shock HED samples plot far apart from the glass sample and are indistinguishable from other polyhedrons.

It is rather difficult to distinguish the various lithologies in the VS versus BS versus R0.55µm band parameter space (Fig. 5D). The CM2 and the Macibini eucrite glass sample are offset from the HEDs. The glass sample does not overlap with the HEDs. However the other impact-melt and shock HED samples neither follow the glass sample nor separable from HEDs.

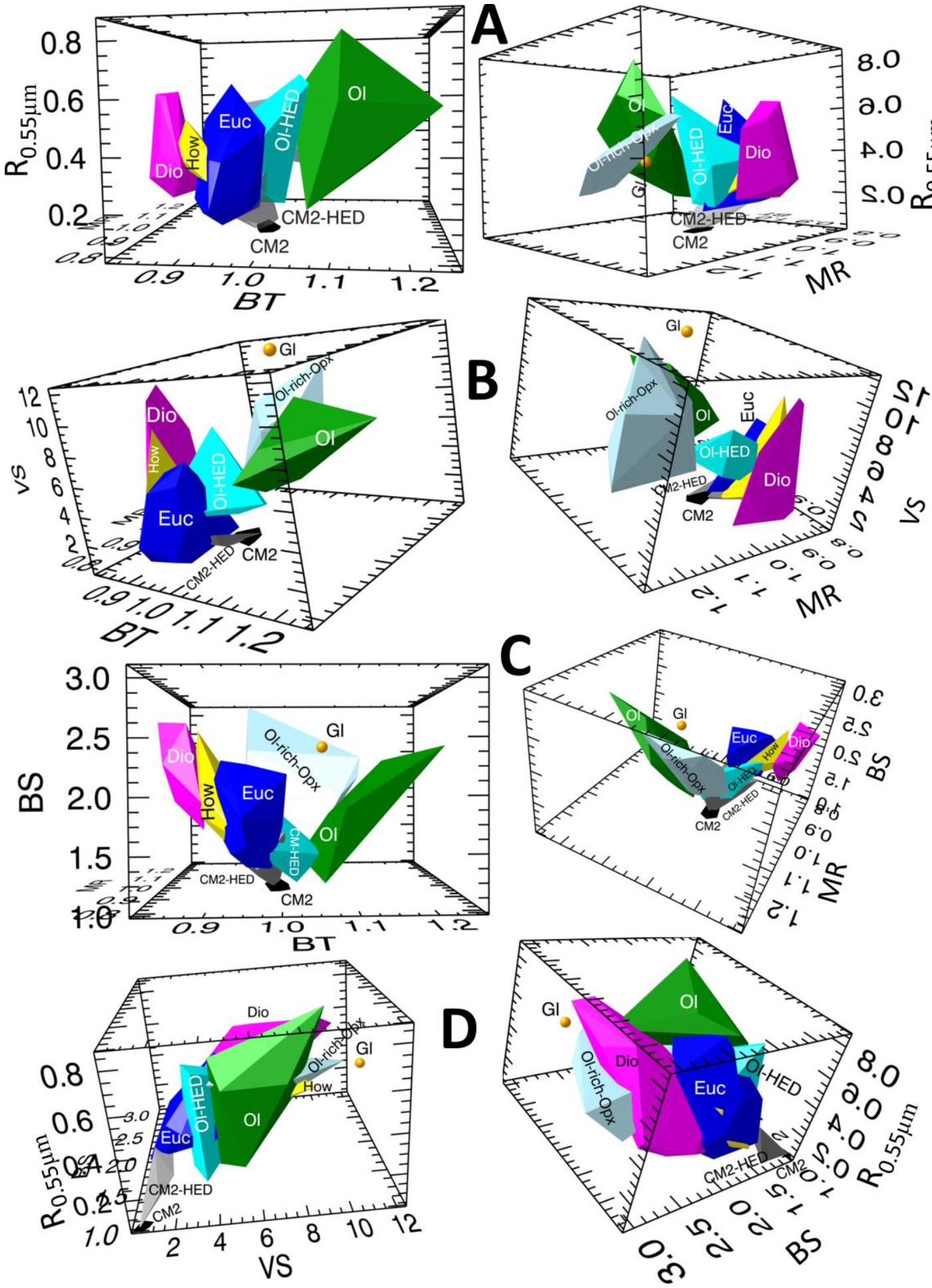

Fig. 5: Different perspective views of three-dimensional band parameter spaces. (A) BT versus MR versus $R_{0.55\mu m}$, (B) BT versus MR versus VS, (C) BT versus MR versus BS, (D) VS versus BS versus $R_{0.55\mu m}$. Each polyhedron represents a particular mineralogy (Euc: eucrite, Dio: diogenite, How: howardite, Ol: olivine, Ol-HED: olivine plus howardite mixtures, and olivine-bearing diogenites, CM2: CM2 chondrites, CM2-HED: Jbilet plus howardite mixture, Murchison plus eucrite mixture, and CM2-bearing howardite). The orange data point represents the Macibini eucrite glass sample.

## 4.5 Implications for Dawn FC at Vesta

The three-dimensional polyhedrons defined from various laboratory spectra are applied to Dawn FC data to enable compositional analysis and mapping. Algorithm in IDL program has been employed to assign FC data points to the respective polyhedrons. Details of the algorithm are provided in Appendix 4.

As a case study, we select Arruntia region in the northern hemisphere of Vesta. This region shows many interesting lithologies, including olivine-rich sites (Ammannito et al., 2013a; Thangjam et al., 2014; Ruesch et al., 2014b; Nathues et al., 2015), impact-melt/glass/shocked or orange material (Le Corre et al., 2013), and dark and bright material (e.g., Thangjam et al., 2014; Ruesch et al., 2014a; Zambon et al., 2014). The Arruntia crater is described as one of the freshest impact craters on Vesta, showing dark and bright ejecta (Ruesch et al., 2014b). We use FC color data (~ 60 m/pixel) obtained during High Altitude Mapping Orbit (HAMO and HAMO 2) for spectral analysis and mapping. The calibration and processing of the FC color data including photometric corrections and error or uncertainty in the data are discussed in Nathues et al. (2014). The errors in the spectral data points are usually less than the size of the symbols displayed in the plots. For investigations of morphological features, the FC clear filter images (~ 18 m/pixel) obtained during Low Altitude Mapping Orbit (LAMO) is used. Figure 6 shows the reflectance and several band parameter images of our study area: (A) reflectance at 0.55 µm ($R_{0.55µm}$), (B) MR (Mid Ratio), (C) VS (Visible Slope in % per 0.1 µm), (D) BS (Band Strength), and (E) BT (Band Tilt). Gray material (mean $R_{0.55µm}$ ~ 0.19) covers the region predominantly, while bright material (mean $R_{0.55µm}$ ~ 0.25) is more abundant than the dark material (mean $R_{0.55µm}$ ~ 0.14). Bright material is found in the ejecta and on the wall of the crater, whereas dark material is locally enriched in a few regions of ejecta and the crater wall. The band parameter values (VS, BS, BT, and MR) do not follow a clear trend of dark, bright and gray material, but the majority of the bright material in the ejecta usually exhibits higher BT, MR, VS values and lower BS values. A slope map is derived from the HAMO-DTM (~ 62 m/pixel) to visualize the relief of this region (Fig. 6F). The Arruntia crater walls show rather steep slopes compared to other craters in this region. The topographic profile of this crater resembles a V-shape without having a pronounced flat crater floor. The distribution of the ejecta material follows the local relief.

For our application, we choose the three-dimensional polyhedrons defined by the BT, MR and $R_{0.55\mu m}$ parameters. The band parameter values of the inflight data are calculated using an IDL algorithm similar to that used for laboratory spectra. Areas with the dominant lithology in this region, i.e., eucrite (Fig. 7A) are mapped. Figure 7B shows those areas (~ 23%) that are assigned to the CM2-HEDs polyhedron where less than 20-30 wt.% of CM2 content is indistinguishable from eucrite/howardite. However, the CM2-rich localities (≥ 20-30 wt.% of CM2, distinguished from eucrite/howardite) are very minor in extent (< 1% of the total area, figure not shown), and this suggests that dark material, particularly CM2, does not contribute to this region significantly. The mapping of lithologies containing olivine mixtures with HEDs is accomplished by a combined polyhedron of Ol-HEDs and Ol-rich-Opx. The Ol-rich-Opx mixtures (>40 wt.% olivine, Appendix 3) are already used to identify and map olivine-rich lithologies by Thangjam et al. (2014) and Nathues et al. (2015). Figure 7C shows areas of olivine mixed with HEDs where less than 40-60 wt.% of olivine content are not separable from eucrite/howardite. There is no unambiguous interpretation for this observation because they can be either eucrite/howardite without olivine or below detection limits. Figure 7D shows olivine-rich areas (≥ 40-60%) that are distinguished from eucrite/howardite. These potential and distinct olivine-rich areas are consistent with the olivine-rich exposures identified by Thangjam et al. (2014) but are more abundant. These exposures cover approximately 4% of the study region. Within a region of 2.5 crater radii, the olivine-rich sites occupy about 14% of the surface, which is 12% more than what Thangjam et al. (2014) found. Most of the olivine-rich sites are on the ejecta near the crater rim, with few sites on the crater wall and the floor. Based on our observations, eucrite covers the majority of the Arruntia region, and therefore, this implies that this region is dominantly eucrite-rich howardite in composition, which is consistent with earlier observations (Thangjam et al., 2014; Ammannito et al., 2013a). Most probably there are no exposures of (nearly) pure diogenite, olivine or CM2-material in this region. The absence of diogenite was also noted by Ammannito et al. (2013a) and Thangjam et al. (2014). We are not able to identify sites consisting mainly of glass or impact-melts because of the paucity of laboratory impact-melt/glass spectral data. Meanwhile, we do not find any visible slope value that is comparable to the Macibini-glass sample.

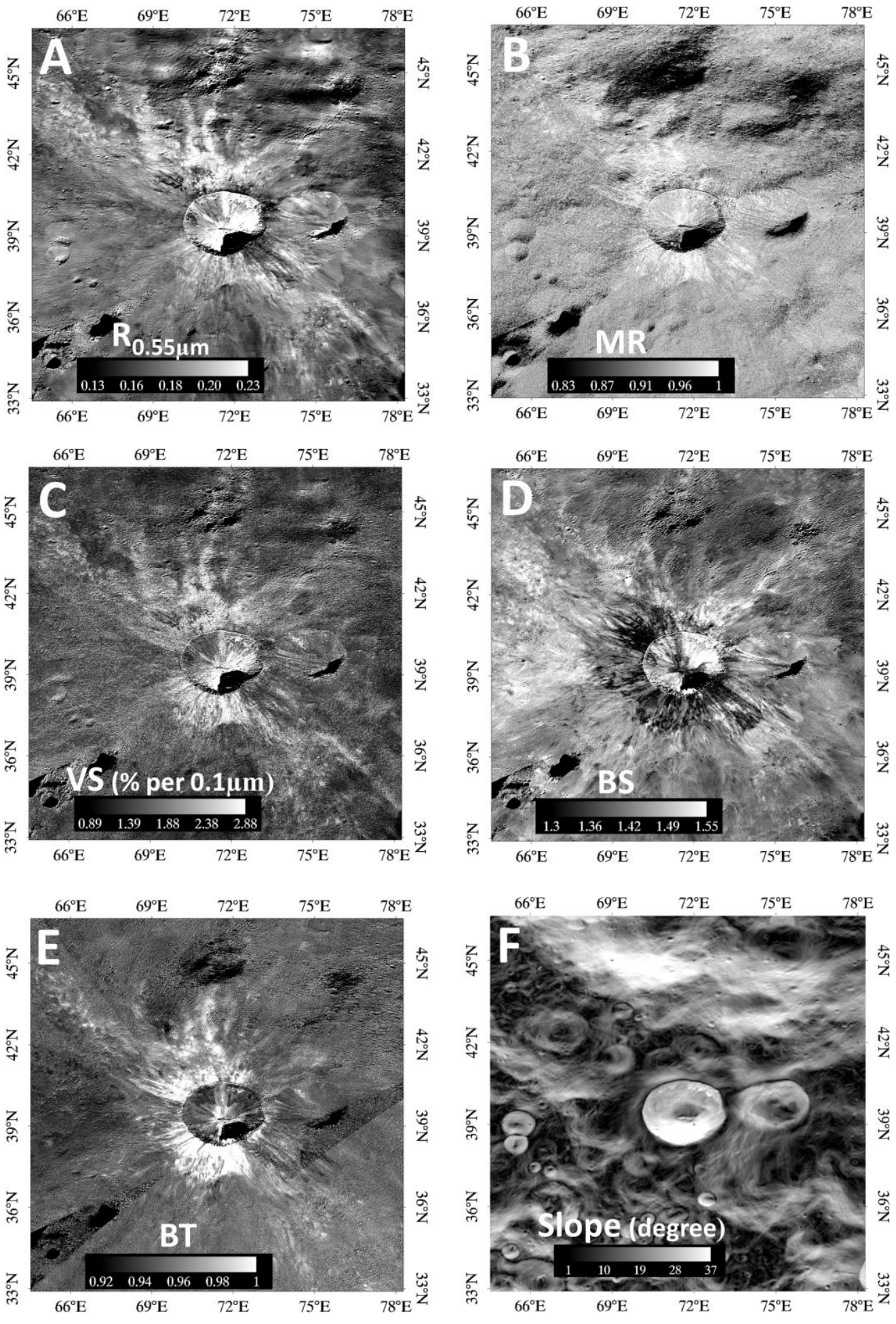

Fig. 6: Arruntia region from FC HAMO data (~ 60 m/pix). (A) Reflectance image at 0.55 µm. Derived band parameters: (B) MR (Mid Ratio), (C) VS (Visible Slope in % per 0.1µm), (D) BS (1-µm band strength), (E) BT (Band Tilt), (F) Surface (topographic) slope in degrees computed from HAMO-DTM (~ 62 m/pixel).

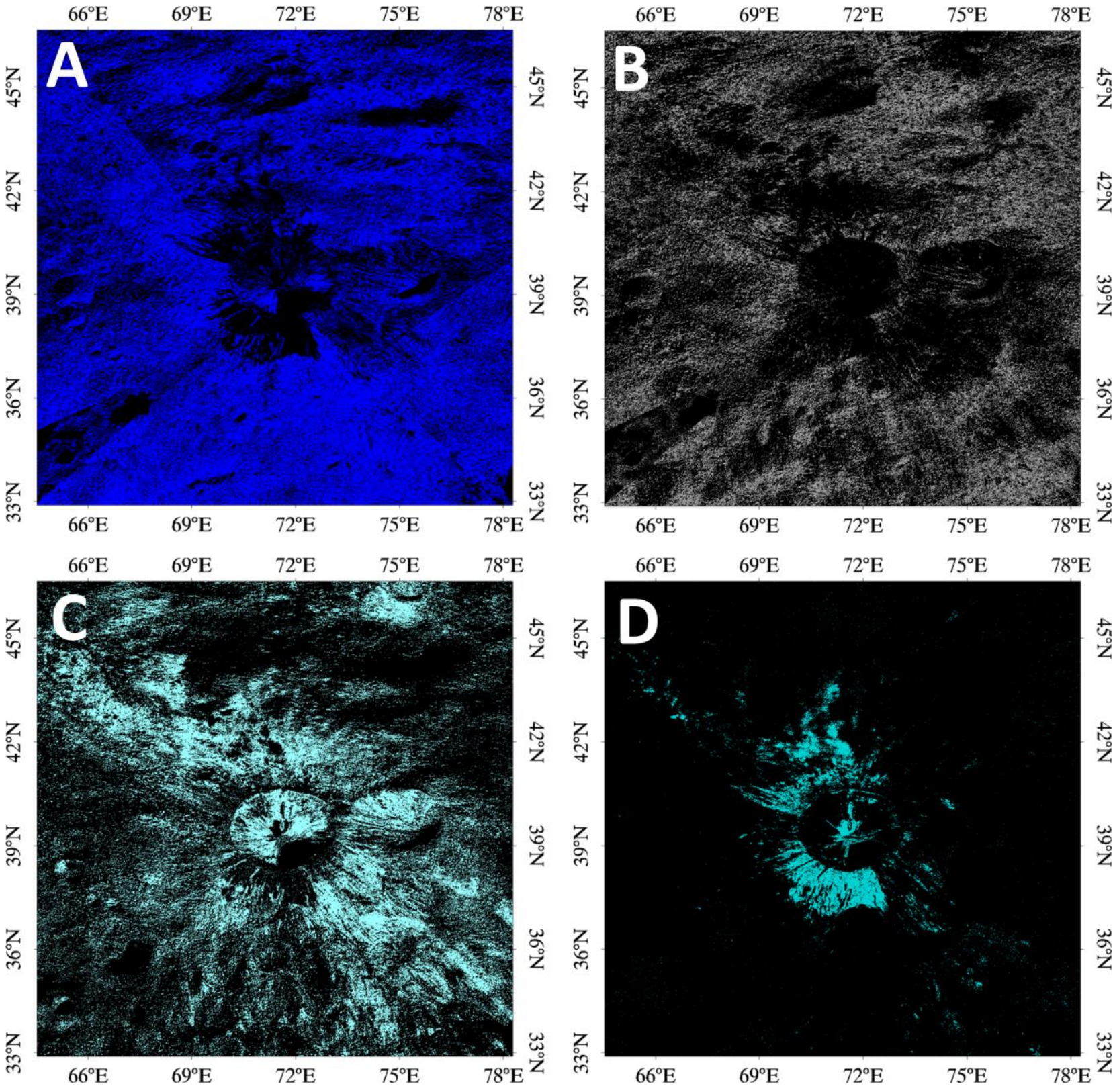

Fig. 7: Mineralogy of Arruntia based on our laboratory polyhedrons of BT versus MR versus $R_{0.55\mu m}$: (A) Eucrites, (B) CM2 plus HED mixtures (≤ 20-30 wt.% of CM2, indistinguishable from eucrites/howardites), (C) Olivine plus HED mixtures (≤ 40-60 wt.% of olivine, indistinguishable from eucrites/howardites), (D) Olivine-rich regions (≥ 40-60 wt.% olivine, distinguished from eucrites/howardites).

## 4.6 Discussion

### 4.6.1 Olivine on Vesta

Olivine on Vesta is currently an unresolved issue, but crucial for understanding the evolutionary aspects (e.g., Clenet et al., 2014; Consolmagno et al., 2015). Despite the predictions of olivine in the Rheasilvia basin (e.g., Gaffey, 1997; Thomas et al., 1997) and an olivine-rich mantle of Vesta (e.g., Righter and Drake, 1997), Dawn observations have not unambiguously detect olivine of mantle origin yet. Most of the olivine-rich sites are detected in the northern hemisphere instead (e.g., Thangjam et al., 2014; Nathues et al., 2015). Because of the lack of significant amount of olivine in the Rheasilvia basin, Clenet et al. (2014) suggest a deeper crust-mantle transition zone of Vesta (up to 80-100 km) which is quite unusual compared to other terrestrial silicate-dominated bodies. The unresolved issue of olivine is one of the arguments by Consolmagno et al. (2015) who claimed that Vesta is probably not an intact and pristine body.

Based on linear spectral unmixing analysis, Combe et al. (2015) argue that the olivine-rich areas identified at Bellicia, Arruntia and Pomponia craters are possible without olivine, i.e., a mixture of low- and high-Ca pyroxenes (hypersthene, pigeonite and diopside). However, the presence of olivine-rich material in the Arruntia region is obvious based on our three-dimensional spectral analysis as well as the earlier works (Ammannito et al., 2013a; Thangjam et al., 2014; Nathues et al., 2015; Ruesch et al., 2014b; Palomba et al., 2015). Spectral modeling by Poulet et al. (2014) suggests that olivine mixed with howardite is likely a ubiquitous component on Vesta. Nathues et al. (2015) and Le Corre et al. (2015) mention that olivine on Vesta is exogenic in origin. However, Cheek and Sunshine (2014) claim that olivine in Bellicia and Arruntia region are of shallow crustal plutonic in origin because they observe localities where Cr-rich pyroxene and olivine occur together. They further mention that the presence of such differentiated plutons on Vesta favor the evolution model of late stage serial magmatism (e.g., Yamaguchi et al., 1997; Mittlefehldt, 1994). However, there is geochemical and microstructural evidence who claimed the presence of potential olivine-bearing mantle material in few olivine-rich diogenites (Tkalcec et al., 2013; Tkalcec and Brenker, 2014) and howardites (Lunning et al., 2015; Hahn et al., 2015). Hahn et al. (2015) claim that they discovered harzburgite clasts of Mg-rich pyroxene plus olivine in howardites that preserved recrystallized textures. These authors hypothesize that the upper mantle of Vesta is harzburgitic in composition.

Thus, the origin of olivine on Vesta is still under debate. Meanwhile, limitations and difficulties in detecting olivine on Vesta's surface is also of concern, because the spectral signature of olivine may be either masked by the presence of pyroxenes (e.g., Beck et al., 2013) or is spatially unresolved by the instruments onboard Dawn (e.g., Beck et al., 2013;

Jutzi et al., 2013). Our spectral analysis technique still leave open the possibility of up to a few tens of weight percent olivine to be present on Vesta that is spectrally undetectable.

### 4.6.2 'Orange material' or shocked/glass/impact-melt

Le Corre et al. (2013) identified sites of 'Orange Material' (OM) mostly in equatorial and northern hemispheric regions including the Arruntia region. They suggest that such material shows red slope in the visible wavelength range. However, visible slope that is comparable with the Macibini-eucrite glass is not found in this work, though it is not necessarily true that all HED glass spectra should show similar characteristics to this sample. Eucrites with significant glass/impact melt components (e.g., LEW 85303, Padvarninkai) are neither distinguishable from the general eucrites nor match the visible slope with the nearly pure Macibini-eucrite glass. On the other hand, other components like olivine do exhibit red visible slope. Therefore, a red spectral slope in the visible wavelength range is not necessarily a unique diagnostic parameter of impact-melt-bearing HEDs. This point is noted by Ruesch et al. (2014b), too. From the morphological observations in the Arruntia crater, Ruesch et al. (2014a) suggest the possibility of the presence of thin impact melt veneers in the ejecta. We investigated morphological features for impact melts in and outside Arruntia using high-resolution clear filter LAMO images (~ 18 m/pixel). Potential impact-melt flow features are found on the crater wall, floor and ejecta (Fig. 8). The HAMO image (~ 60 m/pixel) shows the locations (Fig. 8A) of the putative impact-melt features identified in LAMO images (Fig. 8B-F). Streaks of dark material are observed in the continuous ejecta near the crater rim (Fig. 8B) and also on the inner crater wall. Streaks of dark material on the rim and ejecta (away from the crater) coated with thin veneers of impact melt are one of the common lunar impact-melt features, particularly in small lunar highland craters (e.g., Plescia and Cintala, 2012). The available resolution is not sufficient to resolve the features in fine details as was inspected from Lunar Reconnaissance Orbiter Camera - Narrow Angle Camera (e.g., Plescia and Cintala, 2012). Though the features are difficult to distinguish from ejecta debris, the dark streaks may represent pockets of impact melts ejected during the impact. Such streaks are found mainly in the continuous ejecta near the rim along with the bright material. More typical impact melt flow features are detected on crater wall slopes (Fig. 8C, D). The feature on the northwest flank of the crater wall (Fig. 8C) shows what appear to be pressure ridges and prominent levees. Pressure ridges are seen in lunar impact-melt flows where such features occurred in late stage melt-flows (e.g., Carter et al., 2012). Such ridges are also frequently observed in terrestrial lava flows, where viscous material is compressed due to horizontal and vertical velocity gradients and often develop at distal ends (Fink and Fletcher, 1978; Theilig and Greeley, 1986). Another impact-melt flow feature is observed at the northeastern wall showing faint pressure ridges and prominent levees (Fig. 8D). Both the features (Fig. 8C, D) occur along the steep slopes of the walls, and their widths become narrower and merge downslope as the underlying slope becomes shallower towards the floor. The floor of Arruntia crater is partially in the shade, but weakly illuminated by multi-scattered light, and therefore, it is difficult to inspect the features. The visibility of the floor of the crater has been enhanced in this scene (Fig. 8E). Bright-toned debris covers the floor, but the smooth-textured and comparatively darker material seems to overlie the debris partly. The flow features are not only detected on the floor and the ejecta near the rim, but are also found on the outer rim and in the discontinuous ejecta of the crater (Fig. 8F). The lobate flow features that exhibit a smooth texture and significant difference in spatial distribution of craters represent a potential impact melt veneer deposited on rather a densely-cratered surface. These features extend as continuous and discontinuous lobes. Similar putative impact-melt flow features are identified on Vesta by Williams et al. (2014). Based on numerical modeling and scaling laws, Williams et al. (2014) suggest that impact-melt formation on Vesta is possible though the volumes are less compared with similar-sized lunar craters. On the other hand, Keil et al. (1997) claim that impact-melt deposits on asteroids like Vesta are negligible because of rather low impact velocities compared to the terrestrial planets. It is worth mentioning that impact-melt/shock/glass has been identified in HED meteorites but is not as common and abundant as in lunar and Martian achondrites (e.g., Buchanan et al., 2000; Singerling et al., 2013; Rubin, 2015). Material like impact-melts, shocked- and glassy-components in HEDs may bear different compositional/mineralogical characteristics depending on the intensity of the shocked pressures/temperatures or thermal metamorphism (e.g., Rubin, 2015; Takeda and Graham, 1991), and accordingly they are expected to show different spectral features.

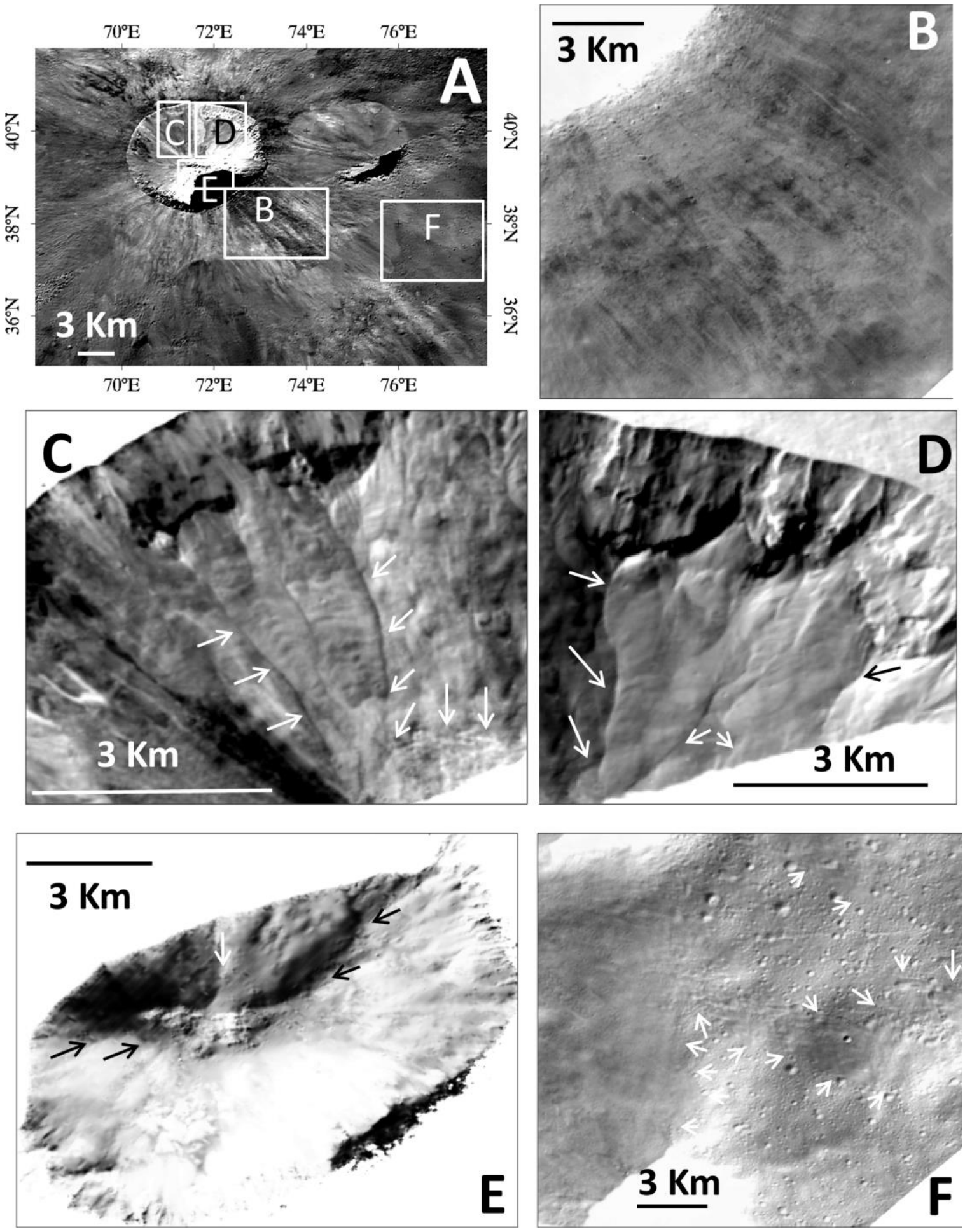

Fig. 8: Potential impact-melt flow features (marked by arrows) observed in Arruntia and its ejecta. North is upwards for all figures. (A) HAMO image at 0.55 µm ( ~ 60 m/pixel) showing the locations of the features identified from LAMO images (~ 18 m/pixel). (B) dark streaks on the bright ejecta near the outer rim, (C, D) lobate flow features on the slopes of inner crater wall, (E) potential impact-melt deposits partly ponded, imparting a darker tone on the brighter debris in the floor of the crater (enhanced image showing a scene that is only illuminated by multi-scattered light), (F) lobate flow-features in the relatively brighter ejecta overlying the densely cratered older surface.

### 4.6.3 Dark and bright material

Vesta is known for showing significant albedo variations across its surface (e.g., Reddy et al., 2012b, c). Dark material is more abundant than bright material (e.g., Palomba et al., 2014; Zambon et al., 2014). Palomba et al. (2014) suggest that dark material is dominantly eucritic or howarditic in composition with significant contributions of exogenous carbonaceous chondritic material. They mentioned that shocked components in HEDs could also be a representative of dark material on Vesta because of their low reflectances. Bright material is interpreted as relatively fresh and recently excavated sub-surface components because of their higher reflectance and deeper band depth values (Zambon et al., 2014). Olivine-rich areas in Bellicia and Arruntia region are noted as a constituent of the bright material (Zambon et al., 2014). They note that the bright material units of Arruntia region are mixed with dark material. However, Palomba et al. (2014) does not mention the presence of dark material in and around the Arruntia region though they defined at least 15% lower reflectance than the local average as one of the criteria to identify this material. For Arruntia region, we observe that dark, bright and gray regions show average reflectance ($R0.55\mu m$) of approximately 0.14, 0.19, and 0.25, respectively. The variation of reflectance of the gray material with the bright and dark material are about 31% and 26%, respectively, while the variation of bright to the dark material is about 66%. Many of the bright areas on the inner walls, and few ejecta of the crater display rather a high reflectance (~ 0.4 or more). The CM2-rich exposures (> 20-30% of CM2 content) are found in very few localized areas. The existence of plenty of areas with lesser amounts of CM2-material (less than 20-30 wt.%) is possible, mixed with eucrites/howardites (Fig. 7B). The olivine-rich and possibly olivine-poor areas also display higher reflectance than the surrounding areas (Fig. 7C, D).

## 4.7 Summary and future work

We introduce and apply an innovative three-dimensional spectral approach (three band parameter space with polyhedrons) that better characterizes the heterogeneous surface compositions compared to the generally used two-dimensional spectral approach (two band parameter space). Several lithologic units are mapped using the FC color data based on the laboratory polyhedrons defined from various spectra of HEDs, olivine, CM2, and their mixtures. According to this study, we find that eucrite-rich howardite is the dominant composition in the Arruntia region. Olivine-rich exposures are found in rather large spatial extents compared to the earlier studies. Spectral features in FC color data do not allow unique identification of impact-melt/shock/glass materials; however, morphological features investigated using higher resolution FC clear filter images identified potential impact melt flow features in this region. Nearly pure CM2, olivine, glass and diogenite exposures are likely not present in this region. This three-dimensional spectral analysis can be applied to the entire Vestan surface to map the compositional heterogeneity. The applicability of this approach to the hyperspectral datasets such as the VIR data as well as to other planetary bodies are yet to be proven.

## 4.9 Acknowledgement

We are thankful to the RELAB and USGS spectral library for the spectra used in this work. G.T. is grateful to Ladislav Rezac, Nagaraju Krishnappa, Jayant Joshi, Dick Jackson and Megha Bhatt for their help in IDL/MATLAB/python programming. G.T. would like to heartily thank many colleagues from TU/Clausthal (Cornelia Ambrosi, Silke Schlenczek, Dietlind Nordhausen), TU/Hannover (Harald Behrens), and MPS (Walter Goetz, Harald Steininger, Henning Fischer) for their help during the preparation of the samples/mixtures. G.T. is thankful to Pierre Beck for providing spectra of olivine-rich diogenites NWA 5480 and NWA 4223. E.A.C. thanks the Canadian Space Agency, the Canada Foundation for Innovation, the Manitoba Research Innovations Fund, and NSERC for supporting the establishment and operation of HOSERLab at the University of Winnipeg. We would like to acknowledge the two anonymous reviewers for their thorough review and constructive comments that helped to improve the manuscript.

## 4.10 Appendix

### 4.10.1 Description of the meteorite specimens used in this study

The meteorite Dar al Gani 779 (DaG 779) has been classified as a howardite with a weathering grade W1 and a shock stage S2 (Grossman, 2000). The modal mineralogy is calculated by using the mineral data (electron microprobe) and whole rock XRF analyzes at TUC. Optical inspection in plane and crossed polarized light reveals distinct clasts in a fine-grained matrix. The majority of the clasts are of diogenite composition, which is dominated by orthopyroxene or, low calcium pyroxene. In size, they are not larger than 5 mm. Few lithic clasts of eucritic composition are present. Mineral analyzes show non-uniform composition of pyroxenes. Their composition ranges from a more enstatitic (~$Wo_{4\pm3}En_{66\pm5}Fs_{30\pm5}$) to pigeonite (~$Wo_{7\pm1}En_{36\pm2}Fs_{57\pm2}$). With increasing Ca-content, there is augite ($\sim Wo_{43\pm1}En_{29\pm1}Fs_{29\pm2}$) as well as rare diopsides. Plagioclase feldspars range from $An_{70}$ to $An_{94}$ with negligible K-feldspar component. Minor minerals are spinel (mostly chromite in composition), ilmenite, olivine ($\sim Fo_{67}$), a silica phase, troilite and rare Ca-phosphate. This howardite contains 77 wt.% orthopyroxene, 8 wt.% plagioclase, 7 wt.% clinopyroxene, 2 wt.% spinel and minor content (≤1 wt.%) of olivine, silica phase, troilite, ilmenite, apatite/whitlockite and rare terrestrial weathering products, such as Ca-carbonates and iron-oxides/-oxy-hydroxides. In DaG 779, a number of secondary processes typical for howardites are observed: 1) recrystallization after melting due to an impact, especially in eucritic clasts; 2) rapidly cooled glasses with quench crystals; 3) metasomatic alteration reactions producing secondary augite, a silica phase, and FeS from pigeonite plus sulfur-rich fluid.

The meteorite North West Africa 1929 (NWA 1929) has been classified as a howardite (Connolly et al., 2006). They find that this specimen consists of 72 vol.-% cumulate eucrite clasts ($Fs_{45–40}Wo_{7–20}$; plagioclase: $An_{91.2–95.3}$; metal: Ni = 0.97 wt.-%, Cr = 0.87 wt.-%), 14 vol.-% diogenite clasts (pyroxene: $Fs_{43–54}Wo_{2.5–3.6}$; Fe/Mn = 36, 37), 6 vol.-% melt (glass) clasts, and 8 vol.-% subophitic clasts. The eucritic clasts look like coarse-grained gabbroic rocks (Connolly et al., 2006). Solid state recrystallization of pyroxene and plagioclase occurs throughout the sample, showing localized melt pockets/veins within clasts (Connolly et al., 2006). NWA 1929 yields a bulk density of 2.91 g/cm$^3$ (McCausland and Flemming, 2006). Howardite NWA 2698 and eucrite NWA 2690 are probably paired with NWA 1929 (Connolly et al., 2006).

The chondrite Jbilet-Winselwan is categorized as a carbonaceous meteorite, Type II (CM2) with shock stage S0 and weathering grade W1 (Meteoritical Bulletin 102, in preparation). Hewins and Garvie (Meteoritical Bulletin 102, in preparation) find that the specimen they analyzed contains chondrules/fragments of Types I and II that includes barred olivine-porphyritic olivine of formerly metal-rich and olivine-pyroxene Type I chondrules, and forsterite relict-grained Type II chondrules. They reported that most of the chondrules sizes are around 200 µm, but range up to 1.2 mm, while a few CAIs are 800 µm in size. They also noted that powder X-ray diffraction analysis show serpentine, smectites, and tochilinite. Hewins (Meteoritical Bulletin 102, in preparation) analyzed olivine composition ranging from $Fa_{0.98\pm0.44}$ ($Fa_{25-40}$) to $Fa_{2.6\pm1.5}$ ($Fa_{40-61}$). Russell et al. (2014) report that Jbilet-Winselwan experienced only minor aqueous alteration though a few severely altered portions is also seen. They observe rather clear chondrules in a fine-grained matrix. Whole rock analyzes chemical data are not yet available.

Olivine used here for mixtures with howardite and CC is from the Mineral Collection of Clausthal University of Technology, their origin is unknown but probably from reworked mantle xenoliths similar to those of Navajo Co. AZ, USA deposits. Individual grains (3 to 7 mm in size) are of gem stone quality with very rare inclusions of chromite.

### 4.10.2 Chemical whole rock data of two howardites and an olivine.

The result of Jbilet-Winselwan is not yet available. Data from XRF based on fused glass diluted by $Li_2B_4O_7$ (1:5).

| Whole-rock composition | | | |
|---|---|---|---|
| **Wt. %** | **DaG 779** | **NWA 1929** | **Olivine** |
| **$SiO_2$** | 47.79 | 46.22 | 38.92 |
| **$TiO_2$** | 0.16 | 0.67 | 0.01 |
| **$Al_2O_3$** | 3.46 | 11.56 | 0.02 |
| **$Al_2O_3$** | 18.30 | 19.32 | 10.07 |
| **MnO** | 0.51 | 0.52 | 0.11 |
| **MgO** | 20.17 | 7.31 | 50.93 |
| **CaO** | 3.28 | 10.02 | 0.07 |
| **$Na_2O$** | 0.13 | 0.43 | - |
| **$K_2O$** | 0.02 | 0.05 | - |
| **Cr (ppm)** | 7438.00 | 13.00 | 100.00 |
| **Co (ppm)** | 19.00 | 4.00 | - |
| **Ni (ppm)** | 33.00 | 29.00 | 2700.00 |
| **∑ (%)** | 94.57 | 96.15 | 101.04 |

## 4.10.3 Spectra/samples and band parameters (values)

| Euc | Size (μm) | BT | MR | R0.555 | BS | VS | Source |
|---|---|---|---|---|---|---|---|
| A-87272 | 0-25 | 0.97 | 0.89 | 0.49 | 2.09 | 5.74 | Relab |
| A-881819 | 0-25 | 0.93 | 0.88 | 0.48 | 1.82 | 5.14 | Relab |
| ALH-78132 | 0-25 | 0.92 | 0.86 | 0.37 | 1.63 | 4.16 | Relab |
| ALH-78132 | 25-45 | 0.91 | 0.87 | 0.22 | 2.15 | 3.40 | Relab |
| ALH-78132 | 45-75 | 0.91 | 0.91 | 0.17 | 2.33 | 2.74 | Relab |
| ALH85001 (dry-sieved) | 0-25 | 0.91 | 0.90 | 0.51 | 1.68 | 3.16 | Relab |
| ALH85001 (wet-sieved) | 0-25 | 0.92 | 0.89 | 0.42 | 1.56 | 2.32 | Relab |
| ALHA76005 | 0-25 | 0.95 | 0.85 | 0.43 | 1.52 | 4.23 | Relab |
| ALHA76005 | 125-250 | 0.96 | 0.92 | 0.18 | 1.78 | 1.74 | Relab |
| ALHA76005 | 250-500 | 0.97 | 0.90 | 0.18 | 1.65 | 1.45 | Relab |
| ALHA76005 | 25-45 | 0.94 | 0.85 | 0.29 | 1.79 | 3.75 | Relab |
| ALHA76005 | 45-75 | 0.94 | 0.85 | 0.21 | 1.99 | 2.91 | Relab |
| ALHA76005 | 75-125 | 0.95 | 0.89 | 0.19 | 1.88 | 2.21 | Relab |
| ALHA76005 | 0-250 | 0.96 | 0.89 | 0.23 | 1.66 | 2.37 | Relab |
| ALHA81001 | 0-125 | 0.93 | 0.81 | 0.20 | 1.49 | 1.20 | Relab |
| ALHA81001 | 0-45 | 0.95 | 0.84 | 0.33 | 1.35 | 2.58 | Relab |
| ALHA81011 | 0-125 | 1.00 | 0.86 | 0.24 | 1.71 | 1.69 | Relab |
| Bereba | 0-25 | 0.96 | 0.84 | 0.39 | 1.58 | 2.50 | Relab |
| Bouvante | 0-500 | 0.99 | 0.85 | 0.19 | 1.53 | 0.77 | Relab |
| Bouvante | 0-25 | 0.96 | 0.78 | 0.27 | 1.56 | 3.49 | Relab |
| Bouvante | 0-250 | 0.98 | 0.81 | 0.20 | 1.61 | 1.26 | Relab |
| Bouvante | 0-44 | 0.96 | 0.80 | 0.16 | 1.69 | 1.62 | Relab |
| Cachari | 0-25 | 0.96 | 0.86 | 0.43 | 1.97 | 6.40 | Relab |
| EET87520 | 0-45 | 1.01 | 0.88 | 0.49 | 2.19 | 3.91 | Relab |
| EET87542 | 0-25 | 0.98 | 0.89 | 0.30 | 1.28 | 4.11 | Relab |
| EET90020 | 0-25 | 0.98 | 0.88 | 0.52 | 1.97 | 6.83 | Relab |
| EET92003 | 0-125 | 0.93 | 0.90 | 0.31 | 2.20 | 4.04 | Relab |
| EETA79005 | 0-25 | 0.93 | 0.84 | 0.38 | 1.73 | 5.03 | Relab |
| EETA79005 | 0-250 | 0.94 | 0.89 | 0.26 | 1.81 | 2.82 | Relab |
| EETA79006 | 0-125 | 0.94 | 0.86 | 0.34 | 1.78 | 3.56 | Relab |
| GRO95533 | 0-25 | 0.97 | 0.87 | 0.51 | 1.81 | 3.38 | Relab |
| Jonzac | 0-25 | 0.94 | 0.81 | 0.38 | 1.99 | 4.25 | Relab |
| Juvinas | 0-25 | 0.95 | 0.86 | 0.42 | 1.77 | 3.08 | Relab |
| Juvinas | 125-250 | 0.98 | 0.99 | 0.13 | 2.02 | 1.15 | Relab |
| Juvinas | 25-45 | 0.95 | 0.91 | 0.27 | 2.19 | 2.99 | Relab |
| Juvinas | 45-75 | 0.96 | 0.95 | 0.20 | 2.26 | 2.44 | Relab |
| Juvinas | 75-125 | 0.97 | 0.97 | 0.15 | 2.17 | 1.71 | Relab |
| LEW85303 | 0-25 | 0.98 | 0.83 | 0.31 | 1.56 | 3.30 | Relab |
| LEW87004 | 0-25 | 0.94 | 0.84 | 0.35 | 1.63 | 4.37 | Relab |
| Macibini-clast3 | 0-63 | 0.96 | 0.87 | 0.21 | 1.71 | 1.61 | Relab |
| Millbillillie (C1MB69) | 0-25 | 0.98 | 0.93 | 0.29 | 1.95 | 1.32 | Relab |
| Millbillillie (C2MB69) | 0-25 | 0.98 | 0.93 | 0.29 | 2.01 | 1.53 | Relab |
| Millbillillie (CAMB69) | 0-25 | 0.96 | 0.87 | 0.43 | 1.72 | 1.97 | Relab |
| Millbillillie (C3MB69) | 0-25 | 0.97 | 0.85 | 0.43 | 1.74 | 2.12 | Relab |
| Millbillillie | 25-45 | 0.95 | 0.88 | 0.31 | 2.25 | 2.39 | Relab |
| Millbillillie | 45-75 | 0.96 | 0.89 | 0.25 | 2.26 | 1.92 | Relab |
| Millbillillie | 75-125 | 0.97 | 0.91 | 0.23 | 2.19 | 1.48 | Relab |
| Millbillillie | 0-80 | 0.97 | 0.89 | 0.38 | 1.44 | 1.86 | Relab |
| Millbillillie | 0-75 | 0.98 | 0.90 | 0.40 | 1.64 | 2.49 | Relab |
| Moore_County | 0-25 | 0.95 | 0.90 | 0.48 | 2.29 | 4.01 | Relab |
| PCA82501 | 0-125 | 0.98 | 0.86 | 0.26 | 1.42 | 2.40 | Relab |
| PCA82502 | 0-25 | 0.95 | 0.80 | 0.35 | 1.67 | 3.81 | Relab |
| PCA91006 | 0-125 | 0.98 | 0.91 | 0.35 | 1.90 | 3.77 | Relab |
| PCA91007 | 0-125 | 0.95 | 0.77 | 0.22 | 1.82 | 2.60 | Relab |
| PCA91078 | 0-45 | 0.99 | 0.83 | 0.28 | 1.80 | 2.78 | Relab |
| Padvarninkai | 0-25 | 0.95 | 0.84 | 0.35 | 1.73 | 2.48 | Relab |
| Padvarninkai | 25-45 | 0.96 | 0.91 | 0.26 | 1.95 | 1.93 | Relab |
| Pasamonte | 0-25 | 0.96 | 0.80 | 0.37 | 1.82 | 5.63 | Relab |
| Serra-de-Mage | 0-25 | 0.94 | 0.97 | 0.65 | 1.72 | 4.32 | Relab |
| Stannern | 0-25 | 0.95 | 0.83 | 0.35 | 1.55 | 2.80 | Relab |
| Stannern | 25-45 | 0.95 | 0.85 | 0.22 | 1.83 | 2.14 | Relab |
| Y-74450 | 0-25 | 0.95 | 0.87 | 0.32 | 1.54 | 3.01 | Relab |
| Y-74450 | 25-45 | 0.94 | 0.88 | 0.21 | 2.01 | 2.85 | Relab |
| Y-74450 | 45-75 | 0.95 | 0.92 | 0.18 | 2.18 | 2.14 | Relab |
| Y-74450 | 75-125 | 0.96 | 0.94 | 0.16 | 2.10 | 1.89 | Relab |
| Y-792510 | 0-25 | 0.97 | 0.85 | 0.43 | 1.87 | 3.59 | Relab |
| Y-792769 | 0-25 | 0.96 | 0.82 | 0.31 | 1.54 | 3.13 | Relab |
| Y-793591 | 0-25 | 0.96 | 0.82 | 0.37 | 1.55 | 3.80 | Relab |
| Y-82082 | 0-125 | 0.99 | 0.84 | 0.36 | 1.52 | 5.71 | Relab |
| **Dio** | **Size (μm)** | **BT** | **MR** | **R0.555** | **BS** | **VS** | **Source** |
| A-881526 | 0-25 | 0.82 | 1.01 | 0.54 | 2.65 | 10.98 | Relab |
| ALHA77256 | 0-25 | 0.84 | 0.95 | 0.48 | 2.23 | 8.48 | Relab |
| Aioun-el-Atrous | 0-25 | 0.86 | 0.97 | 0.62 | 2.54 | 6.42 | Relab |
| EETA79002 | 0-25 | 0.84 | 0.91 | 0.38 | 1.93 | 2.66 | Relab |
| EETA79002 | 125-250 | 0.87 | 1.01 | 0.22 | 2.28 | 1.30 | Relab |
| EETA79002 | 250-500 | 0.88 | 1.00 | 0.20 | 2.16 | 1.29 | Relab |
| EETA79002 | 25-45 | 0.81 | 0.96 | 0.28 | 2.73 | 2.36 | Relab |
| EETA79002 | 45-75 | 0.82 | 1.02 | 0.24 | 2.64 | 2.31 | Relab |
| EETA79002 | 75-125 | 0.85 | 1.06 | 0.22 | 2.55 | 1.75 | Relab |
| Ellemeet | 0-25 | 0.83 | 0.91 | 0.28 | 2.30 | 2.31 | Relab |
| GRO95555 | 0-25 | 0.84 | 0.95 | 0.61 | 2.51 | 10.17 | Relab |
| Johnstown | 0-25 | 0.87 | 0.94 | 0.40 | 1.75 | 4.23 | Relab |
| Johnstown | 25-45 | 0.83 | 1.07 | 0.35 | 2.57 | 4.55 | Relab |
| Johnstown | 0-75 | 0.83 | 1.06 | 0.50 | 2.70 | 7.27 | Relab |
| LAP91900 | 0-25 | 0.83 | 0.95 | 0.58 | 2.52 | 6.68 | Relab |
| Tatahouine | 0-25 | 0.82 | 0.97 | 0.61 | 2.61 | 5.33 | Relab |
| Y-74013 | 0-25 | 0.89 | 0.89 | 0.32 | 1.66 | 4.95 | Relab |
| Y-74013 | 25-45 | 0.85 | 0.93 | 0.29 | 2.42 | 4.62 | Relab |
| Y-75032 | 0-25 | 0.89 | 0.87 | 0.33 | 1.93 | 2.41 | Relab |
| Y-75032 | 25-45 | 0.86 | 0.92 | 0.20 | 2.70 | 2.82 | Relab |
| **How** | **Size (μm)** | **BT** | **MR** | **R0.555** | **BS** | **VS** | **Source** |
| Binda | 0-25 | 0.87 | 0.95 | 0.51 | 2.63 | 5.85 | Relab |
| Bununu | 0-25 | 0.93 | 0.88 | 0.41 | 1.72 | 2.52 | Relab |
| EET83376 | 0-25 | 0.93 | 0.84 | 0.42 | 1.71 | 4.87 | Relab |
| EET87503 (wet-sieved) | 0-150 | 0.93 | 0.88 | 0.26 | 1.65 | 1.91 | Relab |
| EET87503 | 0-25 | 0.93 | 0.88 | 0.39 | 1.56 | 2.32 | Relab |
| EET87503 (wet-sieved) | 0-25 | 0.94 | 0.87 | 0.32 | 1.43 | 2.75 | Relab |
| EET87503 | 125-250 | 0.96 | 0.96 | 0.19 | 1.73 | 0.91 | Relab |
| EET87503 | 250-500 | 0.97 | 0.95 | 0.18 | 1.64 | 0.75 | Relab |
| EET87503 | 25-45 | 0.93 | 0.89 | 0.30 | 1.75 | 1.90 | Relab |
| EET87503 | 45-75 | 0.93 | 0.94 | 0.23 | 1.90 | 1.14 | Relab |
| EET87503 | 75-125 | 0.95 | 0.97 | 0.20 | 1.87 | 1.11 | Relab |
| EET87513 | 0-25 | 0.94 | 0.87 | 0.34 | 1.50 | 3.10 | Relab |
| Frankfort | 0-25 | 0.90 | 0.85 | 0.37 | 1.95 | 2.01 | Relab |
| GRO95535 | 0-25 | 0.92 | 0.87 | 0.38 | 1.65 | 3.06 | Relab |
| GRO95574 | 0-125 | 0.92 | 0.90 | 0.29 | 1.83 | 1.81 | Relab |
| Kapoeta | 0-25 | 0.93 | 0.88 | 0.32 | 1.60 | 2.34 | Relab |
| Le-Teilleul | 0-25 | 0.89 | 0.87 | 0.40 | 2.09 | 3.35 | Relab |
| Petersburg | 0-25 | 0.94 | 0.86 | 0.32 | 1.67 | 9.11 | Relab |
| QUE94200 | 0-25 | 0.88 | 0.89 | 0.43 | 1.82 | 3.20 | Relab |
| QUE97001 | 0-150 | 0.87 | 0.95 | 0.31 | 2.16 | 3.00 | Relab |
| Y-7308 | 0-25 | 0.89 | 0.90 | 0.44 | 2.22 | 4.56 | Relab |
| Y-790727 | 0-25 | 0.91 | 0.86 | 0.36 | 1.80 | 6.25 | Relab |
| Y-791573 | 0-25 | 0.90 | 0.87 | 0.37 | 1.78 | 6.49 | Relab |
| NWA1949 | 0-63 | 0.95 | 0.79 | 0.29 | 1.58 | 4.11 | **This work** |
| DaG779 | 0-63 | 0.90 | 0.90 | 0.38 | 1.91 | 3.26 | **This work** |
| **CM-HED** | **Size (μm)** | **BT** | **MR** | **R0.555** | **BS** | **VS** | **Source** |
| NWA1949-10-Jbilet90 | 0-63 | 0.99 | 1.00 | 0.04 | 1.00 | 0.44 | **This work** |
| NWA1949-20-Jbilet80 | 0-63 | 0.99 | 0.99 | 0.05 | 1.01 | 0.46 | **This work** |
| NWA1949-30-Jbilet70 | 0-63 | 0.99 | 0.98 | 0.05 | 1.02 | 0.51 | **This work** |
| NWA1949-40-Jbilet60 | 0-63 | 0.99 | 0.97 | 0.05 | 1.03 | 0.57 | **This work** |
| NWA1949-50-Jbilet50 | 0-63 | 0.98 | 0.95 | 0.06 | 1.08 | 0.64 | **This work** |
| NWA1949-60-Jbilet40 | 0-63 | 0.98 | 0.94 | 0.07 | 1.10 | 0.79 | **This work** |
| NWA1949-70-Jbilet30 | 0-63 | 0.98 | 0.92 | 0.09 | 1.13 | 0.99 | **This work** |
| NWA1949-80-Jbilet20 | 0-63 | 0.97 | 0.90 | 0.12 | 1.18 | 1.31 | **This work** |
| NWA1949-90-Jbilet10 | 0-63 | 0.97 | 0.86 | 0.15 | 1.30 | 2.01 | **This work** |
| DAG779-10-Jbilet90 | 0-63 | 0.99 | 1.00 | 0.05 | 1.01 | 0.41 | **This work** |
| DAG779-20-Jbilet80 | 0-63 | 0.98 | 0.99 | 0.06 | 1.03 | 0.47 | **This work** |
| DAG779-30-Jbilet70 | 0-63 | 0.98 | 0.98 | 0.06 | 1.04 | 0.52 | **This work** |
| DAG779-40-Jbilet60 | 0-63 | 0.98 | 0.98 | 0.07 | 1.07 | 0.58 | **This work** |
| DAG779-50-Jbilet50 | 0-63 | 0.97 | 0.97 | 0.08 | 1.11 | 0.62 | **This work** |
| DAG779-60-Jbilet40 | 0-63 | 0.96 | 0.96 | 0.09 | 1.14 | 0.71 | **This work** |
| DAG779-70-Jbilet30 | 0-63 | 0.95 | 0.94 | 0.12 | 1.23 | 0.84 | **This work** |
| DAG779-80-Jbilet20 | 0-63 | 0.94 | 0.93 | 0.14 | 1.32 | 1.01 | **This work** |
| DAG779-90-Jbilet10 | 0-63 | 0.92 | 0.91 | 0.20 | 1.52 | 1.26 | **This work** |
| Millbillillie95-Murchison5 | 0-45 | 1.00 | 0.90 | 0.31 | 1.63 | 1.98 | HOSERLab |
| Millbillillie90-Murchison10 | 0-45 | 1.00 | 0.91 | 0.27 | 1.50 | 1.12 | HOSERLab |
| Millbillillie80-Murchison20 | 0-45 | 1.00 | 0.92 | 0.19 | 1.36 | 0.89 | HOSERLab |
| Millbillillie70-Murchison30 | 0-45 | 1.00 | 0.93 | 0.15 | 1.26 | 0.73 | HOSERLab |
| Millbillillie40-Murchison60 | 0-45 | 1.00 | 0.94 | 0.12 | 1.20 | 0.63 | HOSERLab |
| Millbillillie50-Murchison50 | 0-45 | 0.99 | 0.95 | 0.10 | 1.14 | 0.55 | HOSERLab |
| Millbillillie40-Murchison60 | 0-45 | 0.99 | 0.96 | 0.08 | 1.10 | 0.50 | HOSERLab |
| PRA04401 | 0-45 | 0.99 | 0.96 | 0.09 | 1.17 | 0.62 | HOSERLab |
| PRA04401 | 45-90 | 1.00 | 1.00 | 0.07 | 1.42 | 0.41 | HOSERLab |
| PRA04401 | 90-125 | 1.00 | 1.02 | 0.07 | 1.30 | 0.51 | HOSERLab |
| **Olivine-HED** | **Size (μm)** | **BT** | **MR** | **R0.555** | **BS** | **VS** | **Source** |
| NWA1949-10-Olivine90 | 0-63 | 1.06 | 0.97 | 0.65 | 1.36 | 3.33 | **This work** |
| NWA1949-20-Olivine80 | 0-63 | 1.03 | 0.93 | 0.56 | 1.37 | 3.06 | **This work** |
| NWA1949-30-Olivine70 | 0-63 | 1.02 | 0.90 | 0.49 | 1.38 | 3.52 | **This work** |
| NWA1949-40-Olivine60 | 0-63 | 1.01 | 0.89 | 0.46 | 1.38 | 3.72 | **This work** |
| NWA1949-50-Olivine50 | 0-63 | 1.00 | 0.87 | 0.42 | 1.41 | 3.72 | **This work** |
| NWA1949-60-Olivine40 | 0-63 | 0.98 | 0.85 | 0.39 | 1.43 | 3.95 | **This work** |
| NWA1949-70-Olivine30 | 0-63 | 0.98 | 0.84 | 0.35 | 1.48 | 3.86 | **This work** |
| NWA1949-80-Olivine20 | 0-63 | 0.97 | 0.82 | 0.32 | 1.52 | 4.07 | **This work** |
| NWA1949-90-Olivine10 | 0-63 | 0.96 | 0.82 | 0.32 | 1.53 | 3.92 | **This work** |
| DAG779-10-Olivine90 | 0-63 | 1.04 | 1.00 | 0.69 | 1.48 | 2.88 | **This work** |
| DAG779-20-Olivine80 | 0-63 | 1.02 | 0.98 | 0.62 | 1.49 | 2.91 | **This work** |
| DAG779-30-Olivine70 | 0-63 | 0.99 | 0.97 | 0.55 | 1.54 | 2.91 | **This work** |
| DAG779-40-Olivine60 | 0-63 | 0.97 | 0.96 | 0.49 | 1.57 | 2.73 | **This work** |
| DAG779-50-Olivine50 | 0-63 | 0.96 | 0.95 | 0.48 | 1.63 | 2.93 | **This work** |
| DAG779-60-Olivine40 | 0-63 | 0.94 | 0.93 | 0.46 | 1.65 | 3.05 | **This work** |
| DAG779-70-Olivine30 | 0-63 | 0.92 | 0.93 | 0.43 | 1.78 | 3.03 | **This work** |
| DAG779-80-Olivine20 | 0-63 | 0.92 | 0.92 | 0.42 | 1.84 | 3.04 | **This work** |
| DAG779-90-Olivine10 | 0-63 | 0.90 | 0.91 | 0.38 | 1.91 | 2.81 | **This work** |
| NWA5480 | | 0.91 | 1.05 | 0.35 | 2.30 | 7.58 | Beck et al. 2011 |
| NWA4223 | | 0.95 | 0.99 | 0.16 | 1.39 | 2.96 | Beck et al. 2011 |
| NWA6013 | 0-45 | 0.95 | 0.92 | 0.31 | 1.69 | 4.00 | HOSERLab |
| NWA2968 | 0-45 | 1.02 | 0.97 | 0.16 | 1.06 | 3.66 | HOSERLab |
| **Olivine** | **Size (µm)** | **BT** | **MR** | **R0.555** | **BS** | **VS** | **Source** |
| Fo11 | 0-60 | 1.10 | 0.93 | 0.27 | 1.85 | 7.20 | USGS |
| Fo18 | 0-60 | 1.07 | 1.03 | 0.35 | 1.76 | 8.01 | USGS |
| Fo29 | 0-60 | 1.16 | 1.03 | 0.33 | 2.34 | 9.02 | USGS |
| Fo41 | 0-60 | 1.10 | 0.99 | 0.36 | 1.80 | 7.75 | USGS |
| Fo51 | 0-60 | 1.12 | 1.00 | 0.38 | 1.85 | 6.60 | USGS |
| Fo60 | 0-60 | 1.14 | 1.01 | 0.46 | 1.98 | 7.83 | USGS |
| Fo66 | 0-60 | 1.17 | 1.01 | 0.44 | 2.28 | 7.93 | USGS |
| Fo89 | 0-60 | 1.10 | 0.96 | 0.58 | 1.40 | 5.94 | USGS |
| Fo91 | 0-60 | 1.26 | 1.01 | 0.55 | 2.50 | 8.36 | USGS |
| Fo10 | 0-45 | 1.03 | 1.03 | 0.58 | 1.30 | 4.56 | Relab |
| Fo20 | 0-45 | 1.08 | 1.06 | 0.68 | 1.57 | 4.03 | Relab |
| Fo30 | 0-45 | 1.11 | 1.08 | 0.70 | 1.81 | 6.25 | Relab |
| Fo40 | 0-45 | 1.07 | 0.96 | 0.36 | 1.24 | 5.76 | Relab |
| Fo50 | 0-45 | 1.11 | 1.05 | 0.57 | 1.85 | 7.40 | Relab |
| Fo60 | 0-45 | 1.12 | 1.02 | 0.49 | 1.92 | 8.85 | Relab |
| Fo70 | 0-45 | 1.06 | 0.90 | 0.17 | 1.13 | 5.38 | Relab |
| Fo80 | 0-45 | 1.08 | 0.96 | 0.27 | 1.34 | 5.53 | Relab |
| Fo90 | 0-45 | 1.07 | 0.96 | 0.19 | 1.26 | 4.20 | Relab |
| Fo91 | 0-38 | 1.07 | 1.01 | 0.60 | 1.27 | 7.54 | HOSERLab |
| Fo90 | 0-63 | 1.11 | 1.04 | 0.88 | 1.57 | 9.52 | **This work** |
| **CM** | **Size (µm)** | **BT** | **MR** | **R0.555** | **BS** | **VS** | **Source** |
| ALH84044 | 0-125 | 0.98 | 0.95 | 0.05 | 0.99 | 0.87 | Relab |
| LEW87148 | 0-125 | 0.99 | 0.96 | 0.05 | 1.04 | 0.69 | Relab |
| GRO85202 | 0-125 | 0.99 | 0.97 | 0.05 | 1.04 | 0.89 | Relab |
| ALHA77306 | 0-125 | 0.98 | 0.97 | 0.05 | 1.05 | 1.03 | Relab |
| LEW87022 | 0-125 | 0.99 | 0.96 | 0.04 | 0.98 | 0.56 | Relab |
| EET87522 | 0-125 | 0.99 | 1.00 | 0.03 | 0.98 | 0.27 | Relab |
| MAC88100 | 0-125 | 1.00 | 1.00 | 0.03 | 1.03 | 0.30 | Relab |
| LON94101 | 0-125 | 0.99 | 0.98 | 0.04 | 1.03 | 0.58 | Relab |
| Y-791191 | 0-125 | 0.99 | 0.97 | 0.05 | 1.00 | 0.47 | Relab |
| Y-791824 | 0-125 | 0.99 | 0.97 | 0.04 | 1.07 | 0.68 | Relab |
| A-881458 | 0-125 | 0.99 | 0.99 | 0.05 | 1.00 | 0.48 | Relab |
| A-881955 | 0-125 | 0.99 | 1.00 | 0.05 | 1.04 | 0.79 | Relab |
| A-881280 | 0-125 | 1.01 | 0.97 | 0.05 | 0.99 | 0.61 | Relab |
| A-881594 | 0-125 | 1.00 | 0.99 | 0.04 | 1.00 | 0.16 | Relab |
| Y-82054 | 0-125 | 1.01 | 1.01 | 0.04 | 1.00 | 0.41 | Relab |
| MET00639 | 0-75 | 1.00 | 0.98 | 0.03 | 0.99 | 0.03 | Relab |
| Cold-Bokkeveld | 0-75 | 0.97 | 1.00 | 0.05 | 1.01 | 0.33 | Relab |
| Mighei | 0-75 | 0.98 | 0.99 | 0.05 | 0.99 | 0.60 | Relab |
| Jbilet | 0-63 | 0.99 | 1.00 | 0.04 | 0.99 | 0.39 | **This work** |
| Murchison | 0-40 | 0.98 | 0.96 | 0.04 | 0.99 | 0.64 | HOSERLab |
| Murchison | 0-45 | 0.99 | 0.99 | 0.05 | 0.98 | 0.37 | HOSERLab |
| A-881955 | 0-125 | 0.99 | 1.00 | 0.05 | 1.04 | 0.78 | HOSERLab |
| LEW87016 | 0-125 | 0.98 | 1.04 | 0.05 | 1.07 | 0.54 | HOSERLab |
| Y791191 | 0-125 | 1.00 | 0.97 | 0.05 | 1.00 | 0.45 | HOSERLab |
| Y74662 | 0-125 | 1.00 | 1.00 | 0.05 | 1.03 | 0.31 | HOSERLab |
| Mighei-bulk | 0-40 | 0.99 | 0.99 | 0.04 | 0.99 | 0.37 | HOSERLab |
| **Olivine-(rich)-Opx mixture** | **Size (µm)** | **BT** | **MR** | **R0.555** | **BS** | **VS** | **Source** |
| Olivine50-Opx50 | 0-38 | 0.93 | 1.09 | 0.56 | 1.85 | 4.33 | HOSERLab |
| Olivine50-Opx50 | 38-53 | 0.94 | 1.15 | 0.43 | 2.48 | 8.45 | HOSERLab |
| Olivine50-Opx50 | 63-90 | 0.93 | 1.29 | 0.38 | 3.11 | 4.42 | HOSERLab |
| Olivine50-Opx50 | 90-125 | 0.96 | 1.26 | 0.27 | 2.85 | 3.09 | HOSERLab |
| Olivine60-Opx40 | 0-38 | 0.96 | 1.07 | 0.57 | 1.72 | 4.20 | HOSERLab |
| Olivine60-Opx40 | 38-53 | 0.93 | 1.15 | 0.44 | 2.50 | 8.40 | HOSERLab |
| Olivine60-Opx40 | 63-90 | 0.95 | 1.27 | 0.37 | 2.83 | 8.46 | HOSERLab |
| Olivine60-Opx40 | 90-125 | 0.97 | 1.28 | 0.26 | 2.89 | 6.19 | HOSERLab |
| Olivine70-Opx30 | 0-38 | 0.99 | 1.05 | 0.59 | 1.55 | 4.09 | HOSERLab |
| Olivine70-Opx30 | 38-53 | 0.98 | 1.12 | 0.43 | 2.26 | 8.69 | HOSERLab |
| Olivine70-Opx30 | 63-90 | 1.04 | 1.19 | 0.39 | 2.54 | 9.75 | HOSERLab |
| Olivine70-Opx30 | 90-125 | 1.05 | 1.16 | 0.27 | 2.36 | 6.60 | HOSERLab |
| Olivine80-Opx20 | 0-38 | 1.02 | 1.03 | 0.59 | 1.41 | 3.95 | HOSERLab |
| Olivine80-Opx20 | 38-53 | 1.03 | 1.09 | 0.42 | 2.06 | 8.57 | HOSERLab |
| Olivine80-Opx20 | 63-90 | 1.06 | 1.16 | 0.40 | 2.55 | 10.51 | HOSERLab |
| Olivine80-Opx20 | 90-125 | 1.10 | 1.10 | 0.27 | 2.20 | 7.03 | HOSERLab |
| Olivine90-Opx10 | 0-38 | 1.04 | 1.02 | 0.59 | 1.34 | 7.52 | HOSERLab |
| Olivine90-Opx10 | 38-53 | 1.10 | 1.05 | 0.41 | 1.96 | 8.82 | HOSERLab |
| Olivine90-Opx10 | 63-90 | 1.13 | 1.10 | 0.42 | 2.39 | 11.34 | HOSERLab |
| Olivine90-Opx10 | 90-125 | 1.13 | 1.08 | 0.25 | 2.16 | 6.48 | HOSERLab |
| **Melt/glass/shock** | **Size (μm)** | **BT** | **MR** | **R0.555** | **BS** | **VS** | **Source** |
| Padvarninkai (+melt) | 0-25 | 0.99 | 0.89 | 0.17 | 1.22 | 1.25 | Relab |
| Padvarninkai (+melt) | 25-45 | 1.00 | 0.91 | 0.14 | 1.19 | 0.81 | Relab |
| LEW85303 (+melt) | 0-25 | 0.98 | 0.83 | 0.31 | 1.56 | 3.30 | Relab |
| Macibini-cl.3-melt (glass) | | 1.06 | 1.05 | 0.37 | 2.50 | 12.42 | Relab |
| JaH626-shock | 250-500 | 1.08 | 0.87 | 0.12 | 1.56 | 0.52 | HOSERLab |
| JaH626-shock | 45-90 | 1.11 | 0.78 | 0.15 | 1.81 | 1.42 | HOSERLab |
| JaH626-shock | 90-125 | 1.10 | 0.83 | 0.11 | 1.79 | 0.71 | HOSERLab |
| JaH626-shock | <250 | 1.12 | 0.77 | 0.20 | 1.60 | 1.77 | HOSERLab |
| JaH626-shock | <45 | 1.11 | 0.77 | 0.28 | 1.63 | 3.06 | HOSERLab |

## 4.10.4 Finding FC data points that belong to a particular polyhedron

We use an expanding volume approach i.e., an algorithm in IDL programming language to test whether a data point does or does not belong to that particular polyhedron. The volume of the polyhedron is computed using an array of vertices (the band parameter values) and a connectivity array defined through the convex hull method. For example, $\text{Volume}_{\text{lab_polyhedron}}$ represents the volume using the arrays of $\text{Vertices}_{\text{lab_polyhedron}}$ and $\text{Connectivity}_{\text{lab_polyhedron}}$ i.e.,

$$\text{Vertices}_{\text{lab_polyhedron}} = \begin{bmatrix} x_1 & y_1 & z_1 \\ x_2 & y_2 & z_2 \\ \cdots & \cdots & \cdots \\ \cdots & \cdots & \cdots \\ \cdots & \cdots & \cdots \\ x_n & y_n & z_n \end{bmatrix}, \quad \text{Connectivity}_{\text{lab_polyhedron}} = \begin{bmatrix} a_1 & b_1 & c_1 & d_1 \\ a_2 & b_2 & c_2 & d_2 \\ \cdots & \cdots & \cdots & \cdots \\ \cdots & \cdots & \cdots & \cdots \\ \cdots & \cdots & \cdots & \cdots \\ a_n & b_n & c_n & d_n \end{bmatrix}$$

New polyhedrons are created with each and every pixel of the FC data appending the band parameter values to the existing array of vertices. The volume of the new polyhedron is calculated using the new array of vertices and the new connectivity array. Suppose that, for a particular FC data point, $Volume_{new_polyhedron}$ represents the new volume using $Vertices_{new_polyhedron}$ and $Connectivity_{new_polyhedron}$ i.e.,

$$Vertices_{new_polyhedron} = \begin{bmatrix} x_1 & y_1 & z_1 \\ x_2 & y_2 & z_2 \\ \cdots & \cdots & \cdots \\ \cdots & \cdots & \cdots \\ \cdots & \cdots & \cdots \\ x_n & x_n & x_n \\ x_{FC} & y_{FC} & z_{FC} \end{bmatrix}, \; Connectivity_{new_polyhedron} = \begin{bmatrix} p_1 & q_1 & r_1 & s_1 \\ p_2 & q_2 & r_2 & s_2 \\ \cdots & \cdots & \cdots & \cdots \\ \cdots & \cdots & \cdots & \cdots \\ \cdots & \cdots & \cdots & \cdots \\ p_n & q_n & r_n & s_n \end{bmatrix}$$

For each and every pixel in the FC dataset, the volumes are compared. If the new volume (of the new polyhedron) is less than the volume of the laboratory polyhedron, i.e., $Volume_{new_polyhedron} < Volume_{lab_polyhedron}$, then the FC data point is inside the laboratory polyhedron.

## 4.11 Reference

Ammannito, E., et al., 2013a. Olivine in an unexpected location on Vesta's surface. Nature 504, 122-125.

Ammannito, E., et al., 2013b. Vestan lithologies mapped by the visual and infrared spectrometer on Dawn. Meteorit. Planet. Sci. 48, 2185-2198.

Beck P., Barrat J. A., Grisolle F., Quirico E., Schmitt B., Moynier F., Gillet P., and Beck C. 2011. NIR spectral trends of HED meteorites: Can we discriminate between the magmatic evolution, mechanical mixing and observation geometry effects? Icarus 216:560–571.

Beck, A.W., et al., 2013. Challenges in detecting olivine on the surface of 4 Vesta. Meteorit. Planet. Sci. 48, 2166-2184.

Beck, A.W., McSween H. Y., 2010. Diogenites as polymict breccias composed of orthopyroxenite and harzburgite. Meteorit. Planet. Sci. 47, 850-872.

Beck, P., Barrat J. A., Grisolle F., Quirico E., Schmitt B., Moynier F., Gillet P., and Beck C. 2011. NIR spectral trends of HED meteorites: Can we discriminate between the magmatic evolution, mechanical mixing and observation geometry effects? Icarus 216:560–571.

Binzel, R., Gaffey, M., Thomas, P., 1997. Geologic mapping of Vesta from 1994 Hubble space telescope images. Icarus 103, 95–103. doi:10.1006/icar.1997.5734.

Binzel, R.P., Xu, S., 1993. Chips off of asteroid 4 Vesta: Evidence for the parent body of basaltic achondrite meteorites. Science 260, 186-191.

Bobrovnikoff, N.T., 1929. The spectra of minor planets. Lick Observatory Bulletin 14 407, 18-27.

Buchanan, P.C., et al., 2000. The South African polymict eucrite Macibini. Meteorit. Planet. Sci. 35, 1321–1331.

Buchanan, P.C., et al., 2014. Effects of varying proportions of glass on reflectance spectra of HED polymict breccias. Lunar Planet. Sci. 45 (abstract # 1525).

Buratti, B.J., et al., 2013. Vesta, vestoids, and the HED meteorites: interconnections and differences based on Dawn Framing Camera observations. J. Geophys. Res. Plan. 118, 1991–2003.

Burbine, T.H. et al., 2001. Vesta, Vestoids, and the howardite, eucrite, diogenite group: Relationships and the origin of spectral differences. Meteorit. Planet. Sci. 36, 761–781.

Bus, S.J., Binzel, R.P., 2002. Phase II of the small main-belt asteroid spectroscopic survey. A feature-based taxonomy. Icarus 158, 146–177.

Carter, L.M. et al., 2012. Initial observations of lunar impact melts and ejecta flows with the Mini-RF radar. J. Geophys. Res. 117, E00H09.

Cheek, L.C., and Sunshine, J.M. 2014. Evidence of differentiated near-surface plutons on Vesta in integrated Dawn color images and spectral datasets (abstract #2735). Asteroids, Comets, Meteors Conference.

Clenet, H., et al., 2014. A deep crust–mantle boundary in the Asteroid 4 Vesta. Nature 511, 303–306. http://dx.doi.org/10.1038/nature13499.

Cloutis, E. A., et al., 2013. Spectral reflectance properties of HED meteorites + CM2 carbonaceous chondrites: Comparison to HED grain size and compositional variations and implications for the nature of low-albedo features on Asteroid 4 Vesta. Icarus 223, 850-877.

Combe, J-P., et al., 2015. Composition of the northern regions of Vesta analyzed by the Dawn mission, Icarus, 259, 53–71.

Connolly, et al., 2006. The Meteoritical Bulletin, No. 90, 2006 September. Meteorit. Planet. Sci. 41, 9, 1383-1418.

Consolmagno, G.J., et al., 2015. Is Vesta an intact and pristine protoplanet? Icarus, 254, 190–201. doi:10.1016/j.icarus.2015.03.029.

De Sanctis, M.C., et al. 2011. The VIR spectrometer. Space Sci. Rev. 163, 329-369.

De Sanctis, M.C., et al., 2012a. Spectroscopic characterization of mineralogy and its diversity across Vesta. Science 336, 697-700.

De Sanctis, M.C., et al., 2012b. Detection of widespread hydrated materials on Vesta by the VIR imaging spectrometer on board the Dawn mission. Astrophys. J. 758, L36–L41.

Dhingra, D., 2008. Exploring links between crater floor mineralogy and layered lunar crust. Adv. Space Res. 42, 275–280.

Doressoundiram, A., et al., 2008. Color Properties and Trends of the Transneptunian Objects., in The Solar System Beyond Neptune, ed. M. A. Barucci, H. Boehnhardt, D. P. Cruikshank, & A. Morbidelli (Univ. of Arizona Press, Tucson), 91-104.

Duffard, R., Lazzaro D., and De Leon, J., 2005. Revisiting spectral parameters of silicate-bearing meteorites. Meteorit. Planet. Sci. 40, 445–459.

Dumas, C., and Hainaut, O.R., 1996. Ground-based mapping of the asteroid 4 Vesta. The Messenger 84, 13-16.

Feierberg, M.A., Drake, M.J., 1980. The meteorite-asteroid connection: the infrared spectra of eucrites, shergottites, and vesta. Science 209, 805–807. doi:10.1126/science.209.4458.805.

Filacchione, G., et al., 2012. Saturn’s icy satellites and rings investigated by Cassini – VIMS: III. Radial compositional variability. Icarus 220, 1064–1096. doi:10.1016/j.icarus.2012.06.040.

Fink, J.H., and Fletcher R.C., 1978. Ropy pahoehoe: Surface folding of a viscous fluid, J. Volcanol. Geotherrr. Res., 4, 151-170, 1978.

Gaffey M.J., 1997. Surface lithologic heterogeneity of asteroid 4 Vesta. Icarus 127, 130-157.

Gehrels, T., 1967. Minor planets. I. The rotation of Vesta. Astron. J. 72, 929-938.

Grossman, J. N., 2000. The Meteoritical Bulletin, No. 84, 2000 August. Meteorit. Planet. Sci. 35, A199-A225.

Hahn, T.M., McSween, H.Y., and Taylor, L.A., 2015. Vesta’s Missing Mantle: Evidence from new harzburgite components in Howardites. Lunar Planet. Sci. 46. Abstract #1964.

Herrin, J.S., et al., 2011. Carbonaceous chondrite-rich howardites: The potential for hydrous lithologies on the HED parent. Lunar Planet. Sci. 42. Abstract #2806.

Hicks, M.D., et al., 2014. Spectral diversity and photometric behavior of main-belt and near-Earth Vestoids and (4) Vesta: A study in preparation for the Dawn encounter. Icarus 235, 60-74.

Hiroi, T., and Pieters, C.M., 1998. Origin of vestoids suggested from the space weathering trend in the visible reflectance spectra of HED meteorites and lunar soils, in Antarct. Meteorite Res., edited by T. Harasawa, pp. 165-172, National Institute of Polar Research., Tokyo, Japan.

Hiroi, T., et al., 1995. Grain Sizes and Mineral Compositions of Surface Regoliths of Vesta-like Asteroids. Icarus 115, 374–386. doi:10.1006/icar.1995.1105.

Isaacson, P.J., and Pieters, C.M., 2009. Northern Imbrium Noritic anomaly. J. Geophys. Res. 114, E09007.

Ivanov, B.A., Melosh, H.J., 2013. Two-dimensional numerical modeling of the Rheasilvia impact formation. J. Geophys. Res. Planets 118, 1545-1557.

Izawa, M.R.M., et al. Effects of viewing geometry, aggregation state, and particle size on reflectance spectra of the Murchison CM2 chondrite in preparation for the Dawn encounter with Ceres. Icarus (accepted).

Jaumann, R., et al., 2012. Vesta's Shape and Morphology. Science 336, 687-690.

Jutzi, M., Asphaug, E., 2011. Mega-ejecta on asteroid Vesta. Geophys. Res. Lett. 38, L01102. http://dx.doi.org/10.1029/2010GL045517.

Jutzi, M., et al., 2013. The structure of the asteroid 4 Vesta as revealed by models of planet-scale collisions. Nature 494, 207-210.

Keil, K., 2002. Geological history of asteroid 4 Vesta: The "Smallest Terrestrial Planet". In Asteroids III, ed. by W. Bottke, A. Cellino, P. Paolicchi, R. P. Binzel (University of Arizona Press, Tucson, 2002), 573-584.

Keil, K., Stoeffler, D., Love, S. G. and Scott, E. R. D. (1997). Constraints on the role of impact heating and melting in asteroids. Meteoritics and Planetary Science 32, 349-363.

Le Corre, L., et al. 2013. Olivine vs. Impact Melt: Nature of the Orange Material on (4) Vesta from Dawn Observations. Icarus 226, 1568–1594.

Le Corre, L., et al., 2015. Exploring exogenic sources for the olivine on Asteroid (4) Vesta. Icarus, 258, 483-499.

Li, J.-Y., et al., 2010. Photometric mapping of asteroid (4) Vesta's southern hemisphere with Hubble Space Telescope. Icarus 208, 238-251.

Longobardo, A., et al., 2014. Photometric behavior of spectral parameters in Vesta dark and bright regions as inferred by the Dawn VIR spectrometer. Icarus 240, 20–35. doi:10.1016/j.icarus.2014.02.014.

Lunning, N.G., McSween, H.Y., Tenner, T.J., and Kita, N.T., 2015. Olivine and pyroxene from the mantle of asteroid 4 Vesta. Earth Planet. Sci. Let. 418, 126–135.

Luu, J.X. and Jewitt, D.C., 1990. CCD Spectra of Asteroids I. Near-Earth and 3:1 Resonance Asteroids. Astron. J. 99, 1985-2011.

McCausland, P.J.A., and Flemming, R.L., 2006. Preliminary bulk and grain density measurements of Martian, HED and other achondrites. Lunar Planet. Sci. 37 (abstract # 1574).

McCord T. B., Adams J. B., Johnson T. V. 1970. Asteroid Vesta: Spectral reflectivity and compositional implications. Science 178:745-747.

McCord, T.B., et al., 2012. Dark material on Vesta: Adding carbonaceous volatile-rich materials to planetary surfaces. Nature 491, 83-86.

McFadden, L.A., Pieters, C., McCord, T.B., 1977. Vesta – The first pyroxene band from new spectroscopic measurements. Icarus 31, 439–446.

McSween, H.Y., et al., 2014. Dawn's exploration of Vesta's south pole basin – where's the mantle? 75th Annual Meteoritical Society Meeting, (abstract # 5020).

McSween, H.Y., Mittlefehldt D.W., Beck A.W., Mayne, R.G., McCoy T.J., 2011. HED meteorites and their relationship to the geology of Vesta and the Dawn mission. Space Sci. Rev. 163, 141-174.

Meteoritical Bulletin, no. 102, in preparation (2013) http://www.lpi.usra.edu/meteor/metbull.php?code=57788

Mittlefehldt, D.W., 1994. The genesis of diogenites and HED parent body petrogenesis. Geochim. Cosmochim. Acta 58, 1537-1552.

Mittlefehldt, D.W., 2015. Asteroid (4) Vesta: I. The howardite-eucrite-diogenite (HED) clan of meteorites. Chemie Erde-Geochem. 75, 2, 155–183.

Mittlefehldt, D.W., McCoy, T.J., Goodrich, C.A., Kracher A., 1998. Non-chondritic meteorites from asteroidal bodies. In Planetary Materials, vol. 36 (ed. J. J. Papike). Mineralogical Society of America, Chantilly, Virginia, pp. 4-1 to 4-195.

Moskovitz, N.A., Willman, M., Burbine, T.H., Binzel, R.P., Bus, S.J., 2010. A spectroscopic comparison of HED meteorites and V-type asteroids in the inner main-belt. Icarus 208, 773–788.

Nathues, A. et al., 2015. Exogenic olivine on Vesta from Dawn Framing Camera color data. Icarus 258, 467-482.

Nathues, A., 2000. Spectroscopic study of Eunomia asteroid family. Ph.D. dissertation, University of Berlin, Germany.

Nathues, A., et al., 2012. Identification of Vesta surface units with principal component analysis by using Dawn Framing Camera imagery. 43rd Lunar Planet. Sci. Conference (abstract#1779).

Nathues, A., et al., 2014. Detection of serpentine in exogenic carbonaceous chondrite material on Vesta from Dawn FC data. Icarus 239, 222-237.

Palomba, E., et al. 2014. Composition and mineralogy of dark material deposits on Vesta. Icarus, 240, 58–72.

Palomba, E., et al., 2015. Detection of new olivine-rich locations on Vesta. Icarus 258, 120-134.

Pieters, C.M., et al., 2001. Rock types of South Pole-Aitken basin and extent of basaltic volcanism, J. Geophys. Res., 106, 28,001-28,022.

Pieters, C.M., et al., 2012. Distinctive space weathering on Vesta from regolith mixing processes. Nature 491, 79-82.

Pinilla-Alonso, N., et al., 2011. Iapetus surface variability revealed from statistical clustering of a VIMS mosaic: The distribution of $CO_2$. Icarus, 215, 75–82.

Plescia, J.B., Cintala, M., 2012. Impact melt in small lunar highlands craters. J Geophys Res 177:E00H12. doi:10.1029/2011JE003941.

Poulet, F., Ruesch, O., Langevin, Y., Hiesinger, H., 2014. Modal mineralogy of the surface of Vesta: Evidence for ubiquitous olivine and identification of meteorite analogue. Icarus 253, 364-377.

Prettyman, T.H., et al., 2012. Elemental mapping by Dawn reveals exogenic H in Vesta's regolith. Science 338, 242–246.

Reddy, V. et al., 2012c. Delivery of dark material to Vesta via carbonaceous chondritic impacts. Icarus 221, 544-559.

Reddy, V., et al., 2012a. Photometric, spectral phase and temperature effects on Vesta and HED meteorites: Implications for Dawn mission. Icarus 217, 153-168.

Reddy, V., et al., 2012b. Color and albedo heterogeneity of Vesta from Dawn. Science 336, 700-704.

Reddy, V., Gaffey, M. J., Kelley, M. S., Nathues, A., Li J-Y., Yarbrough, R., 2010. Compositional heterogeneity of asteroid 4 Vesta's southern hemisphere: Implications for the Dawn mission. Icarus 210, 693-706.

Righter, K., Drake, M.J., 1997. A magma ocean on Vesta: Core formation and petrogenesis of eucrites and diogenites. Meteorit. Planet. Sci. 32, 929-944.

Roig, F., Gil-Hutton, R., 2006. Selecting candidate V-type asteroids from the analysis of the Sloan Digital Sky Survey colors. Icarus 183 (2), 411–419.

Rubin, A.E., 2015. Maskelynite in asteroidal, lunar and planetary basaltic meteorites: An indicator of shock pressure during impact ejection from their parent bodies, Icarus (in press), doi:http://dx.doi.org/10.1016/j.icarus.2015.05.010.

Ruesch, O. et al., 2014b. Detections and geologic context of local enrichments in olivine on Vesta with VIR/Dawn data. J. Geophys. Res.: Planets 119, 2078–2108. http://dx.doi.org/10.1002/2014JE004625.

Ruesch, O., et al. 2014a. Geologic map of the northern hemisphere of Vesta based on Dawn Framing Camera (FC) images, Icarus 244, 41-59. doi:10.1016/j.icarus.2014.01.035.

Russell, C.T., et al., 2012. Dawn at Vesta: testing the protoplanetary paradigm. Science 336, 684-686.

Russell, C.T., et al., 2013. Dawn completes its mission at 4 Vesta. Meteorit. Planet Sci. 48, 2076-2089.

Russell, S.S., et al., 2014. The Jbilet Winselwan carbonaceous chondrite 1. Mineralogy and petrology: strengthening the link between CM and CO meteorites? 77th Annual Meteoritical Society Meeting (abstract # 5253).

Ruzicka, A., Grossman, J., Bouvier, A., Herd, C. D. K. and Agee, C. B. (2015), The Meteoritical Bulletin, No. 102. Meteoritics & Planetary Science, 50: 1662. doi: 10.1111/maps.12491.

Schaefer, M., et al., 2014. Imprint of the Rheasilvia impact on Vesta – Geologic mapping of quadrangles Gegania and Lucaria. Icarus 244, 60–73.

Sierks, H., et al., 2011. The Dawn Framing Camera. Space Sci. Rev. 163, 263-327.

Singerling, S.A., et al., 2013. Glasses in howardites: Impact melts or pyroclasts?. Meteorit. Planet. Sci. 48, 715-729.

Takeda, H., Graham, A.L., 1991. Degree of equilibration of eucrite pyroxenes and thermal metamorphism of the earliest planetary crust. Meteorit. Planet. Sci. 26, 129–134.

Thangjam, G. et al., 2014. Olivine-rich exposures at Bellicia and Arruntia craters on (4) Vesta from Dawn FC. Meteorit. Planet. Sci. 49, 1831–1850.

Thangjam, G., et al., 2013. Lithologic mapping of HED terrains on Vesta using Dawn Framing Camera color data. Meteorit. Planet. Sci. 48, 2199-2210.

Theilig, E., and Greeley R., 1986. Lava flows on Mars: Analysis of small surface features and comparisons with terrestrial analogs, Proc. Lun. Planet. Sci. Conf. 17th, Part I, J. Geophys. Res., 91, suppl., E193-E206, 1986.

Thomas, P.C., et al., 1997. Impact excavation on asteroid 4 Vesta: Hubble Space Telescope results. Science 277, 1492-1495.

Tkalcec, B.J., and Brenker, F., 2014. Plastic deformation of olivine-rich diogenites and implications for mantle processes on the diogenite parent body. Meteorit. Planet. Sci. 49, 1202–1213.

Tkalcec, B.J., Golabek, G.J., Brenker, F.E., 2013. Solid-state plastic deformation in the dynamic interior of a differentiated asteroid. Nature Geosci. 6, 93-97.

Tompkins, S., and Pieters, C.M., 1999. Mineralogy of the lunar crust: Results from Clementine. Meteorit. Planet. Sci. 34, 25-41.

Wee, B.S., Yamaguchi, A., Ebihara, M., 2010. Platinum group elements in howardites and polymict eucrites: Implications for impactors on the HED parent body. Lunar Planet. Sci. 41. Abstract #1533.

Williams, D.A., et al., 2014. Lobate and flow-like features on asteroid Vesta. Planet. Space Sci. 103, 24-35.

Yamaguchi, A., Taylor, G.J., Keil, K., 1997. Metamorphic history of the eucritic crust of 4 Vesta. J. Geophys. Res. 102, 13381-13386.

Zambon, F., et al., 2014. Spectral analysis of the bright materials on the asteroid Vesta. Icarus 240, 73-85.

# Curriculam Vitae

Guneshwar Thangjam

## Personal Details:

**Date of Birth:** 1985

**Nationality:** Indian

**Languages:** Manipuri, English, Hindi

**Marital Status:** Married

## Education:

| | |
|---|---|
| **02.2012 – 09. 2015** | PhD student<br>International Max Planck Research School for Solar System Science (IMPRS), enrolled at Clausthal University of Technology (TUC) |
| **04.2006 – 09.2008** | M.Sc. (Geology)<br>Bundelkhand University, Jhansi, Uttar Pradesh, India |
| **08.2002 – 12.2005** | B.Sc. (Physics, Chemistry, Maths)<br>Guru-Nanak Khalsa College, Yamuna Nagar, Haryana, India |

## Research Experience:

| | |
|---|---|
| **Feb, 2012 – Sept, 2015** | Mineralogy and Geology of (4) Vesta using Dawn Framing Camera data |
| (PhD student) | Max-Planck-Institut für Sonnensystemforschung, Germany |
| **Nov, 2009 – Jan, 2012** | Geology/mineralogy of Moon using Chandrayaan-I data |
| (Research Fellow) | Space Applications Centre, Indian Space Research Organisation, India |
| **March 2009 – Nov, 2011** | Landslide & neotectonics along the Himalayan Main Boundary Thrust (MBT), Dehradun, India |
| (Research Fellow) | Bundelkhand University, Jhansi, Uttar Pradesh, India |
| **May 2008 – July 2008** | Geologic mapping and petrographic analysis of tectonic fabrics of uranium-bearing exposures in Saladipura-Kotri tract, Rajasthan, India |
| (M.Sc. dissertation) | Atomic Minerals Directorate for Exploration and Research, Jaipur, India |

# Scientific contributions during PhD

## Publications (peer-reviewed)

**2015**

**Guneshwar Thangjam**, Andreas Nathues, Kurt Mengel, Michael Schäfer, Martin Hoffmann, Edward A. Cloutis, Paul Mann, Christian Müller, Thomas Platz, Tanja Schäfer. Three-dimensional spectral analysis of ccompositional heterogeneity at Arruntia crater on (4) Vesta using Dawn FC. **Icarus** (accepted in November 2015, doi:10.1016/j.icarus.2015.11.031).

Nathues, A., Hoffmann, M., Schaefer, M, Le Corre, L., Reddy, V., Platz, T., Cloutis, E. A., Christensen, U., Kneissl, T., Li, J.-Y., Mengel, K., Schmedemann, N., Schaefer, T., Russell, C. T., Applin, D. M., Buczkowski, D. L., Izawa, M. R. M., Keller, H. U., O'Brien, D. P., Pieters, C. M., Raymond, C. A., Ripken, J., Schenk, P. M., Schmidt, B. E., Sierks, H., Sykes, M. V., **Thangjam, G. S.**, Vincent, J.-B.,. Sublimation in bright spots on Ceres. **Nature** (accepted in September 2015, published in Dec, 2015, doi:10.1038/nature15754).

Matthew Izawa R. M., Tanja Schäfer, Valerie B. Pietrasz, Paul Mann, Edward A. Cloutis, Andreas Nathues, Kurt Mengel, Michael Schäfer, **Guneshwar Thangjam**, Martin Hoffmann, Kimberly T. Tait, D. M. Applin. Effects of viewing geometry, aggregation state, and particle size on reflectance spectra of the Murchison CM2 chondrite in preparation for the Dawn encounter with Ceres. **Icarus** (accepted in October, 2015, doi:10.1016/j.icarus.2015.10.029).

Andreas Nathues, Martin Hoffmann, Michael Schäfer, **Guneshwar Thangjam**, Lucille Le Corre, Vishnu Reddy, Ulrich Christensen, Kurt Mengel, Holger Sierks, Jean-Baptist Vincent, Edward A. Cloutis, Christopher T. Russell, Tanja Schäfer, Pablo Gutierrez-Marques, Ian Hall, Joachim Ripken, Irene Büttner (2015). Exogenic olivine on Vesta from Dawn Framing Camera color data. **Icarus**, 258, 467-482, doi:10.1016/j.icarus.2014.09.045.

**2014**

**Guneshwar Thangjam**, Andreas Nathues, Kurt Mengel, Martin Hoffmann, Michael Schäfer, Vishnu Reddy, Edward A. Cloutis, Ulrich Christensen, Holger Sierks, Lucille Le Corre, Jean-Baptiste Vincent, Christopher T. Russell (2014). Olivine-rich exposures at Bellicia and Arruntia craters on (4) Vesta from Dawn FC. **Meteoritics and Planetary Science**, 49, 10, 1831-1850, DOI: 10.1111/maps.12356.

Andreas Nathues, Martin Hoffmann, Edward A. Cloutis, Michael Schäfer, Vishnu Reddy, Ulrich Christensen, Holger Sierks, **Guneshwar Singh Thangjam**, Lucille Le Corre, Kurt Mengel, Jean-Baptist Vincent, Christopher T. Russell, Tom Prettyman, Nico Schmedemann, Thomas Kneissl, Carol Raymond, Pablo Gutierrez-Marques, Ian Hall, Irene Büttner (2014). Detection of serpentine in exogenic carbonaceous chondrite material on Vesta from Dawn FC data. **Icarus**, 239, 222-237. DOI: 10.1016/j.icarus.2014.06.003.

Michael Schäfer, Andreas Nathues, David A. Williams, David W. Mittlefehldt, Lucille Le Corre, Debra L. Buczkowski, Thomas Kneissl, **Guneshwar S. Thangjam**, Martin Hoffmann, Nico Schmedemann, Tanja Schäfer, Jennifer E. C. Scully, Jian-Yang Li, Vishnu Reddy, W. Brent Garry, Katrin Krohn, R. Aileen Yingst, Robert W. Gaskell, Christopher T. Russell (2014). Imprint of the Rheasilvia Impact on Vesta – Geologic Mapping of Quadrangles Gegania and Lucaria. **Icarus**, 244, 60-73, DOI: 10.1016/j.icarus.2014.06.026.

**2013**

**Guneshwar Thangjam**, Vishnu Reddy, Lucille Le Corre, Andreas Nathues, Holger Sierks, Harald Hiesinger, Jian-Yang Li, Juan A. Sanchez, Christopher T. Russell, Robert Gaskell, and Carol Raymond (2013). Lithologic mapping of HED terrains on Vesta using Dawn Framing Camera color data. **Meteoritics and Planetary Science**, 48, 11, 2199-2210, DOI: 10.1111/maps.12132.

# Publications (conference abstracts)

**2015**

**Thangjam**, G., Nathues, A., Mengel, K., Hoffmann, M., Schäfer, M., Mann., P., Cloutis, E., A., Müller, C., Platz, T., Behrens, H., Schäfer, T. 2015. Compositional heterogeneity of (4) Vesta from Dawn FC using a 3-dimensional spectral approach: a case study in Arruntia region. European Planetary Science Conference, EPSC2015-522 (oral).

Nathues, A., M. Hoffmann, M. Schäfer, L. Le Corre, V. Reddy, T. Platz, C. T. Russel, J.-Y. Li, E. Ammanito, I. Buettner, U. Christensen, I. Hall, M. Kelley, P. Gutiérrez Marqués, T. B. McCord, L. A. McFadden, K. Mengel, S. Mottola, D. O'Brien, C. Pieters, C. Raymond, J. Ripken, J. Ripken, T. Schäfer, P. Schenk, H. Sierks, M. V. Sykes, **G. S. Thangjam,** F. Tosi, J.-B. Vincent. 2015. DAWN Framing Camera results from Ceres orbit. European Planetary Science Conference, EPSC2015-450.

Platz, T., A. Nathues, M. Schäfer, M. Hoffmann, T. Kneissl, N. Schmedemann, J.-B. Vincent, I. Büttner, P. Gutierrez-Marques, J. Ripken, C.T. Russell, T. Schäfer, and **G.S. Thangjam.** 2015. Dawn Framing Camera: Morphology and morphometry of impact craters on Ceres. European Planetary Science Conference, EPSC2015-466.

Hoffmann M., Nathues, A., M. Schäfer, T. Platz, C. T. Russell, J.-B. Vincent, M. V. Sykes, J.Y. Li, V. Reddy, L. Le Corre, L. A. McFadden, P. Schenk, C. Pieters, P. Gutiérrez Marqués, J. Ripken, T. Schäfer, K. Mengel, **G. Thangjam,** H. Sierks, U. Christensen, I. Buettner, I. Hall. 2015. Active regions on 1 Ceres in Dawn Framing Camera colours European Planetary Science Conference, EPSC2015-474.

Schäfer M., A. Nathues, M. Hoffmann, T. Schäfer, M. R. M. Izawa, L. Le Corre, E. A. Cloutis, **G. S. Thangjam,** T. Platz, K. Mengel, C. T. Russell, V. Reddy, J. Ripken, P. Gutierrez-Marques, I. Büttner, I. Hall, H. Sierks, and U. Christensen. 2015. Dawn Framing Camera Color Mosaics of Ceres. European Planetary Science Conference, EPSC2015-488.

Platz T., A. Nathues, M. Hoffmann, M. Schäfer, D.A. Williams, S.C. Mest, D.A. Crown, M.V. Sykes, J.-Y. Li, T. Kneissl, N. Schmedemann, O. Ruesch, H. Hiesinger, H. Sizemore, I. Büttner, P. Gutierrez-Marques, J. Ripken, C.A. Raymond, C.T. Russell, T. Schäfer, **G. S. Thangjam.** 2015. Putative volcanic landforms on Ceres. European Planetary Science Conference, EPSC2015-915.

Nathues, A., Sykes, M. V., Büttner, I., Buczkowski, D. L., Carsenty, U., Castillo-Rogez, J., Christensen, U., Gutiérrez Marqués, P., Hall, I., Hoffmann, M., Jaumann, R., Joy, S., Keller, H. U., Kersten, E., Krohn, K., Li, J.-Y., Marchi, S., Matz, K.-D., McCord, T. B., McFadden, L. A., Mengel, K., Mertens, V., Mottola, S., Neumann, W., Mastrodemos, N., O'Brien, D. P., Otto, K., Pieters, C., Pieth, S., Polanskey, C., Preusker, F., Rayman, M. D., Raymond, C., Reddy, V., Ripken, J., Roatsch, T., Russell, C. T., Schäfer, M., Schäfer, T., Schenk, P., Schmedemann, N., Scholten, F., Schröder, S. E., Schulzeck, F., Sierks, H., Smith, D., Stephan, K., **Thangjam, G.**, Weiland, M., Williams, D., Zuber, M. 2015. Dawn Framing Camera Clear Filter Imaging on Ceres Approach. 46th Lunar and Planetary Science Conference, abstract#2069.

Nathues, A., Hoffmann, M., Schaefer, M., Russell, C. T., Schaefer, T., Mengel, K., Reddy, V., **Thangjam, G. S.**, Sierks, H., Christensen, U., Sykes, M. V., Li, J.-Y., Hiesinger, H., Le Corre, L., Gutiérrez Marqués, P., Buettner, I., Hall, I., Ripken, J., Dawn Science Team. 2015. Framing Camera Color Filter Imaging on Ceres Approach. 46th Lunar and Planetary Science Conference, abstract#1957.

Hoffmann, Martin, Nathues, Andreas, Schäfer, Michael, **Thangjam, Guneshwar**, Le Corre, Lucille, Vishnu, Reddy, Christensen, Ulrich, Mengel, Kurt, Sierks, Holger, Vincent, Jean-Baptist, Cloutis, Edward A., Russell, Christopher T., Schäfer, Tanja, Gutierrez-Marques, Pablo, Hall, Ian, Ripken, Joachim, Büttner, Irene. 2015. Exogenous Olivine on Vesta. American Astronomical Society, DPS meeting #46, #415.03.

Schäfer, M., Nathues, A., Hoffmann, M., Schäfer, T., **Thangjam, G. S.**, Mengel, K., Russel, C. T., Sierks, H., 2015. Comparison of Spectral Parameters for HED Discrimination with Dawn Data. 46th Lunar and Planetary Science Conference, abstract#2004.

**2014**

**Thangjam, G.**, Nathues, A., Mengel, K., Hoffmann, M., Schaefer, M., Reddy, V., Cloutis, E. A., Christensen, U., Sierks, H., Le Corre, L., Vincent, J. B., Russell, C. T., Raymond, C. 2014. Olivine Rich Exposures in Bellicia and Arruntia Craters on Vesta Using Dawn FC. 45th Lunar and Planetary Science Conference, abstract#1755 (poster).

Hoffmann, M., Nathues, A., Schäfer, M., Christensen, U., Sierks, H., **Thangjam, G. S.,** Russell, C. T., Raymond, C. A., Gutierrez-Marques, P. 2014. Dawn FC Color Data: Results of Advanced Processing for Vesta. 45th Lunar and Planetary Science Conference, abstract#1737.

Nathues, A., Hoffmann, M., Schäfer, M., **Thangjam, G.**, Reddy, V., Cloutis, E. A., Mengel, K., Christensen, U., Sierks, H., Le Corre, L., Vincent, J. B., Russell, C. T., Raymond, C. 2014. Distribution of Potential Olivine Sites on the Surface of Vesta by Dawn FC. 45th Lunar and Planetary Science Conference, abstract#1740.

Schäfer, M., Nathues, A., Hoffmann, M., Cloutis, E. A., Reddy, V., Christensen, U., Sierks, H., **Thangjam, G. S.**, Le Corre, L., Mengel, K., Vincent, J. B., Russel, C. T., Schmedemann, N., Kneissl, T., Raymond, C. 2014. Serpentine in Exogenic Carbonaceous Chondrite Material on Vesta Detected by Dawn FC. 45th Lunar and Planetary Science Conference, abstract#1745.

**2013**

Reddy, V., Li, J.-Y., Le Corre, L., Russell, C. T., Scully, J. E. C., Nathues, A., Park, R., Gaskell, R., Holger, S., Gaffey, M. J., Raymond, C., **Thangjam, G. S.**, McFadden, L. A. 2013. 44th Lunar and Planetary Science Conference, abstract#1040.

**2012**

**Thangjam G.,** L. Le Corre, V. Reddy, A. Nathues, M. Hoffmann, and H. Sierks. 2012. Comparative spectral analysis of HED meteorites and Dawn Framing Camera data. Rocks`n' Stars Conference, Göttingen (Poster).

**Thangjam G.,** et al., 2012. Compositional And lithological variations of Vesta from Dawn Framing Camera. 44th DPS Meeting (abstract withdrawn due to funding/registration problem).

# Acknowledgment

I am deeply grateful to my PhD supervisors, Dr. Andreas Nathues and Prof. Kurt Mengel, without whom my thesis would not have been completed. It is my great pleasure to work as a part of NASA’s Dawn mission Framing Camera team at MPS. A special thanks to Dr. Andreas Nathues for giving me this opportunity; and a special acknowledgment to Prof. Kurt Mengel for his kind helps not only for my thesis/PhD work but also for my family, as well as the great hospitality for my TU/Clausthal visits/works. The help and encouragement of my PhD supervisors in developing scientific thoughts including skills in writing and presenting scientific publications are worth mentioning. They are the ones who made me to explore the best of my scientific knowledge about the geology and evolution of (4) Vesta and (1) Ceres.

It is a nice pleasure to work together sharing ideas and skills among the FC team members, especially with Dr. Martin Hoffmann, Dr. Michael Schaefer, Tanja Schaefer, Dr. Thomas Platz, and Jan Kallisch. The fruitful discussions and comments/suggestions from all colleagues and co-authors are deeply acknowledged, especially from Prof. Edward A. Cloutis. I also would like to thank Dr. Vishnu Reddy, Dr. Lucille Le Corre, Prof. Harald Hiesinger, Ottaviano Ruesch, Juan Sanchez and Karsten Schindler for their warm help and company. I am very grateful to many colleagues/staffs from TU/Clausthal (Cornelia Ambrosi, Silke Schlenczek, Dietlind Nordhausen, Dr. Kai H. Schmidt, Dr. Karl W. Strauß, Dr. Thomas Schirmer, Christian Müller, and many others), TU/Hannover (Harald Behrens), and MPS (Dr. Walter Goetz, Dr. Harald Steininger, and Henning Fischer) for their help during the preparation of the samples/mixtures for spectral measurements. Thanks to Tanja Schaefer, Dr. Michael Schaefer, Ina Voβ, Sonja Fetkenheuer, Bianca Schwarz, Ute Seute, and many others for helping me in completing many administrative and official documentations/formalities during thesis submission. Thanks to Jan Kallisch for his kind interpretation of my thesis summary in German.

I am grateful to International Max Planck Research School for Solar System Science (IMPRS) for accepting me as a PhD student, and to be a part of the renowned research institute, Max-Planck-Institute for Solar System Research (MPS). I am thankful to Dr. Sonja Schuh, Dr. Dieter Schmitt (Late), Prof. Sami Solanki, Prof. Ulrich Christensen for their help during my PhD studentship. The enormous help by the administrative staff at MPS is really a nice pleasure. Especially, I would like to thank Mrs. Fee von Saltzwedel, Mrs. Petra Fahlbusch and Mrs. Rudolph Claudia for their help in need always. Thanks to Daniel Masse, Lukas Stark, and Sarah-Patricia Wedemeyer for their help in solving problems related to my computer. I am also thankful to Prof. Wolfgang Pfau/TUC, Prof. Hans-Jürgen Gursky/TUC, and Prof. W. van Berk/TUC for kindly approving/reviewing my doctoral thesis.

My gratitude to all my IMPRS/MPS colleagues/friends, especially Nagaraju and his family, Ladislav and his family, Jayant, Jisesh, Goutam, Megha, Chaitanya, Girjesh, Navdeep, Sanjiv, Sandip, Anusha, Rakesh, Ankit, Nafiseh, Atefeh, David and many others who shared pleasant memories.

Special thanks to Prof. Robert G. Strom (Professor Emeritus, Lunar and Planetary Lab, University of Arizona) and Dr. Phani R. Rajasekhar (Scientist, SAC/ISRO) for providing recommendation letters while applying for PhD position at MPS. Thanks to Prof. J. N. Goswami (the then Director, Physical Research laboratory/PRL, India), Dr. M. Bidyananda (Associate Professor, Manipur University, India), A. S. Kiran Kumar (Chairman, Indian Space Research Organization/ISRO, and the then Associate Director, Space Applications Centre/SAC-ISRO), Dr. Prakash Chauhan (Head, Planetary Sciences & Marine Optics Division, SAC/ISRO), and my colleagues and friends from SAC/ISRO and all those who helped and encouraged me to pursue PhD.

I heartily thank my entire family and I feel blessed to have the support and love of my brothers and sister, my parents, my sister-in-law and my brother-in-law. Especially, the eternal love and nurture of my mother (Mema Thangjam) and my eldest brother (Ibotombi Thangjam) are worth mentioning, and I would like to give the credit for this work to them. My loving wife Jhanji deserves a special acknowledgment for her understanding, affection and unconditional support for me always and ever. She helped me a lot in preparing the texts and checking table of contents, references, page settings, etc.

Last but not the least, I am deeply thankful to those who helped some time or another in my life to realize this moment. I wish I could visit those blessed moments! Life is indeed a very beautiful journey, with many ups and downs, and with many bonding bound to restrictions. Well, life has to keep moving…

*“ The woods are lovely, dark, and deep,*
*But I have promises to keep,*
*And miles to go before I sleep,*
*And miles to go before I sleep ”*

(From ‘Stopping by Woods on a Snowy Evening’ by Robert Frost)

## Guneshwar Thangjam: Mineralogy and Geology of asteroid (4) Vesta from Dawn Framing Camera

This thesis presents insight into the surface compositional heterogeneity and geology of asteroid Vesta. Lithologic mapping using color data from the Framing Camera (FC) onboard the NASA Dawn spacecraft revealed that majority of the surface is howarditic in composition (Thangjam et al. 2013). An important outcome of this thesis is identification of olivine-rich exposures for the first time using FC color data (Thangjam et al., 2014). Another significant contribution from this thesis is introduction of an innovative three-dimensional approach of spectral analysis to study surface compositional heterogeneity of Vesta (Thangjam et al., in press). Besides the spectral information, a revised petrologic evolution model is also presented to investigate the ongoing problem of missing olivine-rich mantle of Vesta as well as the petrogenesis of the HED meteorites (howardite, eucrite, diogenite clan of stony achondrites).